\documentclass[reprint,nofootinbib,showpacs]{revtex4}
\usepackage{graphicx}  
\usepackage{dcolumn}   
\usepackage{bm}        
\usepackage{amssymb}
\usepackage{hyperref}
\usepackage{multirow}
\usepackage{epsfig}
\usepackage{cases}
\usepackage{placeins}
 \usepackage{epstopdf}
\begin{document}

\title{
New physics footprints in the angular distribution of $B_s\to D_s^*(\to D_s\gamma,\,D_s\pi)\, \tau\, \nu$ decays}

\author{Nilakshi~Das${}$}
\email{nilakshi\_rs@phy.nits.ac.in} 
\author{Rupak~Dutta${}$}
\email{rupak@phy.nits.ac.in}
\affiliation{National Institute of Technology Silchar, Silchar 788010, India\\}
\begin{abstract}
Hints of lepton flavor universality violation observed in various flavor ratios such as $R_D$, $R_{D^*}$, $R_{J/\psi}$,
$P_\tau^{D^*}$ and $F_L^{D^*}$ in $B\to D^{(*)}\,\ell \, \nu$ and $B_c\to J/\psi \, \ell\, \nu$ charge current decays have opened new 
avenues to search for indirect evidences of beyond the standard model physics. Motivated by these anomalies, we perform a detailed angular 
analysis of $B_s\to D_s^*(\to D_s\gamma,\,D_s\pi)\, \ell\, \nu$ decays that proceed via similar $b \to c\, \ell\, \nu$ quark level transition.
We use the most general effective Hamiltonian for $b\to \,c\, l\nu$ process and give predictions of several $q^2$ and $\cos\theta$ dependent 
observables for the $B_s\to D_s^*(\to D_s\gamma,\,D_s\pi)\, \ell\, \nu$ decays in the standard model and in the presence of various real and 
complex new physics couplings. The results pertaining to this decay are competent to address the anomalies in the charge current 
sector.

\end{abstract}


\maketitle

\section{Introduction}
Lepton flavor universality that treats the three generations of charged leptons $(e,\mu,\tau)$
to be identical except the differences in their masses in the weak decays of flavor changing processes
has exposed the possibility of new physics~(NP) which lies beyond the Standard Model~(SM).
The hunt of new physics lies not just at the frontiers of the lepton flavor violating decays at the collider experiments
but also in various other phenomena such as matter-antimatter asymmetry of the universe, dark matter, neutrino mass, mass hierarchy 
problem and so on. The $B$ factories, since their inception, have been instrumental in exploring NP.
The $B$ factories have literally witnessed the breaking of lepton flavor universality in $b\,\to\,c\,l\,\nu$ 
charged current and $b\,\to\,s\,l^+\,l^-$ neutral current transition decays.
Although the results of several decay modes revealed the signature of lepton flavor universality violation, none of them are statistically 
significant to account for the evidence of new physics. The future upgradation of LHC with improved precision and with more number of new 
measurements can reduce the systematic error in the existing measurements and at the same time the efforts to study various similar decay 
modes eventually add up to tackle the possible new physics puzzle in semileptonic $B$ decays.
In the present context, we limit ourself to discuss the anomalies in the $b\,\to\,c\,l\,\nu$ charged current
quark level transitions.

\begin{itemize}
 \item \textbf{Anomalies in $R_{D}$}: The ratio of branching ratio $R_{D}$ for the decay mode $B\,\to \, D\, l\, \nu$ is defined as
  \begin{equation}
  R_{D}=\frac{\mathcal{B}\,(B \,\to \, D\, \tau \, \nu_{\tau})}{\mathcal{B}\,(B\,\to D\, \{e/\mu\}\,\nu_{(e/\mu)}}\,.
 \end{equation}
A very precise SM prediction of $R_D\,=\,0.299 \pm 0.003$ and $R_D=0.300\pm0.008$~\cite{Lattice:2015rga, Na:2015kha, Aoki:2016frl,
Bigi:2016mdz,Bernlochner:2017jka,Jaiswal:2017rve} was reported using the $B \to D$ form factors obtained in lattice QCD approach.
In 2016, FLAG working group predicated the most accurate SM results of $R_D=0.300\pm0.008$ by combining two lattice QCD results with the 
experimental form factor of $B\to D\, l\,\nu$ obtained from BABAR (2010)~\cite{BaBar:2009zxk} and BELLE (2016)~\cite{Belle:2015pkj}.
In 2012, for the first time BABAR collaboration experimentally measured the value of the ratio of branching to be 
$R_D=0.440\pm0.058\pm0.042$~\cite{BaBar:2012obs}. This measurement was 
found to be deviated from the theoretical prediction at $2.6\sigma$ level.
Later, BELLE collaboration in 2015~\cite{Belle:2015qfa} measured the value to be $R_D =\,0.375\pm 0.064 \pm 0.026$. Similarly
in the Mariond 2019, the BELLE collaboration announced the updated measurement in $R_D$ and reported it to be 
$R_D=0.307 \pm 0.037\pm 0.016$~\cite{Abdesselam:2019dgh}.
Although it is consistent with it's previous measurement, the average of all the three measurements obtained
from the HFLAV still deviates at $1.4\sigma$ from the SM expectation~\cite{Lattice:2015rga, Na:2015kha, Aoki:2016frl,Bigi:2016mdz,
Bernlochner:2017jka,Jaiswal:2017rve}. 
Although the deviation from the SM prediction is decreased from $2.6\sigma$ to $1.4\sigma$, the tension between theory and experiment still 
exists.  
 
\item \textbf{Anomalies in $R_{D^*}$}: 
The ratio of branching ratio $R_{D^*}$ for the decay mode $B\,\to \, D^*\, l\, \nu$ is defined as
\begin{equation}
R_{D^*}=\frac{\mathcal{B}\,(B \,\to \, D^*\, \tau \, \nu_{\tau})}{\mathcal{B}\,(B\,\to D^*\, \{e/\mu\}\,\nu_{(e/\mu)}}\,.
\end{equation}
The first SM prediction of $R_{D^*} = 0.252 \pm 0.003$ was reported in Ref.~\cite{Fajfer:2012vx}. Several New calculations have become available since 
$2017$~\cite{Bernlochner:2017jka,Bigi:2017jbd,Jaiswal:2017rve}. Although there are differences in the evaluation of the theoretical uncertainty, all the new calculations are found to be in 
very good agreement with each other. They are more robust and are
consistent with the old predictions for $R_{D^*}$ as well. The arithmetic average obtained by HFLAV is $R_{D^*}=0.258 \pm 0.005$~\cite{Bernlochner:2017jka,Bigi:2017jbd,Jaiswal:2017rve}. 
As of $B \to D^*$ lattice QCD form factors are concerned, at present only some unquenched calculations 
at the zero recoil exists from the Fermilab Lattice and MILC Collaborations~\cite{Bernard:2008dn,Bailey:2014tva}. The non zero recoil 
calculations for the $B \to D^*$ form factors are limited by the availability of computational resources and the efficient algorithms.
First experimental measurement of $R_{D^{*}} = 0.332\pm0.024\pm0.018$ was reported by BABAR collaboration~\cite{BaBar:2013mob} and it was found to be 
deviated at $2.7\sigma$ from the SM predication. Later in $2015$, $2016$ and $2017$, BELLE collaboration measured the value of 
$R_{D^*}$ to be $0.293\pm0.038 \pm 0.015 $~\cite{Belle:2015qfa}, $0.302\pm0.030 \pm 0.011 $~\cite{Belle:2016dyj} and $0.270\pm0.035^{+0.028}_{-0.025} $~\cite{Belle:2017ilt}, respectively. 
Similarly, in the year $2015$ and $2017$, LHCb collaboration also measured the value of $R_{D^*}$ to be $0.336 \pm 0.027\pm 0.030$~\cite{LHCb:2015gmp}
and $0.291 \pm 0.019\pm 0.029$~\cite{LHCb:2017rln} , respectively. The recent update of $R_{D^*}$ measurement from the BELLE collaboration~\cite{Belle:2019gij} 
announced in the Mariond 2019 is $R_{D^*}= 0.283\pm 0.018 \pm 0.014$. At present, the average of various measurements of $R_{D^*}$
from HFLAV still deviates from the SM expectation at the level of $2.5\sigma$.

\item \textbf{Anomalies in $R_{J/\psi}$} :
The ratio of branching ratio $R_{J/\psi}$ for the decay mode $B_c\,\to \, J/\psi\, l\, \nu$ is defined as
\begin{equation}
 R_{J/\psi}=\frac{\mathcal{B}\,(B_c\, \to  J/\psi \, \tau \,\bar{\nu}_{\tau})}{\mathcal{B}\,(B_c \,\to \, J/\psi \, \{e/\mu\}\
 \bar{\nu}_{(e/\mu)})}\,.
\end{equation}
The SM prediction of $R_{J/\psi}$ can be found in the Refs.~\cite{Ivanov:2000aj,Ebert:2003cn,AbdElHady:1999xh,Wen-Fei:2013uea,Hsiao:2016pml,
Dutta:2017xmj,Dutta:2017wpq}.
In addition, the authors in Ref.~\cite{Cohen:2018dgz} provide the SM bound to be $R_{J/\psi}\in [0.20, 0.39]$ at $95\%$ confidence level. 
Very recently, the HPQCD collaboration reported the first lattice QCD results of $R_{J/\psi}$ and reported it to be 
$0.2582\pm 0.0038$~\cite{Harrison:2020gvo}. 
The experimental measurement of $R_{J/\psi}$ from the LHCb collaboration in 2017 has reported the value of $R_{J/\psi}=0.71 \pm 0.17\pm 0.18$.
This measurement of $R_{J/\psi}$ deviates from the SM prediction at $1.8\sigma$ level.

\item \textbf{Anomalies in $P_{\tau}^{D^*}$ and $F_L^{D^*}$} :
The $\tau$ polarization fraction and the longitudinal polarization fraction of $D^*$ meson in $B\to D^*\,\tau\,\nu$ decays 
are defined as
\begin{eqnarray}
&&P_{\tau}^{D^*} = \frac{\Gamma^{+}(B\,\to\, D^*\,\tau\,\bar{\nu}_{\tau}) - \Gamma^{-}(B\,\to\, D^*\,\tau\,\bar{\nu}_{\tau})}
{\Gamma(B\,\to\, D^*\,\tau\,\bar{\nu}_{\tau})}\,, \qquad\qquad
F_L^{D^*}=\frac{\Gamma\,(B\,\to\, D_L^*\,\tau\,\bar{\nu}_{\tau})}{\Gamma\,(B\,\to\, D^*\,\tau\,\bar{\nu}_{\tau})}\,.
\end{eqnarray}
The measured value of the $\tau$ polarization fraction $P_{\tau}^{D^*} = -0.38\pm0.51^{+0.21}_{-0.16}$~\cite{Hirose:2016wfn, Hirose:2017dxl}
deviates from the SM prediction of~$0.497\pm 0.013$~\cite{Tanaka:2012nw} at $1.6\sigma$ level.
Similarly, for $F_L^{D^*}$, the measured value $F_L^{D^*}=0.60 \pm 0.08 \pm 0.035$~\cite{Abdesselam:2019wbt} deviates from
the SM expectation of~$0.46\pm 0.04$~\cite{Alok:2016qyh} at $1.5\sigma$ level.
 \end{itemize}
 
So far till date there have been several model independent and model dependent NP analysis on $b\to c l \nu$ decays.
We report here an incomplete list of various literature~\cite{Sakaki:2014sea,Biancofiore:2013ki,Freytsis:2015qca,Dutta:2015ueb,
Bhattacharya:2016zcw,Colangelo:2016ymy,Dutta:2016eml,Alok:2017qsi,Azatov:2018knx,Bifani:2018zmi, Huang:2018nnq,Hu:2018veh,Feruglio:2018fxo,
Jung:2018lfu,Datta:2017aue, Bernlochner:2018kxh,Alok:2018uft,Dutta:2018zqp,Dutta:2018jxz,Fajfer:2012jt,Crivellin:2012ye,Li:2016vvp,
Bhattacharya:2016mcc,Leljak:2019fqa,Becirevic:2019tpx,Rajeev:2018txm,Dutta:2018vgu,Colangelo:2018cnj,Bardhan:2016uhr,Li:2018lxi,Gomez:2019xfw,
Alok:2019uqc,Rajeev:2019ktp,Dutta:2019wxo,Yan:2019hpm,Popov:2019tyc,
Azizi:2019aaf,Mu:2019bin,Azizi:2019tcn,Colangelo:2019axi,Altmannshofer:2017poe,Rui:2016opu}. Recently, in Ref
.~\cite{Blanke:2018yud,Blanke:2019qrx}, the authors calculate  the best fit values of vector, scalar and tensor
 NP couplings  in $1D$ and $2D$ scenarios by fitting the experimental measurements of $R_{D^{(*)}}$,
 $P_{\tau} ^{D_s^*}$ and $F_L^{D_s^*}$ by considering the correlation between the observable
 $R_D\,-\,  R_{D^*}$. Similarly, in Ref~\cite{Becirevic:2019tpx}, the authors obtained the best fit values of NP Wilson coefficients~(WC)
by considering the experimental values of $R_{D}-R_{D^*}$ in a bayesian statistical approach assuming 
complex NP WCs. Moreover, in Ref~\cite{Murgui:2019czp}, the authors perform a global fit 
of NP WCs by considering $R_{D^{(*)}}$, $P_{\tau}^{D^*}$,  $F_{L}^{D^*}$ and differential $q^2$ 
 distribution of $B \to D\tau\nu$  and $B \to D^*\tau\nu$ decays. In addition the authors consider the constraints coming from the 
branching fraction of $B_c\, \to \tau \bar{\nu}_{\tau}$ decays.
 
The SM analysis of $B_s\,\to\, D^{*}_{s}\,l\,{\nu}$ decays has been performed by several authors using the form factors obtained in 
the constituent quark meson (CQM) model~\cite{Zhao:2006at}, the QCD sum rule~\cite{Azizi:2008vt, Bayar:2008cv}, the light cone sum rule (LCSR)
~\cite{Li:2009wq,Bordone:2019guc}, the covariant light-front quark model (CLFQM)~\cite{Li:2010bb},  
the instantaneous Bethe-Salpeter equation~\cite{Chen:2011ut,Zhou:2019stx}, the lattice 
QCD at zero recoil point~\cite{Harrison:2017fmw}, the perturbative QCD approach~\cite{Fan:2013kqa,Sahoo:2019hbu}, the BGL 
parameterization of lattice QCD data.~\cite{Cohen:2019zev} and the relativistic quark model (RQM) based
on the quasi-potential approach~\cite{Faustov:2012mt}. In Ref.~\cite{Das:2019cpt}, the authors perform a model independent analysis
of NP effects in $B_s\,\to\, D^{*}_{s}\,l\,{\nu}$ decays by using the RQM form factors of Ref.~\cite{Faustov:2012mt}. They, however, treat 
$D_s^*$ meson to be stable and did not consider any further decay of $D_s^*$ to $D_s\gamma$ or $D_s\pi$.
 
In the present paper, we use the most general effective Lagrangian in the presence of NP and perform a detail angular analysis of 
$B_s\to D_s^*(\to\, D_s\gamma, D_s\pi) \,l\,\nu$ decays using the lattice QCD 
form factor results in the full $q^2$ range. Among the two decay channels, the probability of $D_s^*$ going to $D_s\gamma$ is $93\%$, whereas,
for $D_s^*\to D_s\pi$, it is $5\%$. In this analysis we treat the NP WCs to be both real and complex. 
We give prediction of the branching fraction, longitudinal polarization fraction of $D_s^*$ meson, forward backward
asymmetry  and several angular observables pertinent to $B_s\to D_s^*(\to\, D_s\gamma, D_s\pi) \,l\,\nu$ decays. 

Study of this decay channel is well motivated for several reasons. from the experimental point of view, very recently, LHCb collaboration has 
provided a complementary information regarding the CKM matrix element $V_{cb}$ using this decay channel. Similarly, LHCb collaboration has 
also reported the measured shape of the normalized differential decay distribution with respect to $q^2$. It will allow to make a direct 
comparison between the experimental measurements with its theoretical values. Moreover, BELLE collaboration is accumulating large data samples
which will help in measuring the branching fractions to a very good precision. A total of $(6.53\pm 0.66)\times 10^6$
 $B_s\,\bar{B}_s$ pair is obtained at the BELLE detector~\cite{Aaij:2020hsi,Aaij:2020xjy} at electron-positron 
 collider KEKB  asymmetric energy. In BELLE-II the statistics will be increased by a factor of $40$, and in the next 
 decade the datas are expected to be more than $50$ times. Hence a precise measurement of observables pertaining to $B_s\to D_s^*l\nu$ decays
 may be feasible in near future which eventually will be crucial to reveal the evidence of lepton flavor universality violation in $B$ meson 
decays. At the same time, from theoretical point of 
view, very recently in 2021, first lattice QCD results for $B_s\,\to\, D^{*}_{s}$ form factors have been reported by
the HPQED collaboration~\cite{Harrison:2021tol}. From the lattice QCD point of view, the $B_s\,\to\, D^{*}_{s}$ form factors have an advantage 
over the $B\to D^*$ form factors mainly for two reasons. First, the $B_s\to D_s^*$ does not contain the valance $u/d$ quarks. Secondly,
the $D_s^*$ meson can be treated as stable as there is no Zweing-allowed strong two body decays because of its very narrow width.
 
This paper is organized as follows. In section~\ref{theory}, we start with the most general effective weak Lagrangian for
$b \to c\,l\,\nu$ decays in the presence of vector, scalar and tensor NP operators. We also report the relevant formula for all
the observables pertaining to $B_s\to D_s^*(\to\, D_s\gamma, D_s\pi) \,l\,\nu$ decays in section~\ref{theory}. In section~\ref{results}, We 
discuss the results obtained in the SM and in the presence of three different NP scenerios. Finally, we conclude with a brief 
summary of our results in section~\ref{conclusions}.

\section{Theoretical Framework}
\label{theory}
In the presence of NP, the effective weak Lagrangian for the $b \to c\,l\,\nu$ transition decays at renormalization scale $\mu = m_b$, can be 
written as~\cite{Bhattacharya:2011qm, Cirigliano:2009wk} 
\begin{eqnarray}
\mathcal L_{\rm eff} &=&
-\frac{4\,G_F}{\sqrt{2}}\,V_{cb}\,\Bigg\{(1 + g_{V_{L}})\,\bar{l}_L\,\gamma_{\mu}\,\nu_L\,\bar{c}_L\,\gamma^{\mu}\,b_L +
g_{V_{R}}\,\bar{l}_L\,\gamma_{\mu}\,\nu_L\,\bar{c}_R\,\gamma^{\mu}\,b_R \nonumber \\
&&+
g_{S_{L}}\,\bar{l}_R\,\nu_L\,\bar{c}_R\,b_L +
g_{S_{R}}\,\bar{l}_R\,\nu_L\,\bar{c}_L\,b_R + 
g_{T_{L}}\,\bar{l}_R\,\sigma_{\mu\nu}\,\nu_L\,\bar{c}_R\,\sigma^{\mu\nu}\,b_L \Bigg\} + {\rm h.c.}\,,
\label{ham_dspi}
\end{eqnarray}
where, $G_F$ is the Fermi coupling constant and $V_{cb}$ is the Cabibbo~kobayashi Maskawa~(CKM) matrix element. The vector, scalar, and tensor 
type NP interactions denoted by $g_{V_{L,\,R}}$, $g_{S_{L,\,R}}$, and $g_{T_{L}}$ NP couplings are associated with left handed neutrinos. We 
have not considered the right handed neutrino interactions in our analysis.

\subsection{Angular decay distribution of $B_s\to \, D_s^*(\to D_s\gamma)\, l\, \nu$ decay mode}
The four body differential decay distribution for the $B_s\to D_s^*(\to D_s\gamma)\, l\, \nu$ decay can be expressed in terms of the angular 
coefficients as~\cite{Colangelo:2021dnv} 

\begin{eqnarray}
 \frac{d^4\Gamma(B\to D_s^*(\to D_s\gamma)\, l\, \nu)}{dq^2\,d\cos\theta_l\, d\cos\theta_{D_s}\, d\phi} &=& \mathcal{N}_\gamma P_{D_s*}
 {\Big(1-\frac{m_l^2}{q^2}\Big)}^2 \nonumber
 \Big\{ I_{1s}\sin^2\theta_{D_s}+ I_{1c} (3 +\cos2\theta_{D_s}) 
  +(I_{2s}\sin^2\theta_{D_s}+ I_{2c} (3 +  \\  \nonumber 
 && \cos2\theta_{D_s})\cos2\theta_l+ I_{3}\sin^2\theta_{D_s} \sin^2\theta_l 
 \cos2\phi 
 \,+\,I_{4}\,\sin2\theta_{D_s}\, \sin2\, \theta_l \, \cos\phi +  I_5\\  \nonumber
  &&\, \sin2\theta_{D_s}\, \sin\theta_l\, \cos\phi  + (I_{6s}\, \sin^2\theta_{D_s} 
 +I_{6c}(3 + cos2\theta_{D_s}) 
 cos\theta_l)+ I_7\, \sin2\theta_{D_s}\,\\  
 &&  \sin\theta_l\, \sin\phi
 + I_8 \sin2\theta_{D_s}\,\sin2\theta_l \sin\phi
+ I_9 \sin^2\theta_{D_s}\, \sin^2\theta_l\, \sin 2\phi\Big)\,, 
 \label{dsg_sm_eq1}
\end{eqnarray}
where the three momentum vector of the $D_s^*$ meson and the normalization constant are defined as
\begin{eqnarray}
&&|p_{D_s^*}|=\sqrt{\lambda(m_{B_s}^2,\, m_{D_s^*}^2, q^2)}/2m_{B_s}\,,\qquad\qquad
 N_{\gamma}=\frac{3G_F^2|V_{cb}^2|\mathcal{B}(D_s^*\, \to\, D_s\gamma)}{128\, {(2\pi)}^4\,m_{B_s^2}}\,. 
\end{eqnarray}
In the presence of vector, scalar and tensor NP couplings, the angular coefficients $I_i$, where $i = 1,....6$, can be expressed as~\cite{}
\begin{eqnarray}
 I_i&=&{|1+\epsilon_V|}^2 I_i^{SM}\,+\, {|\epsilon_R|}^2\, I_i^{NP,R} + {|\epsilon_P|}^2\, 
 I_i^{NP,T} + 2\,{\rm Re}[\epsilon_R(1+ \epsilon_V^*)] I^{INT,R}_i+\, 2\,{\rm Re}[\epsilon_P(1+\epsilon_V^*)]I_i^{INT,P} \nonumber \\
 && + \, 2\,{\rm Re}[\epsilon_T(1+\epsilon_V^*)]I_i^{INT,T} +2\,{\rm Re}[\epsilon_R\epsilon_T^*] I_i^{INT,RT}\,+\, 2\, {\rm Re}
[\epsilon_P\epsilon_T^*]\,I_i^{INT,PT} + 2\,{\rm Re}[\epsilon_P\epsilon_R^*] I_i^{INT,PR}\,.
\end{eqnarray}
Similarly, the angular coefficients $I_7$, $I_8$ and $I_9$ can be written as 
\begin{eqnarray}
I_7&=&2\,{\rm Im}\,[\epsilon_R\,(1+\epsilon_V^*)\,]\,I_7^{INT,R}\, + 2\,{\rm Im}\,[\epsilon_P\,(1\,+ \, \epsilon_V^*)\,]\, 
I_7^{INT,P} + 2\, Im\, [\epsilon_T\,(1\,+\,\epsilon_V^*)]\, I_7^{INT,T} \nonumber \\
&&+
2{\rm Im}\,[\epsilon_R\epsilon_T^*]\,I_7^{INT,RT}\, +
 \, 2\,{\rm Im} \,[\epsilon_P \, \epsilon_T^*]\, I_7^{INT,PT} +2\,{\rm Im} \,[\epsilon_P \, \epsilon_R^*]\, I_7^{INT,RT}\,, \nonumber \\
I_{(8/9)}&=&2\, {\rm Im} \, [\epsilon_R\, (1\,+\, \epsilon_V^*)]\, I_{(8/9)}^{INT,R}\,, 
\end{eqnarray}
where 
 \begin{eqnarray}
  \epsilon_V\,= g_{V_{L}},\hspace{0.5cm}\epsilon_R= g_{V_{R}},\hspace{0.5cm} 
  \epsilon_P\,=\, g_{S_{R}}-g_{S_{L}}\hspace{0.5cm} \epsilon_S=g_{S_{R}}+g_{S_{L}}\hspace{0.5cm}
  \epsilon_T=g_{T_{L}}\,. 
 \end{eqnarray}
Here $I_i^{\rm SM}$ represents the angular coefficients in the SM and all other terms correspond to NP, interference of NP with NP, and 
interference of SM with NP, respectively. We refer to Ref~\cite{Colangelo:2021dnv} for all the omitted details. 

\subsubsection{The $q^2$ dependent observables}
We define several $q^2$ dependent observables for the $B_s\to D_s^*(\to D_s\gamma) l\nu$ decay mode. 
\begin{itemize}
\item The differential branching ratio, the lepton forward-backward asymmetry $A_{FB}^l (q^2)$, the forward-backward Asymmetry of 
transversely polarized $D_s^*$ meson $A_{FB}^T (q^2)$ and the convexity parameter $C_F^l (q^2)$ are defined as~\cite{Becirevic:2019tpx}
\begin{eqnarray}
&& \frac{d\Gamma}{dq^2} (q^2)=\mathcal{N}_{\gamma}|\vec{P}_{D_s^*}|{\Big(1 - \frac{m_l^2}{q^2}\Big)}^2\frac{16}{9}\pi
 {\Big(3I_{1S}+12 \, I_{1c}-I_{2s}-4I_{2c}\Big)}\,,\nonumber \\
&&A_{FB}^l(q^2)= \frac{8\pi}{3}\, \frac{\mathcal{N}_{\gamma}|\vec{P}_{D_s^*}| \,\  {\Big(1 - \frac{m_l^2}{q^2}\Big)}^2
   {\Big(I_{6s}+  4I_{6c}\Big)}}{d\Gamma/dq^2}\,, \qquad\qquad
A_{FB}^T(q^2)= \frac{32\pi}{3}\,\ \frac{\mathcal{N}_{\gamma}|\vec{P}_{D_s^*}|{\Big(1 - \frac{m_l^2}{q^2}\Big)}^2
  {I_{6c}}}{d\Gamma_T/dq^2}\,,\nonumber\\
&&F_L^{D_s^*}(q^2)=\frac{16\pi}{9}\frac{\mathcal{N}_{\gamma}|\vec{P}_{D_s^*}|{\Big(1 - \frac{m_l^2}{q^2}\Big)}^2
 {\Big(3I_{1s}-I_{2s}\Big)}}{d\Gamma/dq^2}\,, \qquad\qquad
 C_{F}^l(q^2)=\frac{32\pi}{3}\frac{\mathcal{N}_{\gamma}|\vec{P}_{D_s^*}|{\Big(1 - \frac{m_l^2}{q^2}\Big)}{\Big(I_{2s}+
4I_{2c}\Big)}}{d\Gamma/dq^2}\,.
\end{eqnarray}
\item The angular observables $A_3\,(q^2)$, $A_4\,(q^2)$, $A_5\,(q^2)$,  $A_{6s}\,(q^2)$, $A_7\,(q^2)$, $A_8\,(q^2)$ and $A_9\,(q^2)$ 
are defined as\cite{Becirevic:2019tpx} 
\begin{eqnarray}
&&A_3(q^2) = \frac{16}{9}\, \frac{\mathcal{N}_{\gamma}\,|\vec{P}_{D_s^*}|\,{\Big(1 -\frac{m_l^2}{q^2}\Big)\,I_3}}{d\Gamma/dq^2}\,, 
\qquad\qquad
A_{4}(q^2)= - \frac{64}{9}\, \frac{\mathcal{N}_{\gamma}|\vec{P}_{D_s^*}|{\Big(1 - \frac{m_l^2}{q^2}\Big)}I_4}{d\Gamma/dq^2}\,, 
\nonumber \\
&&A_{5}(q^2)=-\frac{8\pi}{3}\,\frac{\mathcal{N}_{\gamma}\,|\vec{P}_{D_s^*}|\,{\Big(1 - \frac{m_l^2}{q^2}\Big)I_5}}{d\Gamma/dq^2}\,, 
\qquad\qquad
A_{6s} (q^2)\,=\, - \frac{288\pi}{24}\, \frac{\mathcal{N}_{\gamma}|\vec{P}_{D_s^*}|{\Big(1 - \frac{m_l^2}{q^2}\Big)}I_{6s}}{d\Gamma/dq^2}\,,
\nonumber \\
&&A_{7}(q^2)=-\frac{8\pi}{3}\,\frac{\mathcal{N}_{\gamma}|\vec{P}_{D_s^*}|{\Big(1 - \frac{m_l^2}{q^2}\Big) I_7}}{d\Gamma/dq^2}\,, 
\qquad\qquad
A_{8}(q^2)= \frac{64}{9}\, \frac{\mathcal{N}_{\gamma}|\vec{P}_{D_s^*}|{\Big(1 - \frac{m_l^2}{q^2}\Big)}I_8}{d\Gamma/dq^2}\,, \nonumber\\
&&A_9(q^2) =\frac{16}{9}\,\frac{\mathcal{N}_{\gamma}\,|\vec{P}_{D_s^*}|\,{\Big(1 - \frac{m_l^2}{q^2}\Big)\,I_9}}{d\Gamma/dq^2}\,.
\end{eqnarray}

\item The ratio of branching fraction is defined as follows 
\begin{eqnarray}
 R_{D_s^*}(q^2) = \frac{d\Gamma/dq^2|_{\tau-mode}}{d\Gamma/dq^2|_{e-mode}}
\end{eqnarray}
\end{itemize}

\subsubsection{The $\cos\theta$ dependent observables}
We also define several $\cos\theta_{D_s}$ and $\cos\theta_l$ dependent observables. They are
\begin{eqnarray}
 F_L(\cos\theta_{D_s})\, &=& \,\frac{\mathcal{N}_{\gamma}\,|\vec{P}_{D_s^*}|\,
 {(1 - \frac{m_l^2}{q^2})}^2\, 2\pi \,\int_{q^2_{min}}^{q^2_{max}} (2I_{1s}\,-\,\frac{2}{3} I_{2s})(1 - \cos^2\theta_{D_s}) \, dq^2}
{\Gamma(B_s\to D_s^*(\to D\gamma)\, l\,\nu)}\hspace{0.3cm}\nonumber \\
 \vspace{1cm}
 F_T(\cos\theta_{D_s})\,&=&\,\frac{\mathcal{N}_{\gamma}\, |\vec{P}_{D_s^*}|\,
 {(1 - \frac{m_l^2}{q^2})}^2 4\pi\, \int_{q^2_{min}}^{q^2_{max}} (2I_{1c}\,-\,\frac{2}{3} I_{2c})(1 + \cos^2\theta_{D_s}) \, dq^2}
{\Gamma(B_s\to D_s^*(\to D\gamma)\, l\,\nu)}\nonumber \\
  \vspace{1cm}
   F_L(\cos\theta_l )\, &=&\,\frac{8\pi}{3} \frac{\mathcal{N}_{\gamma}\,|\vec{P}_{D_s^*}|\,
 {(1 - \frac{m_l^2}{q^2})}^2\, \, \int_{q^2_{min}}^{q^2_{max}} (I_{1s}\
 ,+\, I_{2s} (2\cos^2\theta_l\, - \,1)+I_{6s} \cos\theta_l ) \, dq^2}{\Gamma(B_s\to D_s^*(\to D_s\gamma) l\nu)}\nonumber \\  
  \vspace{1cm}
 F_T(\cos\theta_l )\, &=&\,\frac{32\pi}{3}\frac{\mathcal{N}_{\gamma}\,|\vec{P}_{D_s^*}|\,
 {(1 - \frac{m_l^2}{q^2})}^2\, \, \int_{q^2_{min}}^{q^2_{max}} (I_{1c}\
 ,+\, I_{2c} (2\cos^2\theta_l\, - \,1)+I_{6c} \cos\theta_l ) \, dq^2}{\Gamma(B_s\to D_s^*(\to D_s\gamma) l\nu)}\nonumber \\
 \vspace{1cm}
 A_{FB}^l(\cos\theta_{D_s})\,&=&\,\frac{\mathcal{N}_{\gamma}|\vec{P}_{D_s^*}|
 {(1 - \frac{m_l^2}{q^2})}^2 2\pi \int_{q^2_{min}}^{q^2_{max}}[ (I_{6s}\,+ 2 I_{6c})\,+ (2I_{6c}\,- \, I_{6s})\cos^2_{\theta_{D_s}} ] \,dq^2}
{d\Gamma/ d\cos\theta_{D_s}}\,,
\end{eqnarray}
where
\begin{eqnarray}
 \frac{d\Gamma}{ d\cos\theta_{D_s}}&=&\frac{4\pi}{3}\int_{q^2_{min}}^{q^2_{max}}\mathcal{N}_{\gamma}|\vec{P}_{D_s^*}|
 {(1 - \frac{m_l^2}{q^2})}^2 [(3I_{1s}\,-\, I_{2s}\, + \, 6I_{1c}-2I_{2c})+ 
 (I_{2s}-3I_{1s}+6I_{1c}-2I_{2c})\cos^2_{\theta_{D_s}}]\, dq^2\nonumber
\end{eqnarray}

\subsection{Angular decay distribution of $B_s\to D_s^*(\to D_s\pi)\, l\, \nu$ decay mode}

Starting with the effective Lagrangian of Eq.~\ref{ham_dspi}, the four body differential decay distribution of 
$B_s\to D_s^*(\to D_s\pi)\, l\, \nu$ can be written as follows~\cite{Becirevic:2019tpx,Mandal:2020htr}.
\begin{eqnarray}
  \frac{d^4\Gamma(B\to D_s^*(\to D_s\pi)\, l\, \nu)}{dq^2\,d\cos\theta_l\, d\cos\theta_{D_s}\, d\phi} &=& \frac{9}{32\pi}
 \Big\{ I_{1s}\sin^2\theta_{D_s}\, +  \, I_{1c}\cos^2\theta_{D_s} +
 (I_{2s}\sin^2\theta_{D_s}\,+ \, I_{2c}\cos^2\theta_{D_s})\cos2\theta_l  \nonumber\\
 &&+ (I_{3} \cos2\phi \,  +\,I_9\sin2\phi)  \sin^2\theta_{D_s}\sin^2\theta_l \,+\,(I_{4}\,\cos\phi+ I_8\sin\phi)\sin2\,  \theta_{D_s}  
 \sin2\, \theta_l \,\nonumber \\ 
 &&  +(I_5\,\, \cos\phi  + I_7\,\, \sin \phi)\sin2\theta_{D_s}\, \sin\theta_l )
+ (I_{6s}\, \sin^2\theta_{D_s}  +I_{6c}\cos^2\theta_{D_s}) \cos\theta_l  \Big\}\,,
 \label{dspi_sm_eq1}
\end{eqnarray}
where the angular coefficients are~\cite{Becirevic:2019tpx,Mandal:2020htr}
\begin{eqnarray}
 I_{1c}&=&N_F\,\Big{[}2(1\,+\,\frac{ m_l^2}{q^2})\,(\mathcal{{A}}_0 +4\,{|\mathcal{A}_{T0})|}^2)\,-\, \frac{16m_l}{\sqrt{q^2}}\, {\rm Re}[\mathcal{{A}}_0^L \mathcal{A}_{T0}^L]\, \nonumber
 +\frac{4m_l^2}{q^2}{|\mathcal{A}_{tP}|}^2\Big{]}\\ \nonumber
 I_{1s}&=&N_F\Big{[}\frac{1}{2}(3+\frac{m_l^2}{q^2})\, ({|\mathcal{{A}}_{\perp}^L|}^2
 +{|\mathcal{{A}}_{||}^L|}^2)\, +\, 2(1+\frac{3m_l^2}{q^2})\,({|\mathcal{{A}}_{T\perp}^L|}^2+{|\mathcal{{A}}_{T||}^L|}^2)\, -\, 8\frac{m_l}{\sqrt{q^2}} 
  {\rm Re}[\mathcal{A}_{\perp}^L \mathcal{A}_{T\perp}^{L*}
+\mathcal{A}_{||}^L \mathcal{A}_{T||}^{L*}] \nonumber\\
I_{2c}&=&-2 N_F (1\,-\,\frac{ m_l^2}{q^2})({|\mathcal{A}_{0}^L|}^2\, - \, {|\mathcal{A}_{T0}^L|}^2 )\nonumber\\
I_{2s}&=&\frac{1}{2}N_F \, (1\,-\,\frac{ m_l^2}{q^2})\, \Big( {|\mathcal{A}_{\perp}^L|}^2+
|{\mathcal{A}_{||}^L}|^2)-4\,({|\mathcal{A}_{T\perp}^L|}^2 \, +\, {|\mathcal{A}_{T||}^L|}^2) \nonumber \\\nonumber\\
I_{3}&=&N_F\, (1\,-\, \frac{m_l^2}{q^2})\,\Big{(} {|\mathcal{A}_{\perp}^L|}^2
-{|\mathcal{A}_{||}^L|}^2 - 4 \,( {|\mathcal{A}_{T\perp}^L|}^2 \,+\, {|\mathcal{A}_{T||}^L|}^2) \Big{)}\nonumber \\
I_4&=&\sqrt{2}N_F\, (1\,-\, \frac{m_l^2}{q^2})\,  {\rm Re}[ \mathcal{A}_0\mathcal{A}_{||}^{L*}\, -\, 4\mathcal{A}_{T0}^L\mathcal{A}_{T||}^{L*}]\nonumber\\
I_5&=&2\sqrt{2}N_F\, \Bigg[  {\rm Re}[(\mathcal{A}_0^L \, - \,2\frac{m_l}{\sqrt{q^2}}\mathcal{A}_{T0}^L)\, (\mathcal{A}_{\perp}^{L*} \, - \,2\frac{m_l}{\sqrt{q^2}}\mathcal{A}_{T\perp}^{L*})] \,-\,\frac{m_l^2}{q^2}\, {\rm Re}[\mathcal{A}_{tP}^{L*} (\mathcal{A}_{||}^{L} \, - \,2\frac{m_l}{\sqrt{q^2}}\mathcal{A}_{T||}^{L}) ]\nonumber \\
I_{6c}&=& N_F\frac{8m_l^2}{q^2}\, {\rm Re}[A_{tP}^{L*}(\mathcal{A}_0^L \, - \,2\frac{\sqrt{q^2}}{m_l}\mathcal{A}_{T0}^{L})]\nonumber\\
I_{6s}&=&4\, N_F\, {\rm Re}[(\mathcal{A}_{||}^L \, - 2\,\frac{m_l}{\sqrt{q^2}}\mathcal{A}_{T||}^L)(\mathcal{A}_{\perp}^{L*}\, -\, 2\frac{m_l}{\sqrt{q^2}} \mathcal{A}_{T\perp}^{L*} )]\nonumber \\
I_7&=&-2\sqrt{2}N_F\Bigg[{\rm Im}[(\mathcal{A}_0^L\,- \, 2\frac{m_l}{q^2} \mathcal{A}_{T0}^L)\,
(\mathcal{A}_{||}^{L*}\,-\, 2\frac{m_l}{q^2}\mathcal{A}_{T||}^{L*}) \,+\,\frac{m_l^2}{q^2}
Im[\mathcal{A}_{tP}^{L*}(\mathcal{A}_{\perp}^L \, - \, 2\frac{q^2}{m_l} \mathcal{A}_{T\perp}^L)]\Bigg]\nonumber \\
I_8&=&\sqrt{2}N_F\, (1\,-\, \frac{m_l^2}{q^2})\, {\rm Im}[\mathcal{A}_0^{L*}\, \mathcal{A}_{\perp}^L \, -\, 4\mathcal{A}_{T0}^{L*}\,\mathcal{A}_{T\perp}^{L}]\nonumber\\
I_9&=&2\,N_F\, (1\,-\, \frac{m_l^2}{q^2}){\rm Im}[\mathcal{A}_{||}^L\mathcal{A}_{\perp}^{L*}\,
-\, 4\mathcal{A}_{T||}^{L}\mathcal{A}_{T\perp}^{L*}]\,,
\end{eqnarray}
with
\begin{eqnarray}
N_F=\frac{G_F^2{|V_{cb}|}^2     }{2^73\pi^3m_{B_s}^3   }q^2\lambda_{D_s^*}^{1/2}{(1\,-\, \frac{m_l^2}{q^2})}^2\mathcal{B}(D_s^*\to D_s\pi)\,.
\end{eqnarray}
The longitudinal, transverse and time like component of amplitude $A_{T0,T\perp,T||}^{L}$, written in terms of NP couplings, are taken from
Ref.~\cite{Mandal:2020htr}.  We refer to Ref~\cite{Mandal:2020htr} for the omitted details.
\subsubsection{The $q^2$ dependent observables}

\begin{itemize}
\item The differential branching ratio, the lepton forward-backward asymmetry $A_{FB}^l (q^2)$, the forward-backward Asymmetry of 
transversely polarized $D_s^*$ meson $A_{FB}^T (q^2)$ and the convexity parameter $C_F^l (q^2)$ can be defined as~\cite{Becirevic:2019tpx}
\begin{eqnarray}
&&\frac{d\Gamma}{dq^2}(q^2)=\frac{1}{4}{\Big(6I_{1S}+3I_{1c}-2I_{2s}-I_{2c}\Big)}\,, \qquad\qquad
  A_{FB}^l(q^2)=\frac{3}{8}\frac{\Big(I_{6c}+ 2I_{6s}\Big)}{d\Gamma/dq^2}\,, \nonumber\\
&&A_{FB}^T(q^2)=\frac{3}{8}\frac{{I_{6s}}}{d\Gamma_T/dq^2}\,, \qquad\qquad
   C_F^l(q^2)=\frac{6}{8}\frac{(2I_{2c}+4I_{2s})}{d\Gamma/dq^2 }\,.
\end{eqnarray}
\item The angular observables $A_3\,(q^2)$, $A_4\,(q^2)$, $A_5\,(q^2)$, $A_{6s}\,(q^2)$, $A_7\,(q^2)$, $A_8\,(q^2)$ and $A_9(q^2)$ 
can be defined as~\cite{Becirevic:2019tpx}
\begin{eqnarray}
 A_3\,(q^2)\,&=&\,\frac{1}{2\pi}\frac{I_3}{d\Gamma/dq^2}\hspace{1cm}
 A_4\,(q^2)\,=\,- \frac{2}{\pi}\frac{I_4}{d\Gamma/dq^2}\nonumber\\
A_5\,(q^2)\,&=&\, - \frac{3}{4}\frac{I_5}{d\Gamma/dq^2}\hspace{1cm}
A_{6s}\,(q^2)\,=\,-\frac{27}{8}\frac{I_{6s}}{d\Gamma/dq^2}\nonumber\\
 A_7\,(q^2)\,&=&\, - \frac{3}{4}\frac{I_7}{d\Gamma/dq^2}\hspace{1cm}
A_8\,(q^2)\,=\,\frac{2}{\pi}\frac{I_8}{d\Gamma/dq^2}\nonumber\\
A_9\,(q^2)\,&=&\,\frac{1}{2\pi}\frac{I_9}{d\Gamma/dq^2}\,.
\end{eqnarray}
\end{itemize}

\subsubsection{The $\cos\theta$ dependent observables}
The $\cos\theta_{D_s}$ and $\cos\theta_l$ dependent observables can be defined as follows~\cite{Colangelo:2018cnj}.
\begin{eqnarray}
 F_L(\cos\theta_{D_s})\,&=&\,\frac{9}{16}\frac{\int_{q^2_{min}}^{q^2_{max}} (2I_{1c}\,-\,\frac{2}{3} I_{2c}) \cos^2\theta_{D_s^*} dq^2}
{\Gamma(B_s\to D_s^* \to D_s^*(D\pi) l\nu)}\nonumber \\
 \vspace{1cm}
  F_T(\cos\theta_{D_s})\,&=&\,\frac{9}{16}\frac{\int_{q^2_{min}}^{q^2_{max}} (2I_{1s}\,-\,\frac{2}{3} I_{2s}) (1-\cos^2\theta_{D_s}) dq^2}
{\Gamma(B_s\to D_s^* \to D_s^*(D\pi) l\nu)} \nonumber\\
 \vspace{1cm}
 F_L(\cos\theta_l )\,&=&\,\frac{9}{24}\frac{\int_{q^2_{min}}^{q^2_{max}} (I_{1c}+ I_{2c} (2\cos^2\theta_l\, - \,1)+I_{6c} \cos\theta_l)dq^2}
{\Gamma(B_s\to D_s^* \to D_s^*(D\pi) l\nu)}\nonumber\\
 \vspace{1cm}
 F_T(\cos\theta_l )\,&=&\,\frac{9}{12}\frac{\int_{q^2_{min}}^{q^2_{max}} (I_{1s} + I_{2s} (\cos^2\theta_l\, - \,1)+I_{6s} \cos\theta_l)
dq^2}{\Gamma(B_s\to D_s^* \to D_s^*(D\pi) l\nu)}\nonumber\\
 \vspace{1cm}
 A_{FB}^l(\cos\theta_{D_s})\,&=&\,\frac{9}{16}\frac{\int_{q^2_{min}}^{q^2_{max}} ( I_{6s}\,+\, (I_{6c}\,-\, I_{6s})\cos^2_{\theta_{D_s}}) \,
\ dq^2}{d\Gamma/ d\cos\theta_{D_s}}\,,
\end{eqnarray}
where
\begin{eqnarray}
 \frac{d\Gamma}{ d\cos\theta_{D_s}}&=&\frac{9}{24}\int_{q^2_{min}}^{q^2_{max}} [(3I_{1s}\,-\, I_{2s})\, + \, (I_{2s}-3I_{1s}+3I_{1c}-2I_{2c})
\cos^2_{\theta_{D_s}}]\, dq^2\,.
\end{eqnarray}

\section{Results and Discussion}
\label{results}
\subsection{Input Parameters}
In Table~\ref{inputs}, we report all the theory inputs such as the masses of various mesons, leptons, the branching fraction of 
$\mathcal{B}(D_s^*\to D_s\gamma)$,  
$\mathcal{B}(D_s^*\to D_s\pi)$ and mass of $b$ quark and $c$ quark evaluated at renormalization scale $\mu = m_b$~\cite{Zyla:2020zbs}.
The mass parameters are expressed in GeV unit and the $B_s$ meson life time $\tau_{B_s}$ is expressed in second.  
We consider the uncertainties associated with the CKM matrix element $|V_{cb}|$ and the relevant vector and axial vector
form factor inputs $V$, $A_0$, $A_1$ and $A_2$ of Ref~\cite{Harrison:2021tol}. The relevant formula for the form factors pertinent
for our discussion, taken from Ref.~\cite{Harrison:2021tol}, is  
\begin{equation}
F(q^2) = \frac{1}{P(q^2)}\sum_{n=0}^3 a_n z^n(q^2,t_0)\,,
\end{equation}
where $F$ stands for the form factors $V$, $A_0$, $A_1$, $A_2$ and $a_0$, $a_1$, $a_2$, $a_3$ are the z-expansion coefficients. The pole
function $P(q^2)$ and $z(q^2,t_0)$ are defined as
\begin{eqnarray}\label{poleformeq}
P(q^2)=\prod_{M_\mathrm{pole}}z(q^2,M_\mathrm{pole}^2)\,\hspace{1cm}z(q^2,t_0)=\frac{\sqrt{t_+-q^2}-\sqrt{t_+-t_0}}{\sqrt{t_+-q^2}+\sqrt{t_+-t_0}}.
\end{eqnarray}
where, $t_0=(M_{B_s}-M_{D_s^*})^2$, $t_+=(M_{B}+M_{D^*})^2$ and the pole masses are represented by $M_{pole}$. In Table.~\ref{ff}, we 
report the form factor inputs relevant for our analysis. 
The uncertainty associated with these parameters are written within parenthesis. 
\begin{table}[htbp]
\centering
\setlength{\tabcolsep}{6pt} 
\renewcommand{\arraystretch}{1.0} 
\begin{tabular}{|l|l|l|l|l|l|}
\hline
\hline
Parameters  & Values & Parameters  & Values & Parameters  & Values\\
\hline
\hline
$m_{B_s}$    & 5.36677 & $m_{D_s^*}$ & 2.112 & $m_e$ & $0.5109989461\times10^{-3}$ \\
$m_b$        & 4.18          &  $m_c$ & 0.91 & $|V_{cb}|$     & 0.0409(11) \\
$m_B$ &   5.27964   &  $m_{D^*}$    &   2.010      & $m_\tau$ & 1.77682\\
$\mathcal{B}(D_s^*\to D_s\pi)$ &   $5.8\times 10^{-2}$   & $\mathcal{B}(D_s^*\to D_s\gamma)$ &   $93.5\times 10^{-2}$     &&\\
$G_F$ & $1.1663787\times10^{-5}$ &$\tau_{B_s}$ & $1.515\times10^{-12}$ & &\\

\hline
\end{tabular}
\caption{Theory input parameters }
\label{inputs}
\end{table}

\begin{table}[htbp]
\setlength{\tabcolsep}{6pt} 
\renewcommand{\arraystretch}{1.4} 
\begin{tabular}{ |c | c| c| c| c|c c c c |}
\hline
& $a_0$	& $a_1$	& $a_2$	& $a_3$ & $M_{pole}$	&&&\\\hline
${A_0}$&	0.1046(79)&	-0.39(15)&	0.02(98)&	-0.03(1.00)& 6.275 & 6.872 & 7.25&\\
${A_1}$&	0.0536(28)&	0.020(75)&	0.09(81)&	0.10(99)& 6.745 & 6.75&  7.15 & 7.15\\
${A_2}$&	0.051(15)&	0.02(26)&	-0.35(79)&	-0.07(99) & 6.745 & 6.75&  7.15 & 7.15\\
${V}$&	0.102(14)&	-0.27(30)&	-0.007(0.998)&	-3e-05 +- 1 & 6.335 & 6.926&  7.02 & 7.28\\
\hline
\end{tabular}
\caption{Form factor Input Parameters.}
\label{ff} 
\end{table}

We have used the equation of motion to find out the relevant tensor form factors so that 
\begin{eqnarray}
&&T_1(q^2) = \frac{m_b + m_c}{m_{B_s} + m_{D_s^*}}\,V(q^2)\,,\qquad
T_2(q^2) = \frac{m_b - m_c}{m_{B_s} - m_{D_s^*}}\,A_1(q^2)\,,\nonumber \\
&&T_3(q^2) = -\frac{m_b -m_c}{q^2}\Big[m_{B_s}\Big(A_1(q^2) - A_2(q^2)\Big)+m_{D_s^*}\Big(A_2(q^2) + A_1(q^2)-2\,A_0(q^2)\Big)\Big]\,.
\end{eqnarray}

\subsection{SM prediction}
We report the SM central value and the $1\sigma$ uncertainty associated with several observables 
such as the branching ratio~($BR$), the ratio of branching ratio~($R_{D_s^*}$), the forward backward asymmetry~($A_{FB}^l$), the 
convexity parameter~($C_F^l$), the forward backward asymmetry for the transversely polarized $D_s^*$ meson~($A_{FB}^T$), the longitudinal 
polarization fraction of $D_s^*$ meson~($F_L^{D_s^*}$), $A_3$, $ A_4$, $A_5$, $A_{6s}$, $A_7$, $A_8$ and $A_9$ for both $e$ and $\tau$ mode 
in Table~\ref{sm_results}. Our observations are as follows.
\begin{itemize}
\item The branching ratio of $B_s\to D_s^*( \to D_s\pi) l \nu $ mode is found to be of $\mathcal O(10^{-3})$, whereas the branching ratio of 
$B_s\to D_s^*( \to D_s\gamma) l \nu $ decay mode is obtained to be of $\mathcal O(10^{-2})$.   

\item As expected, the central value and the $1\sigma$ uncertainty associated with $R_{D_s^*}$, $A_{FB}^l$, $C_F^l$, $A_{FB}^T$ and 
$F_L^{D_s^*}$ is exactly same for the $B_s\to D_s^*( \to D_s\gamma) l \nu $ and the $B_s\to D_s^*( \to D_s\pi) l \nu $ mode.
 
\item The angular observables such as $A_3$, $ A_4$, $A_5$, $A_{6s}$ are, however, quite different for both the decay modes. The central
values obtained for $A_3$, $ A_4$ and $A_5$ in $B_s\to D_s^*( \to D_s\pi) l \nu $ mode are twice as large as the values obtained in case of
$B_s\to D_s^*( \to D_s\gamma) l \nu $ mode. 
  
\item The angular observables $\ A_7,\,\ A_8$ and $A_9$ are zero in the SM and are non-vanishing only if NP induces a complex contribution to 
the amplitude 
 
\item The ratio of branching ratio  $R_{D_s^*}$ is found to be $0.2430\pm 0.0015$ which is quite similar to the value 
reported in Ref~\cite{Harrison:2021tol}.
\end{itemize}

\begin{table}[htbp]
\centering 
  \setlength{\tabcolsep}{6pt} 
 \renewcommand{\arraystretch}{1.2}
\begin{tabular}{|c|c|c|c|c|c|c|c|}
\hline
Observable &\multicolumn{2}{c|}{$B_s\to D_s^*(\to \,D_s\pi)\, l \nu$ decay mode } &\multicolumn{2}{c|}{$B_s\to D_s^*( \to D_s\gamma) l\nu$ decay mode} \\
\cline{2-5}
 &{ $e$-mode} & \multicolumn{1}{c|}{$\tau$  mode} &
 {$e$  mode}&{$\tau$  mode}  \\
\cline{2-5}
&central value  & Central value  &central value  &central value \\
\hline          
\hline
 {$BR$}         & $( 3.0516\,\pm\,0.0988)\times 10^{-3}$  &   $(0.7415\,\pm\,0.0231)\times 10^{-3}$     &   $(4.9194\,\pm\,0.1593)\times10^{-2}$  & $( 1.1954\pm0.0372)\times10^{-2}$  \\
 \hline
 $A_{FB}^{l}$   & $ -0.2640\, \pm\,0.0031$       &  -0.0896 $\pm$ 0.0020       &   $ -0.2640\, \pm\,0.003$    & -0.0896 $\pm$ 0.0020  \\
 \hline
 $A_{FB}^{T}$   &  $-0.5436\, \pm\ 0.0035$      & $-0.3842\, \pm\, 0.0026$       &  $-0.5436\, \pm\ 0.0035$            & $-0.3842\, \pm\, 0.0026$ \\
\hline
$F_L^{D_s^*}$   & $0.5143\, \pm\,0.0040$         & $0.4482 \, \pm\,0.0015$       &  $0.5143\, \pm\,0.0040$        & $0.4482 \, \pm\,0.0015$ \\
\hline
$A_3$           &  $ -0.0252\, \pm\, 0.0003$     & $-0.0162\, \pm\,0.0001$       &   $0.0126 \, \pm\,  0.0001$ & $ 0.0081 \pm 0.0001$ \\
\hline
 $A_4$          &  $0.1909\pm0.0005$            &    $0.0883 \, \pm\, 0.0001$  & $-0.0954 \, \pm\,   0.0002$    &  $-0.0442 \pm 0.0001$ \\
\hline
 $A_5$          &  $-0.2139\pm0.0019$             &  $-0.2265\pm0.0010$           &$0.1069  \, \pm\,  0.0010$ &  $ 0.1133 \pm 0.0005$ \\
\hline
$A_{6s}$        &  $1.1882\pm 0.0140$            &  $0.9539\pm0.0077$           &$-0.0000\pm0.0000$ &  $-0.5509   \pm 0.0026$ \\
\hline
$C_F^l$         &  $-0.4071\pm0.0091$            & $-0.0550\pm0.0014$           & $-0.4071\pm0.0091$    &      $-0.0550\pm0.0014$ \\
\hline
$A_{7}$         &        0.0000          &  0.0000               & 0.0000         &   0.0000 \\
\hline
$A_{8}$         &       0.0000           &  0.0000             & 0.0000          &   0.0000 \\
 \hline
$A_{9}$         &        0.0000 &   0.0000         & 0.0000 &   0.0000 \\
\hline
 $R_{D_s^*}$ & \multicolumn{2}{c|}{$0.2430\pm0.0015$}  & \multicolumn{2}{c|}{$0.2430\pm0.0015$}   \\
\hline
\end{tabular}
\caption{The central values and the corresponding $1\sigma$ ranges of various observables in the SM.}
\label{sm_results}
\end{table}

\begin{figure}[htbp]
\centering
\includegraphics[width=4cm,height=3cm]{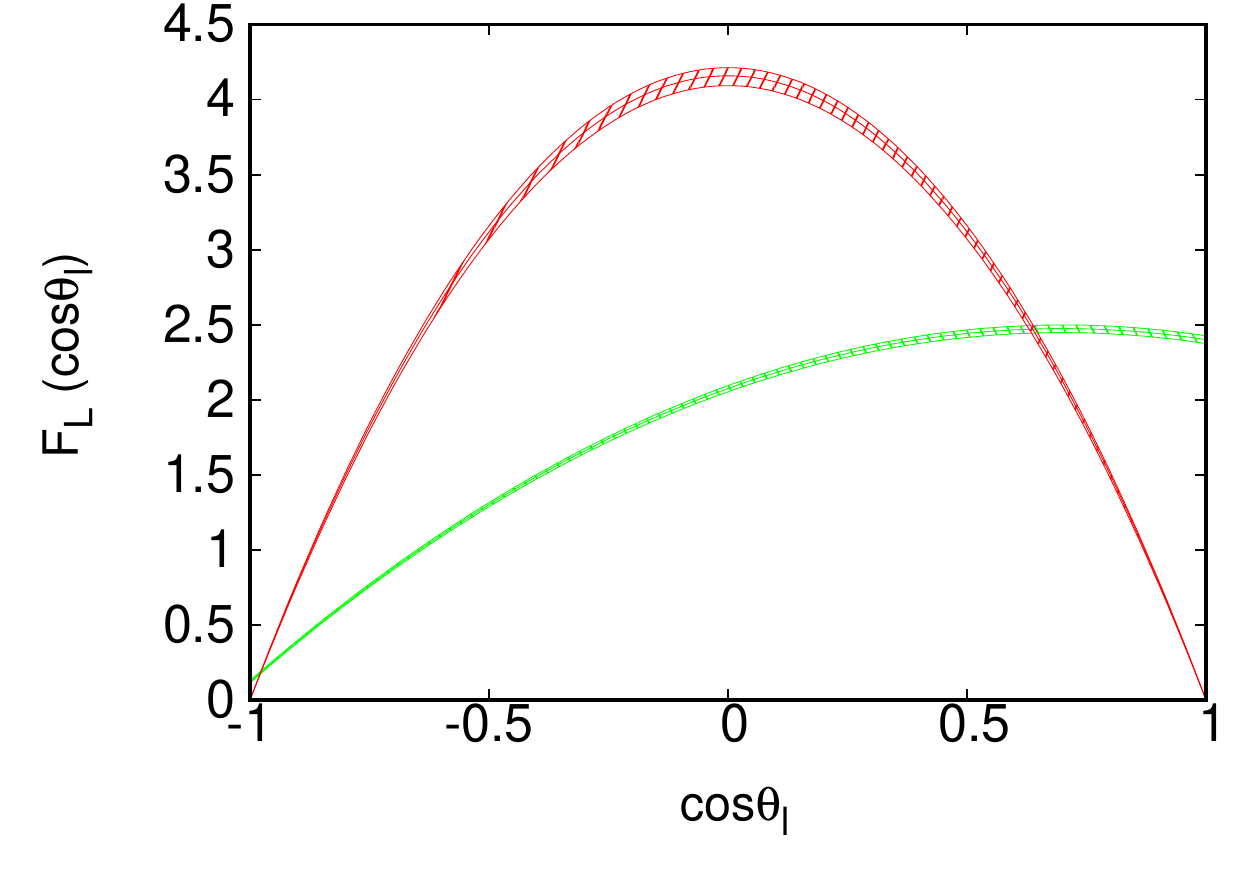}
\includegraphics[width=4cm,height=3cm]{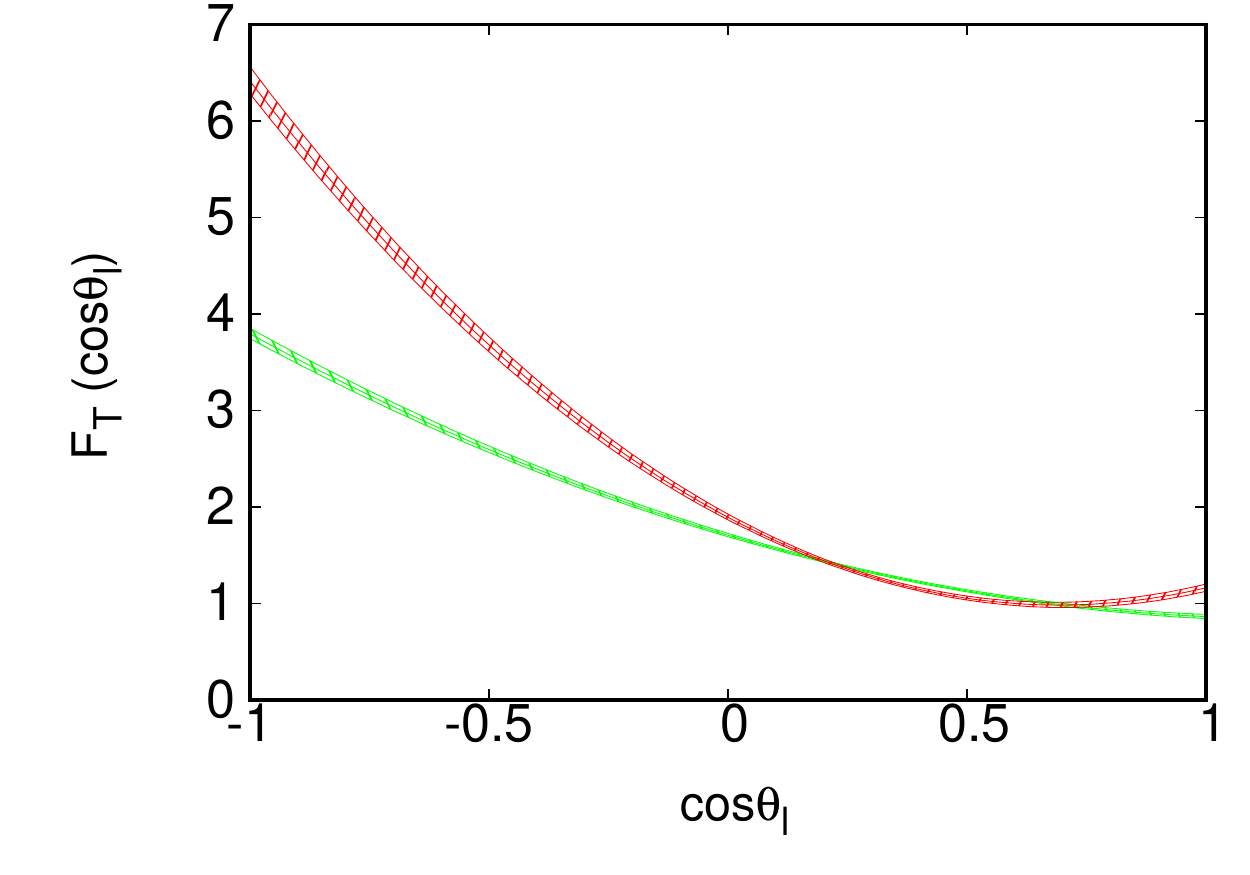}
\includegraphics[width=4cm,height=3cm]{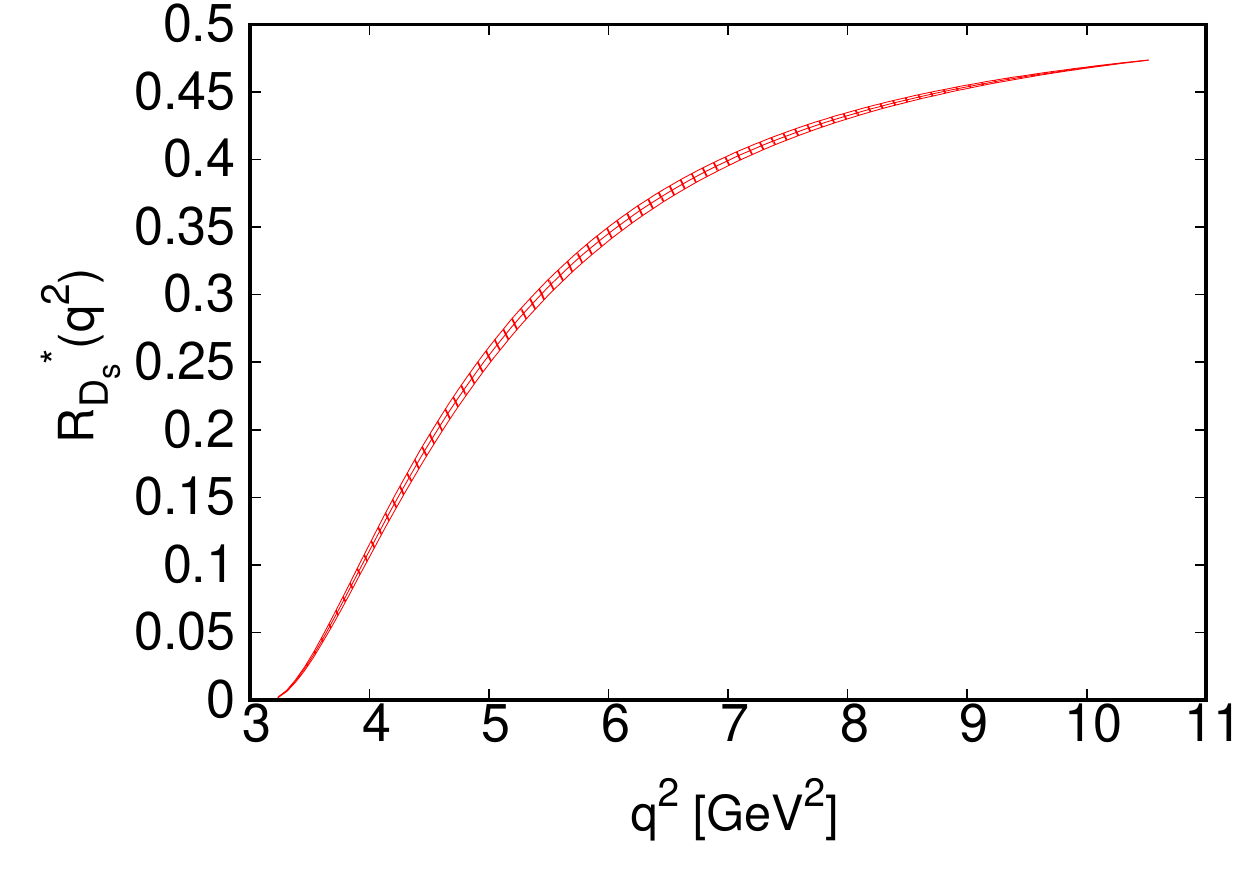}
\includegraphics[width=4cm,height=3cm]{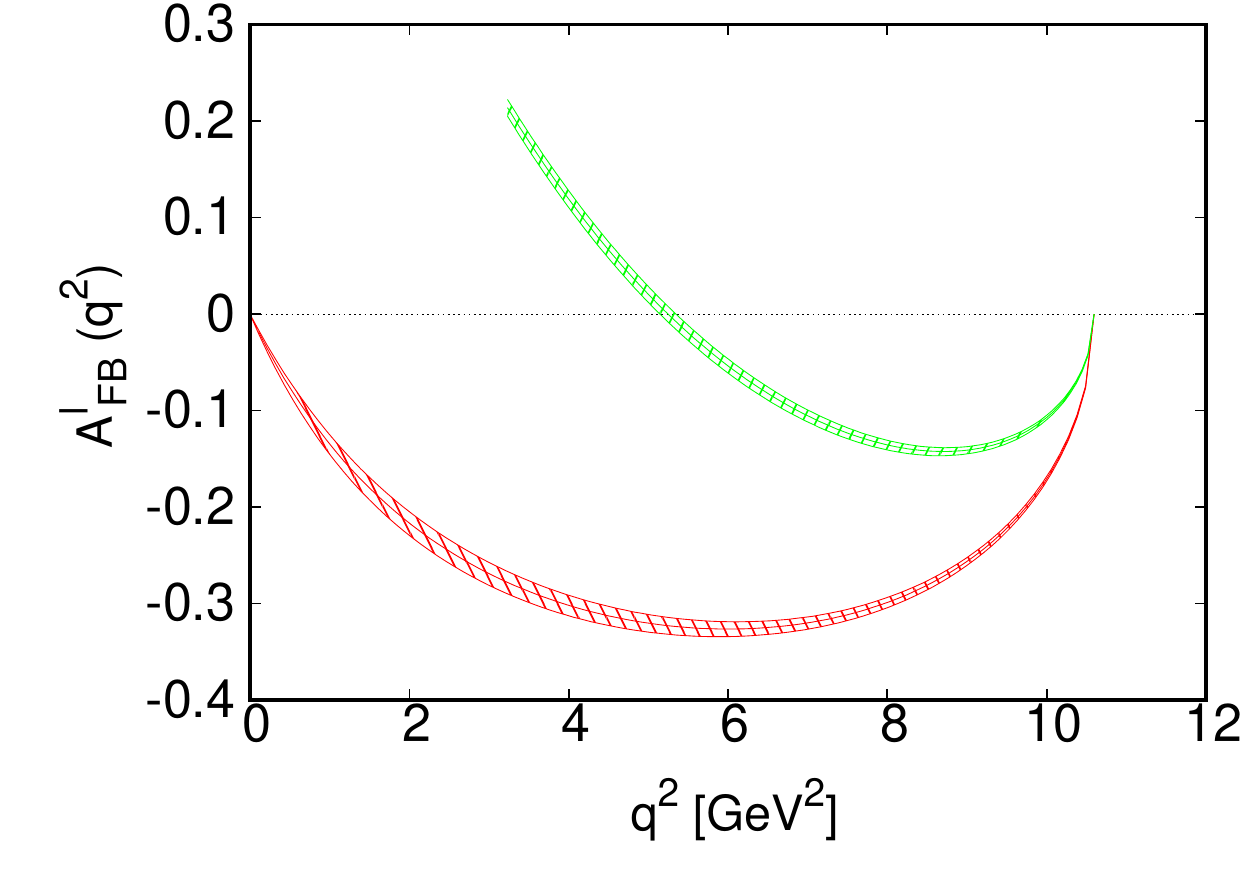}
\includegraphics[width=4cm,height=3cm]{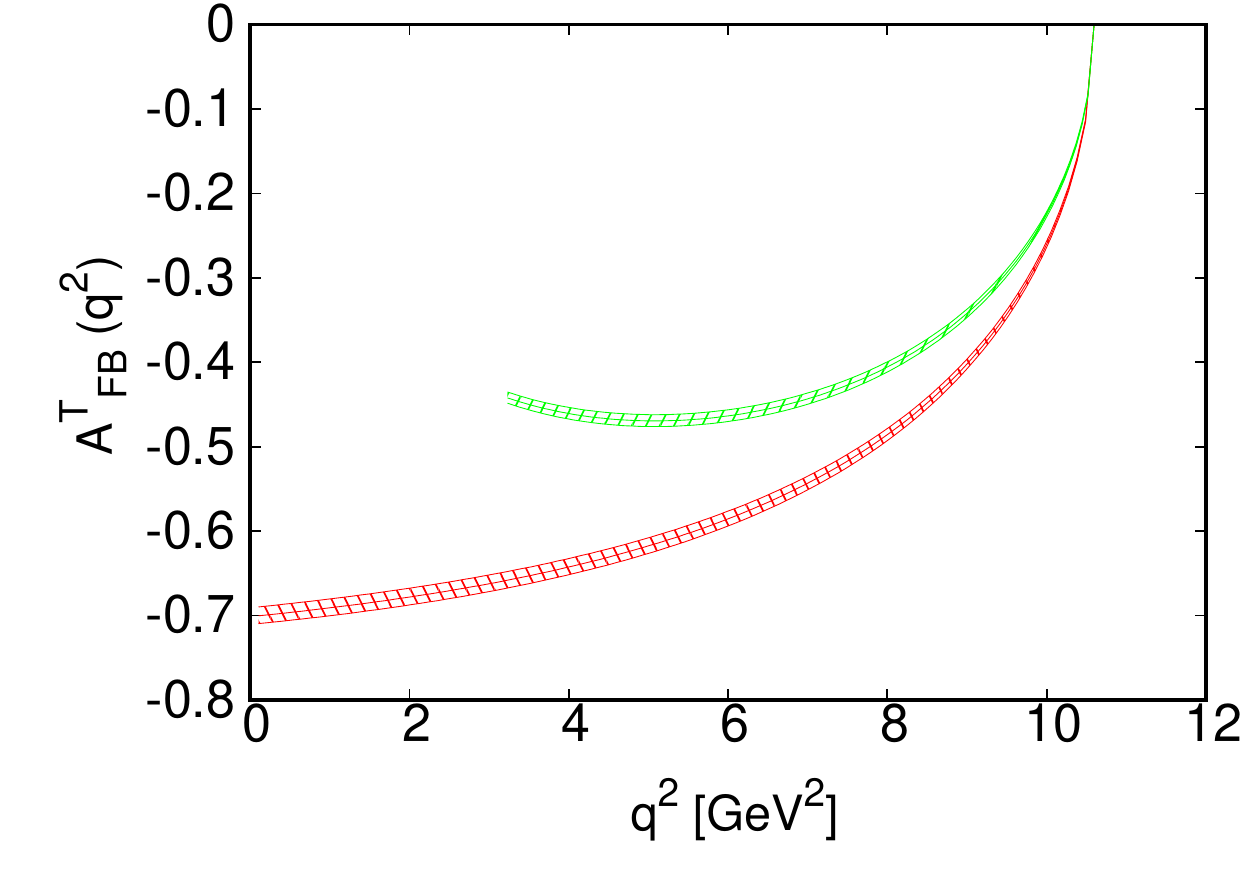}
\includegraphics[width=4cm,height=3cm]{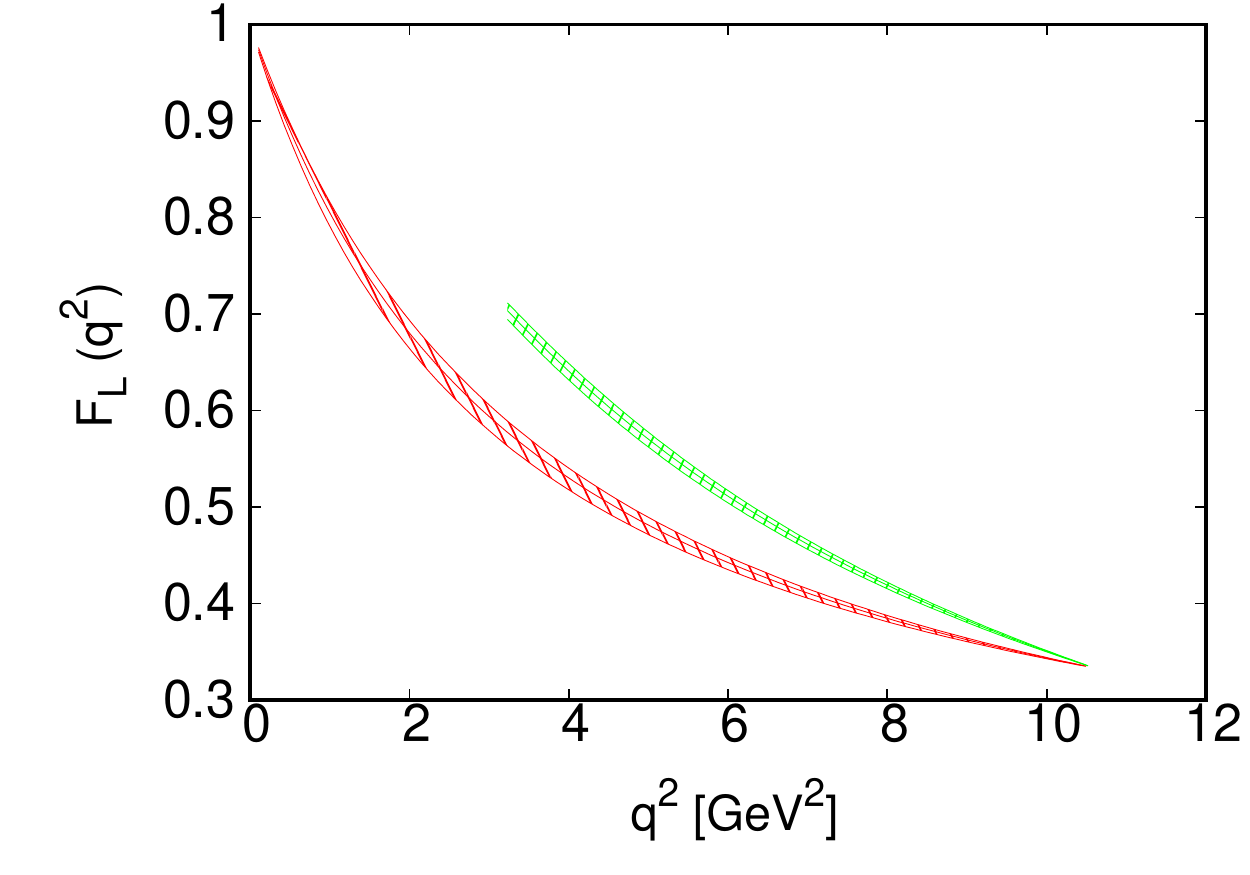}
\includegraphics[width=4cm,height=3cm]{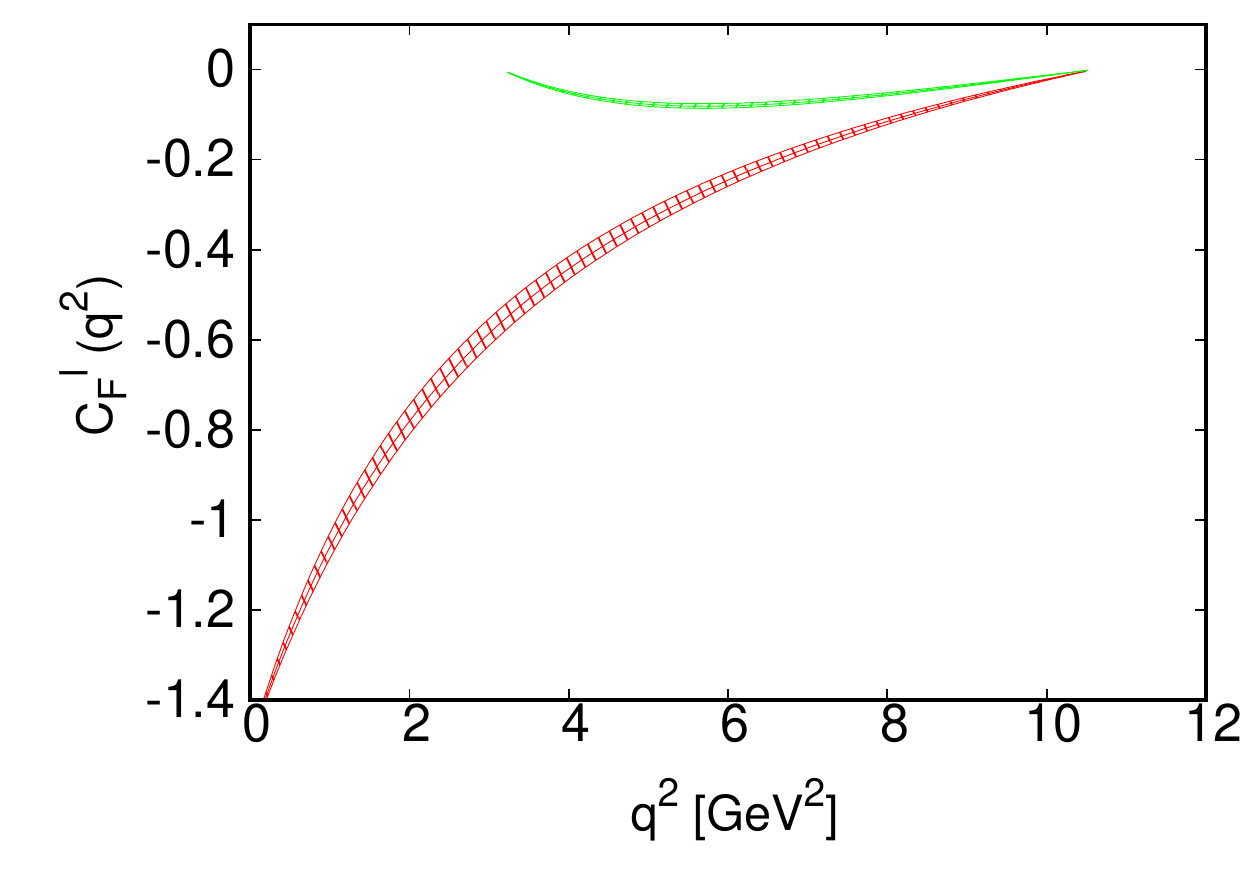}
\caption{$q^2$ and $\cos\theta_l$ dependence of $B_s \to D_s^*\,(\to D_s\gamma, D_s\pi)\,l\nu$ decay observables in the SM for the 
$e$~(red) and the $\tau$~(green) mode. }
\label{sm_plot_same}
\end{figure}

\begin{figure}[htbp]
\centering
\includegraphics[width=4cm,height=3cm]{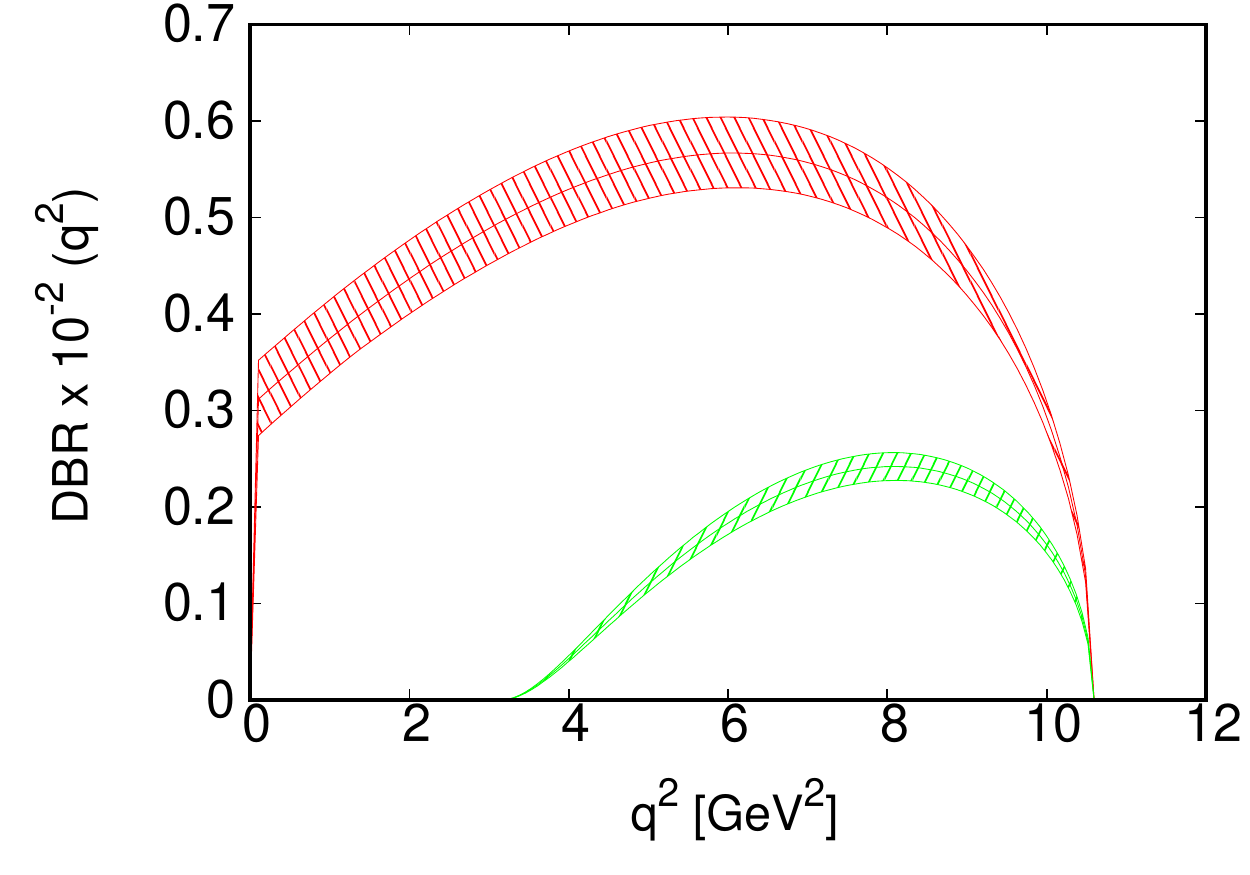}
\includegraphics[width=4cm,height=3cm]{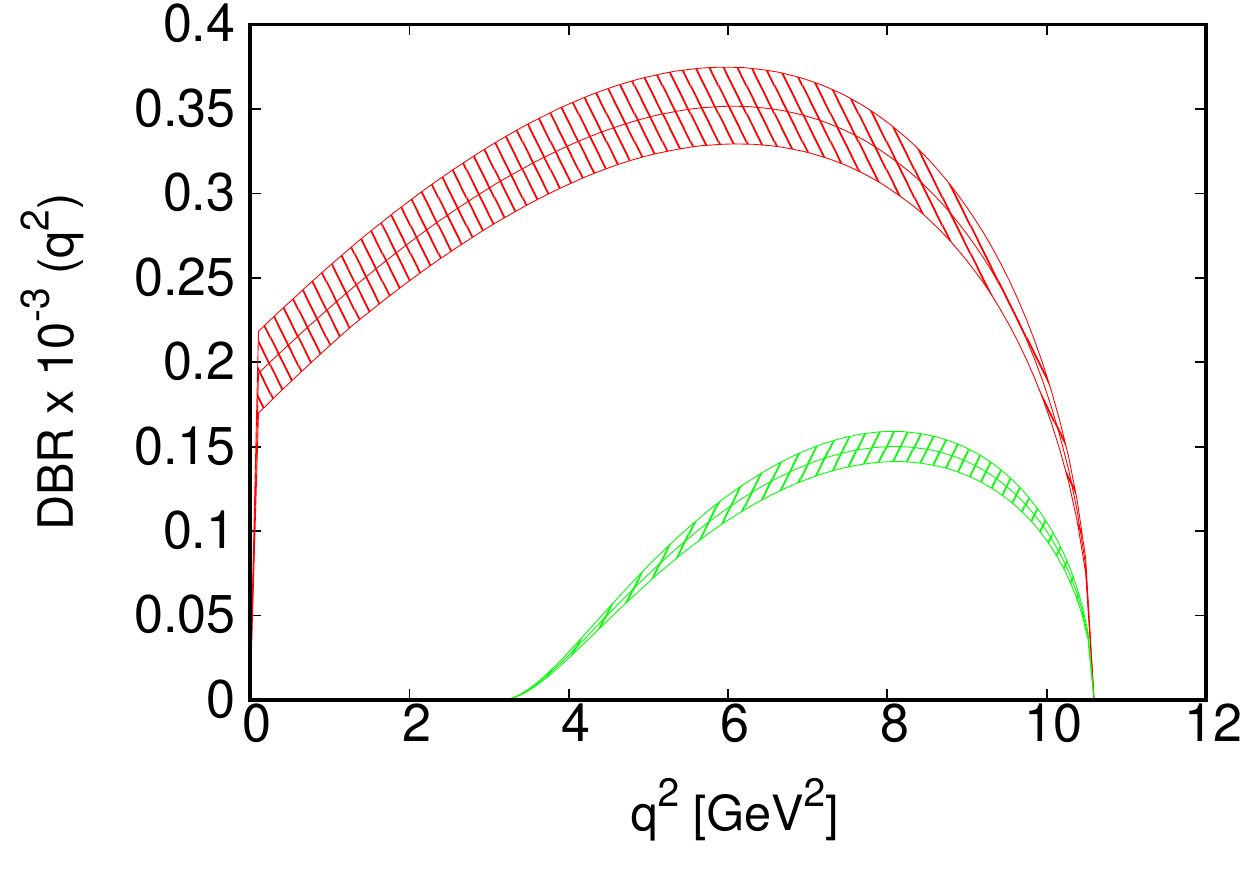}\hspace{1.5cm}
\includegraphics[width=4cm,height=3cm]{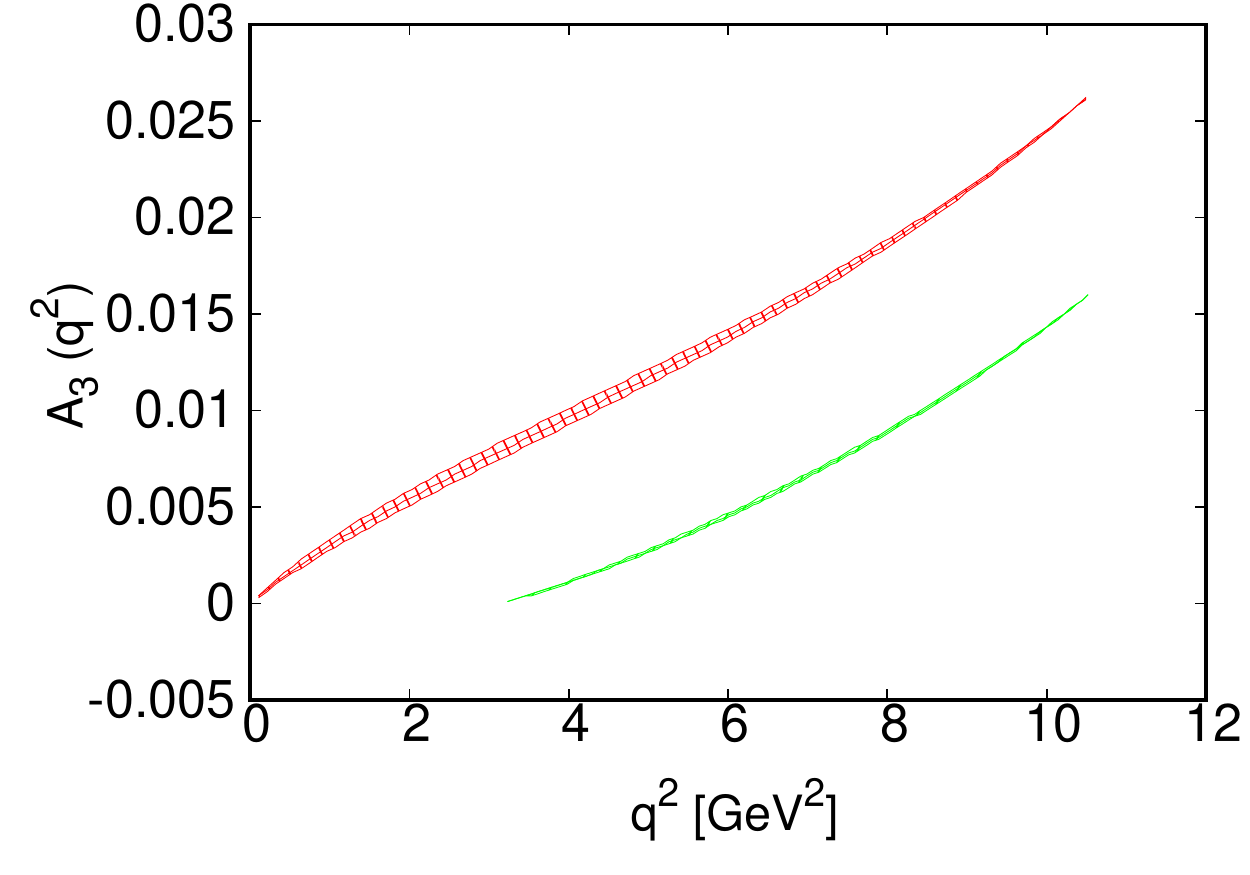}
\includegraphics[width=4cm,height=3cm]{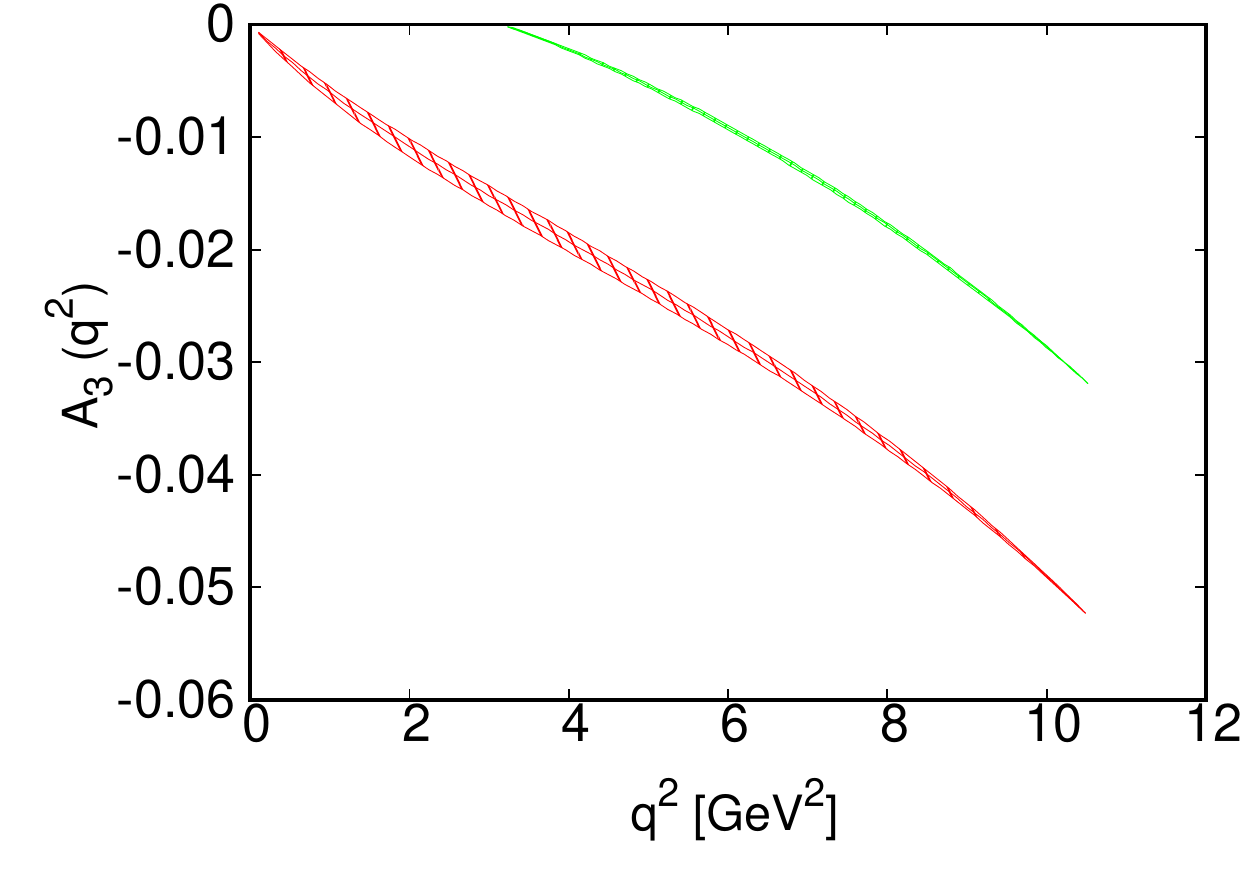}\hspace{1.5cm}
\includegraphics[width=4cm,height=3cm]{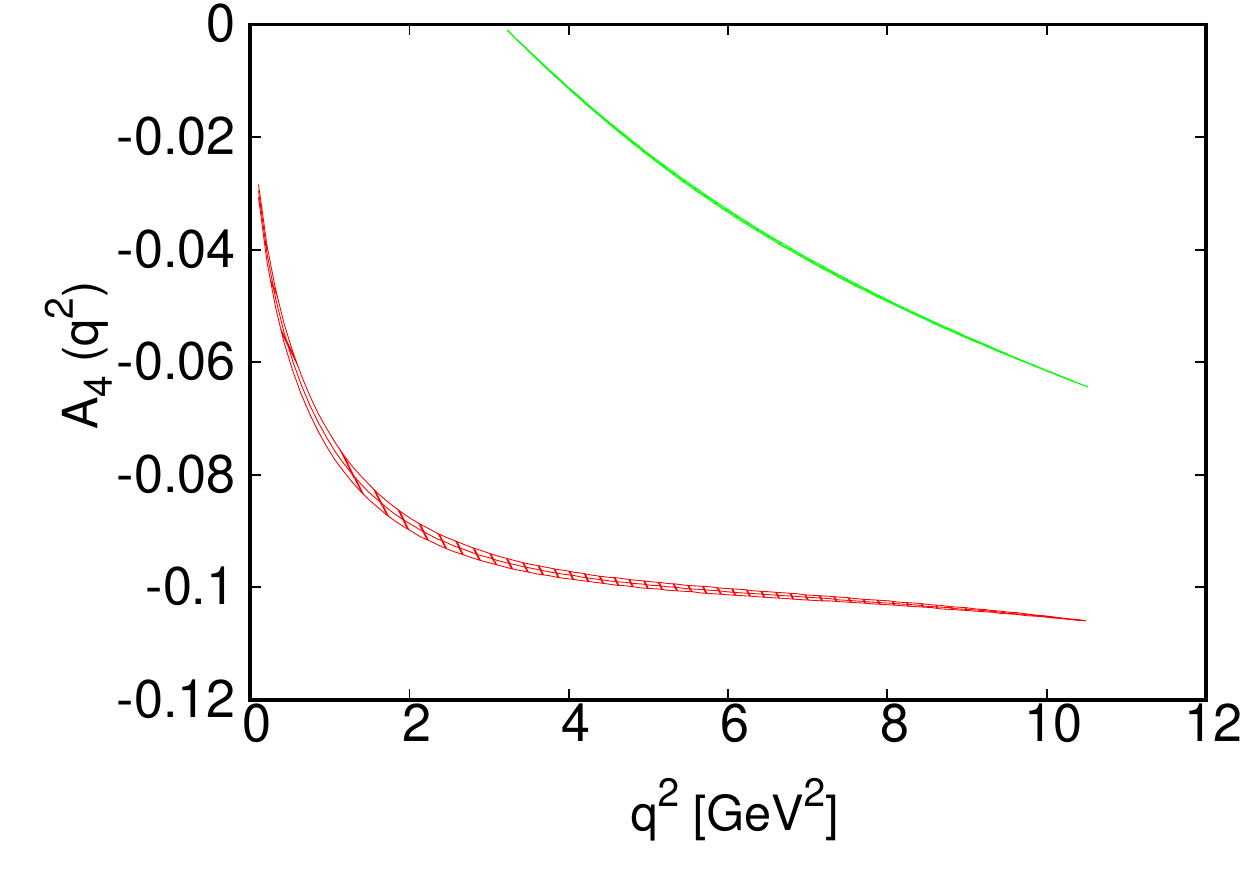}
\includegraphics[width=4cm,height=3cm]{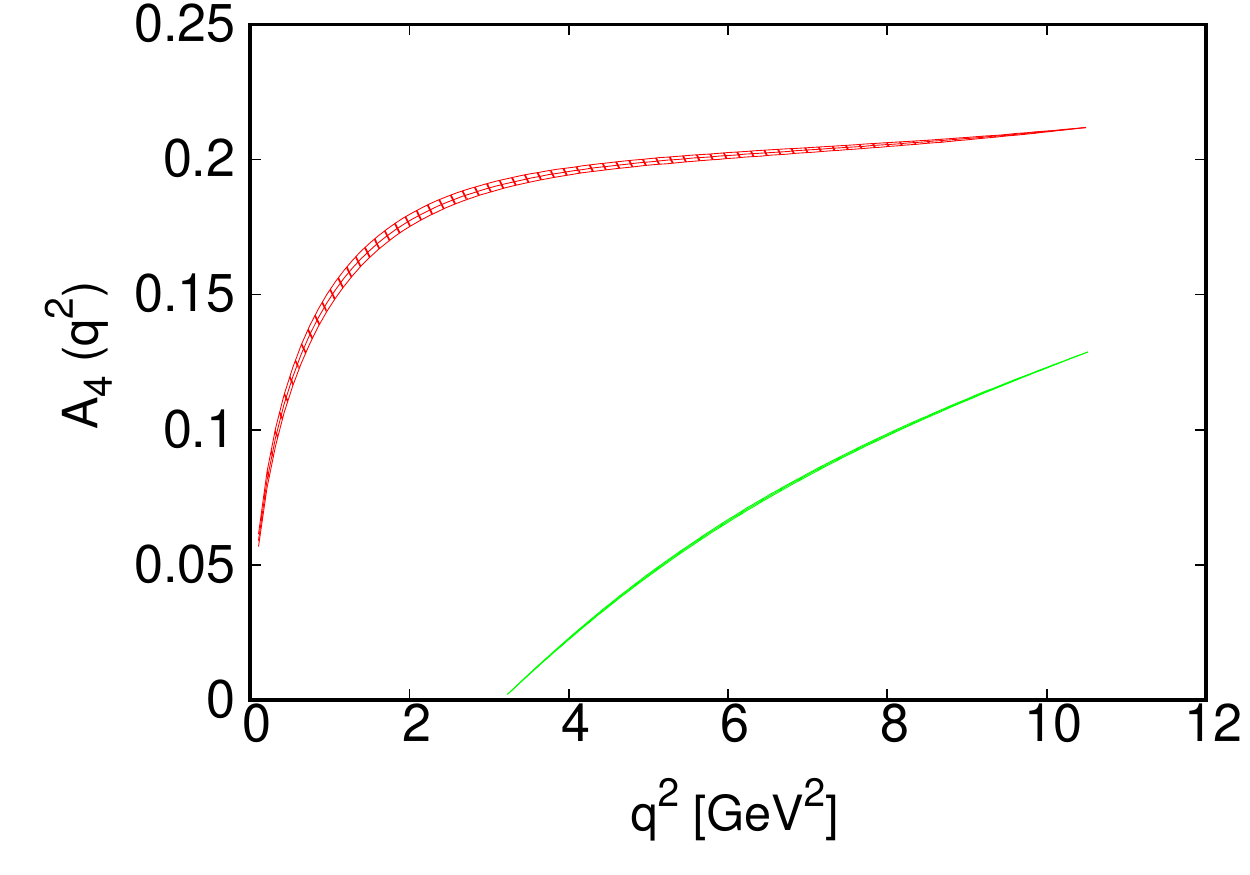}\hspace{1.5cm}
\includegraphics[width=4cm,height=3cm]{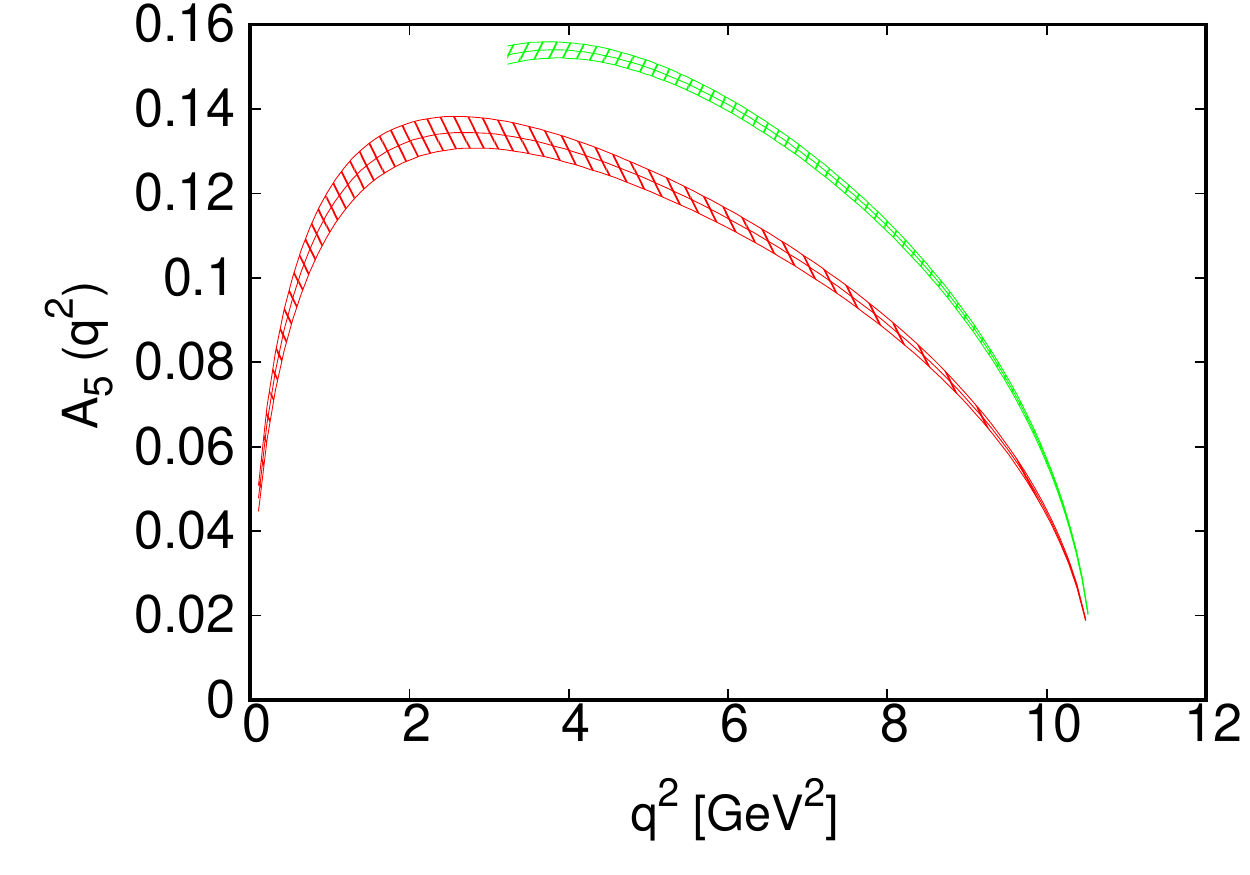}
\includegraphics[width=4cm,height=3cm]{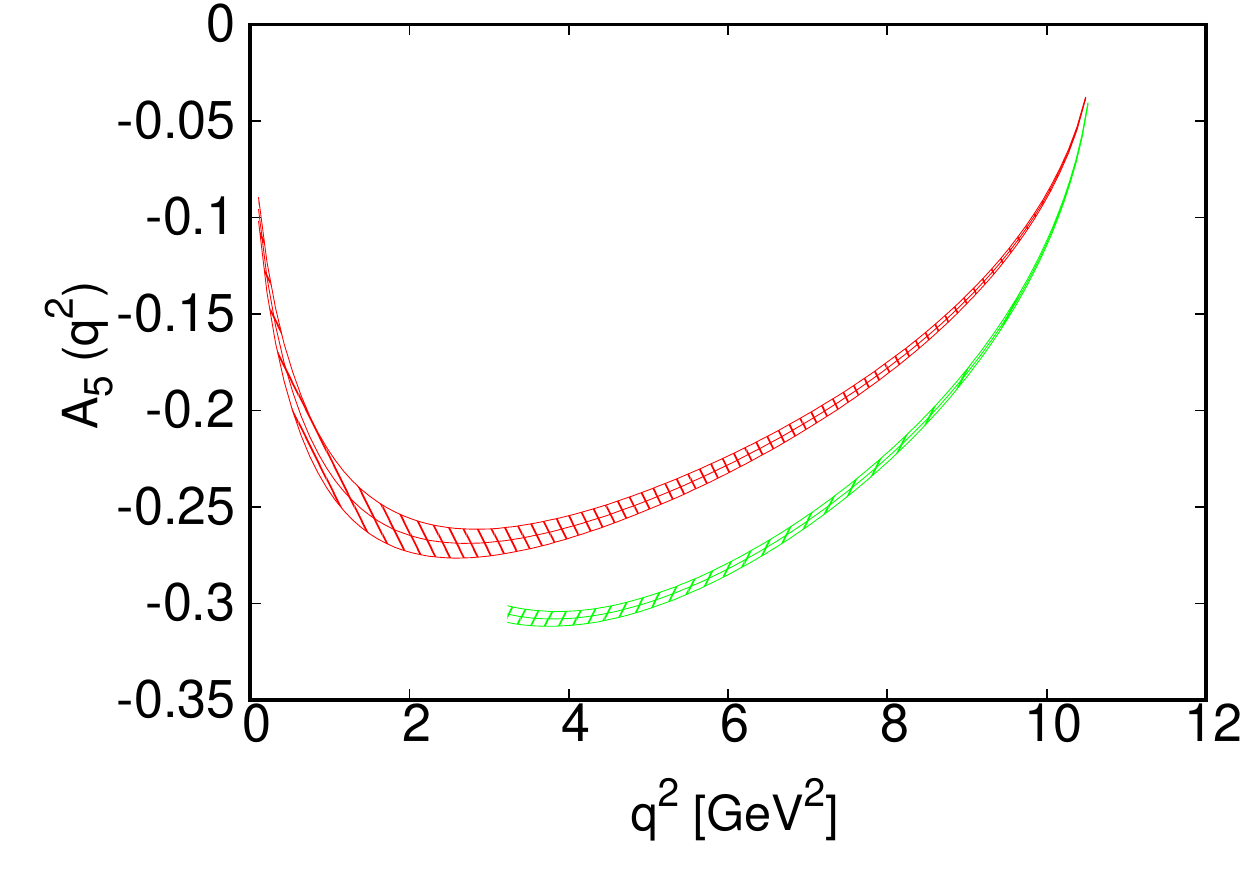}\hspace{1.5cm}
\includegraphics[width=4cm,height=3cm]{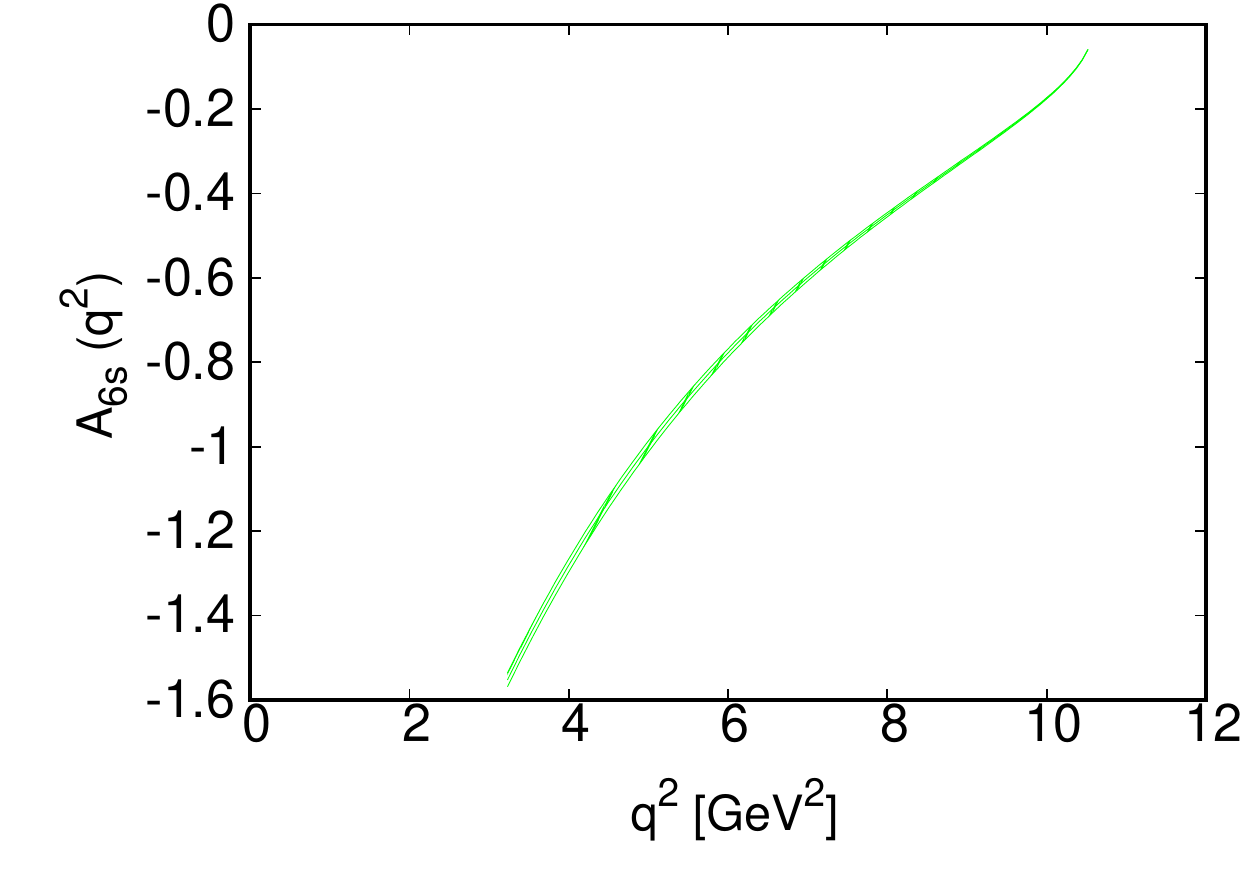}
\includegraphics[width=4cm,height=3cm]{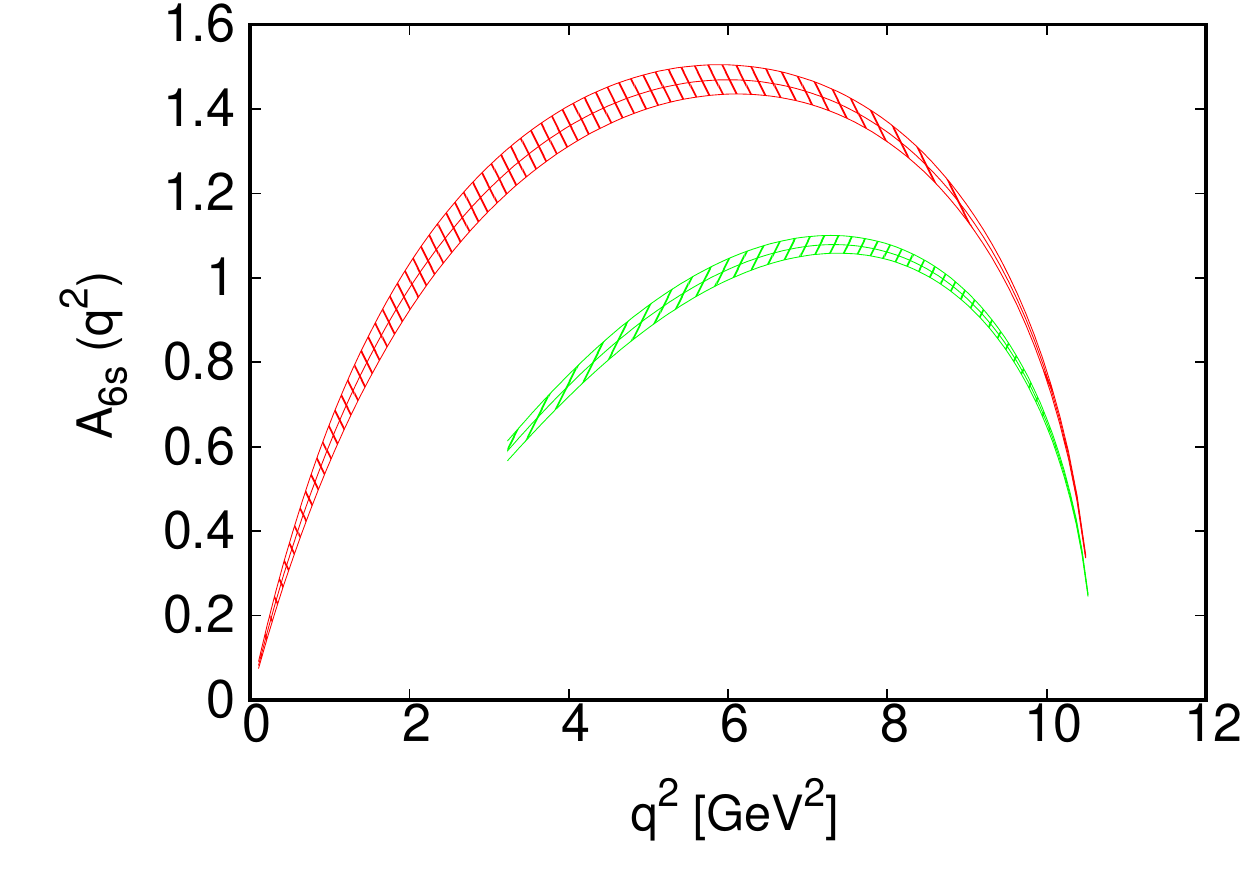}\hspace{1.5cm}
\includegraphics[width=4cm,height=3cm]{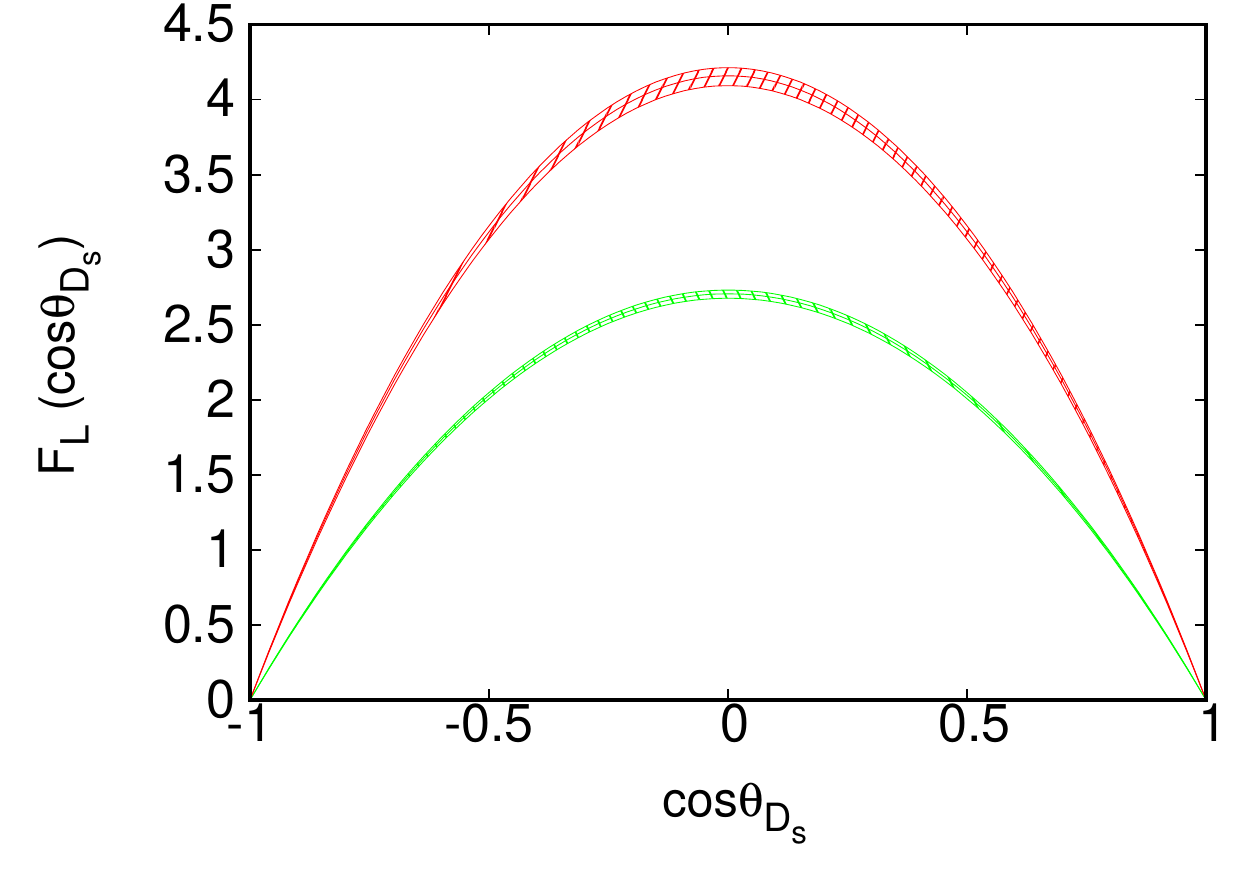}
\includegraphics[width=4cm,height=3cm]{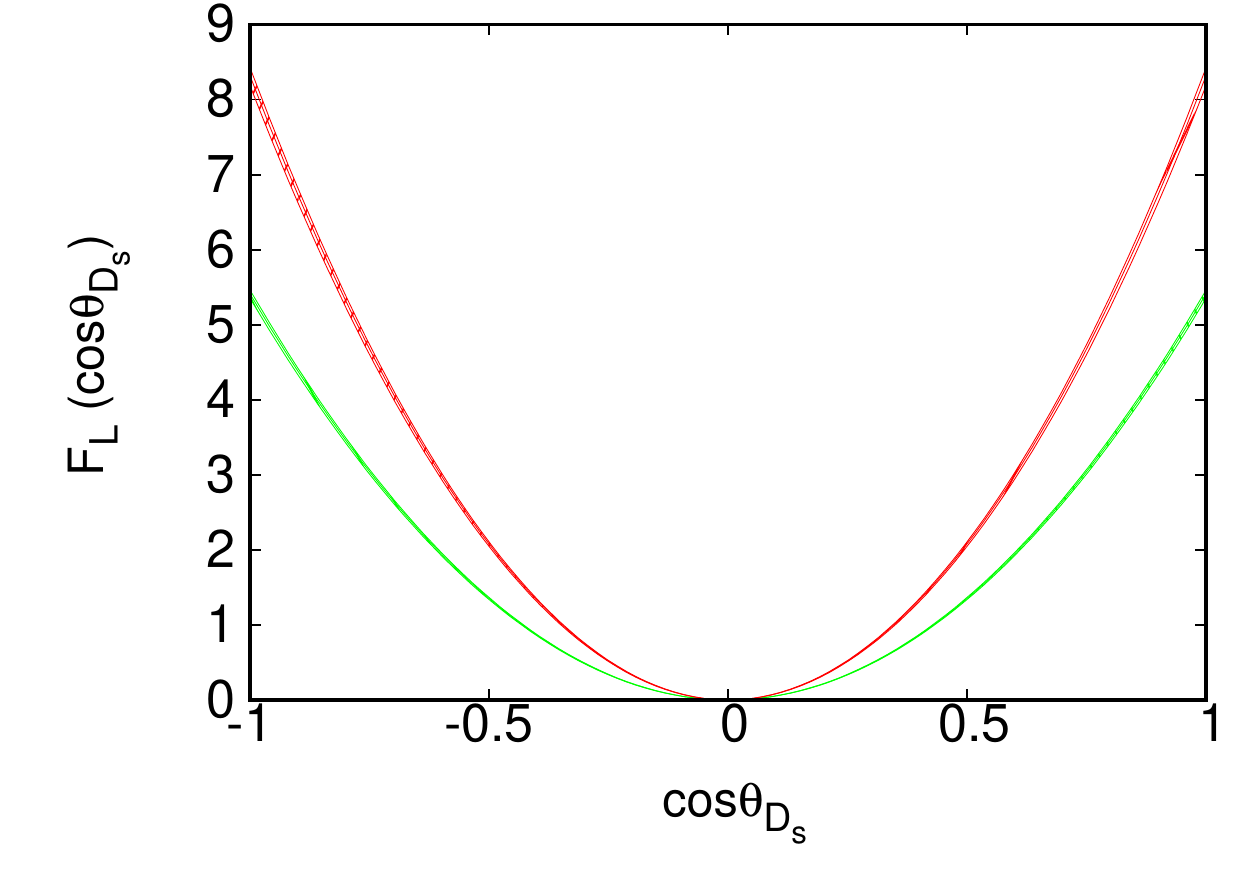}\hspace{1.5cm}
\includegraphics[width=4cm,height=3cm]{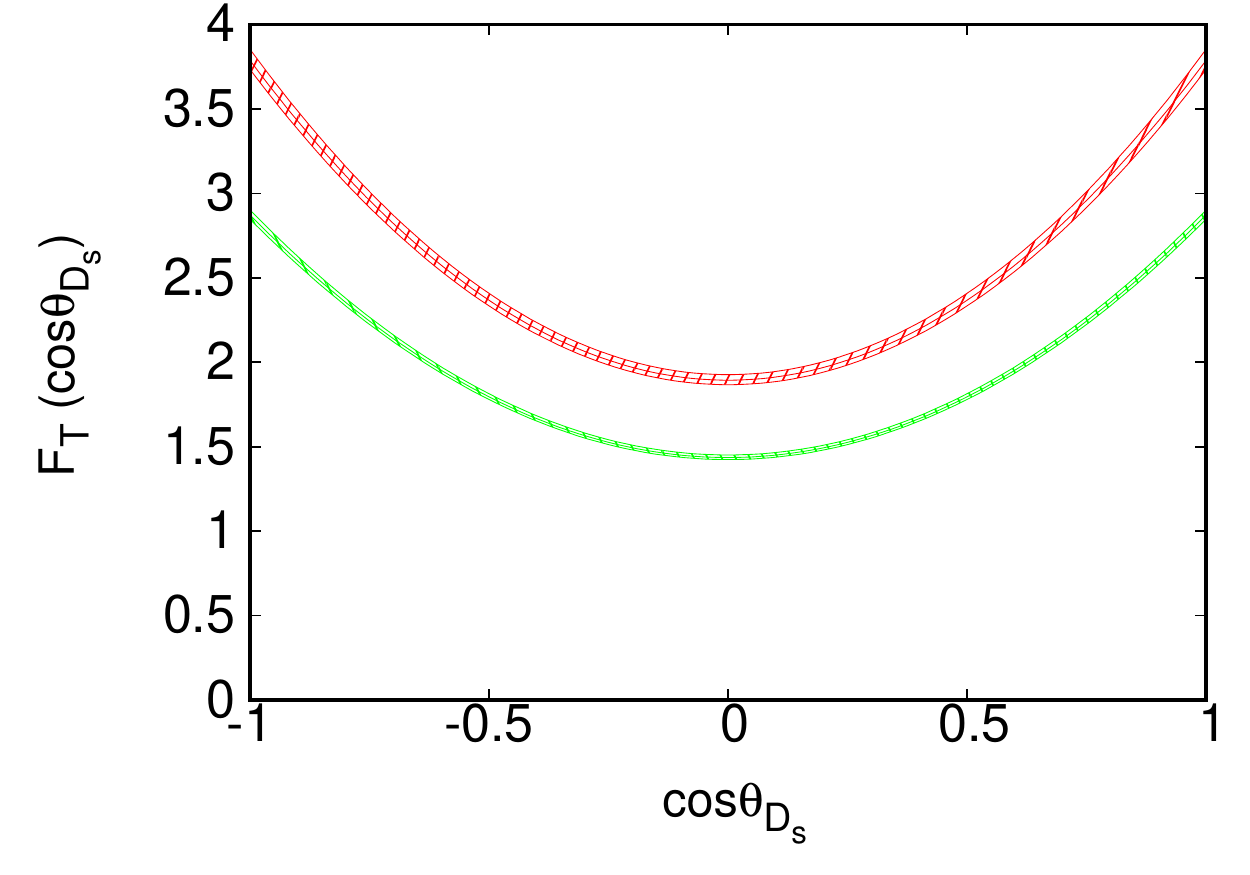}
\includegraphics[width=4cm,height=3cm]{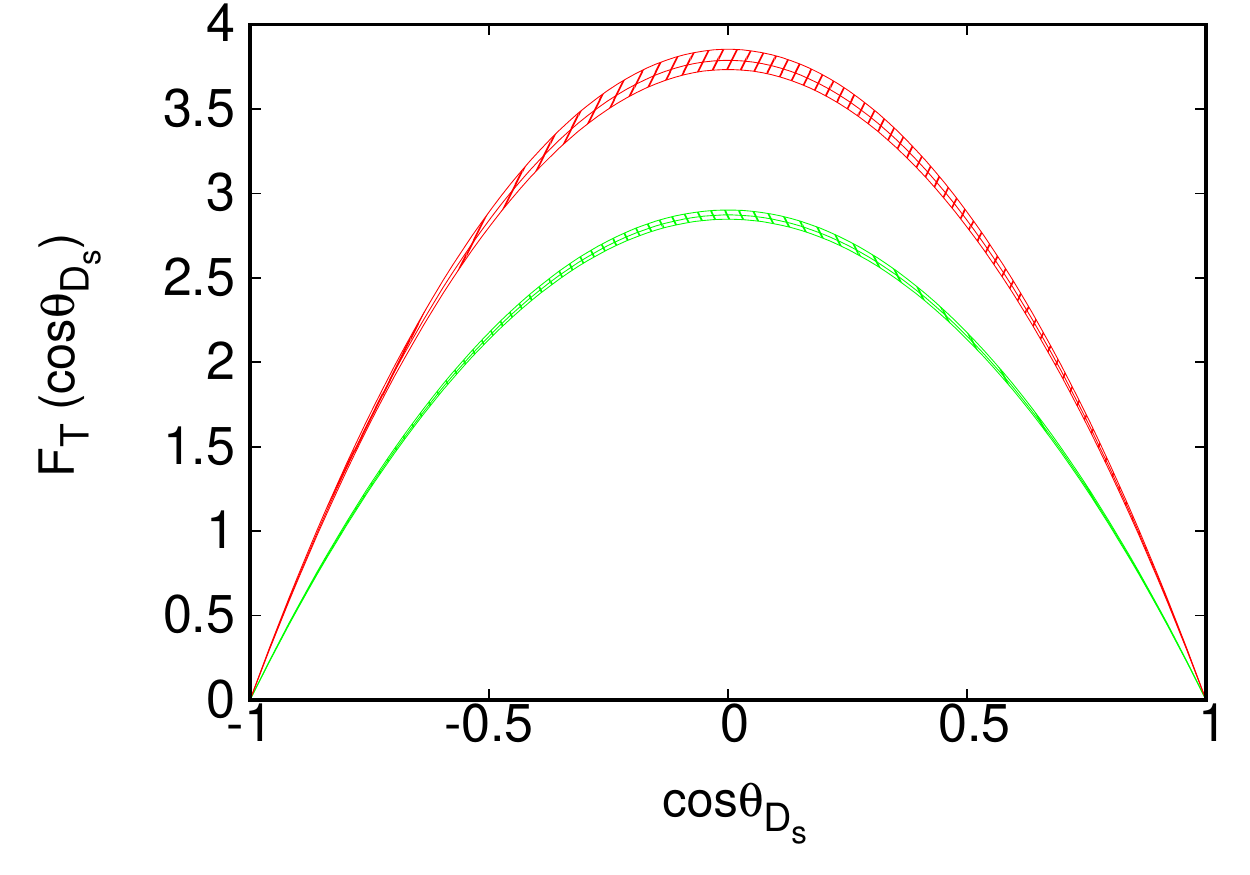}\hspace{1.5cm}
\includegraphics[width=4cm,height=3cm]{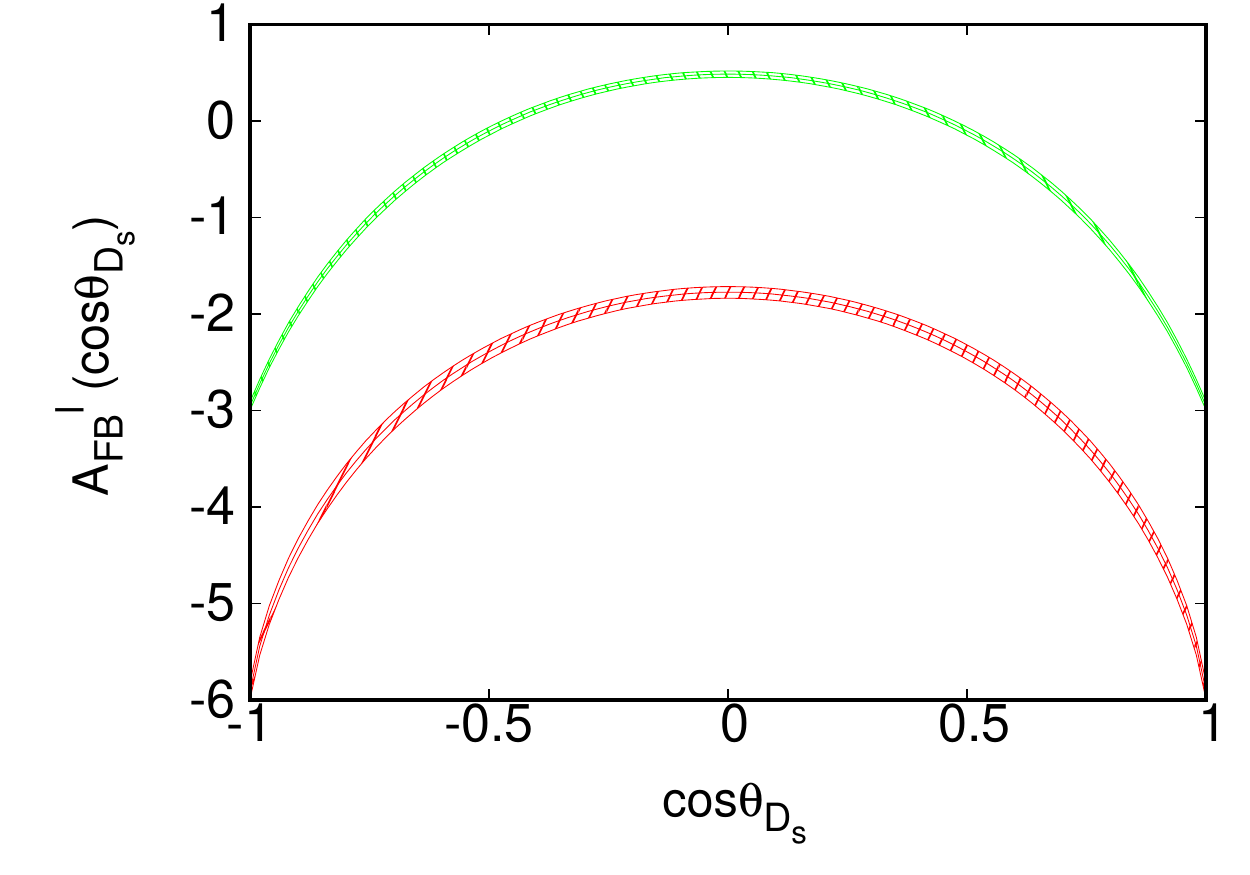}
\includegraphics[width=4cm,height=3cm]{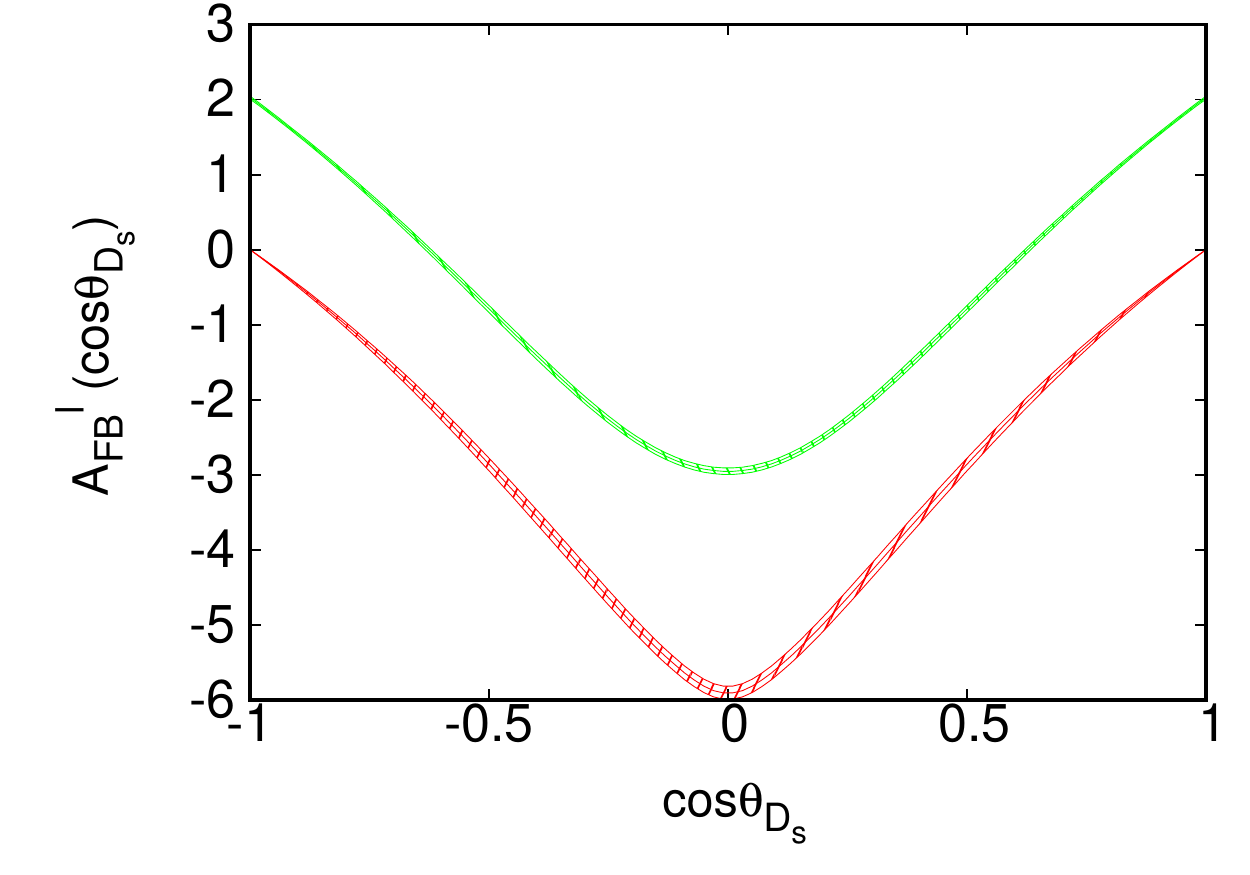}\hspace{1.5cm}
\caption{$q^2$ and $\cos\theta_{D_s}$ dependence of $B_s \to D_s^*\,(\to D_s\gamma, D_s\pi)\,l\nu$ decay observables in the SM for 
the $e$~(red) and the $\tau$~(green) mode. }
\label{sm_plot_diff}
\end{figure}

In Fig\ref{sm_plot_same}, we show several $q^2$ and $\cos\theta_l$ dependent observables such as $R_{D_s^*}(q^2)$, $A_{FB}^l(q^2)$,
$A_{FB}^T(q^2)$, $F_L^{D_s^*}(q^2)$, $C_F^l(q^2)$, $F_L(\cos\theta_l)$ and $F_T(\cos\theta_l)$ for the 
$B_s\to D_s^*(\to D_s \gamma, D_s \pi\, ) \, l \, \nu $ decay mode. It should be mentioned that these observables show exact same behaviour 
for the $D_s\pi$ and the $D_s\gamma$ mode. Here the red color represents the $e$ 
mode and green color represents the $\tau$ mode, respectively. Our main observations are as follows.
\begin{itemize}
\item \textbf{ $A_{FB}^l$ :} We observe a zero crossing of $A_{FB}^{\tau}$ at $q^2=5.25\pm 0.12\,\ \rm GeV^2\rm$.

\item \textbf{ $A_{FB}^T$: } $A_{FB}^T$ is minimum at low $q^2$ and assumes negative values for the whole $q^2$ range in both $e$ mode and 
$\tau$ mode. Moreover, it increases with $q^2$ and becomes zero at $q^2 = q^2_{\rm max}$.

\item \textbf{ $C_F^l$ :} The convexity parameter $C_F^e(q^2)$ is found to be minimum at low $q^2$ and it increases as $q^2$ increases.
At $q^2=q^2_{\rm max}$, it is equal to zero for both $e$ and the $\tau$ mode. 

\item \textbf{ $F_L^{D_s^*}$: } The longitudinal polarization fraction $F_L^{D_s^*}$ is maximum for low value of $q^2$. It gradually decreases
and becomes minimum at $q^2 = q^2_{\rm max}$.

\item \textbf{$F_L^{D_s^*}(\cos{\theta_l})$}:  The distribution is found to be symmetric in case of $e$ mode but not for the $\tau$ mode. This
is due to the presence of lepton mass term in the amplitude. At $\cos\theta_l=0$, $F_L^{D_s^*}(\cos{\theta_l})$ is maximum for $e$ mode,
whereas, for the $\tau$ mode, the maximum occurs at $\cos\theta_l=1$. 

\item \textbf{$F_T^{D_s^*}(\cos{\theta_l})$}: The maximum value of $F_T^{D_s^*}$ is obtained for $\cos\theta_l=-1$ for both $e$ and the $\tau$
mode. it gradually decreases with increasing $\cos\theta_l$ and becomes minimum near $\cos\theta_l=1$.

\end{itemize}
In Fig~\ref{sm_plot_diff}, we display the $q^2$ and $\cos\theta_{D_s}$ dependence of several observables that are different for 
$B_s\to D_s^*(\to D_s \pi) \, l \, \nu $ and $B_s\to D_s^*(\to D_s \gamma) \, l \, \nu $ decay modes. Here the red color represents the $e$ 
mode and green color represents the $\tau$ mode, respectively. Our observations are as follows.  
\begin{itemize}
 
 \item DBR : In case of $B_s\to D_s^*(\to D_s \gamma) \, l \, \nu $ decay mode, the maximum value of ${\rm DBR} = 
(0.567\, \pm\, 0.037)\times10^{-2}$ is observed at $q^2\approx 6.04\,\rm GeV^2\rm$ for the $e$ mode, whereas, 
the maximum value of ${\rm DBR} =(0.241\pm0.015)\times 10^{-2}$ is observed at $q^2\approx 8.28\, \rm GeV^2\rm$ for the $\tau$ mode. 
Similarly, for $B_s\to D_s^*(\to D_s \pi)\, l\, \nu$, the $DBR$ peak of $(0.351\pm 0.023)\times10^{-3}$ is observed at 
$q^2\approx6.15 \rm\, GeV^2\rm$ for e mode and maximum ${\rm DBR} = (0.150\pm 0.010)\times10^{-3}$ is observed at
 $q^2\approx 8.07 \rm\, GeV^2\rm$ for the $\tau$ mode.

\item{$A_3$, $A_4$, $A_5$:} The angular observables $A_i$s obey a strict relation $A_i^{\pi} = -2\,A_i^{\gamma}$ at all 
values of $q^2$ for the $D_s\pi$ and $D_s\gamma$ mode. 

\item{\textbf{$A_{6s}$}} : For the $D_s\gamma$ channel, $A_{6s}$ is observed to be zero for the $e$ mode, whereas, it is minimum at low 
$q^2$ and maximum at high $q^2$ for the $\tau$ mode. It should also be mentioned that value of $A_{6s}$ is negative for the whole $q^2$ range.
For the $D_s \pi$ channel, the maximum of $A_{6s}$ is observed at $q^2 \approx 5.98\, \rm GeV^2$ for the $e$ mode and it is observed at 
$q^2 \approx 7.28\, \rm GeV^2$ for the $\tau$ mode.
  
\item \textbf{$F_L^{D_s^*}(\cos{\theta_{D_s}})$}: The behviour of $F_L^{D_s^*}$ is symmetric about $\cos{\theta_{D_s}}$.
The maximum value of $F_L^{D_s^*}$ is obtained at $\cos\theta_{D_s}=\, 0$ for both $e$ and the $\tau$ mode in the $D_s \gamma$ mode,
whereas, in $D_s \pi$ mode, we observe a minimum at $\cos\theta_{D_s}=\, 0$.

\item \textbf{$F_T^{D_s^*}(\cos{\theta_{D_s}})$} : $F_T^{D_s^*}$ is symmetric in $\cos\theta_{D_s}$ for both $D_s \gamma$ and 
$D_s \pi$ mode. $F_T^{D_s^*}$ is minimum at $\cos\theta_{D_s}=0$, whereas, it is found to be maximum at $\cos\theta_{D_s}=\pm 1$ for
the $D_s \gamma$ mode. For the $D_s \pi$ mode, the maximum, however, occurs at $\cos\theta_{D_s}=0$ and it goes to zero at 
$\cos\theta_{D_s}=\pm 1$.

\item \textbf{$A_{FB}^l (\cos\theta_{D_s})$} : $A_{FB}^l$ is symmetric in $\cos\theta_{D_s}$ for both $D_s \gamma$ and $D_s \pi$ 
modes. For $D_s \gamma$ mode, $A_{FB}^l$ is minimum at $\cos\theta=\pm 1$, whereas, it is maximum at $\cos\theta=0$ for both $e$ and the 
$\tau$ mode. However, for $D_s \pi$ mode, it is completely opposite. $A_{FB}^l$ is maximum at $\cos\theta=\pm 1$ and minimum at 
$\cos\theta=0$ for both $e$ and the $\tau$ cases.
It should also be mentioned that, a zero crossing in $A_{FB}^{\tau} (\cos\theta_{D_s})$ is observed at 
$\cos\theta_{D_s} = \pm0.456\pm0.018$ for the $D_s\gamma$ mode, whereas, the zero crossing point is observed at 
$\cos\theta_{D_s} = \pm 0.626\pm 0.007$ for the $D_s\pi$ mode.
\end{itemize}

\subsection{New physics analysis}
We now proceed to discuss the NP effects on various physical observables in the angular distribution of 
$B_s\to D_s^*(\to D_s \gamma) \, \tau\, \nu $ and $B_s\to D_s^*(\to D_s \pi) \, \tau \, \nu $ decays in a model independent framework. 
We have taken three possible NP scenarios. The best fit values of the NP couplings under each scenarios, taken from recent global fit 
analysis~\cite{Blanke:2018yud,Blanke:2019qrx,Becirevic:2019tpx}, are reported in Table~\ref{global_fit}.

\begin{table}[htbp]
\centering 
 \setlength{\tabcolsep}{6pt} 
 \renewcommand{\arraystretch}{1.8}
\begin{tabular}{|c|c|c|c|c|c|c|}
\hline
\multicolumn{3}{|c|}{New physics scenarios}\\
\hline

Scenerio - I~\cite{Blanke:2018yud,Blanke:2019qrx} & Scenerio - II~\cite{Blanke:2018yud,Blanke:2019qrx}  & Scenerio - III~\cite{Becirevic:2019tpx} \\
\hline
$g_{V_{L}}= 0.07$ & $(g_{V_{L}},\, g_{S_{L}}=-4\,g_{T_L})= (0.10, - 0.04)$ & $g_{V_{L}}=0.07-i 0.16$  \\
$g_{S_{R}}=0.09$ & $(g_{S_{R}},\, g_{S_{L}})= (0.21,-0.15) (Set A)$ or  & $g_{V_{R}}=-0.01-i0.39$ \\
$g_{S_{L}}=0.07$ &$(g_{S_{R}},\, g_{S_{L}}) = (-0.26, -0.61) (Set B)$ & $g_{S_{L}}=0.29-i0.67$ \\
$g_{S_{L}}=4\,g_{T_L}=-0.03$ &  $(g_{V_{L}},\, g_{S_{R}})= (0.08,-0.01)$ & $g_{S_{R}}=0.19+i0.08$\\
 & $(g_{S_{L}}  =  4\,g_{T_{L}}) =  (-0.06+i0.31)$ & $g_{T_{L}}=0.11-i0.18$\\

\hline

\end{tabular}
\caption{Best fit value of NP couplings.}
\label{global_fit}
\end{table}

\subsubsection{Scenerio - I}
In scenerio - I, we choose four different 1D NP hypothesis and the corresponding best fit values of Ref~\cite{Blanke:2018yud,Blanke:2019qrx}
obtained at scale $\mu = 1\,{\rm TeV}$ are reported in Table~\ref{global_fit}. For our analysis, we run these NP couplings down to the 
renormalization scale $\mu=m_b$~\cite{Blanke:2018yud,Blanke:2019qrx}. The effect of these NP couplings on several physical observables 
pertaining to $B_s\to D_s^*(\to D_s\gamma)\,\tau \,\nu$ and $B_s\to D_s^* (\to D_s\pi)\,\tau\,\nu$ decay modes are reported in 
Table~\ref{sc1_dsg_int}.  
 
It is clear from from Table~\ref{sc1_dsg_int} that in the presence of $g_{V_L}$ NP coupling, the branching ratio gets considerable deviations 
from the SM predication. However, no deviation from the SM prediction is observed for observables that are in the form of ratios. The NP 
dependency cancels in these ratios. In the presence of $g_{S_L}$ and $g_{S_R}$ NP couplings, $A_{FB}^{\tau}$ is found to be at more
than $4\sigma$ away from the SM prediction for both $D_s\gamma$ and $D_s\pi$ mode. Similarly, a deviation of around $3.4\sigma$, $4.71\sigma$ 
and $ 1\sigma$ is observed for $F_L^{D_s^*}$ in the presence of $g_{S_L}$, $g_{S_R}$ and $g_{S_L}=4\,g_{T_L}$ NP couplings. Moreover, the
deviation from the SM expectation observed in case of $R_{D_s^*}$ is at the level of $2.05\sigma$ and $3.96$ significance in the presence of $g_{S_R}$ and 
$g_{S_L}=4\,g_{T_L}$ NP couplings respectively, whereas, it is at the level of $15\sigma$ significance for $g_{V_L}$ NP coupling.
The observables $A_3$ and $A_{FB}^T$ show slight deviation from the SM in the presence of $g_{S_L}=4\,g_{T_L}$ NP coupling. As expected, $A_7$, $A_8$ and $A_9$
are all zero and hence we don't report them in Table.~\ref{sc1_dsg_int}.
 
\begin{figure}[htbp]
\centering
\includegraphics[width=4cm,height=3cm]{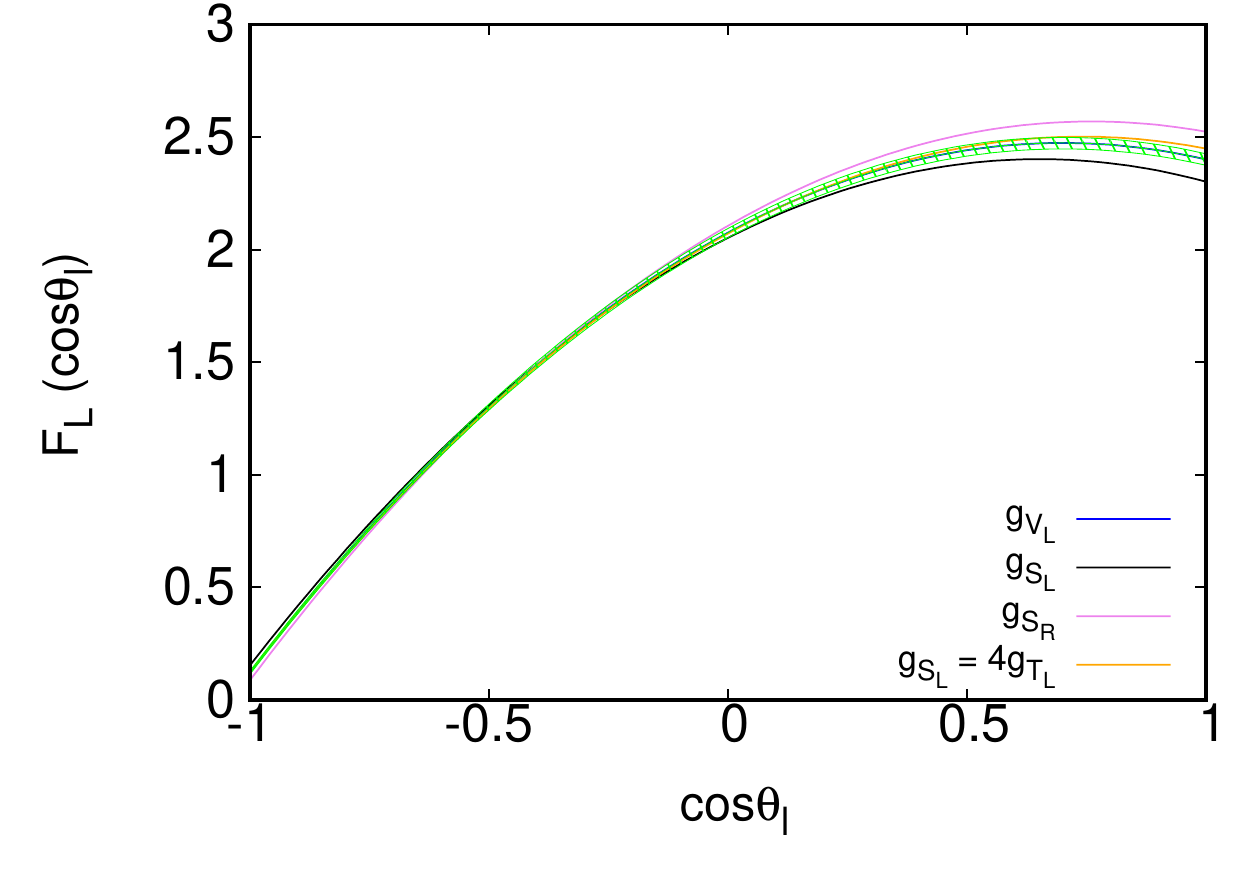}
\includegraphics[width=4cm,height=3cm]{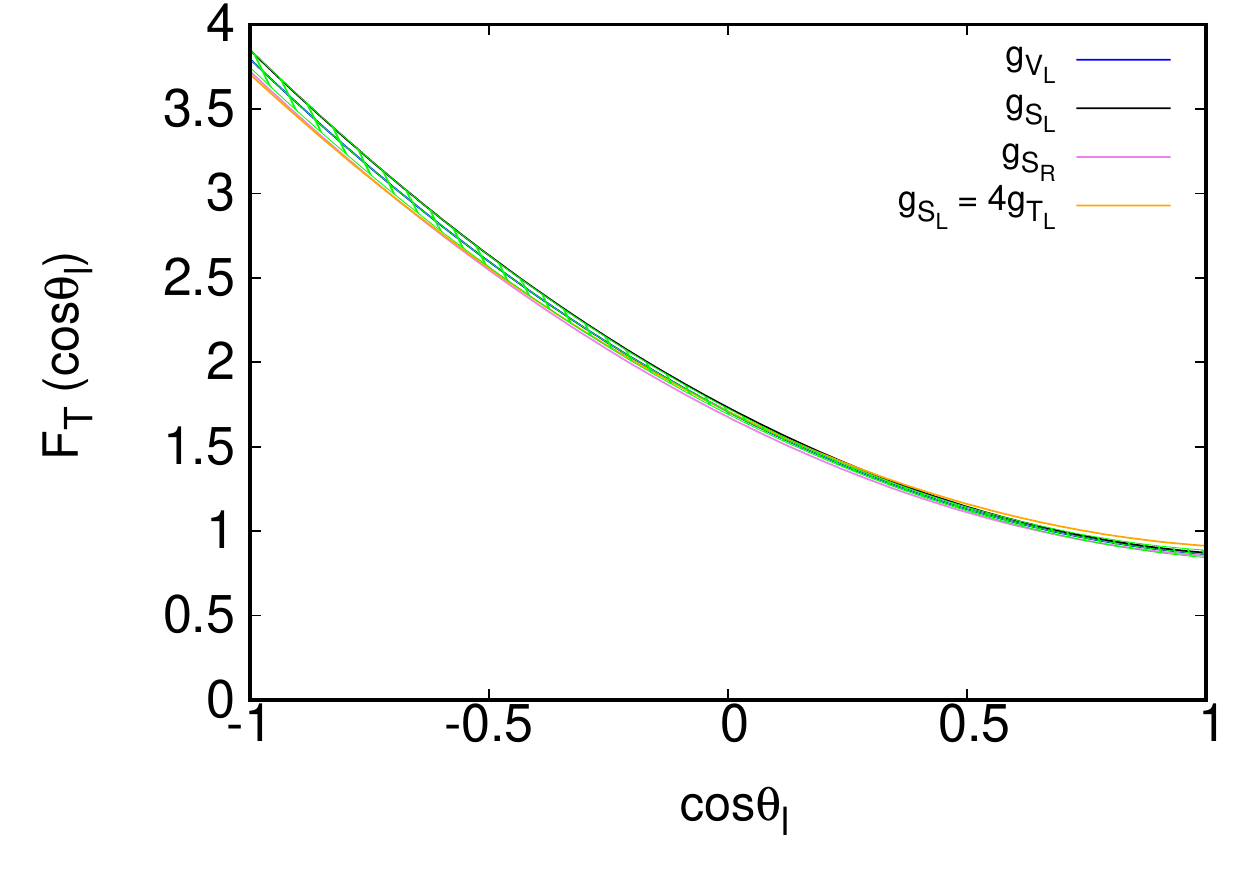}
\includegraphics[width=4cm,height=3cm]{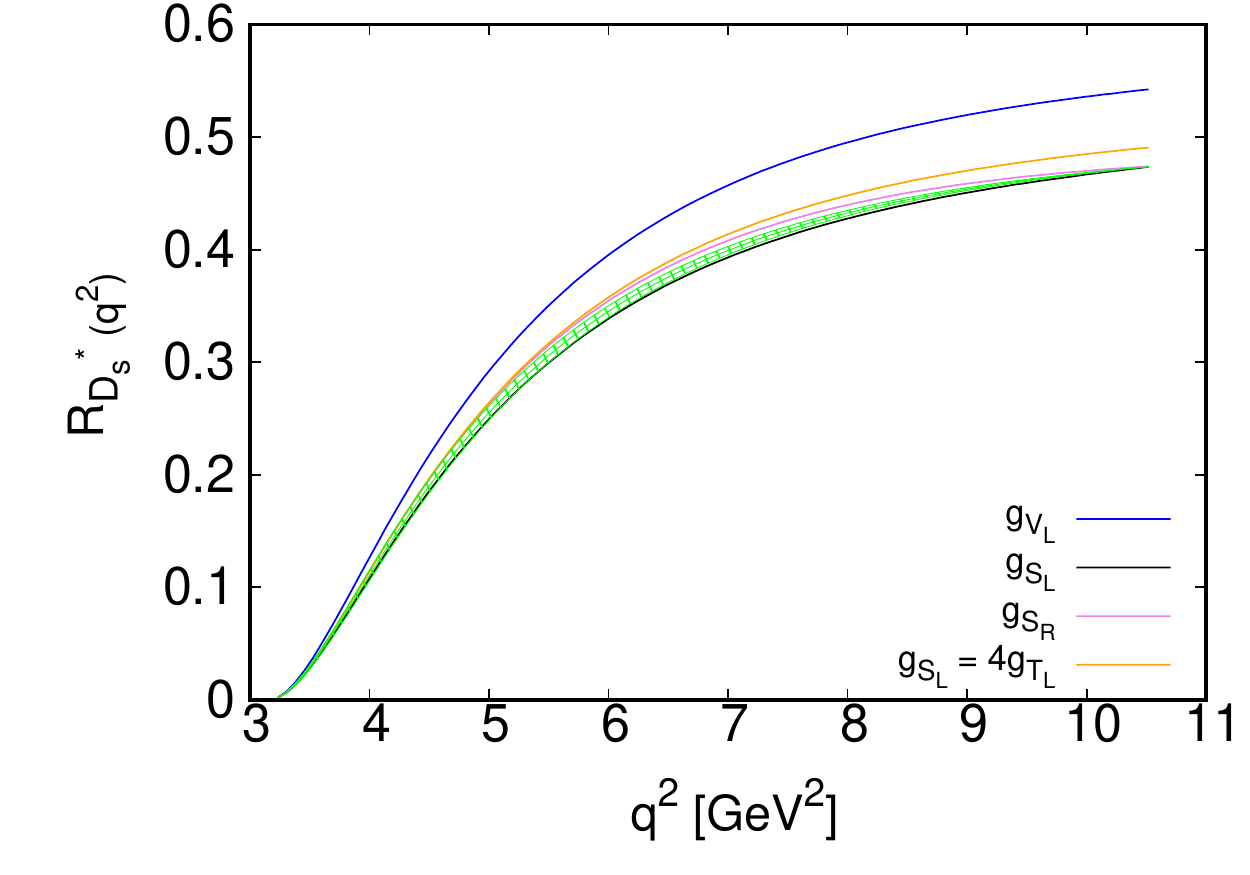}
\includegraphics[width=4cm,height=3cm]{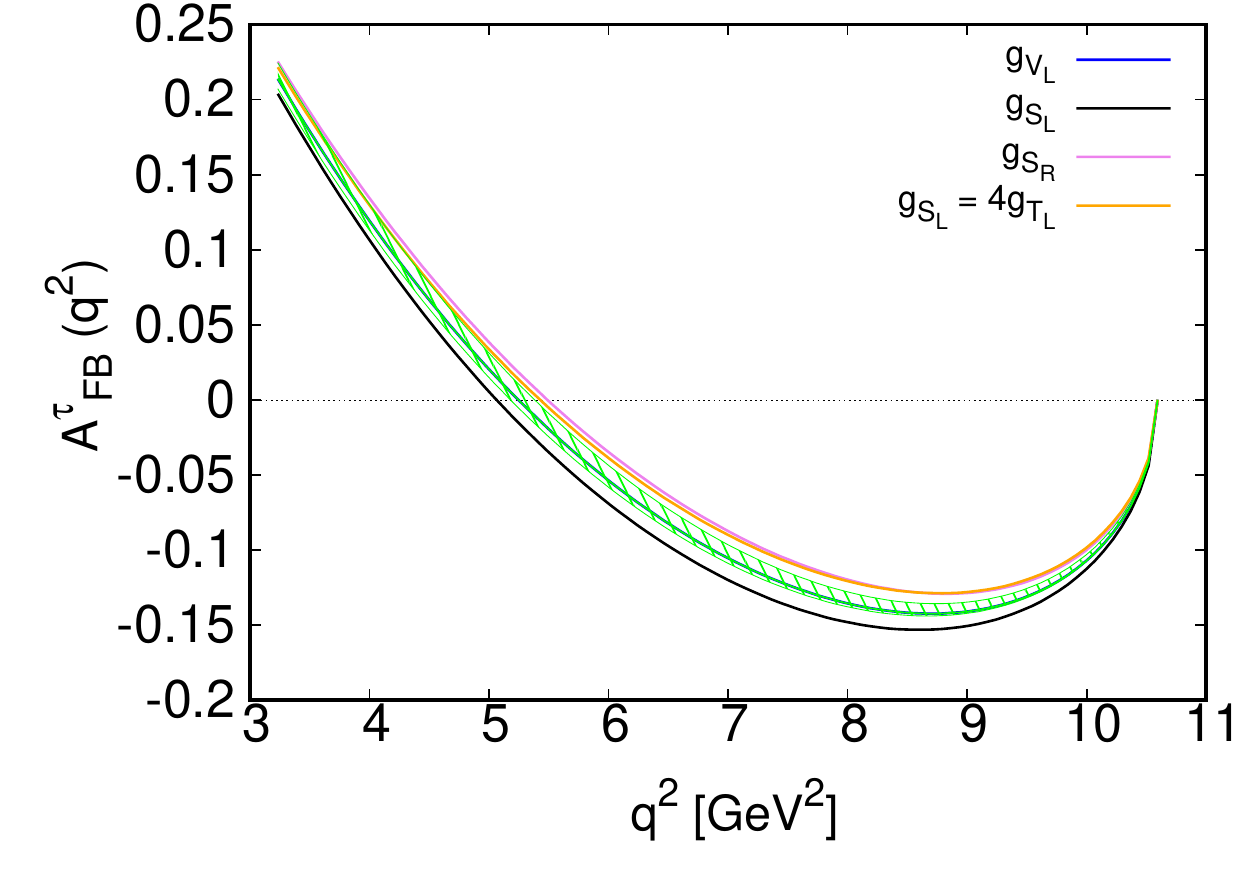}
\includegraphics[width=4cm,height=3cm]{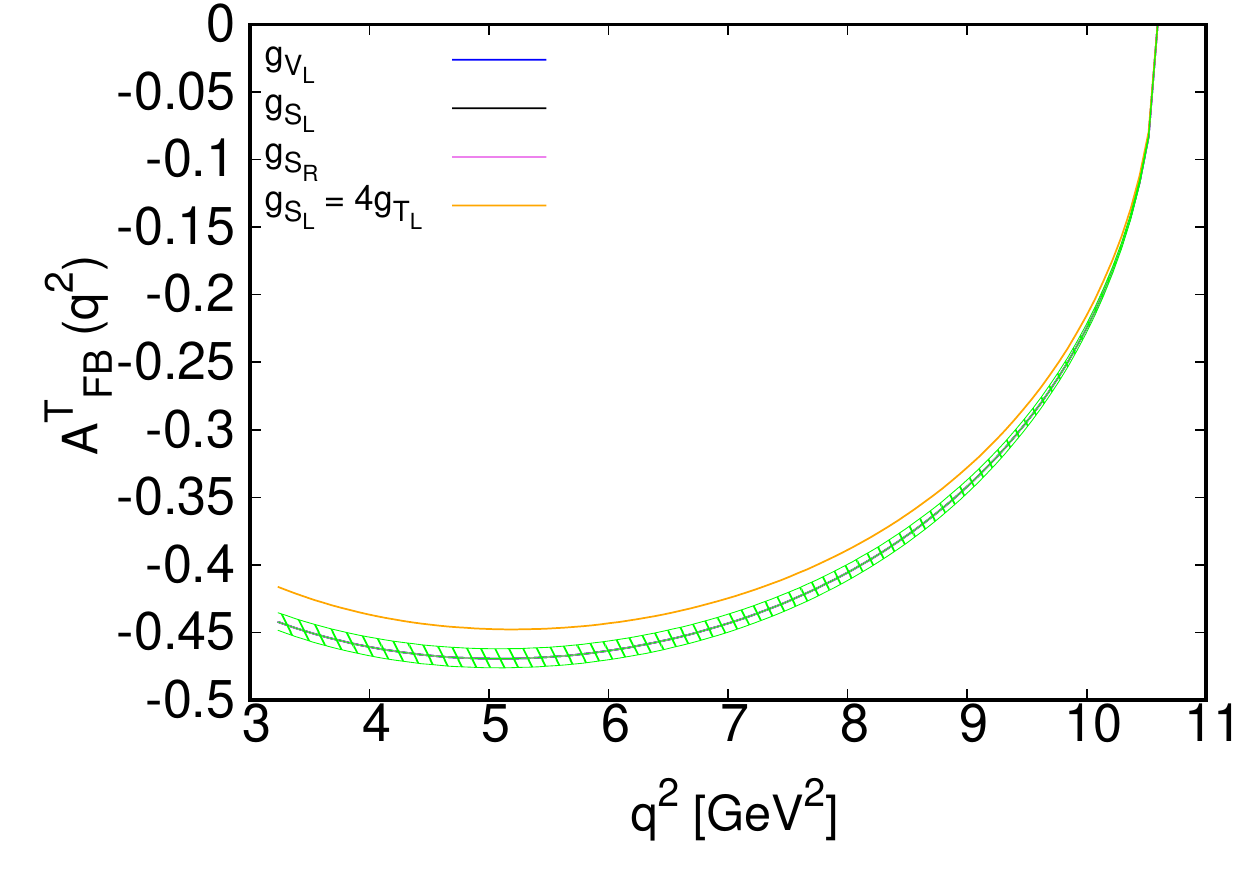}
\includegraphics[width=4cm,height=3cm]{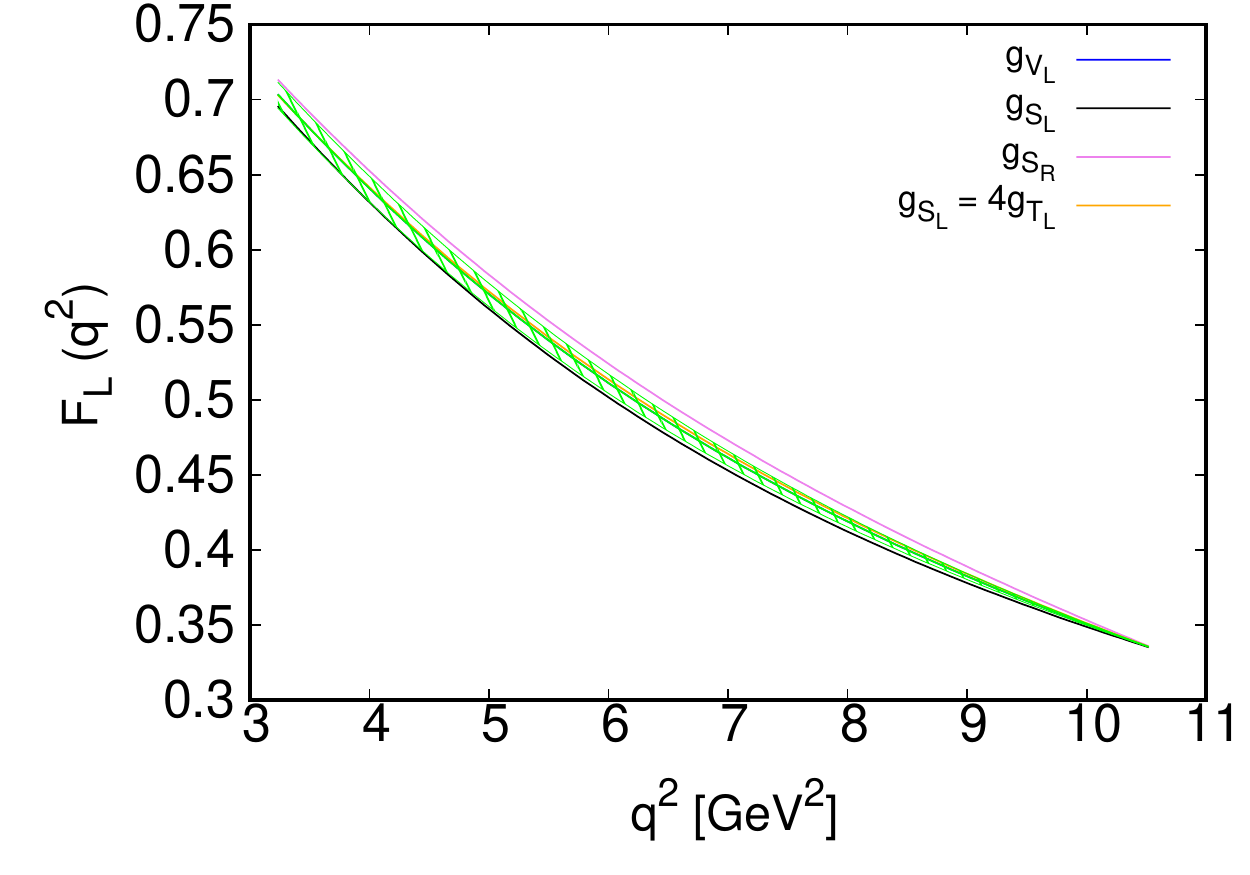}
\includegraphics[width=4cm,height=3cm]{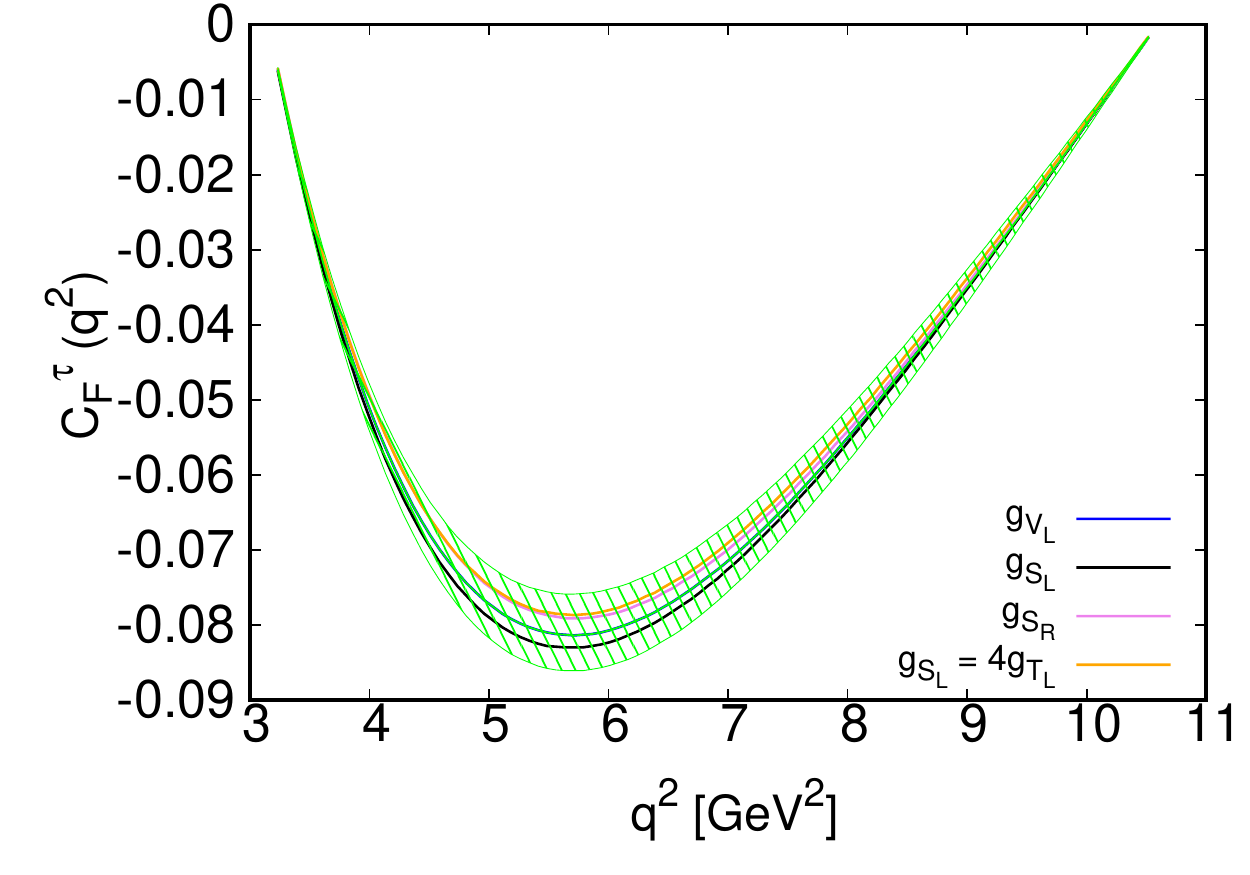}
\caption{The $q^2$ and $\cos\theta_l$ dependence of $B_s\to D_s^*(\to D_s \gamma, D_s\pi )\tau \nu$ decay observables in the SM and in the 
presence of the NP couplings of scenario - I. The SM central line and the corresponding error band are shown with green color. The blue, 
black, violet and orange lines represent the effect of $g_{V_L},\, g_{S_L},\, g_{S_R},\, g_{S_L}=4\,g_{T_L}$ NP couplings, respectively.}
\label{dsg_sc1_same}
\end{figure}

\begin{figure}[htbp]
\centering
\includegraphics[width=4cm,height=3cm]{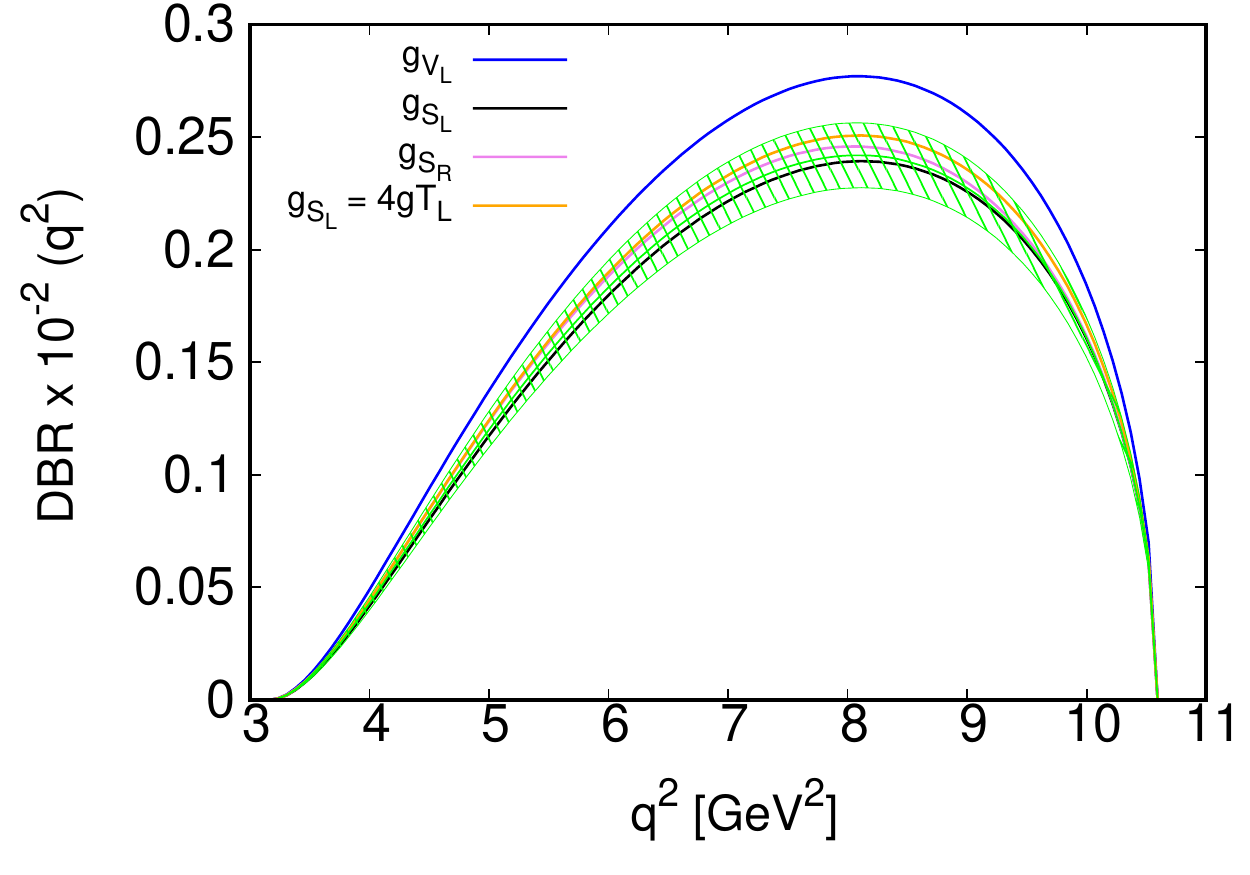}
\includegraphics[width=4cm,height=3cm]{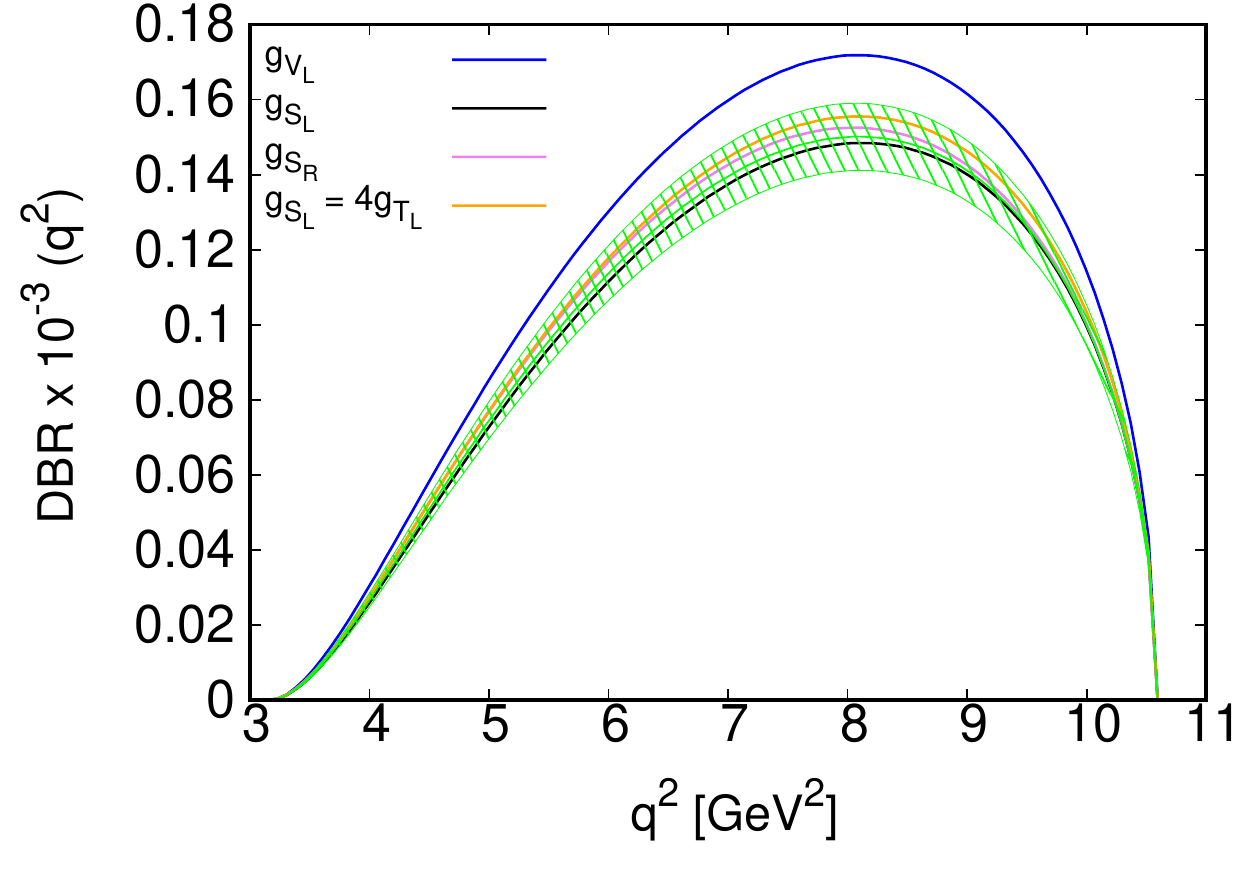}\hspace{1.5cm}
\includegraphics[width=4cm,height=3cm]{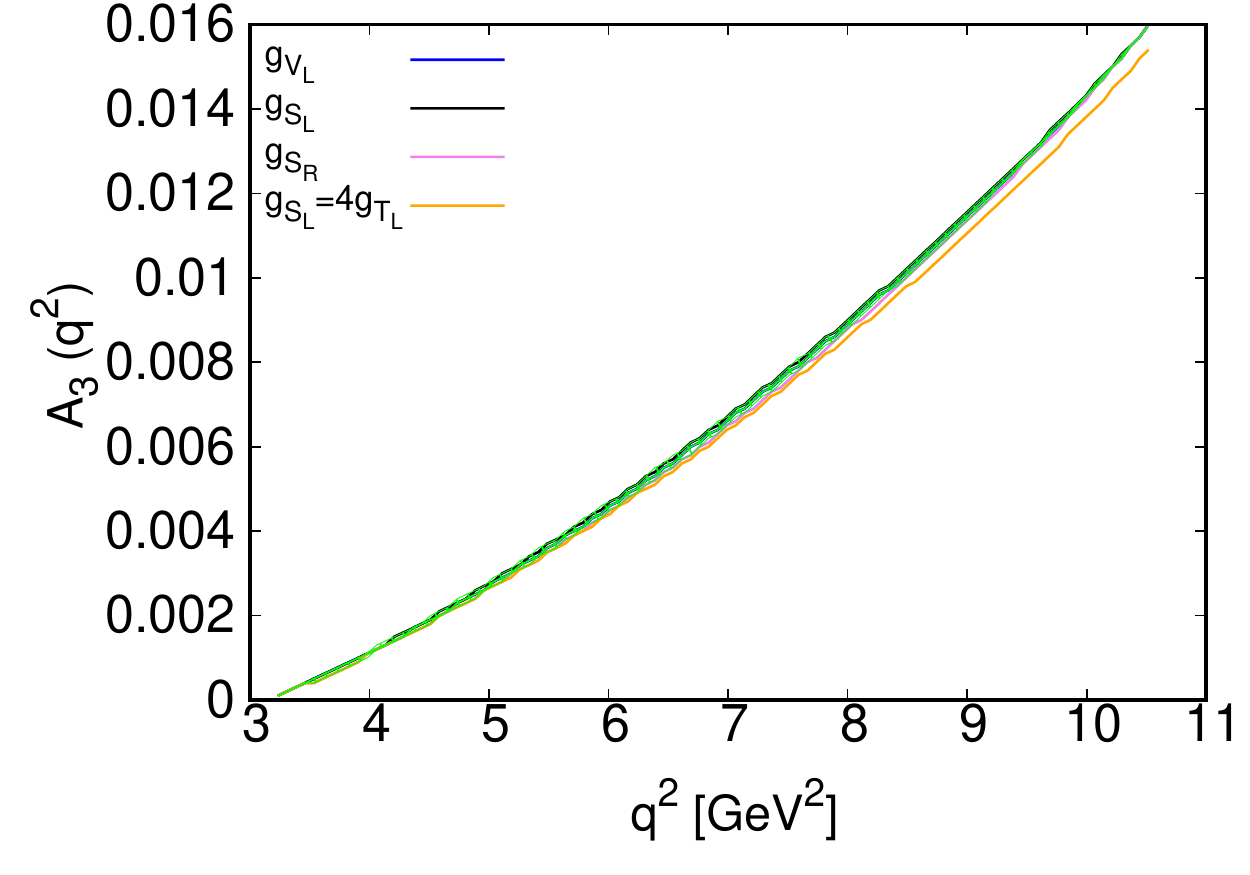}
\includegraphics[width=4cm,height=3cm]{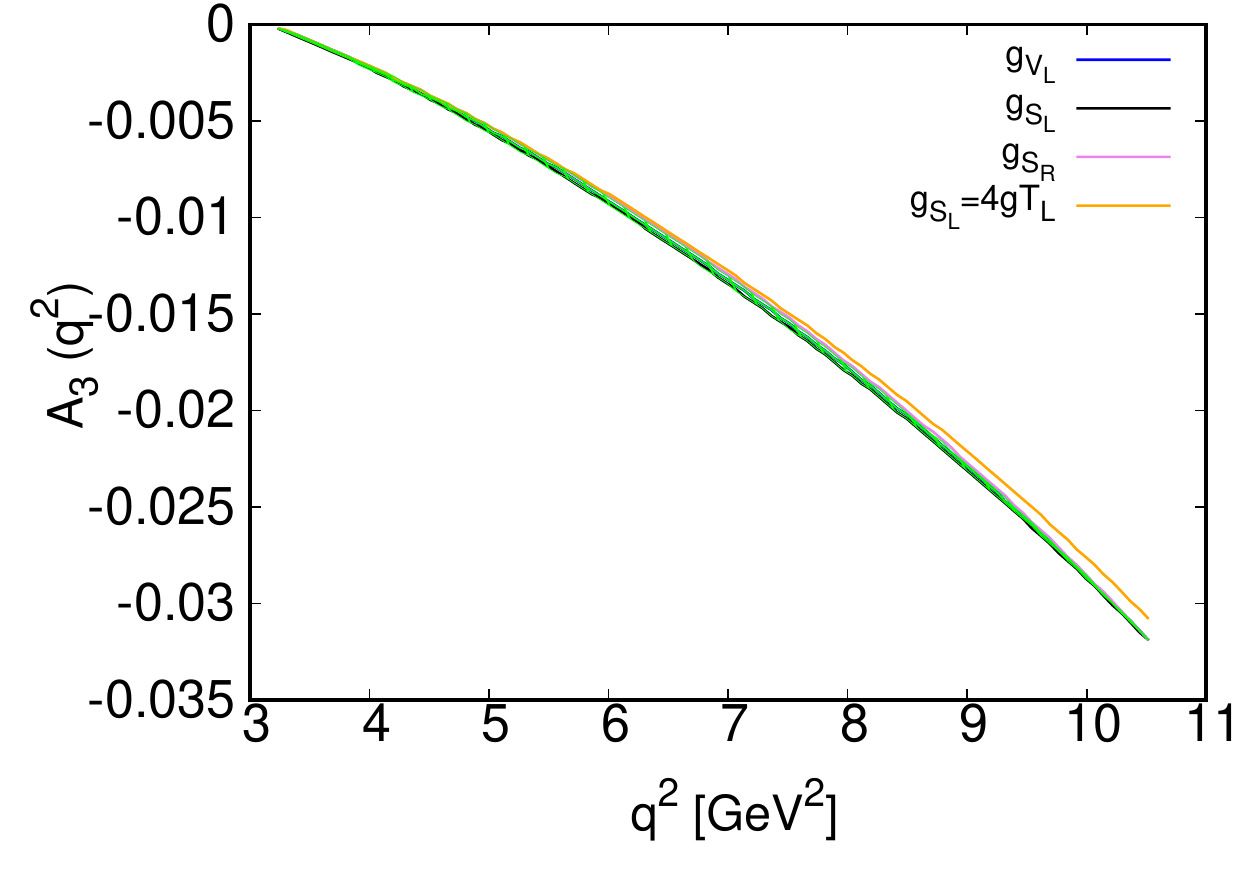}\hspace{1.5cm}
\includegraphics[width=4cm,height=3cm]{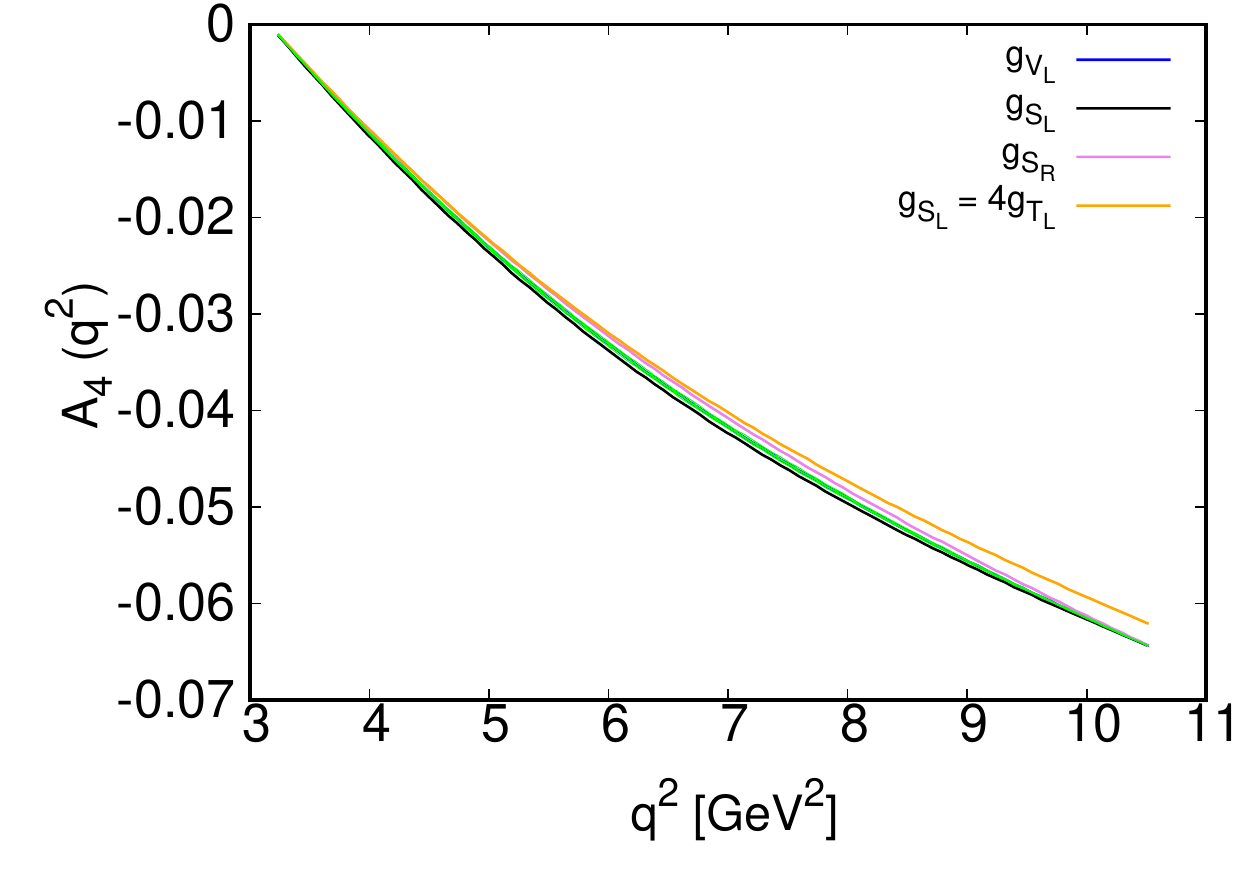}
\includegraphics[width=4cm,height=3cm]{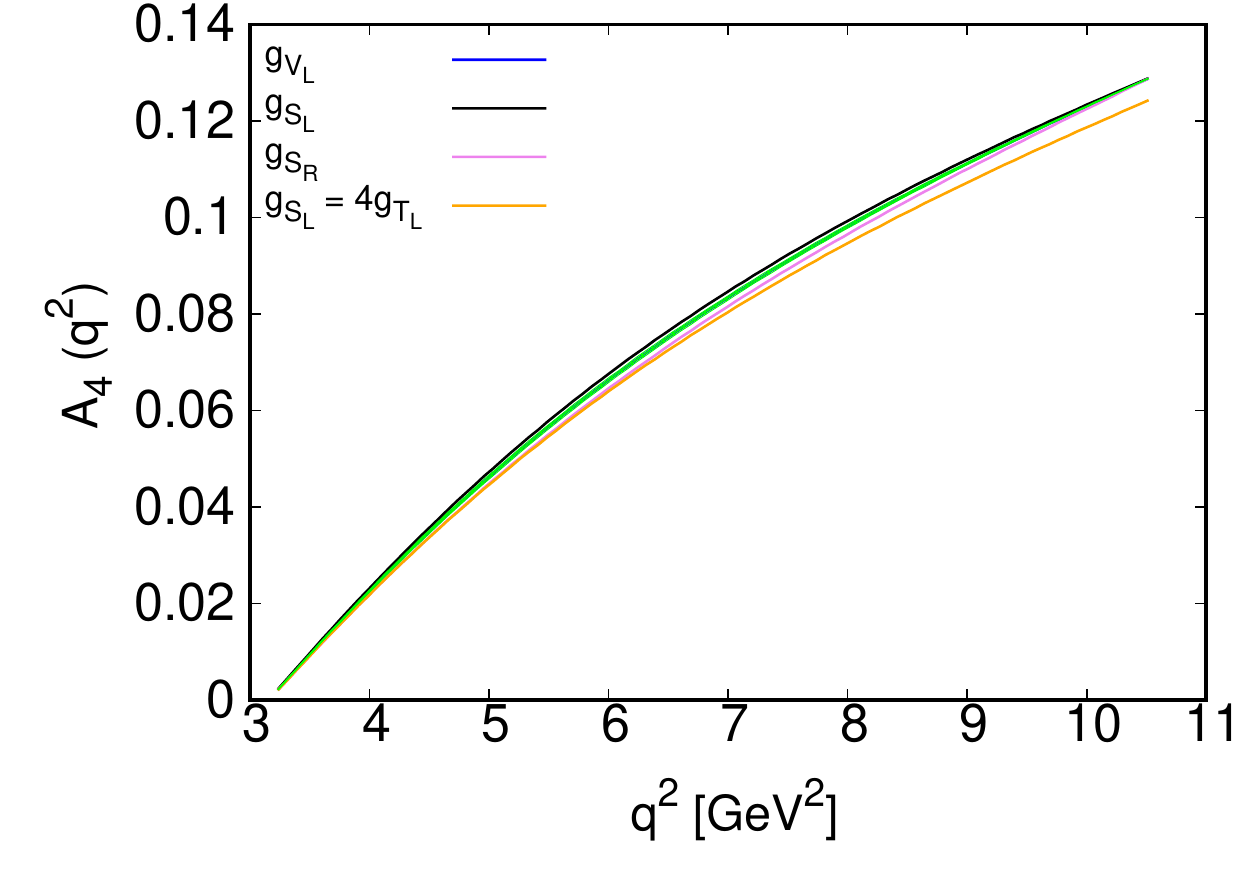}\hspace{1.5cm}
\includegraphics[width=4cm,height=3cm]{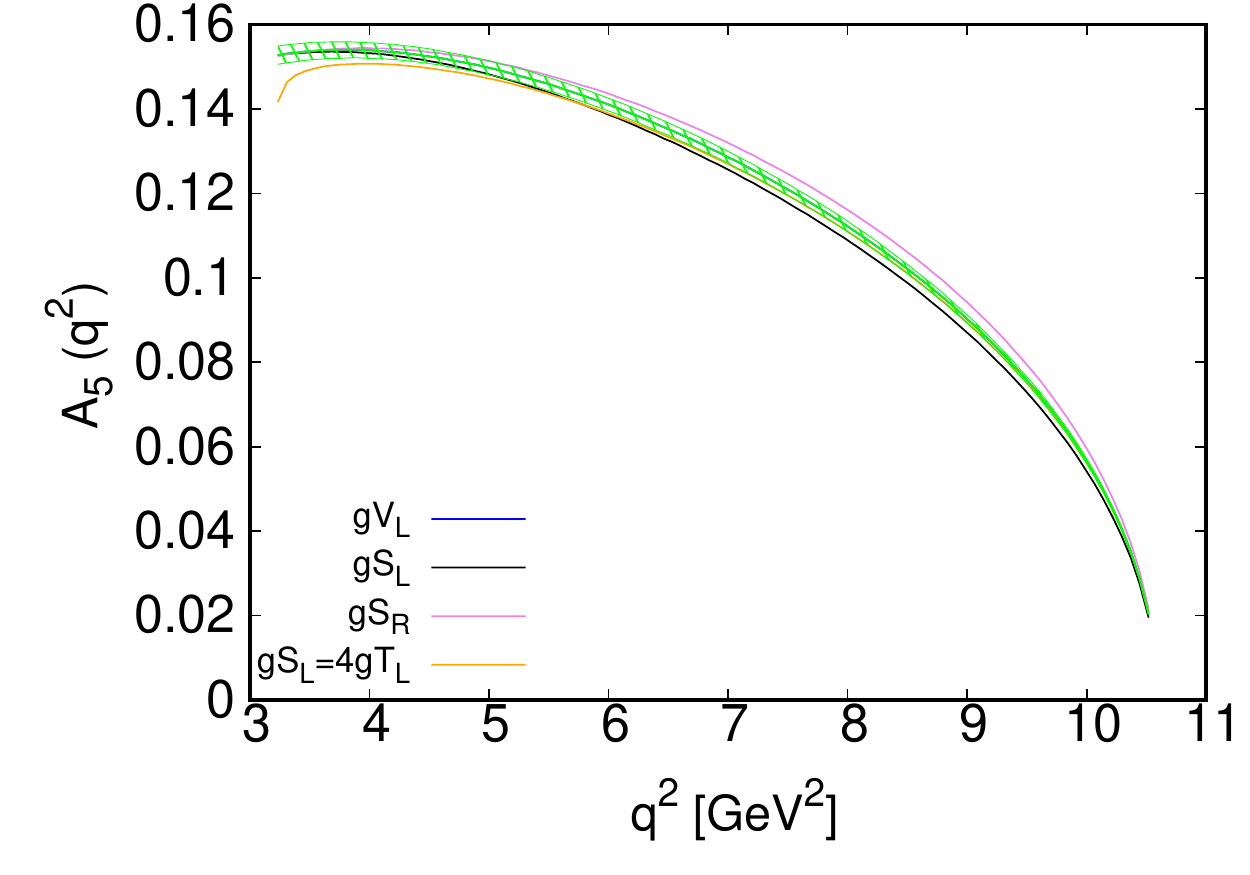}
\includegraphics[width=4cm,height=3cm]{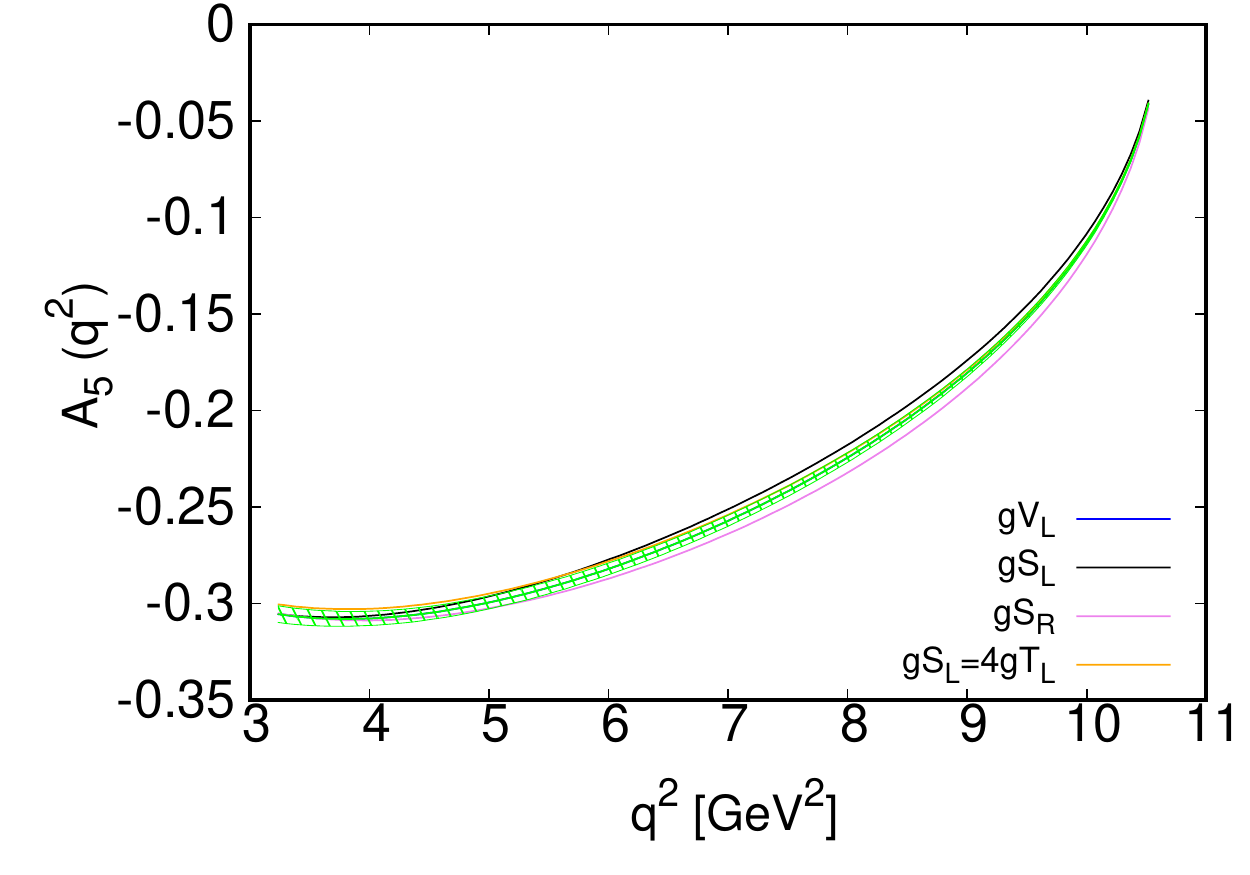}\hspace{1.5cm}
\includegraphics[width=4cm,height=3cm]{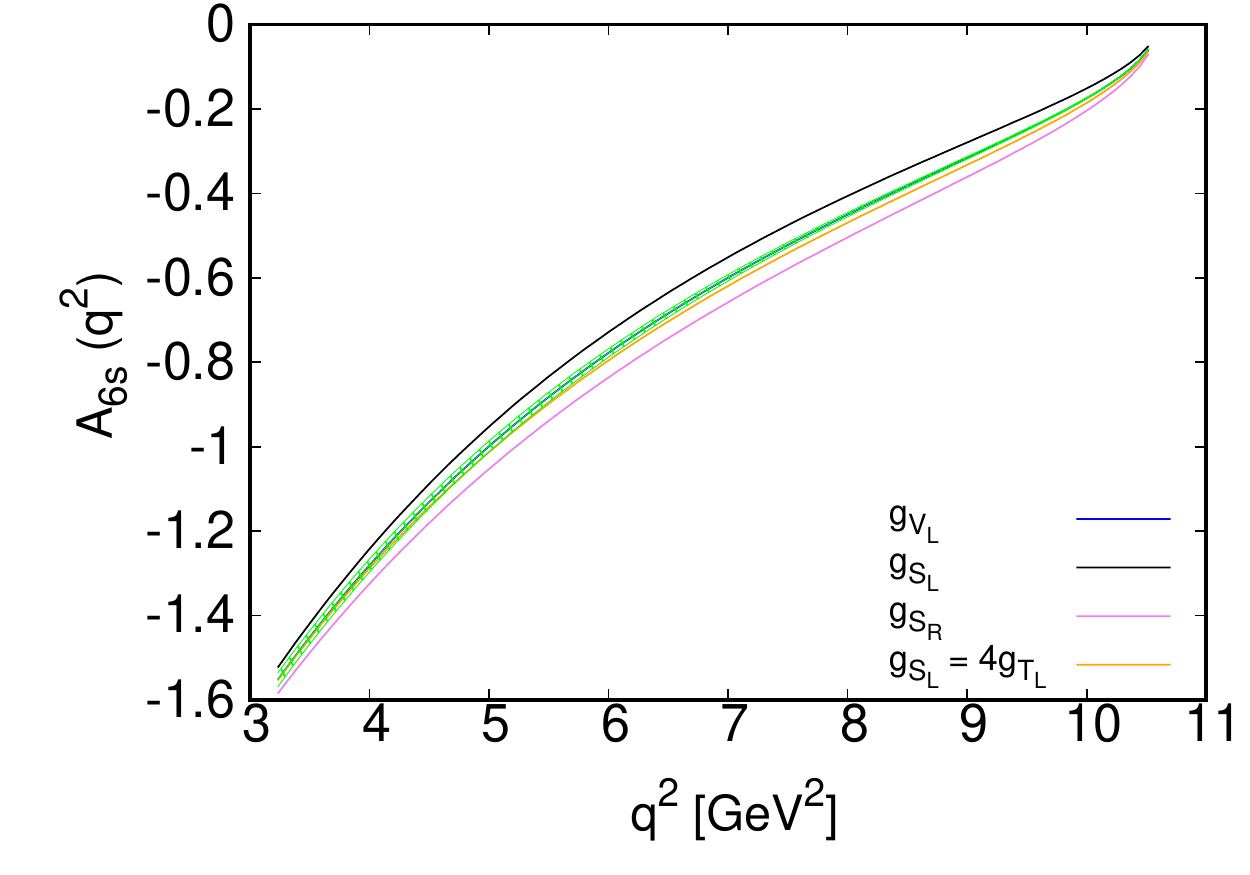}
\includegraphics[width=4cm,height=3cm]{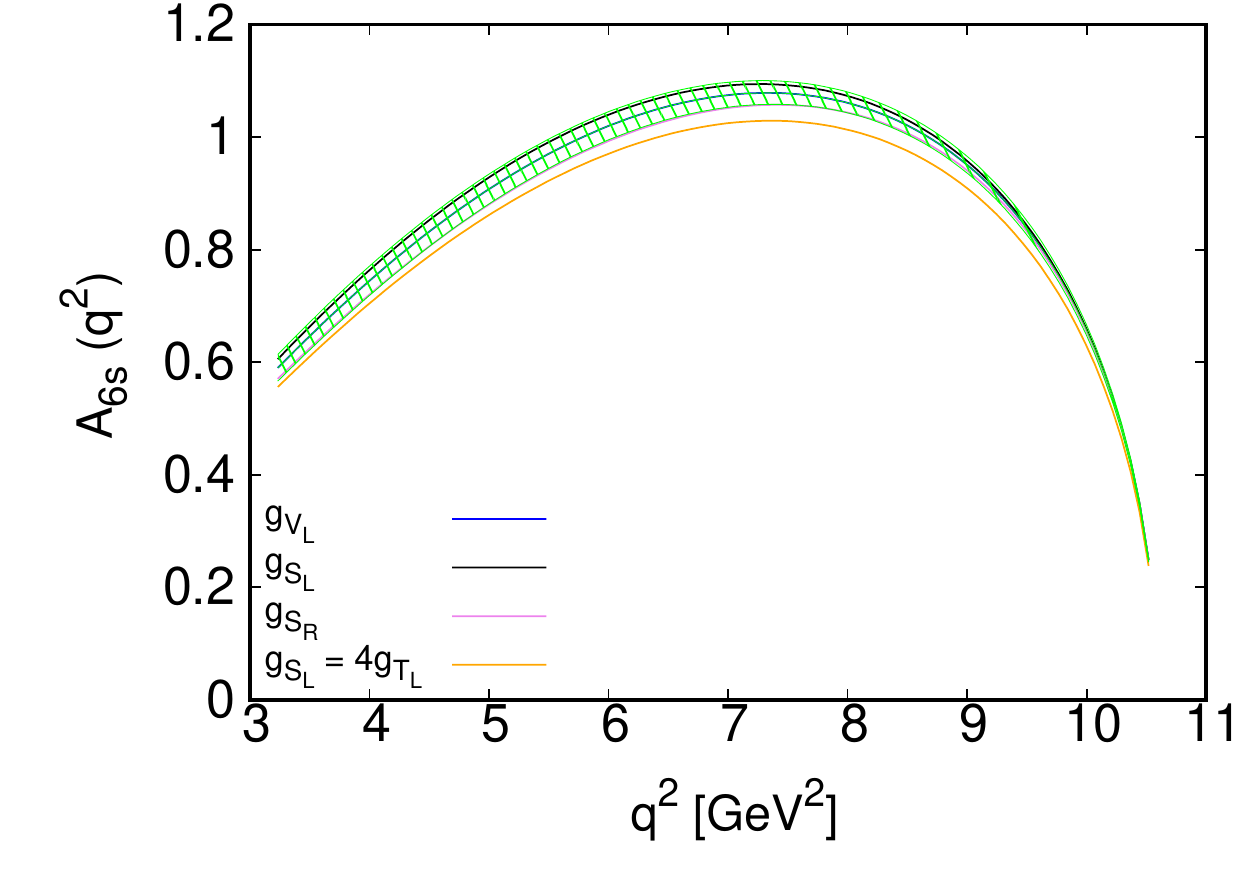}\hspace{1.5cm}
\includegraphics[width=4cm,height=3cm]{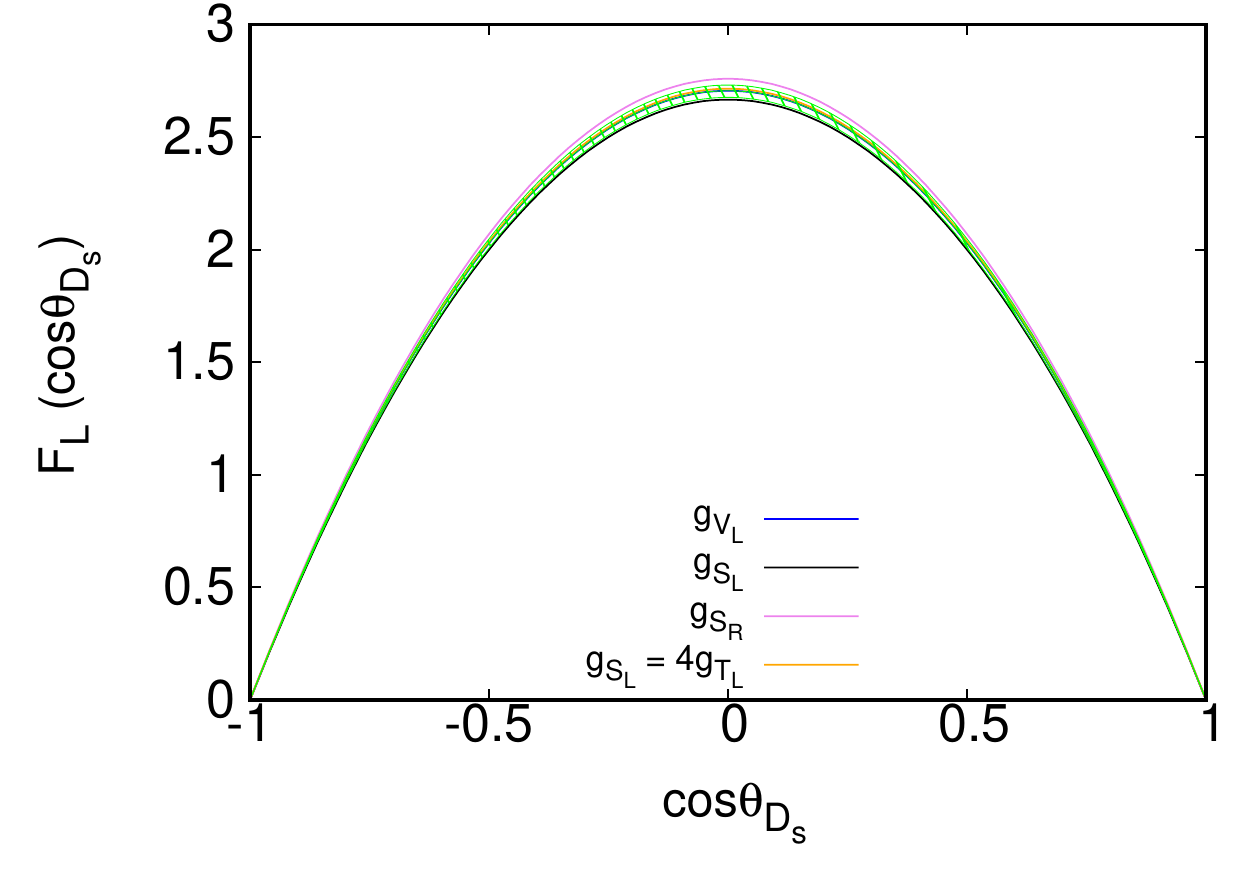}
\includegraphics[width=4cm,height=3cm]{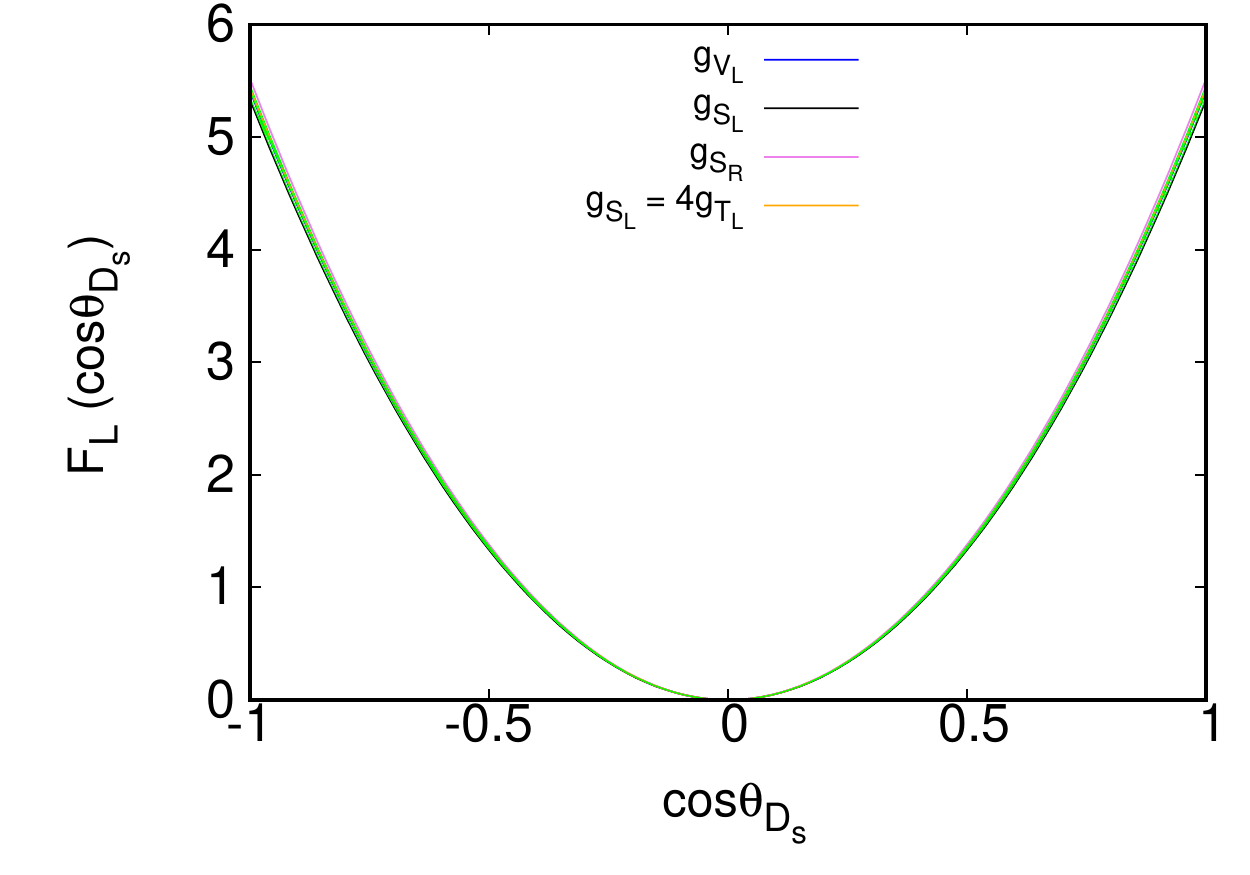}\hspace{1.5cm}
\includegraphics[width=4cm,height=3cm]{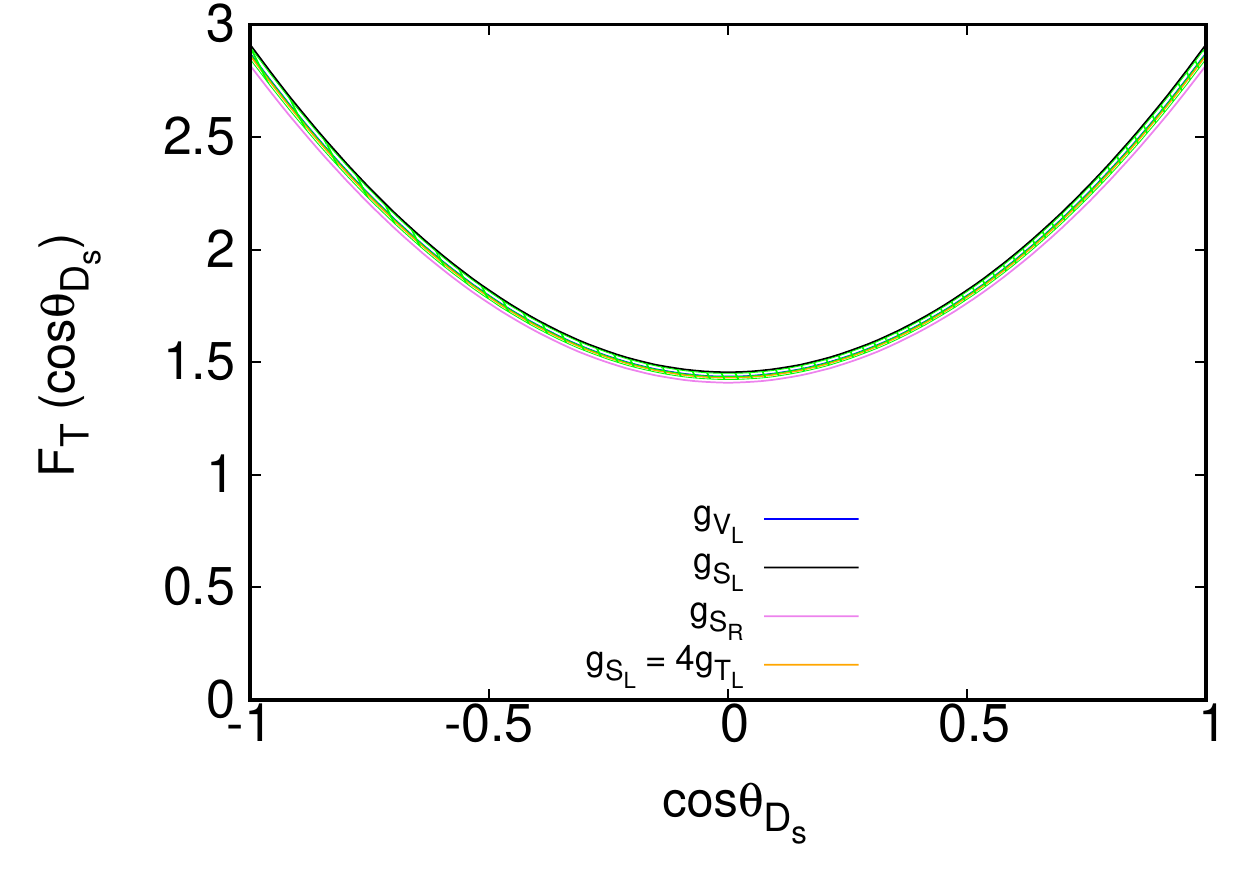}
\includegraphics[width=4cm,height=3cm]{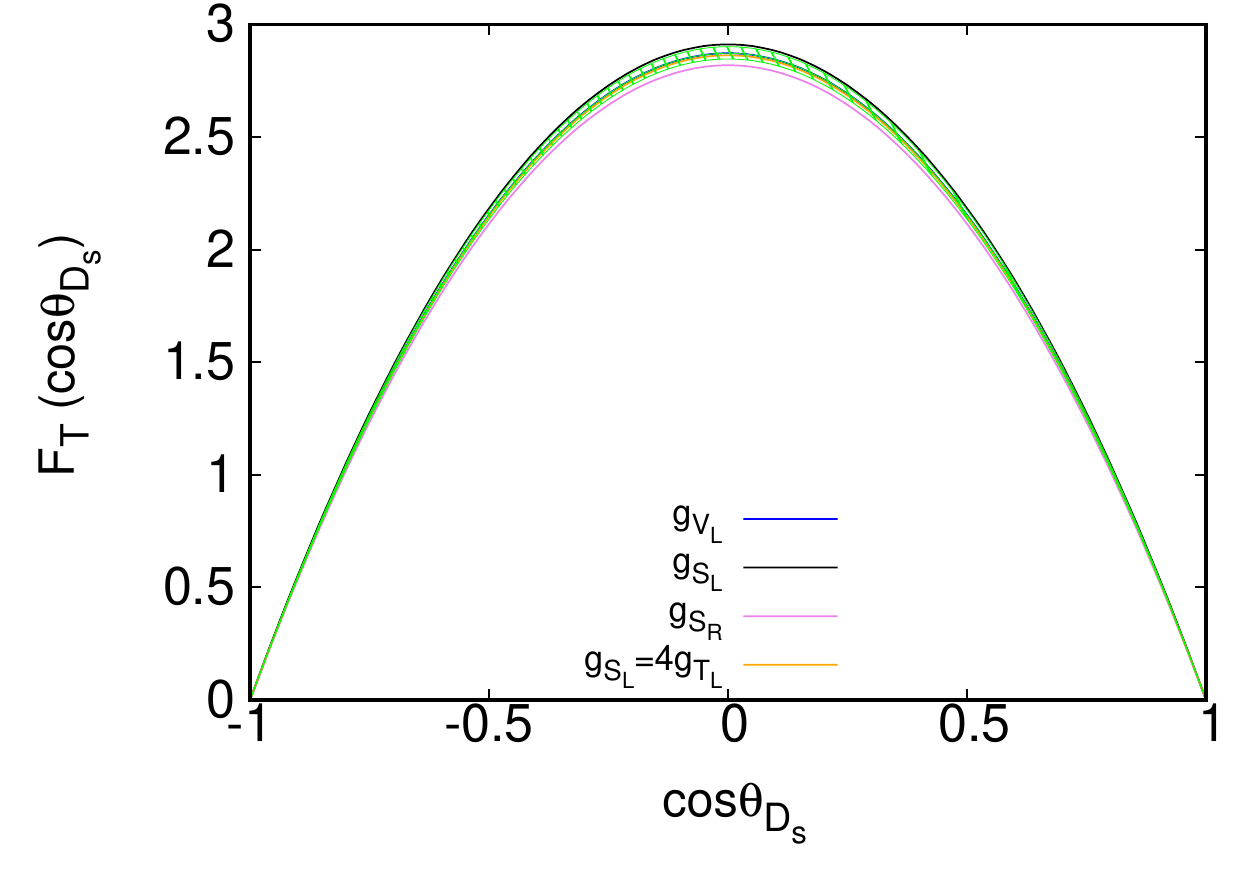}\hspace{1.5cm}
\includegraphics[width=4cm,height=3cm]{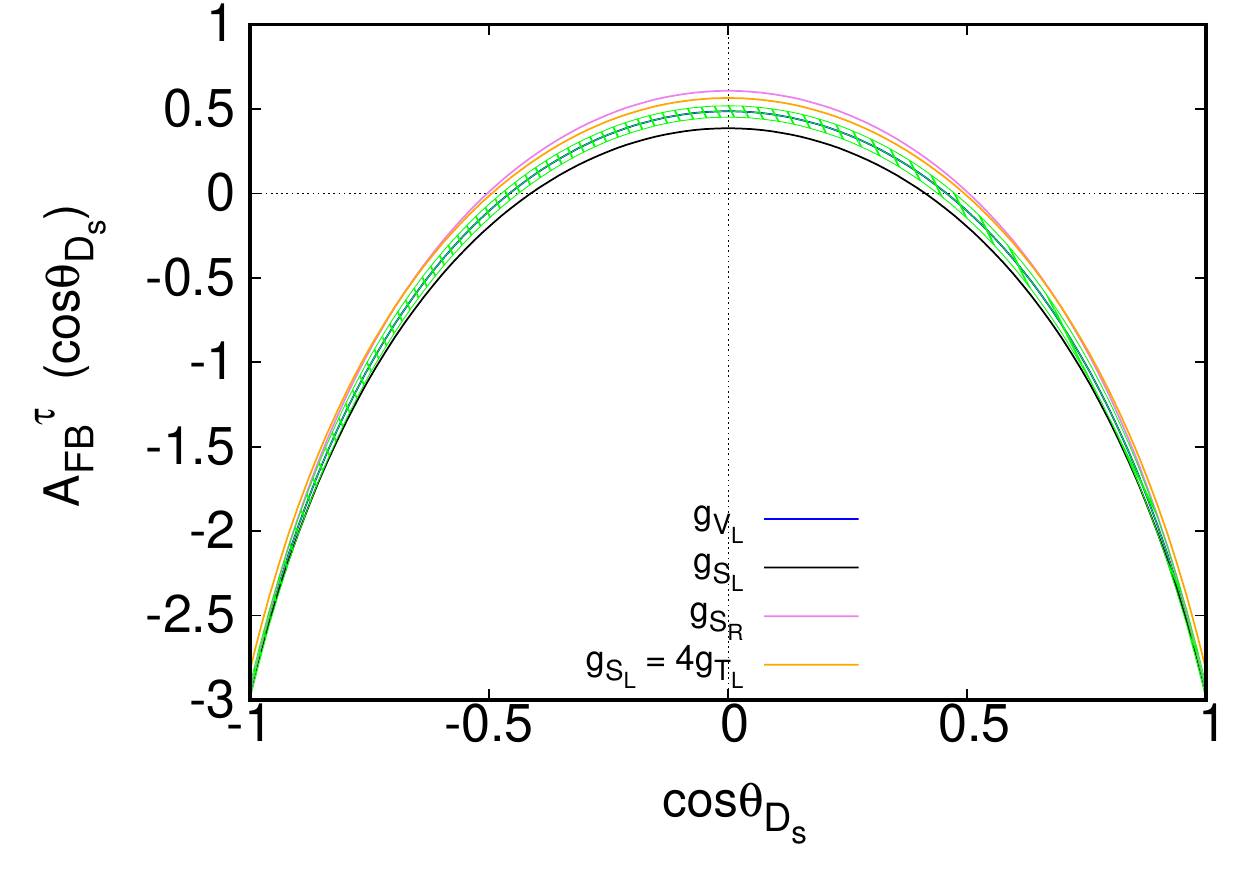}
\includegraphics[width=4cm,height=3cm]{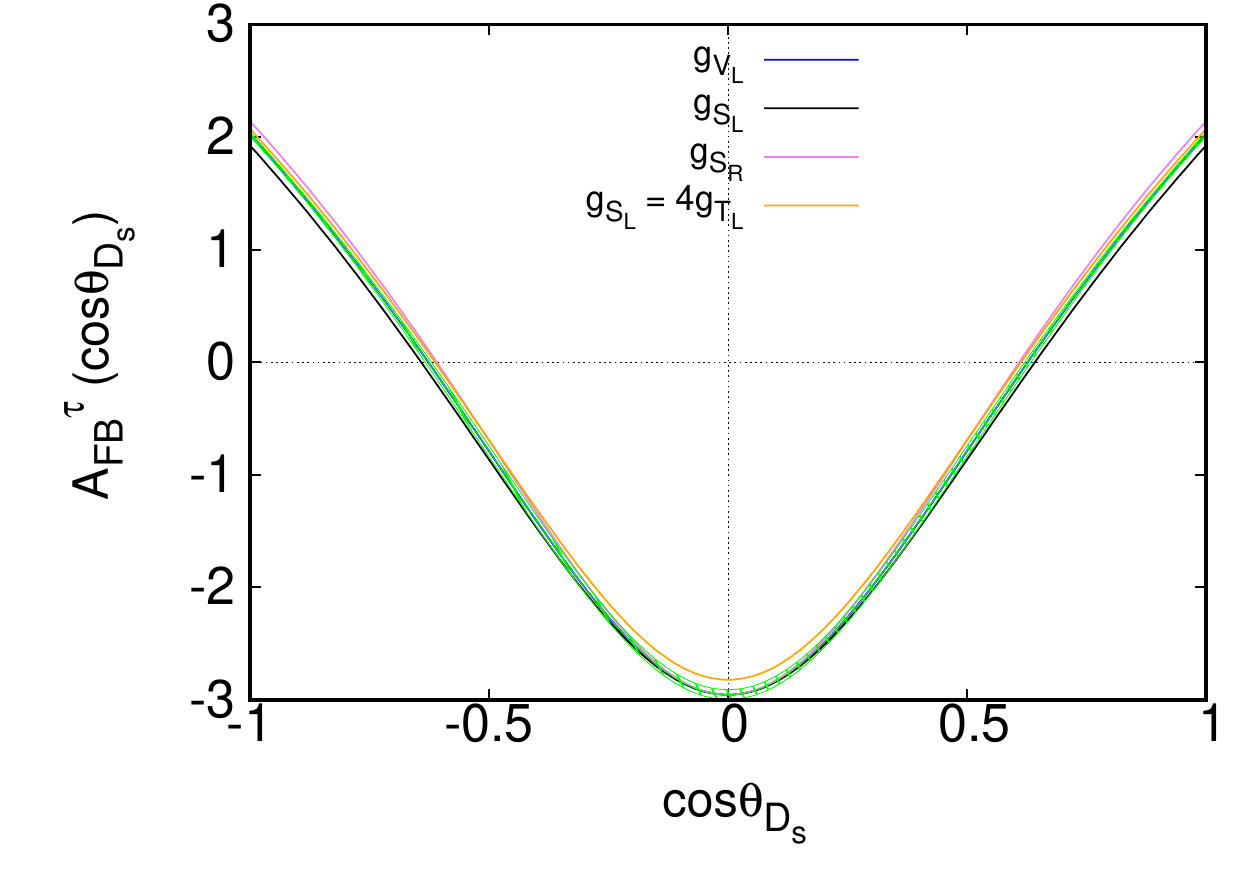}\hspace{1.5cm}
\caption{The $q^2$ and $\cos\theta_{D_s}$ dependence of various physical observable of $B_s\to D_s^*(\to D_s \gamma, D_s\pi )\tau \nu$ in the SM and in the presence of the NP couplings of
scenario - I. The SM central line and the corresponding error band are shown with green color. The  blue, black, violet and orange colors represents the effect of NP coupling $g_{V_L},\,\ g_{S_{L}},\,\ g_{S_R},\,\ g_{S_{L}}=4\,g_{T_L}$  respectively.}
\label{dsg_sc1_diff}
\end{figure}

\begin{table}[htbp]
\centering 
\resizebox{\columnwidth}{!}{
\begin{tabular}{|c|c|c|c|c|c|c|c|c|}
\hline

 &\multicolumn{2}{c|}{$g_{V_L}$}  &\multicolumn{2}{c|}{$g_{S_L}$} & \multicolumn{2}{c|}{$g_{S_R}$} & \multicolumn{2}{c|}{$g_{S_L}=4\,g_{T_L}$} \\
 \cline{2-9}
 & $D_s\gamma$  & $D_s\pi$   & $D_s\gamma$  & $D_s\pi$   & $D_s\gamma$  & $D_s\pi$ & $D_s\gamma$  & $D_s\pi $  \\
\hline
\hline
$BR\times10^{-2}$       & $1.3686 \pm 0.0426$ & $ 0.0849\pm0.0026$  & $1.1798 \pm 0.0367$ & $0.0732 \pm 0.0023$  & $1.2174 \pm 0.0379$ & $0.0755\pm0.0024$ & $1.2382\pm 0.0385$ & $0.0768 \,\pm\,  0.0024 $   \\
\hline
$A_3$                   &  $0.0081 \pm  0.0001$ & $-0.0162 \pm 0.0001$  & $0.0082  \,\pm\,0.0001$& $-0.0164 \pm 0.0001 $    & $  0.0080   \,\pm\, 0.0001$&    $ -0.0159 \,\pm\,0.0001$    & $ 0.0078 \,\pm\,0.0001$ & $ -0.0156\pm\,0.0001$  \\
\hline
$A_4$                   & $-0.0442   \pm  0.0001$  & $ 0.0883 \pm 0.0001$  &  $-0.0448  \pm   0.0001$  & $ 0.0895  \pm 0.0001$   & $-0.0434   \pm 0.0001 $  & $ 0.0867 \,\pm\,0.0001 $  & $ -0.0426  \,\pm\, 0.0001$  &   $ 0.0852  \pm  0.0001 $  \\
\hline
$A_5 $                  & $0.1133  \pm 0.0005$ &  $-0.2265 \pm 0.0010 $  & $ 0.1104  \pm   0.0005$ & $ -0.2208  \pm 0.0010$ & $ 0.1166 \pm  0.0005$ & $ - 0.2333  \,\pm\, 0.0010$    &  $0.1119  \pm 0.0005 $ & $-0.2238 \pm  0.0010$  \\
\hline
$A_{6s}$                & $-0.5509   \pm  0.0026$ & $   0.9539 \,\pm\,0.0077 $   & $-0.5076   \pm   0.0025 $ & $0.9665   \pm  0.0078$  &$ -0.6033   \pm 0.0028$&   $ 0.9366 \,\pm\,0.0076$    &  $-0.5673 \pm 0.0027 $& $0.9098   \pm 0.0075 $ \\
\hline
$R_{D_s^*}$         & \multicolumn{2}{c|}{$0.2782  \pm 0.0018$}    & \multicolumn{2}{c|}{$ 0.2398  \pm  0.0015 $}     & \multicolumn{2}{c|}{$ 0.2475   \pm 0.0016$} & \multicolumn{2}{c|}{$0.2517\pm0.0016$}\\
\hline
$A_{FB}^{\tau}$         & \multicolumn{2}{c|}{$-0.0896  \pm  0.0020$} &  \multicolumn{2}{c|}{$-0.1020\pm 0.0020 $}& \multicolumn{2}{c|}{$-0.0741   \pm 0.0021 $}&  \multicolumn{2}{c|}{ $ -0.0761  \,\pm\, 0.0020 $}    \\
\hline
$A_{FB}^{T}$            & \multicolumn{2}{c|}{$-0.3842   \pm  0.0026$}    & \multicolumn{2}{c|}{$-0.3842 \pm 0.0026 $}   & \multicolumn{2}{c|}{$-0.3842  \pm  0.0026 $}   & \multicolumn{2}{c|}{$-0.3677   \,\pm\,0.0025$}     \\
\hline
$F_L$           & \multicolumn{2}{c|}{$0.4482    \pm 0.0015 $}    & \multicolumn{2}{c|}{$ 0.4409  \pm   0.0015 $}    & \multicolumn{2}{c|}{$ 0.4582   \pm 0.0015 $} &  \multicolumn{2}{c|}{$ 0.4501\,\pm\,0.0015$}   \\
\hline
$C_F^l$                 &\multicolumn{2}{c|}{$-0.0550    \pm 0.0014$}  & \multicolumn{2}{c|}{$ -0.0557  \pm 0.0014 $}    &\multicolumn{2}{c|}{$ -0.0540  \pm  0.0014$ }   &  \multicolumn{2}{c|}{$-0.0531 \,\pm\,0.0013$}    \\
\hline
\end{tabular}}
 \caption{ Prediction of $B_s\to D_s^*(\to D_s\gamma,D_s\pi)\,\tau\,\nu$ decay observables in Scenerio - I.}
\label{sc1_dsg_int}
 \end{table}
In Fig~\ref{dsg_sc1_same} we display the $q^2$ and $\cos\theta_l$  dependence of several physical observables that exhibit same 
behaviour for the $D_s\gamma$ and $D_s\pi$ modes. The contribution coming from $g_{V_L},\, g_{S_L},\, g_{S_R},\, g_{S_L}=4\,g_{T_L}$
NP couplings are represented by blue, black, violet and orange lines, respectively. Our observations are as follows.
\begin{itemize}
\item In case of $F_L(\cos\theta_l)$, a slight deviation from SM expectation is observed at $\cos\theta_l \ge 0.5$ with $g_{S_L}$ and
$g_{S_R}$  NP couplings and they are distinguishable from the SM prediction at slightly more than $1\sigma$
significance. However, for $F_T(\cos\theta_l)$, no such deviation is observed and they all lie within the SM error band.

\item In case of $R_{D_s^*}(q^2)$, maximum deviation is observed in case of $g_{V_L}$ NP coupling and it is clearly distinguishable from the 
SM prediction at more than $3\sigma$ significance at high $q^2$ value.

\item The zero crossing in $A_{FB}^{\tau}(q^2)$ is shifted to lower value of $q^2$ than in the SM with $g_{S_L}$ NP coupling,
whereas, it is found to be shifted to higher value of $q^2$ with $g_{S_R}$ and $g_{S_L}=4\,g_{T_L}$ NP couplings. The zero crossings in 
$A_{FB}^{\tau}$ at $q^2 = 5.06 \,{\rm GeV^2}$, $q^2=5.48 \, {\rm GeV^2}$ and $q^2=5.43\, {\rm GeV^2}$ in the presence of 
$g_{S_L}$, $g_{S_R}$ and $g_{S_L}=4\,g_{T_L}$ NP couplings 
are clearly distinguishable from the SM prediction of $q^2=5.25\pm 0.12\,{ \rm GeV^2}$ at the level of $1.9\sigma$ and $2.3\sigma$
and $1.7\sigma$ significance.

\item At low $q^2$ range, $A_{FB}^{T}(q^2)$ is deviates from the SM predication in the presence of $g_{S_L}=4\,g_{T_L}$ NP coupling. 
In case of  $F_L^{D_s^*}(q^2)$ and $C_F^{\tau}(q^2)$ observables, no significant deviation is observed and they all 
lie within the SM error band. 
 
\end{itemize}
In Fig~\ref{dsg_sc1_diff} we display the $q^2$ and $\cos\theta_{D_s}$ dependence of several physical observables that exhibits different
behaviour for the $D_s\gamma$ and $D_s\pi$ decay modes. Our observations are as follows. 
\begin{itemize}
\item In case of differential branching ratio $DBR\,(q^2)$, the deviation from the SM prediction is more pronounced with $g_{V_L}$ NP coupling
and the peak of the distribution is clearly distinguishable from the SM prediction at the level of $2\sigma$ significance. No such significant
deviation is observed with the rest of the NP couplings and they all lie within the SM error band.

\item The angular observable $A_3$, $A_4$ and $A_5$ are slightly deviated from the SM in the presence of 
$g_{S_{L}}=4g_{T_L}$ NP coupling. Similarly in case of $A_{6s}$, a slight deviation is observed with  $g_{S_{L}}$, $g_{S_{R}}$ and $g_{S_{L}}=4g_{T_L}$ NP coupling for the $D_s\gamma$ mode, whereas, $A_{6s}$ shows slight deviation in the presence of $g_{S_{L}}=4g_{T_L}$ for the $D_s\pi$ mode.

\item The observables $F_L(\cos{\theta_{D_s}})$ and $F_T(\cos{\theta_{D_s}})$ do not show any significant deviation from the SM prediction in
the presence of the NP couplings of scenario - I. 

\item The deviation from the SM prediction observed in case of $A_{FB}^{\tau} \,(\cos\theta_{D_s})$ is more pronounced with $g_{S_L}$, 
$g_{S_R}$ and $g_{S_L}=4\,g_{T_L}$ NP couplings for the $D_s\gamma$ mode. The zero crossing in $A_{FB}^{\tau} \,(\cos\theta_{D_s})$ 
is shifted to $\cos\theta_{D_s}=0.412$, $0.500$ and $0.497$ in the presence of $g_{S_L}$, $g_{S_R}$ and $g_{S_L}=4\,g_{T_L}$ NP
couplings and they are clearly distinguishable from the SM zero crossing of $\cos\theta_{D_s}=\pm 0.456\pm 0.018$ at the level of more than 
$2\sigma$ significance.
Similarly for the $D_s\pi$ mode, $A_{FB}^{\tau} \,(\cos\theta_{D_s})$ shows slight deviation in the presence of $g_{S_L}$, $g_{S_R}$ and 
$g_{S_L}=4\,g_{T_L}$ NP couplings. The zero crossings in $A_{FB}^{\tau} \,(\cos\theta_{D_s})$ observed at 
$\cos\theta_{D_s}=\pm 0.642$, $\pm 0.610$ and $\pm 0.613$ in the presence of $g_{S_L}$, $g_{S_R}$ and $g_{S_L}=4\,g_{T_L}$ NP
couplings are distinguishable from the SM zero crossing of $\cos\theta_{D_s}= \pm 0.626 \pm 0.007$ at the level of $1-2\sigma$ significance.

\end{itemize}

\subsubsection{(Scenerio - II)}
In scenerio-II, we choose four 2D NP hypothesis such as ($g_{V_L}$, $g_{S_{L}}\,=\,-4g_{T_L}$), ($g_{S_R}$, $g_{S_{L}}$)(Set A or Set B), 
($g_{V_L}$, $g_{S_R}$) and ($g_{S_{L}}\,=\,4g_{T_L}$). The best fit values of these NP couplings at $\mu=1\rm TeV$ scale obtained from
Ref.~\cite{Blanke:2018yud,Blanke:2019qrx} are mentioned in the Table~\ref{global_fit}. In our analysis, we run them down to the 
renormalization scale of $\mu = m_b$. In Table~\ref{dsg_sc2}, we report the central values and the corresponding $1\sigma$ range
of several physical observables for both $B_s\to (D_s\gamma)\,\tau\,\nu$ and $B_s\to (D_s\pi)\,\tau\,\nu$ decays in the presence of each $2D$
NP couplings. 

The deviation from the SM prediction observed for $BR$ is more pronounced in the presence of $(g_{V_L}$, $g_{S_{L}}\,=\,-4g_{T_L})$ NP 
coupling and it is clearly distinguishable from the SM prediction at more than $3\sigma$ significance. Similarly, a deviation of around
$2-3\sigma$ is observed with $(g_{S_R},\, g_{S_{L}})$~(setA or Set B) and $(g_{V_L},\,g_{S_R})$ NP couplings.
Significant deviation from the SM prediction is observed for $R_{D_s^*}$ with $(g_{V_L}$, $g_{S_{L}}\,=\,-4g_{T_L})$, $(g_{V_L},\,g_{S_R})$ 
and $(g_{S_R},\, g_{S_{L}})$~(set A or Set B) NP couplings. In case of $F_L^{D_s^*}$, the deviation is more pronounced with 
$(g_{S_R},\, g_{S_{L}})$~(set A or Set B) NP couplings.
The angular observable $A_7$ is found to be non zero in the presence of $(g_{S_{L}}=4g_{T_L})$ complex NP couplings for both $D_s\gamma$ and 
$D_s\pi$ modes. The angular observables $A_8$ and $A_9$ are absent in this scenario-II and hence we do not report them in Table-~\ref{dsg_sc2}.

\begin{figure}[htbp]
\centering
\includegraphics[width=4cm,height=3cm]{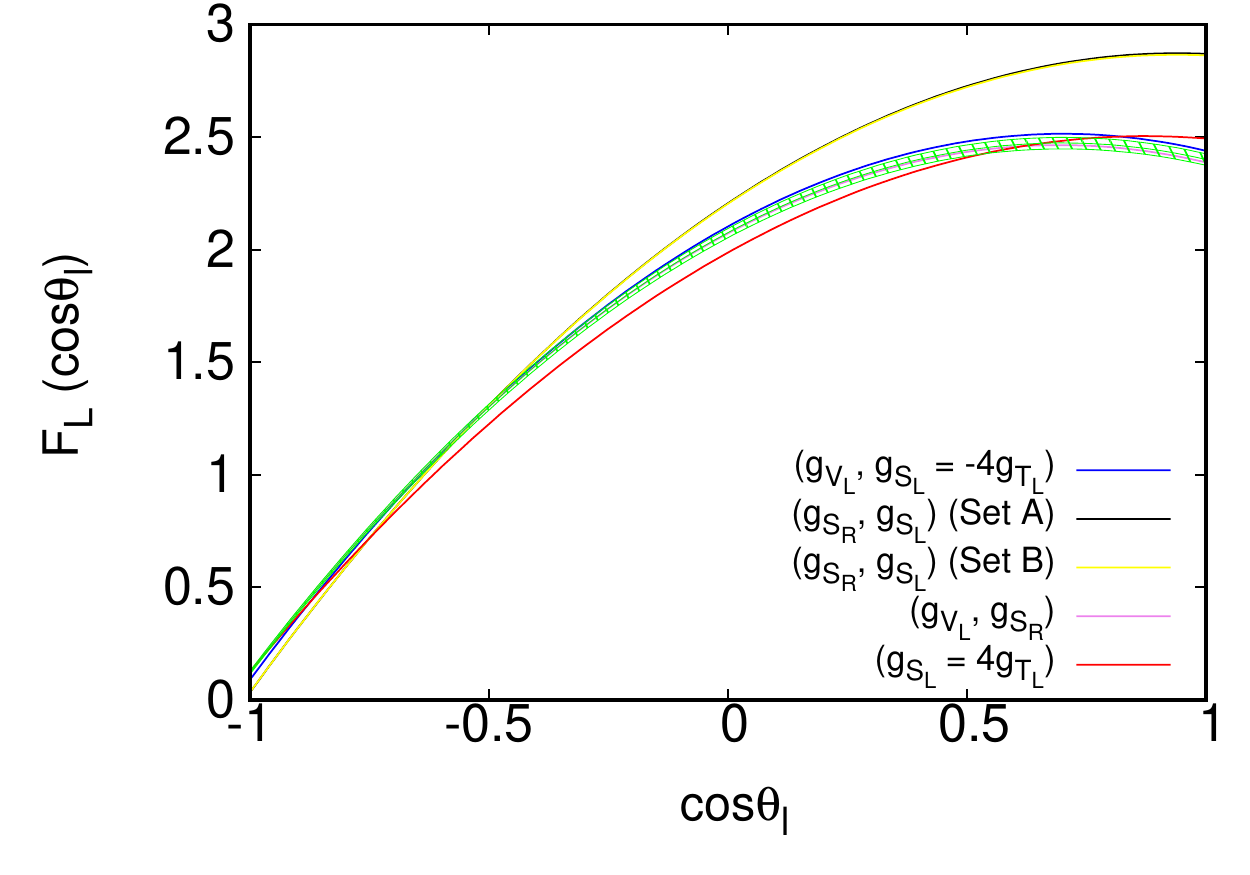}
\includegraphics[width=4cm,height=3cm]{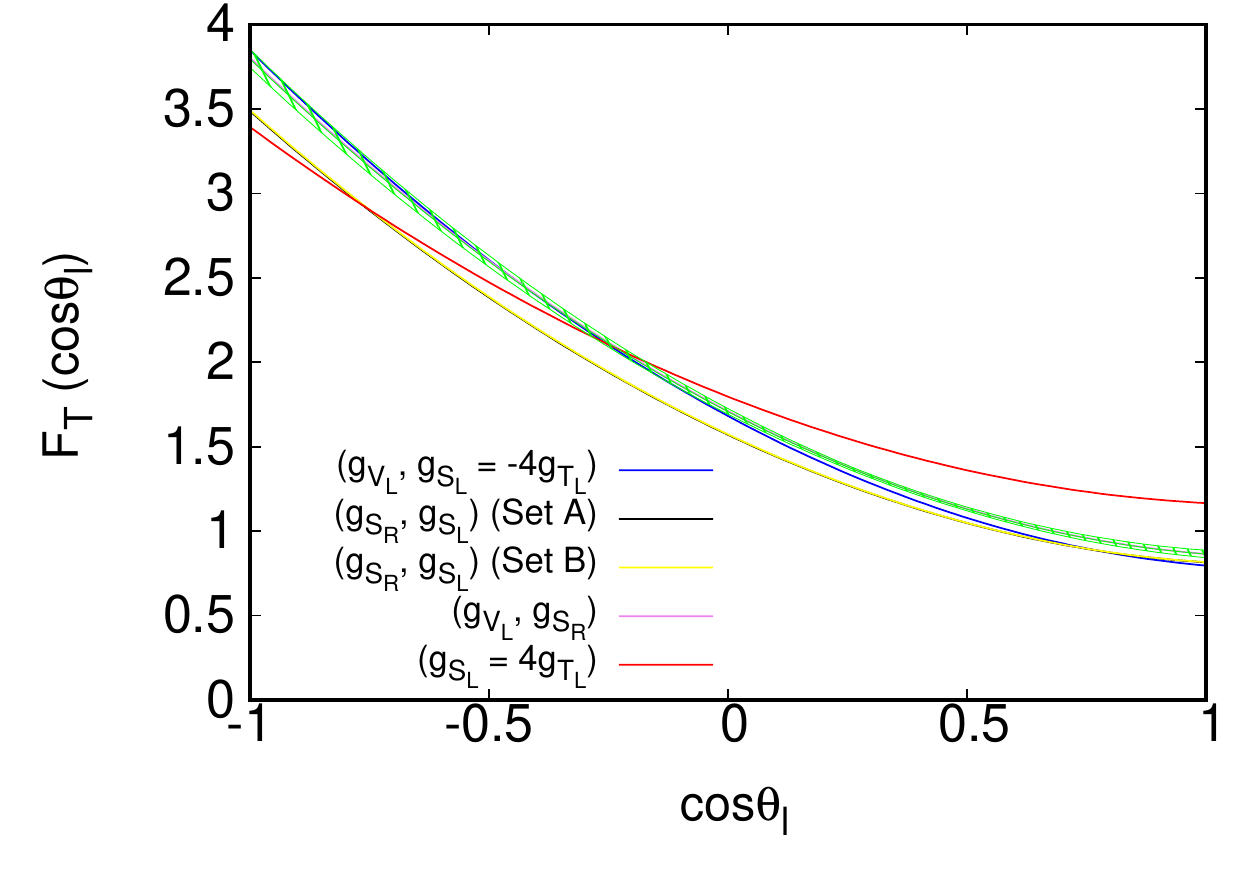}
\includegraphics[width=4cm,height=3cm]{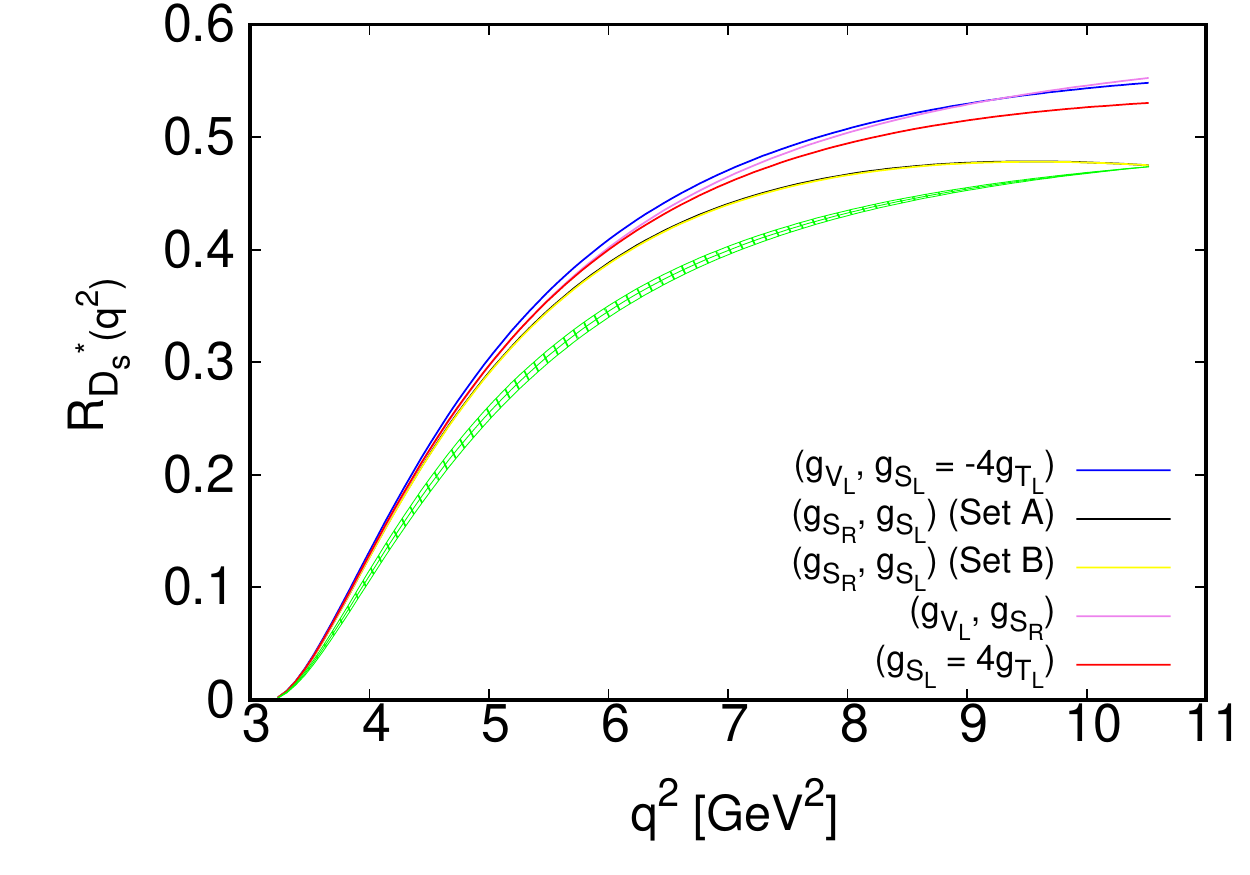}
\includegraphics[width=4cm,height=3cm]{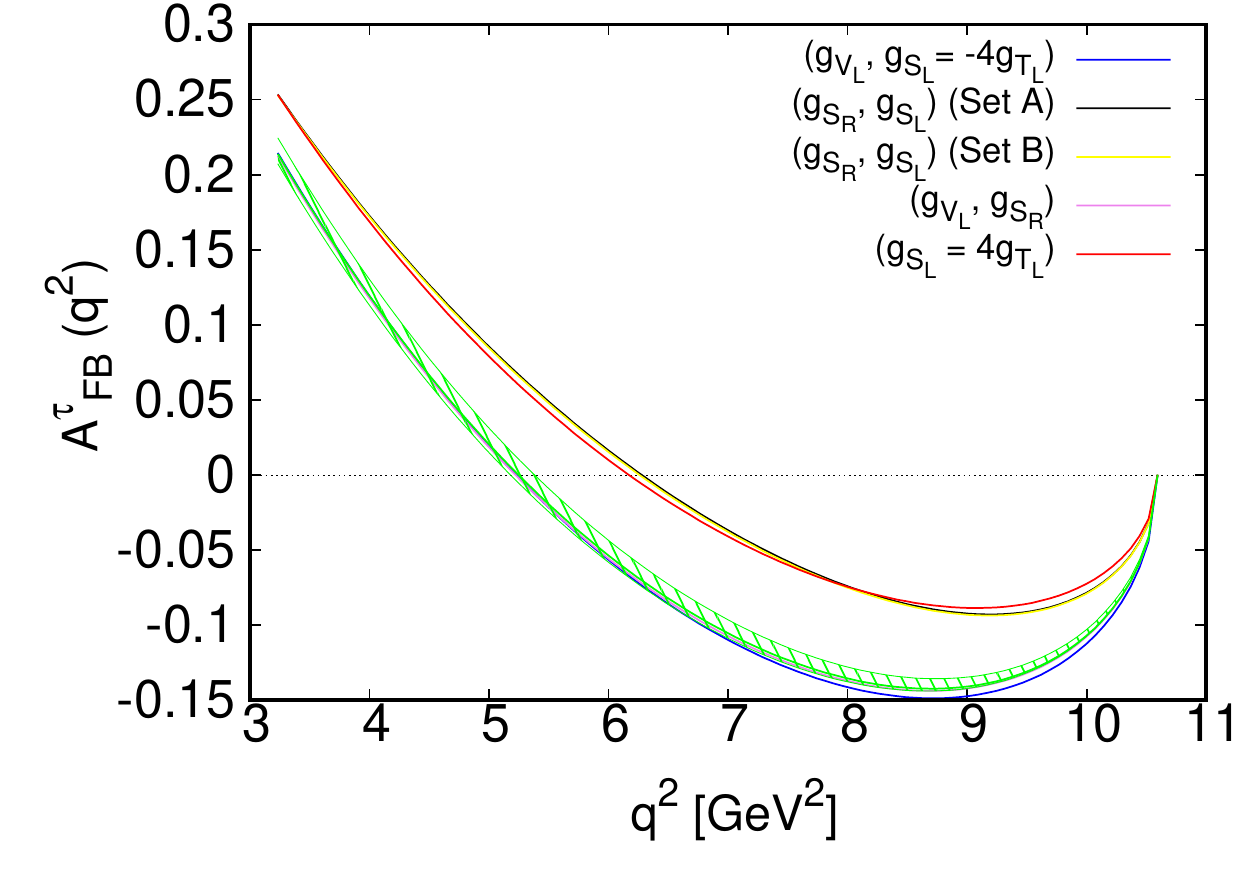}
\includegraphics[width=4cm,height=3cm]{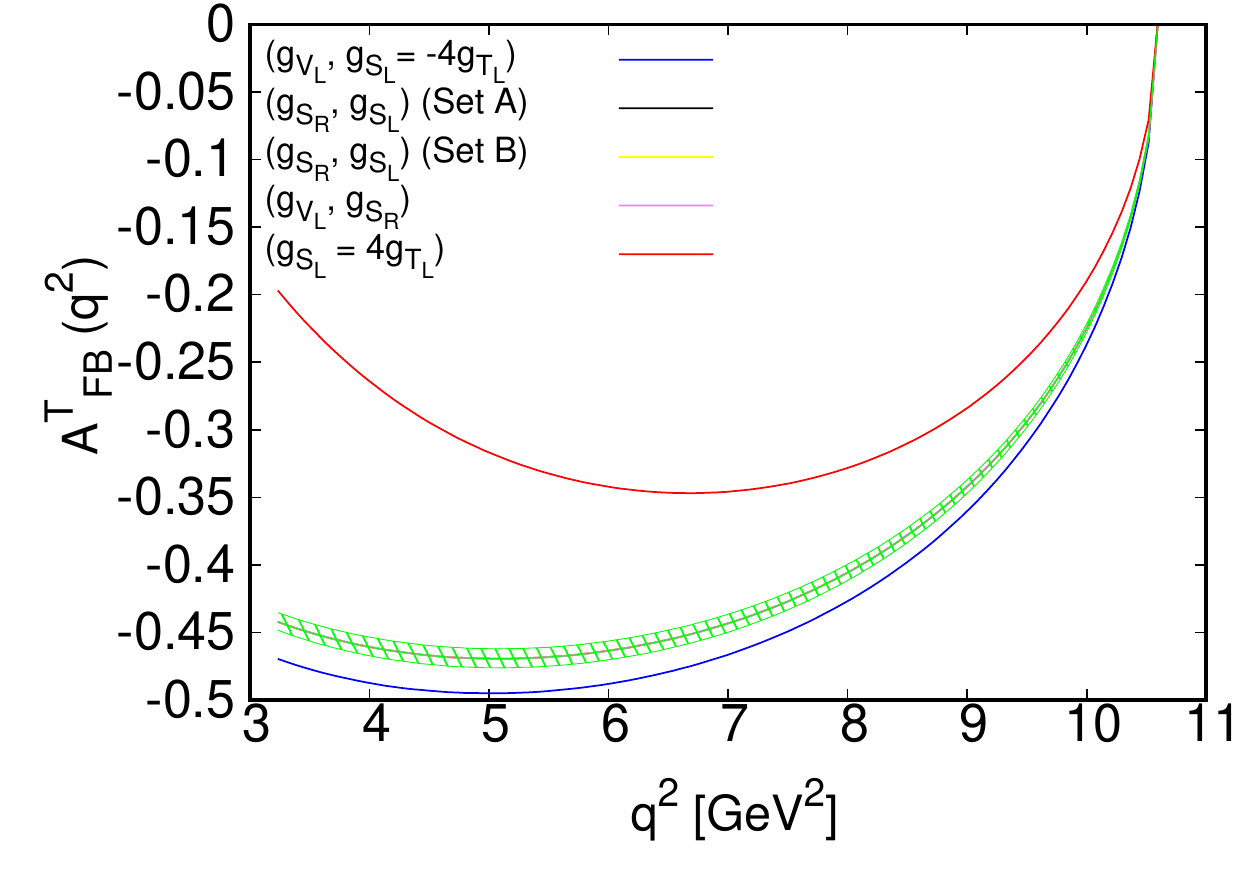}
\includegraphics[width=4cm,height=3cm]{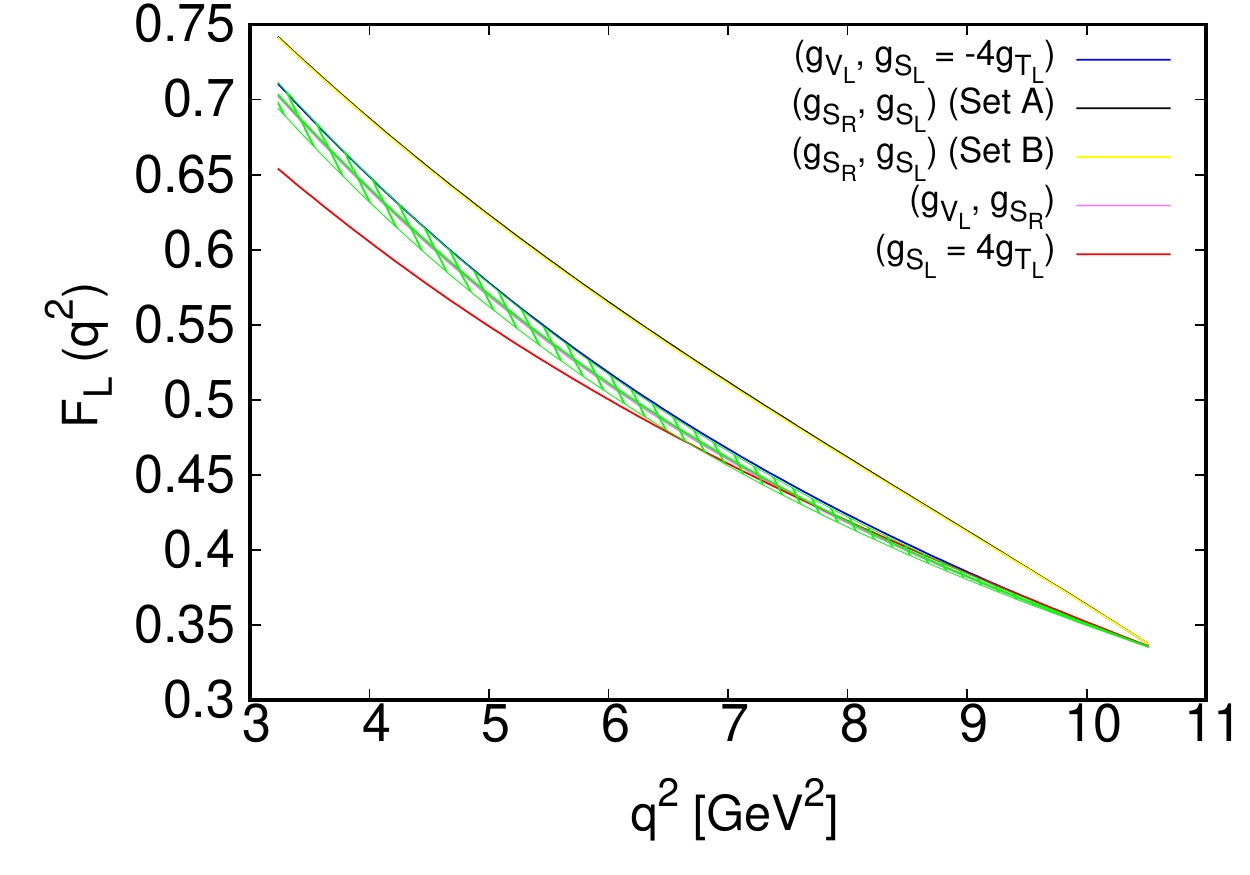}
\includegraphics[width=4cm,height=3cm]{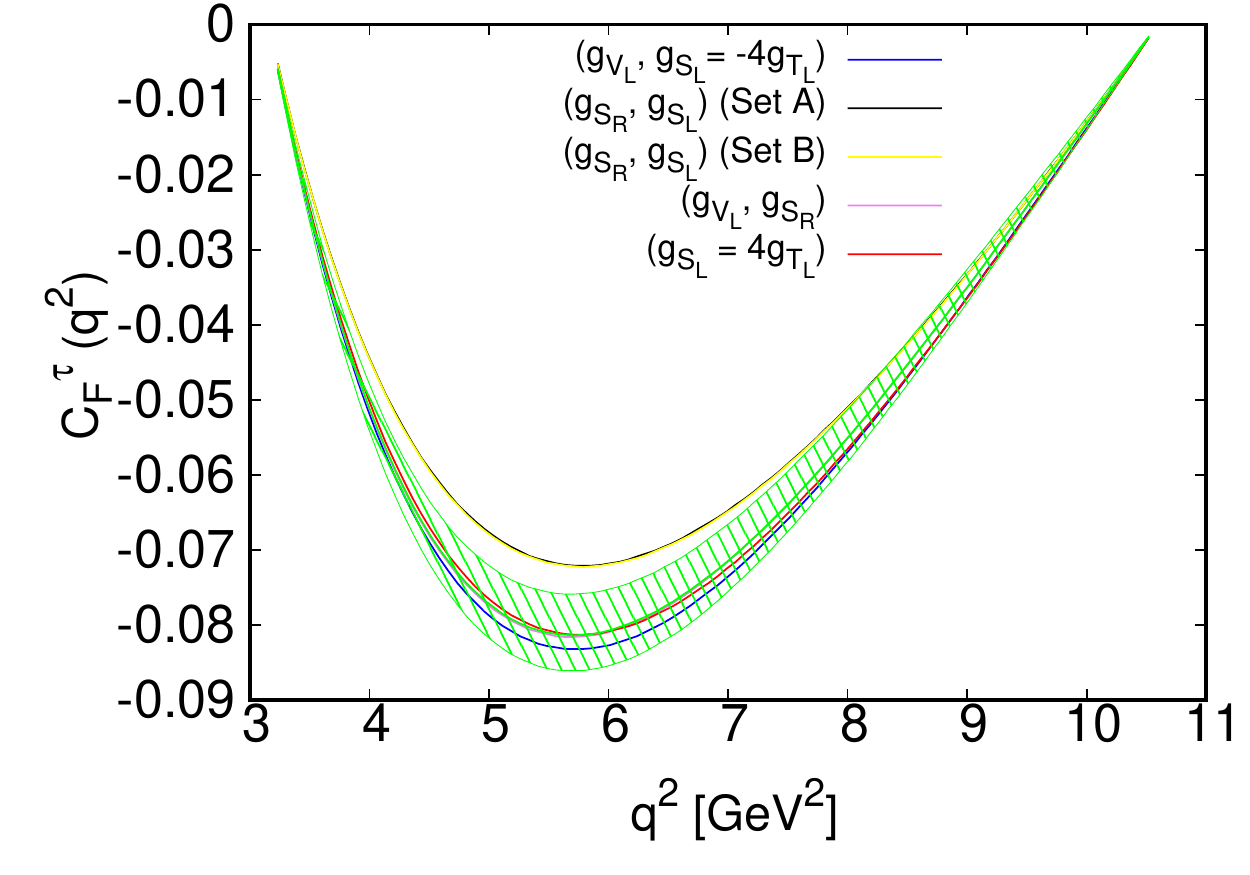}
\caption{The $q^2$ and $\cos\theta_l$ dependence of various physical observable of $B_s\to D_s^*(\to D_s \gamma, D_s\pi )\tau \nu$ in the SM and in the presence of the NP couplings of scenario - II. The SM central line and the corresponding error band are shown with green color. The blue, black, yellow, violet and red color colors represents the effect of NP coupling ($g_{V_L}$, $g_{S_{L}}\,=\,-4g_{T_L}$), ($g_{S_R}$, $g_{S_{L}}$)(Set A), ($g_{S_R}$, $g_{S_{L}}$)(Set B), ($g_{V_L}$, $g_{S_R}$) and $g_{S_{L}}\,=\,4g_{T_L}$  respectively.}
\label{dsg_sc2_same}
\end{figure}

\begin{table}[htbp]
\centering 
  \renewcommand{\arraystretch}{1.3}
\resizebox{18.6cm}{!}{
\begin{tabular}{|c|c|c|c|c|c|c|c|c|c|c|c|c|c|c|c|c|c|}
\hline
&\multicolumn{2}{c|}{ ({ $g_{V_L}$}, $g_{S_{L}}\,=\,-4g_{T_L}$)} & \multicolumn{2}{c|}{($g_{S_R}$, $g_{S_{L}}$)(Set A)} & \multicolumn{2}{c|}{($g_{S_R}$, $g_{S_{L}}$)(Set B)} & \multicolumn{2}{c|}{({ $g_{V_L}$}, $g_{S_R}$)} & \multicolumn{2}{c|}{$g_{S_{L}}=4g_{T_L}$} \\
\cline{2-11}
& $D_s\gamma$  & $D_s\pi$  & $D_s\gamma$  & $D_s\pi$   & $D_s\gamma$ & $D_s\pi $  & $D_s\gamma$  & $D_s\pi $  & $D_s\gamma$ & $D_s\pi $  \\
\hline
\hline
$BR\times 10^{-2}$   & $1.4049 \pm 0.0437$ & $ 0.0872  \,\pm \, 0.0027$ & $ 1.2984\pm0.0404$ &$0.0805\,\pm \,0.0025 $ & $ 1.2963 \pm  0.0404$  & $0.0804  \pm 0.0025$ &  $1.3918 \pm  0.0433$ &$0.0863 \,\pm \, 0.0027 $ & $1.3696\pm  0.0426 $ & $0.0850 \pm 0.0026$\\
\hline
$A_3$                &$0.0083\,\pm\, 0.0001$& $-0.0167  \,\pm \, 0.0001$  & $ 0.0075\,\pm\,0.0001$& $-0.0149\,\pm \,0.000 1$ & $ 0.0075\,\pm\,0.0001 $ & $ -0.0150 \pm   0.0001$&  $0.0081\,\pm\,0.0001$&$-0.0162\,\pm \,0.0001$ &$ 0.0066  \pm 0.0001 $& $-0.0132 \pm 0.0001$\\
\hline
$A_4$                &$-0.0454\,\pm\,0.0001$& $0.0909 \,\pm \, 0.0002 $   & $-0.0407\,\pm\,0.0001 $& $0.0813\,\pm \,0.0002 $ & $-0.0407 \,\pm\, 0.0001$ & $0.0815\,\pm \,0.0002 $ &  $-0.0443\,\pm\,0.0001$ &$0.0885\,\pm \,0.0001 $ & $-0.0359 \pm  0.0001 $& $0.0719 \pm0.0001$\\
\hline  
$A_5 $               &$ 0.1178\,\pm\,0.0005$&  $-0.2356  \,\pm \, 0.0010 $  & $0.1247\,\pm\,0.0004 $& $ -0.2493\,\pm \,0.0009 $ &$0.1245 \,\pm\, 0.0004$ &$ -0.2490\,\pm \,0.0009 $  &  $0.1129\,\pm\,0.0005$ & $ -0.2258\,\pm \, 0.0010 $ &$ 0.1084 \pm  0.0004$& $-0.2168  \pm 0.0008$\\
\hline
$A_{6s}$             &$-0.5736\,\pm\,0.0028$ & $0.9950  \,\pm \,  0.0079 $   & $-0.7422\,\pm\,0.0030$&$0.8782\,\pm \,0.0072 $ & $-0.7394  \,\pm\, 0.0030$&  $0.8796 \pm  0.0072$ &  $-0.5453\,\pm\,0.0026$ &$0.9556\,\pm \,   0.0078 $ & $-0.6032 \pm  0.0027$& $0.7508 \pm0.0061$\\
\hline
$A_{7}$              &  \multicolumn{2}{c|}{0.0000}  &    \multicolumn{2}{c|}{0.0000} &  \multicolumn{2}{c|}{0.0000} & \multicolumn{2}{c|}{0.0000}  &     $0.0086\pm 0.0001$ & $0.0172 \pm 0.0001$\\
\hline
$R_{D_s^*}$      & \multicolumn{2}{c|}{$0.2856\,\pm\,0.0018$}          &   \multicolumn{2}{c|}{$ 0.2639 \,\pm\,0.0018 $} & \multicolumn{2}{c|}{$ 0.2635 \,\pm\,0.0018 $}  &    \multicolumn{2}{c|}{$0.2829 \pm  0.0018$}    & \multicolumn{2}{c|}{$0.2784  \pm  0.0018$}\\
\hline
$A_{FB}^{\tau}$         & \multicolumn{2}{c|}{$-0.0936\pm 0.0021$}  & \multicolumn{2}{c|}{$-0.0302\pm0.0021$} & \multicolumn{2}{c|}{$-0.0311\pm 0.0021$}   & \multicolumn{2}{c|}{$-0.0912\pm   0.0020$} & \multicolumn{2}{c|}{$-0.0328\pm 0.0017$}\\
\hline
$A_{FB}^{T}$         & \multicolumn{2}{c|}{$-0.4047 \,\pm\,0.0026$}  & \multicolumn{2}{c|}{$-0.3842\,\pm\,0.0026$} & \multicolumn{2}{c|}{$-0.3842\,\pm\,0.0026$} & \multicolumn{2}{c|}{$-0.3842\,\pm\,0.0026$} & \multicolumn{2}{c|}{$-0.3000\,\pm\,0.0017$}\\
\hline
$F_L$        & \multicolumn{2}{c|}{$0.4537 \,\pm\, 0.0015$} & \multicolumn{2}{c|}{$0.4920\,\pm\,0.0015$} & \multicolumn{2}{c|}{$ 0.4912 \,\pm\, 0.0015 $}  &   \multicolumn{2}{c|}{$0.4472 \,\pm\,0.0015$} & \multicolumn{2}{c|}{ 0.4438 $\pm$  0.0015}\\
\hline
$C_F^l$              & \multicolumn{2}{c|}{$-0.0567\,\pm\,0.0014$}   &  \multicolumn{2}{c|}{$-0.0506\,\pm\,0.0013 $} & \multicolumn{2}{c|}{$-0.0507  \,\pm\, 0.0013 $} &  \multicolumn{2}{c|}{$-0.0551\,\pm\,0.0014$} & \multicolumn{2}{c|}{$-0.0559 \pm  0.0011$}\\
\hline
\end{tabular}}
 \caption{ Prediction of $B_s\to D_s^*(\to D_s\gamma,D_s\pi)\,\tau\,\nu$ decay observables in Scenerio - II.}
\label{dsg_sc2}
\end{table}
We display the $q^2$ and $\cos\theta_l$  dependence of several physical observables that show same 
behaviour for the $D_s\gamma$ and $D_s\pi$ decay modes in Fig~\ref{dsg_sc2_same}. The blue, black, yellow, violet and red lines represent the 
contribution coming from $(g_{V_L}$, $g_{S_{L}}\,=\,-4g_{T_L})$, $(g_{S_R}$, $g_{S_{L}})$~(Set A), $(g_{S_R}$, $g_{S_{L}})$~(Set B), 
$(g_{V_L}$, $g_{S_R})$ and $(g_{S_{L}}\,=\,4g_{T_L})$ NP couplings, respectively. Our observations are as follows.
\begin{itemize}

 \item Although a slight deviation from the SM prediction is observed for $F_L(\cos\theta_l)$ with $g_{S_{L}}\,=\,4g_{T_L}$ NP coupling, the 
deviation, however, is quite significant in the presence of $(g_{S_R}$, $g_{S_{L}})$~(Set A or Set B) NP couplings. Similarly, 
$F_T(\cos\theta_l)$ is observed to be deviated from the corresponding SM value in the presence of $(g_{S_R}$, $g_{S_{L}})$~(Set A or Set B) 
and $(g_{S_{L}}\,=\,4g_{T_L})$ NP couplings.

\item Although the deviation from the SM prediction for $R_{D_s^*} (q^2)$ is quite significant for all the $2D$ NP couplings, it is more 
pronounced in case of $(g_{V_L}$, $g_{S_{L}}\,=\,-4g_{T_L})$, $(g_{V_L}$, $g_{S_R})$ and $(g_{S_{L}}=4g_{T_L})$ NP couplings and they are
clearly distinguishable from the SM prediction at more than $10\sigma$ significance.

\item The zero crossing in $A_{FB}^{\tau}(q^2)$ is shifted to higher value of $q^2$ than in the SM in the presence of $(g_{S_R}$, $g_{S_{L}})$
~(Set A or Set B) and $(g_{S_{L}}=4g_{T_L})$ NP couplings. The zero crossings of $A_{FB}^{\tau}(q^2)$ at $q^2 = 6.28{\rm GeV^2}$ and 
$q^2 = 6.16{\rm GeV^2}$ in the presence of these NP couplings are clearly distinguishable from the SM prediction of $q^2 = 5.25\pm 0.10 {\rm GeV^2}$ at more
than $8\sigma$ significance. Similarly, for $A_{FB}^T(q^2)$, a significant deviation of more than $10\sigma$ is observed at low $q^2$ in the 
presence of $g_{S_{L}}=4g_{T_L}$ NP coupling.

\item In case of $F_L(q^2)$, although a slight deviation is observed with $(g_{S_{L}}\,=\,4g_{T_L})$ NP coupling, the deviation, however, is more
pronounced in the presence of $(g_{S_R}$, $g_{S_{L}})$~(Set A or Set B) NP couplings. Similarly for $C_F^{\tau}(q^2)$, maximum deviation from the
SM prediction is observed with $(g_{S_R}$, $g_{S_{L}})$~(Set A or Set B) NP couplings.

\end{itemize}

\begin{figure}[htbp]
\centering
\includegraphics[width=4cm,height=3cm]{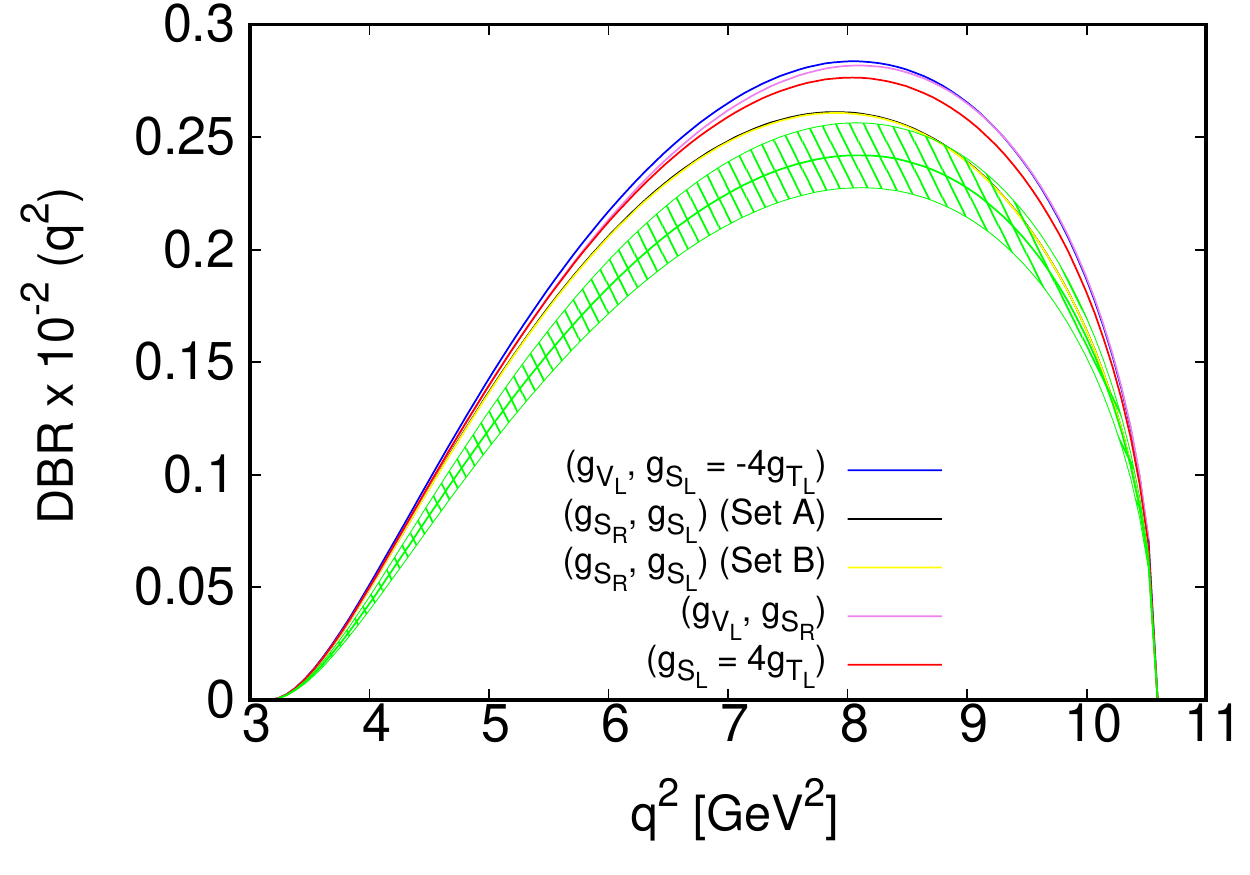}
\includegraphics[width=4cm,height=3cm]{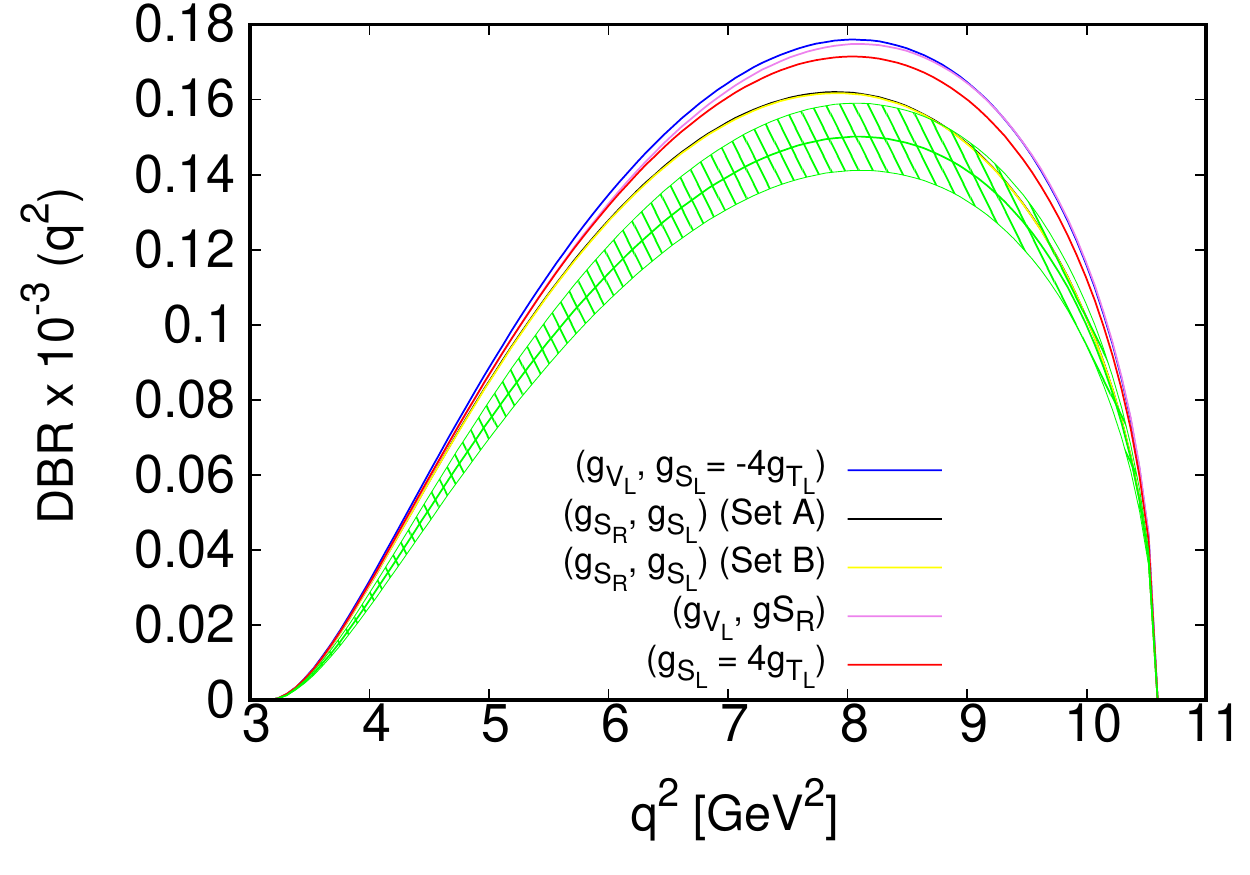}\hspace{1.5cm}
\includegraphics[width=4cm,height=3cm]{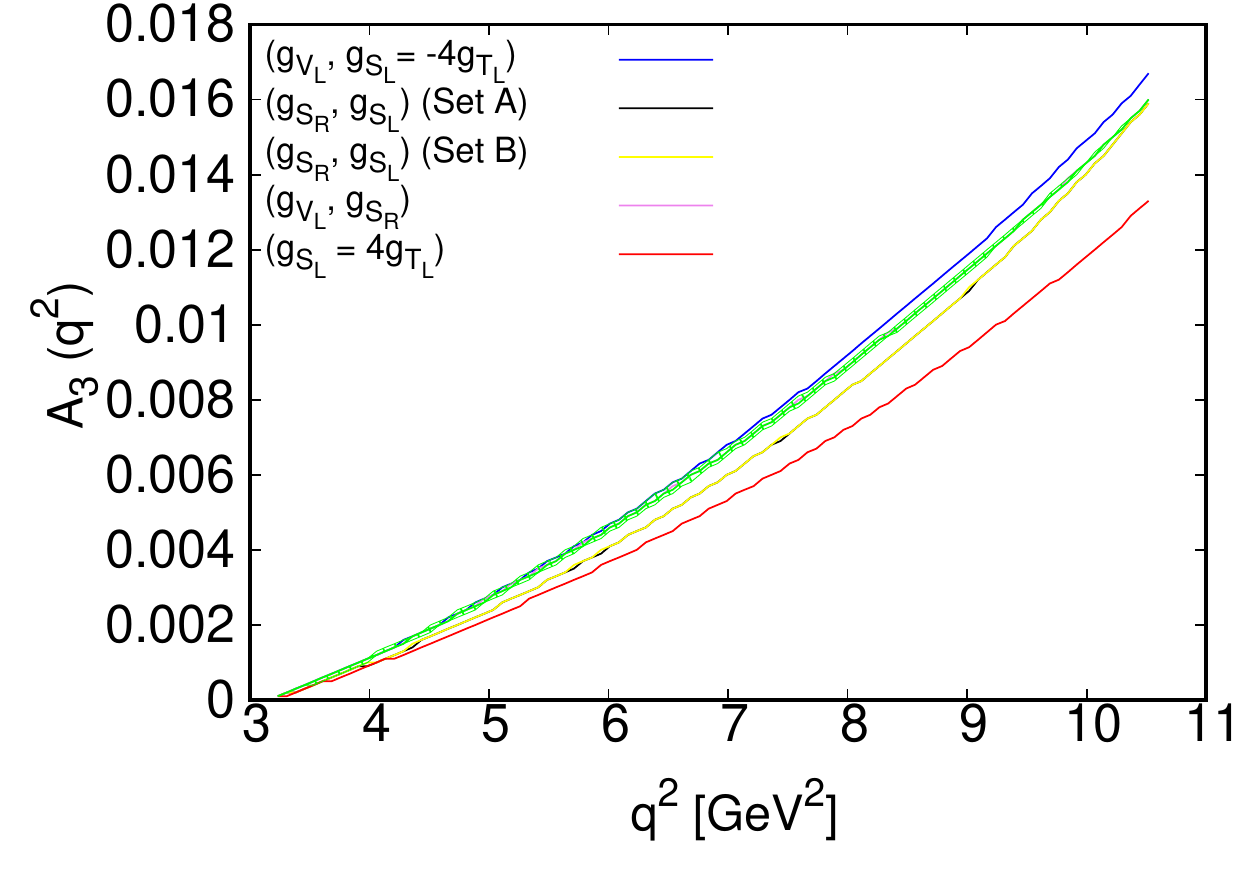}
\includegraphics[width=4cm,height=3cm]{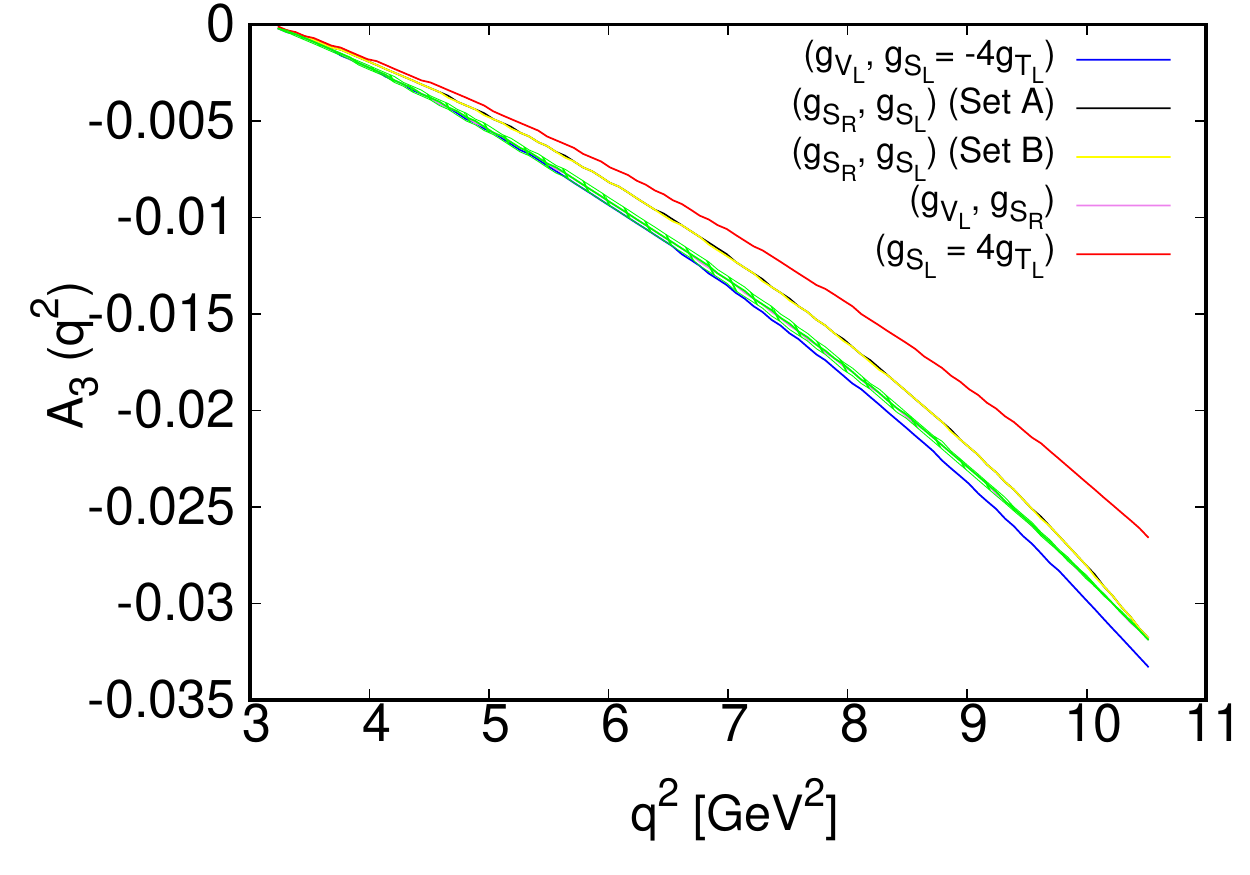}\hspace{1.5cm}
\includegraphics[width=4cm,height=3cm]{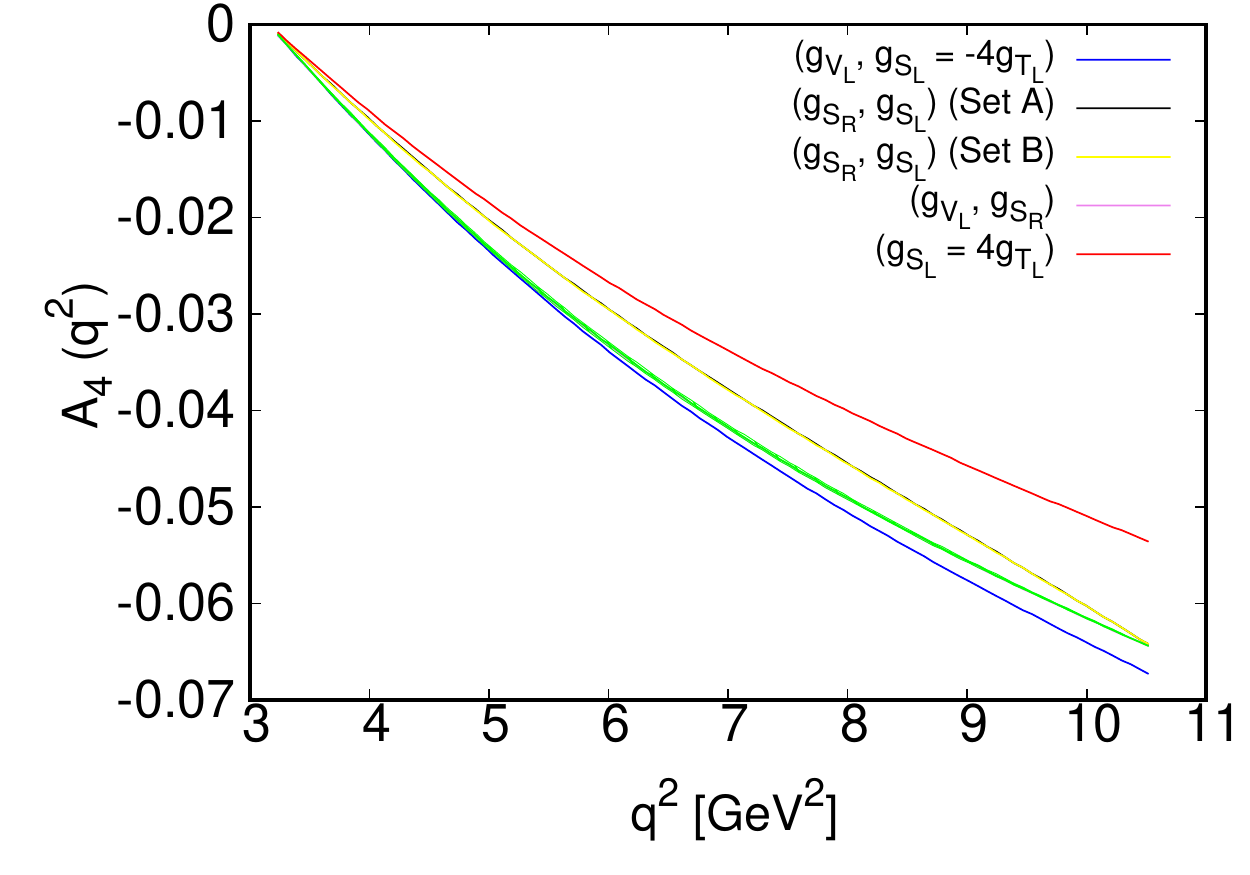}
\includegraphics[width=4cm,height=3cm]{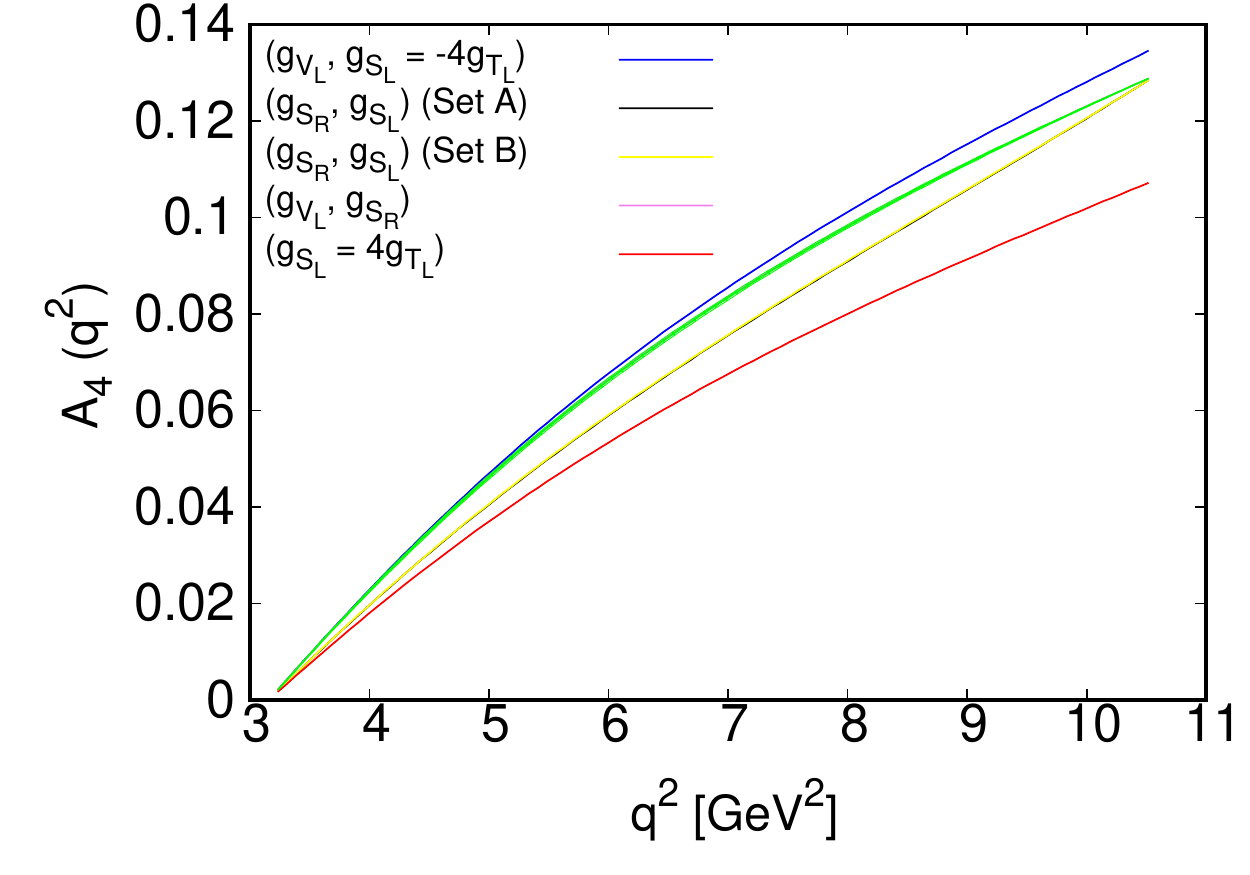}\hspace{1.5cm}
\includegraphics[width=4cm,height=3cm]{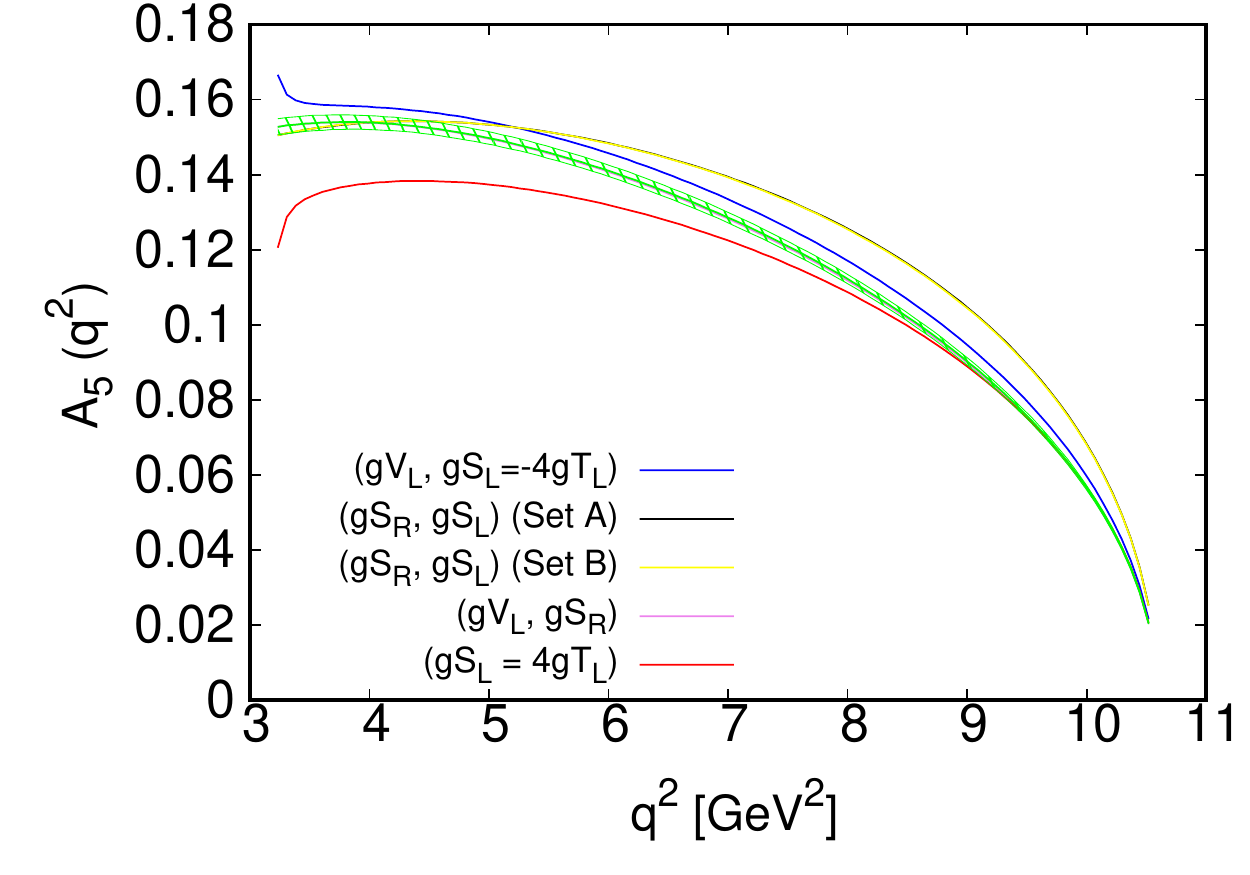}
\includegraphics[width=4cm,height=3cm]{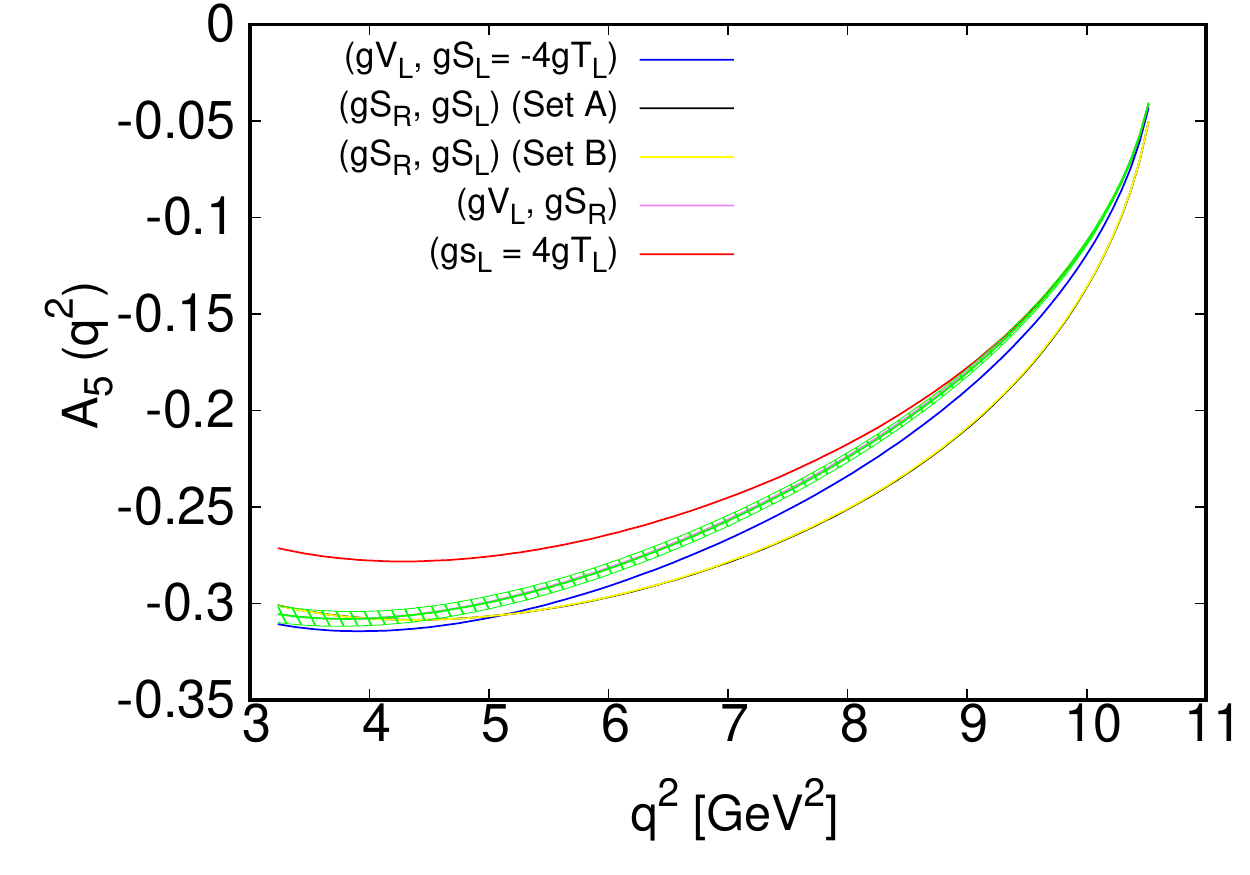}\hspace{1.5cm}
\includegraphics[width=4cm,height=3cm]{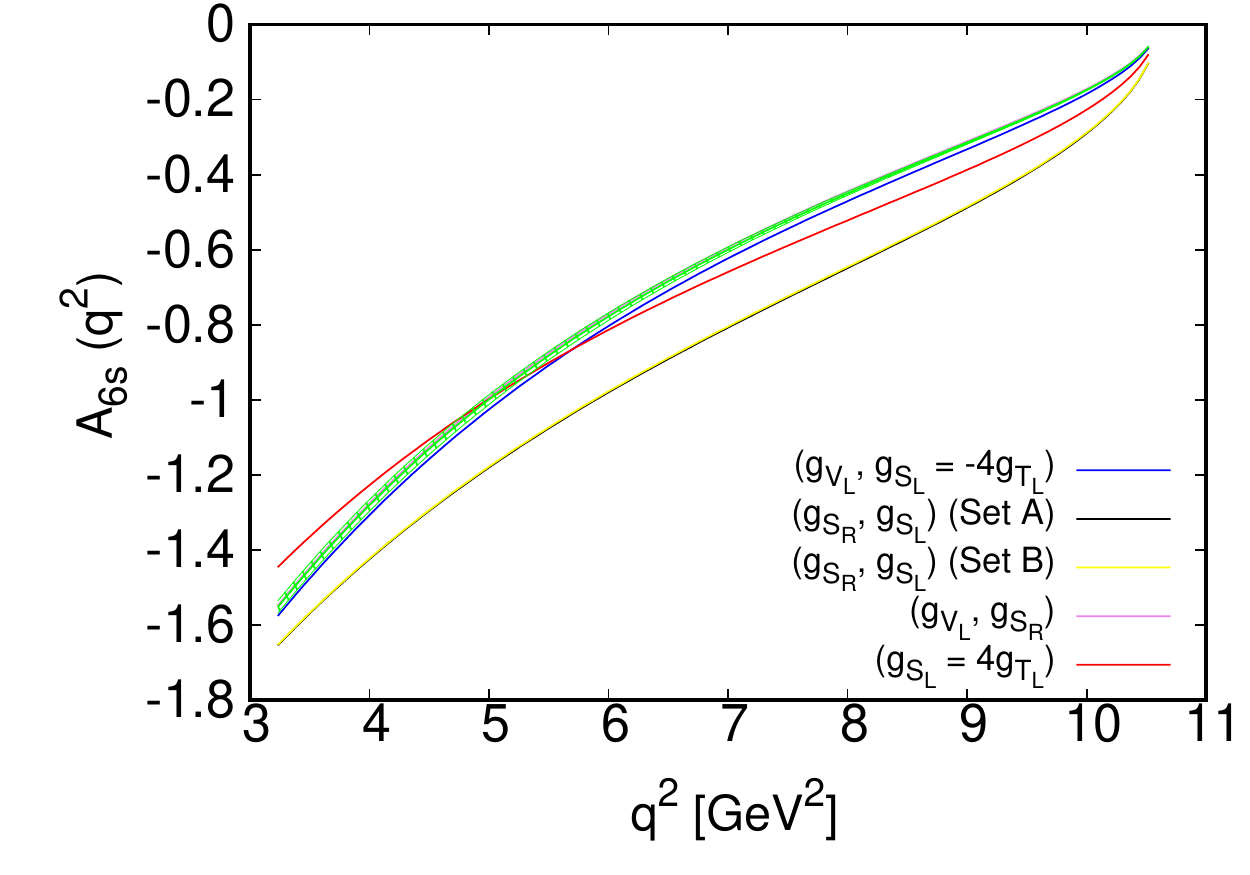}
\includegraphics[width=4cm,height=3cm]{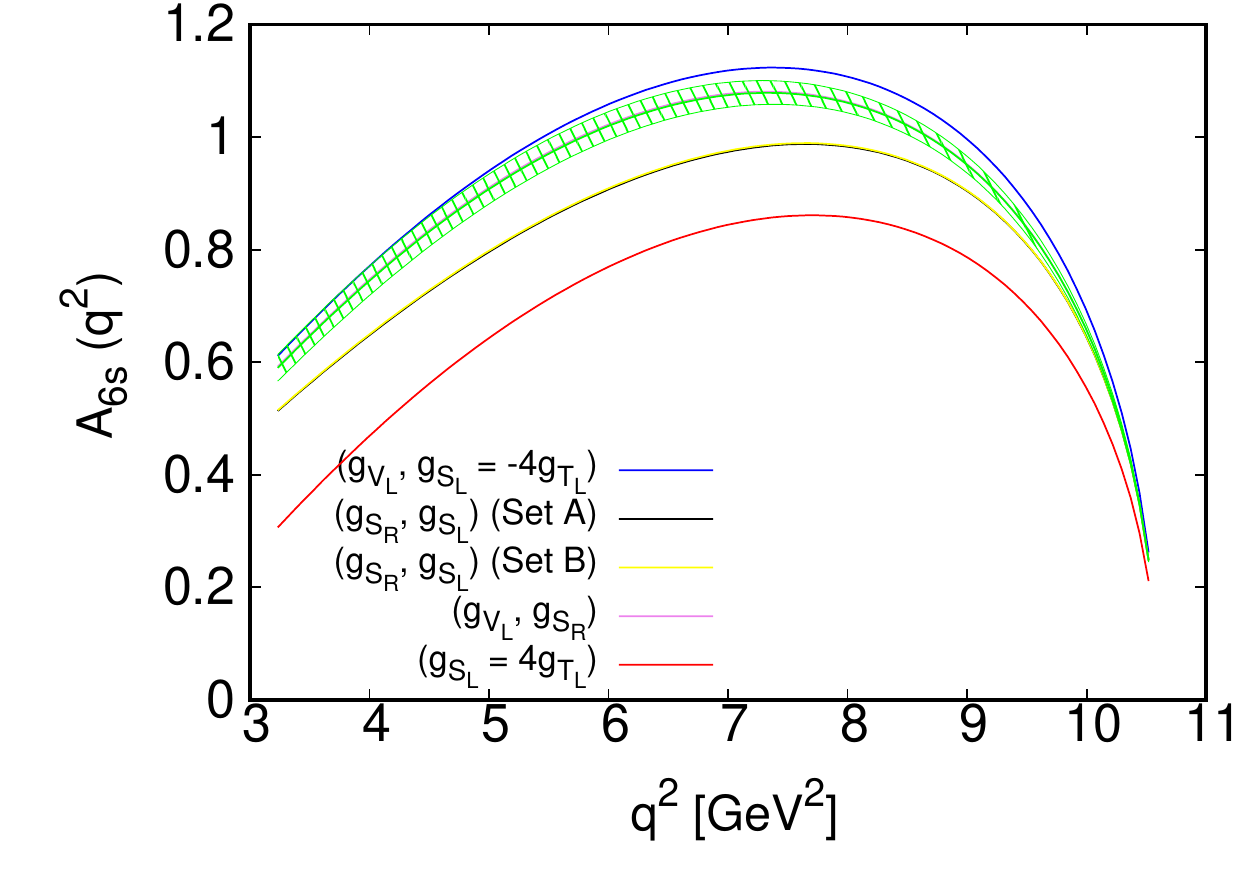}\hspace{1.5cm}
\includegraphics[width=4cm,height=3cm]{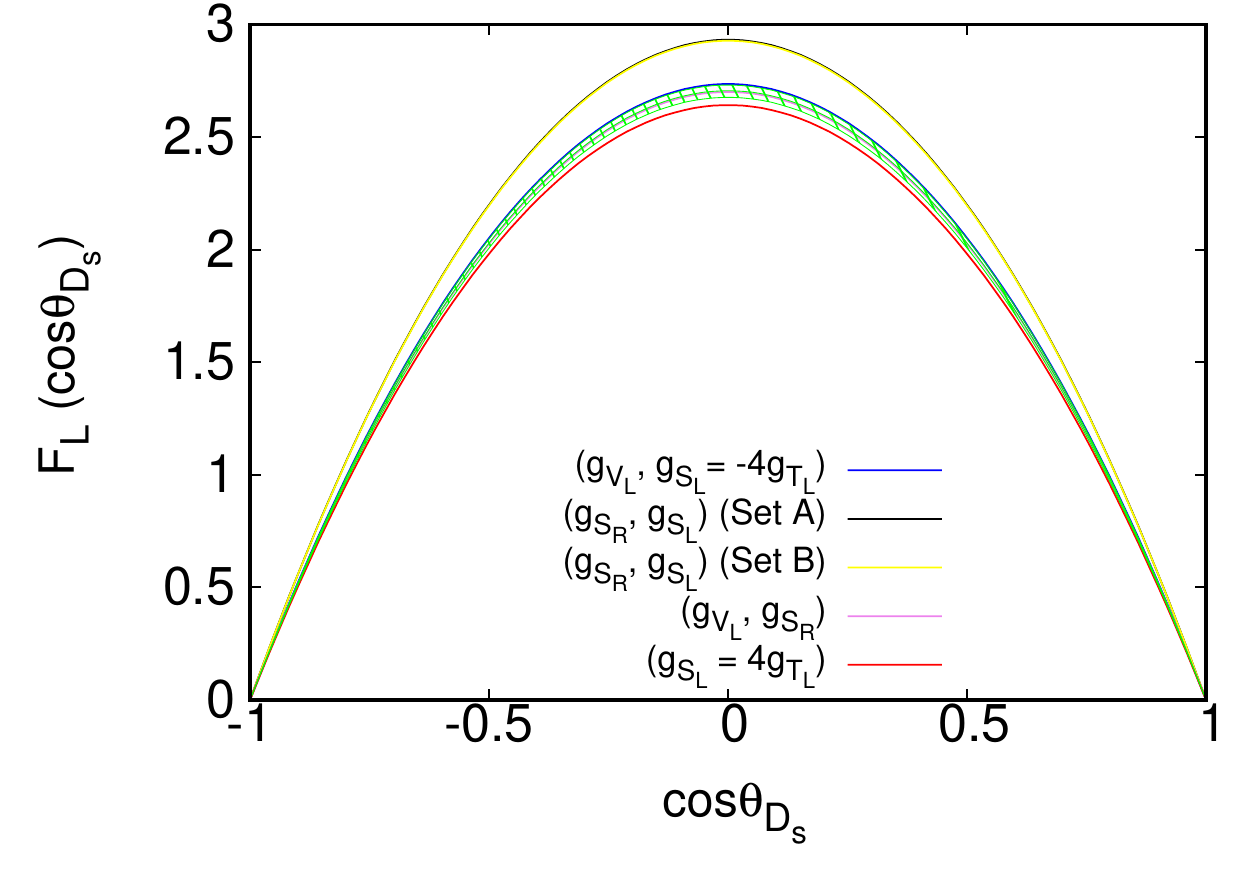}
\includegraphics[width=4cm,height=3cm]{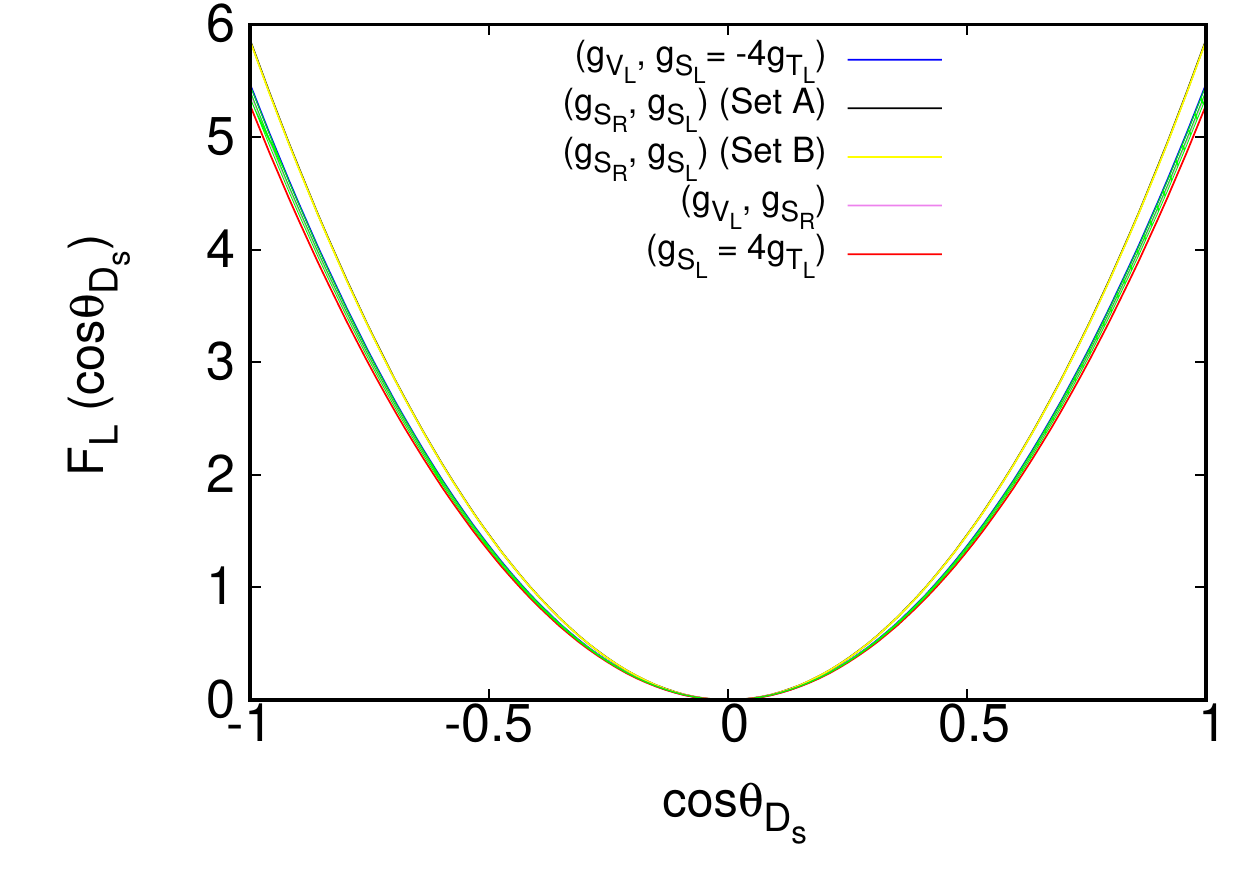}\hspace{1.5cm}
\includegraphics[width=4cm,height=3cm]{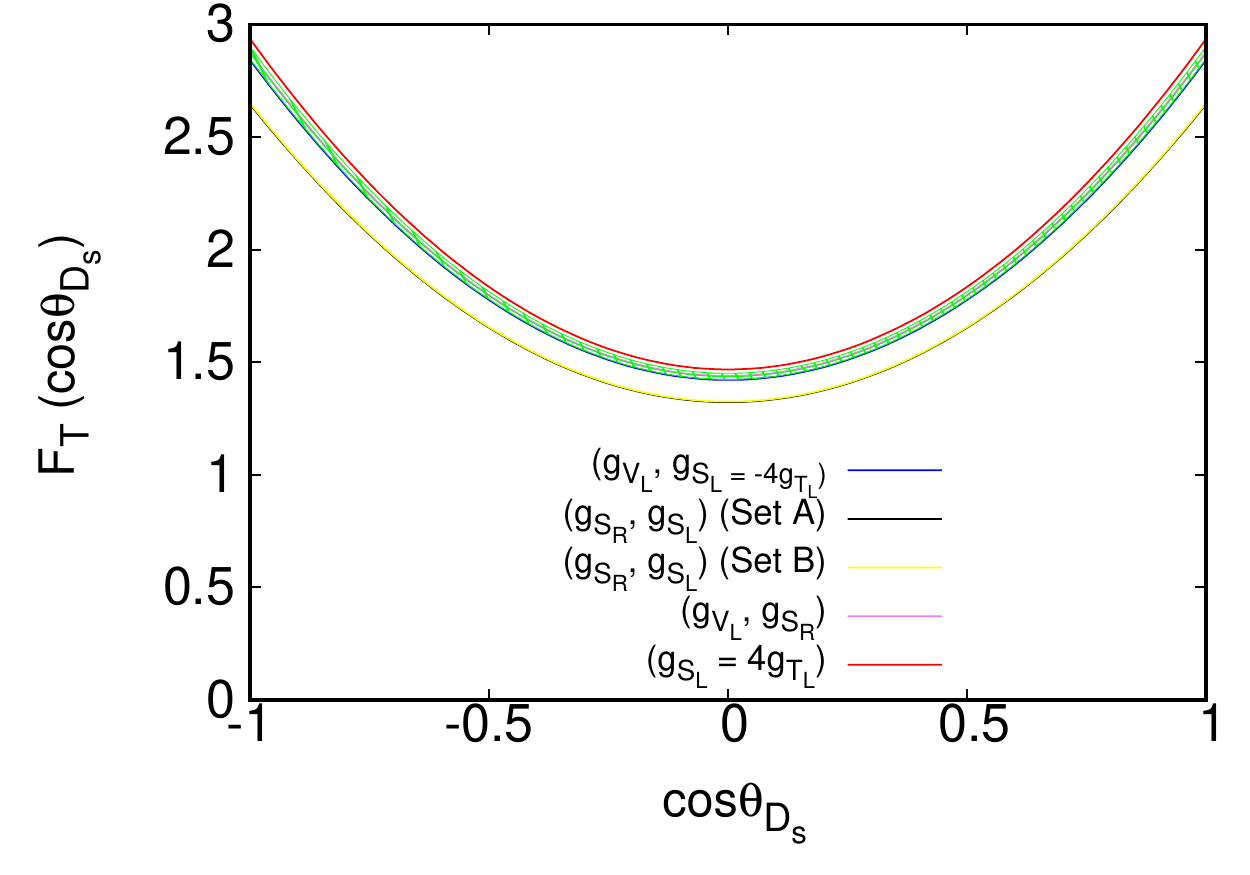}
\includegraphics[width=4cm,height=3cm]{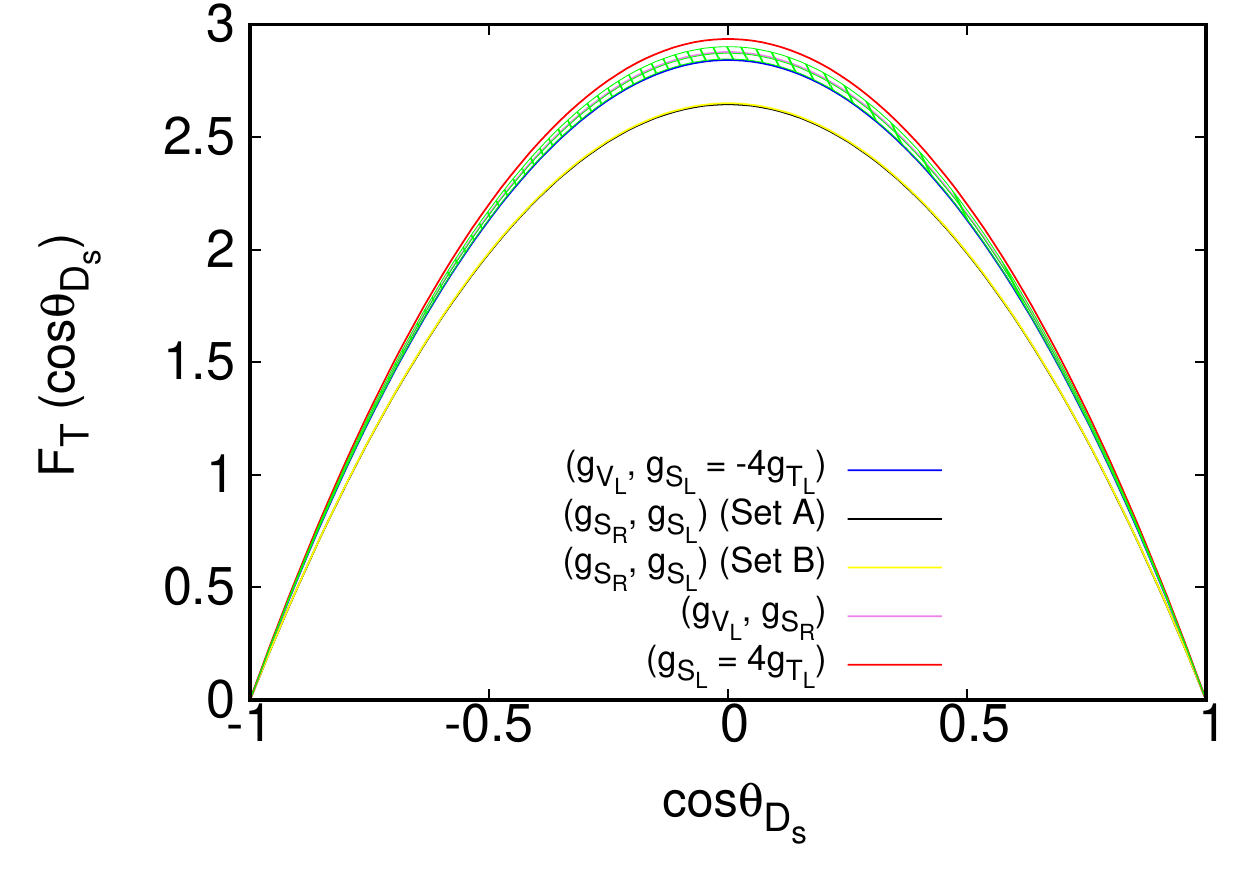}\hspace{1.5cm}
\includegraphics[width=4cm,height=3cm]{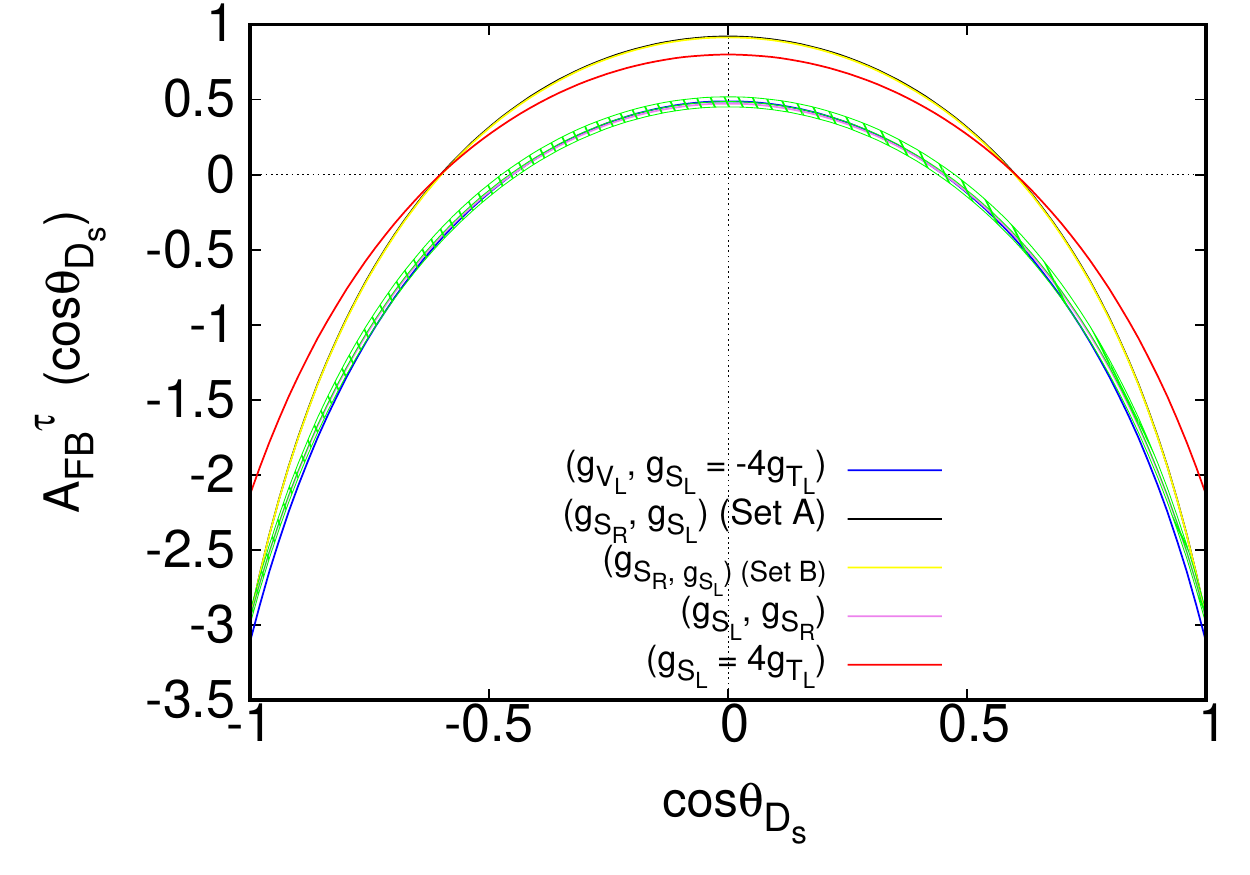}
\includegraphics[width=4cm,height=3cm]{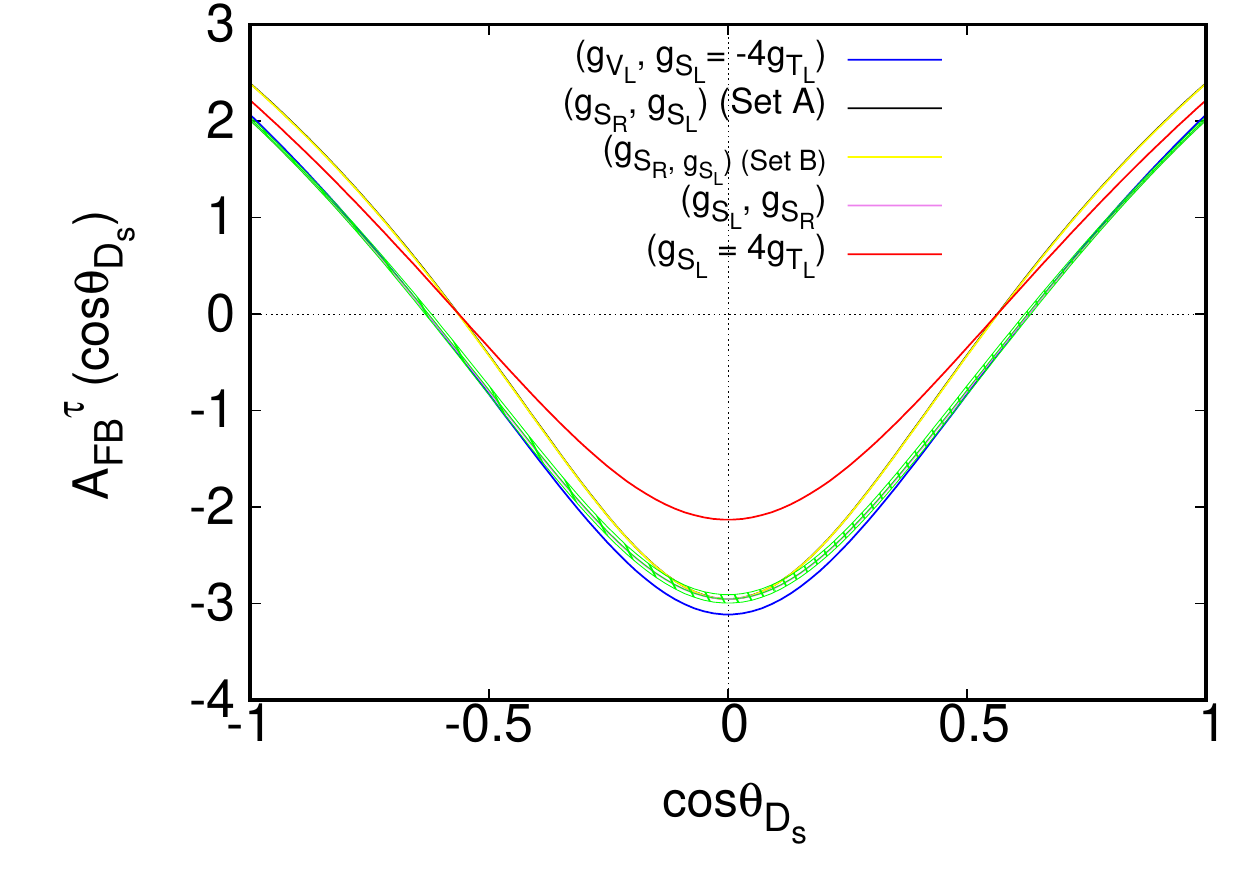}\hspace{1.5cm}
\includegraphics[width=4cm,height=3cm]{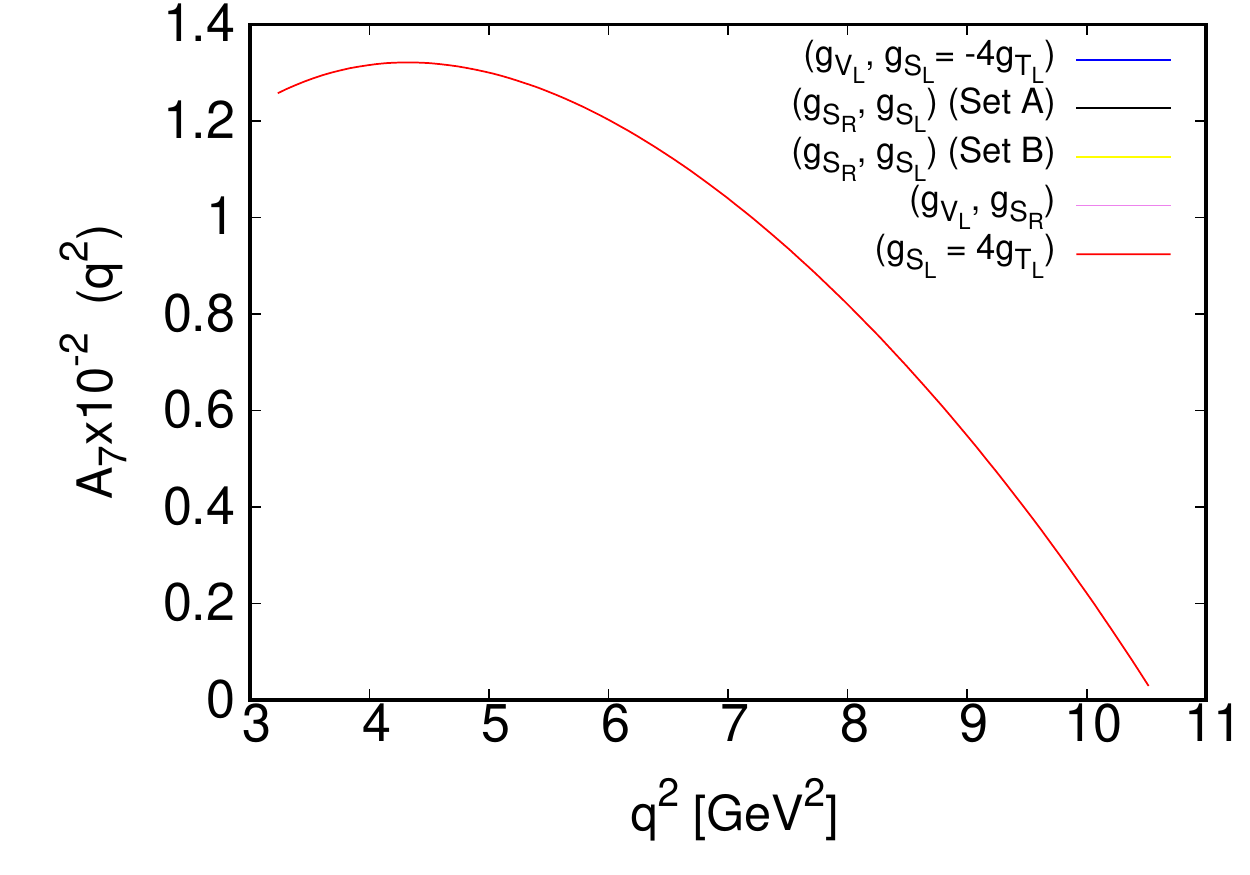}
 \includegraphics[width=4cm,height=3cm]{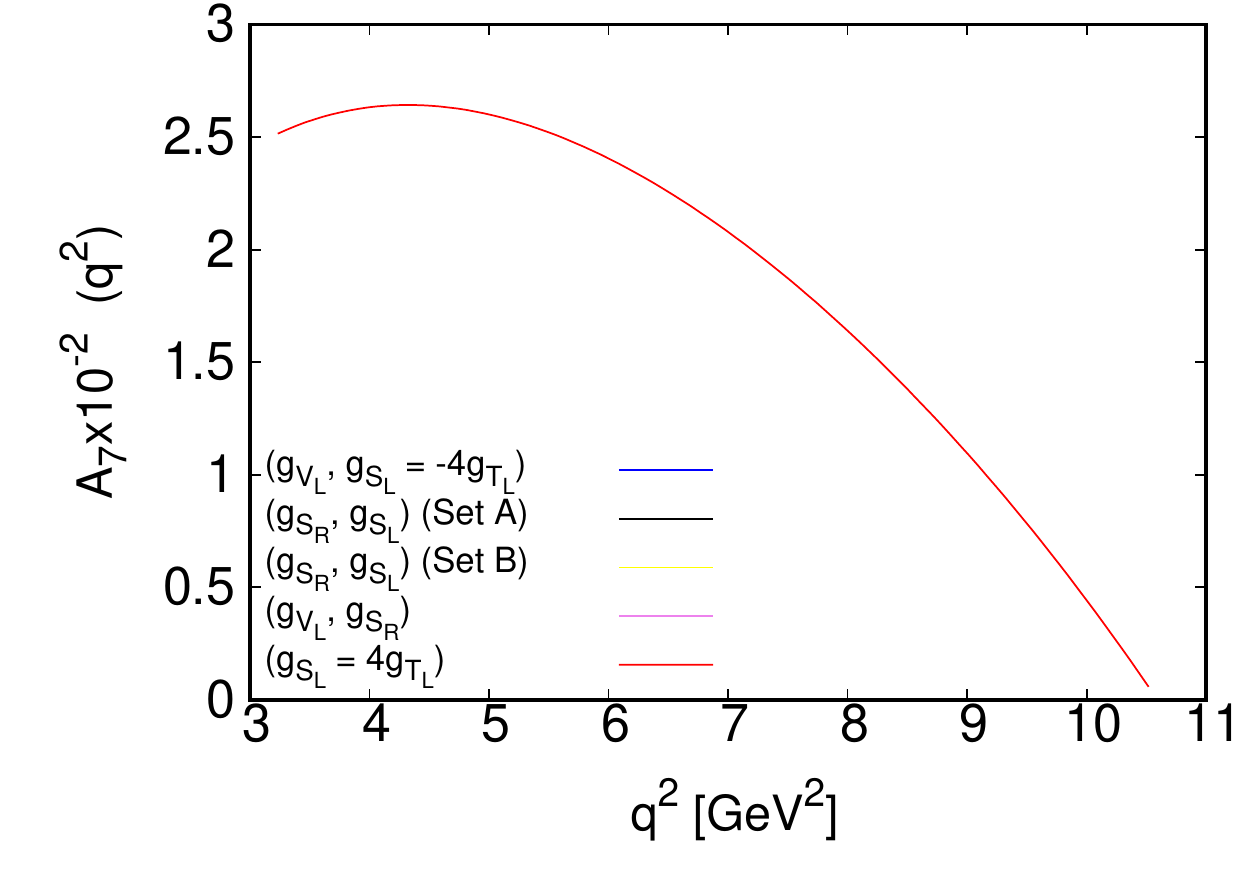}
\caption{The $q^2$ and $\cos\theta_{D_s}$ dependence of various physical observable of $B_s\to D_s^*(\to D_s \gamma, D_s\pi )\tau \nu$ in the SM and in the presence of the NP couplings of scenario - II. The SM central line and the corresponding error band are shown with green color. The blue, black, yellow, violet and red color colors represents the effect of NP coupling ($g_{V_L}$, $g_{S_{L}}\,=\,-4g_{T_L}$), ($g_{S_R}$, $g_{S_{L}}$)(Set A), ($g_{S_R}$, $g_{S_{L}}$)(Set B), ($g_{V_L}$, $g_{S_R}$) and $g_{S_{L}}\,=\,4g_{T_L}$ respectively.}
\label{dsg_sc2_diff}
\end{figure}
The $q^2$ and $\cos\theta_{D_s}$ dependent observables which exhibit different behaviour for $D_s\pi$ and $D_s\gamma$ modes are displayed in 
Fig~\ref{dsg_sc2_diff}. The left panel figures correspond to the $D_s\gamma$ mode and right panel figures correspond to the $D_s\pi$ mode,
respectively. Our observations are as follows.
\begin{itemize}
\item In case of $DBR(q^2)$, although there is deviation from the SM prediction with all NP couplings, the deviation, however, is more
pronounced once the $(g_{V_L}$, $g_{S_{L}}\,=\,-4g_{T_L})$ NP coupling is switched on and it is clearly distinguishable from the SM prediction
at more than $3\sigma$ significance level.

\item For the $A_3$, $A_4$ and $A_5$ observables, the maximum deviation is observed in case of $(g_{S_{L}}\,=\,4g_{T_L})$ NP couplings for both
$D_s\pi$ and $D_s\gamma$ modes. For $A_{6s}$, the maximum deviation is observed with $(g_{S_R}$, $g_{S_{L}})$~(Set A or Set B) for the 
$D_s\gamma$ mode, whereas, for the $D_s\pi$ mode, the maximum deviation is observed with $(g_{S_{L}}\,=\,4g_{T_L})$ NP coupling.
 
\item For the $D_s\gamma$ mode, $F_L(\cos{\theta_{D_s}})$ deviates significantly from the SM prediction at $\cos\theta_{D_s}=0$ in presence 
of $(g_{S_R}$, $g_{S_{L}})$~(Set A or Set B) NP coupling and it is clearly distinguishable from the SM error band, whereas, for the 
$D_s\pi$ mode, $F_L(\cos{\theta_{D_s}})$ shows a significant deviation at $\cos\theta_{D_s}=\pm1$.
In case of $F_T(\cos{\theta_{D_s}})$, the deviation from the SM prediction is more pronounced with $(g_{S_R}$, $g_{S_{L}})$~
(Set A or Set B) NP couplings for both $D_s\pi$ and $D_s\gamma$ modes. 
  
\item For $D_s\gamma$ and $D_s\pi$ mode, $A_{FB}^{\tau}(\cos{\theta_{D_s}})$ deviates significantly from the SM prediction in the presence of 
$(g_{S_R}$, $g_{S_{L}})$~(Set A or Set B) and $(g_{S_{L}}\,=\,4g_{T_L})$ NP couplings.
The zero crossings in $A_{FB}^{\tau}(\cos{\theta_{D_s}})$ at $\cos{\theta_{D_s}} = \pm 0.601$ and $\cos{\theta_{D_s}} = \pm 0.563$ for 
$D_s\gamma$ and $D_s\pi$ modes lie $8\sigma$ away from the SM zero crossing point.

\item We observe a non-zero $q^2$ distribution of $A_7(q^2)$ in the presence of $g_{S_{L}}\,=\,4g_{T_L}$ complex NP couplings.
\end{itemize}

\subsubsection{(Scenerio - III)}

In this scenario, we select five different complex $1D$ NP couplings. 
The best fit values each NP couplings at renormalization scale $\mu=m_b$ obtained from Ref\cite{Becirevic:2019tpx} are reported in 
Table~\ref{global_fit}. In Table~\ref{dsg_sc3}, we report the impact of each NP couplings
on various physical observable in $D_s\gamma$ and $D_s\pi$ decay modes. We see significant deviation of all the observables with these
complex NP couplings.
In the presence of $g_{V_L}$ and $g_{V_R}$ NP couplings, branching ratio deviates from the SM prediction at the level
of $3-6\sigma$ significance. $A_{FB}^{\tau}$ deviates more than $9\sigma$ in the presence of $g_{V_R}$ and $g_{S_{L}}$ NP couplings and 
the observable $A_{FB}^T$ deviates more than $10\sigma$ from the SM expectation in case of $g_{V_R}$ NP coupling. Similarly, the 
longitudinal polarization fraction of $D_s^*$, $F_L$ is found to deviate from the SM value at more than $10\sigma$ significance in the 
presence of $g_{T_L}$ NP coupling for both the decay modes. In case of $R_{D_s^*}$, we observe a considerable deviation of around
$10\sigma$ in the presence of $g_{V_L}$ and $g_{T_L}$ NP couplings. 
Moreover, for $A_3\, A_4,\,{\rm and}\, A_5$ the maximum deviation from the SM prediction is observed with $g_{T_L}$ NP coupling. 
For the angular observable $A_{6s}$, the deviation observed is more pronounced in case of $g_{S_{L}}$, $g_{S_R}$ and $g_{T_L}$ NP couplings
in $D_s\gamma$ mode, whereas, $g_{V_L}$ and $g_{T_L}$ show more significant deviation in case of $D_s\pi$ mode.
A nonzero value of $A_7$ is also observed in the presence of $g_{S_{L}}$, $g_{S_R}$ and $g_{T_L}$ NP couplings. 
The angular observables $A_8$ and $A_9$ assume non zero values once $g_{V_{R}}$ NP coupling is switched on
It should also be 
mentioned that the values of $A_7$, $A_8$ and $A_9$ in $D_s\pi$ mode is twice as large as the values obtained for the $D_s\gamma$ mode.

\begin{table}[htbp]
\centering
\renewcommand{\arraystretch}{1.3}
\resizebox{18.5cm}{!}{
\begin{tabular}{|c|c|c|c|c|c|c|c|c|c|c|c|c|c|c|c|c|c|c|c|}
\hline
&\multicolumn{2}{c|}{ $g_{V_L}$}  & \multicolumn{2}{c|}{ $g_{V_R}$}      &   \multicolumn{2}{c|}{ $g_{S_{L}}$} &   \multicolumn{2}{c|}{$g_{S_R}$} &    \multicolumn{2}{c|}{$g_{T_L}$} \\
\cline{2-11}
& $D_s\gamma$  & $D_s\pi$ & $D_s\gamma$  & $D_s\pi$  & $D_s\gamma$  & $D_s\pi $  & $D_s\gamma$  & $D_s\pi $  & $D_s\gamma$  & $D_s\pi $  \\
\hline
\hline
$BR\times10^{-2}$  & $1.3992\,\pm \,0.0436 $& $0.0868\,\pm\,0.0027 $ &
$1.3980\,\pm \,0.0435 $ &$ 0.0867\,\pm\,0.0027$ &
$  1.1831 \,\pm \,0.0368 $ &$ 0.0734\,\pm\,   0.0023$ &$ 1.2228  \,\pm \, 0.0381$& $0.0759\,\pm\,0.0024 $  &
$ 1.3681\,\pm \,0.0429$ & $ 0.0849\,\pm\,0.0027 $  \\
\hline
$A_3$             & $ 0.0081 \,\pm \,0.0001$& $ -0.0162\,\pm\,0.0001$ & $ 0.0082\,\pm \,0.0001$ & $-0.0163\,\pm\,0.0001 $  &$0.0082 \,\pm \  0.0001  $
&$-0.0164\,\pm\,0.0001 $ &$0.0079\,\pm \,0.0001 $ &  $-0.0159\,\pm\,0.0001$ & $ -0.0019\,\pm \,0.0001$  &  $-0.0038\,\pm\, 0.0001 $\\
\hline
$A_4$             & $-0.0442\,\pm \,0.0001$& $0.0883\,\pm\,0.0001 $  & $-0.0443\,\pm \,0.0001$&  $0.0886\,\pm\,0.0001 $  & $-0.0446 \,\pm \  0.0001$ &  $ 0.0893\,\pm\,0.0001$
&$-0.0432\,\pm \,0.0001 $ & $0.0864\,\pm\,0.0001 $    & $ -0.0112\,\pm \,0.0001$ & $0.0224\,\pm\, 0.0002 $ \\
\hline
$A_5 $            & $0.1133\,\pm \,0.0005$& $- 0.2265\,\pm\,0.0010 $ & $0.0949\,\pm \,0.0004$&$-0.1899\,\pm\,0.0007 $   & $0.1041 \,\pm \  0.0005$ & $-0.2082\,\pm\,0.0010 $
&$0.1173\,\pm \,0.0005 $&   $-0.2346\,\pm\,0.0010 $  & $0.0566\,\pm \,0.0002$&  $-0.1132\,\pm\, 0.0005 $ \\
\hline
$A_{6s}$          & $-0.5509\,\pm \,0.0026$& $ 0.9539\,\pm\,0.0077 $  & $-0.5521\,\pm \,0.0026$& $0.6915\,\pm\,0.0056 $ & $-0.4374 \,\pm \  0.0022 $ &$ 0.9638\,\pm\,0.0078$
&$-0.6141\,\pm \,0.0028$&$0.9325\,\pm\,0.0076 $    & $-0.2392\,\pm \,0.0023$ & $0.3422\,\pm\, 0.0014 $  \\
\hline
$A_{7}$          &   \multicolumn{2}{c|}{0.0000} & $-0.0130\pm0.0001$  &$-0.0260\pm 0.0001$ & $-0.0095\pm0.0001$ &$-0.0190\pm0.0001$ & $-0.0011\pm0.0001$ & $-0.0022\pm0.0001$ &  $-0.0062\pm0.00001$&  $-0.0124\pm0.0001$ \\
\hline
$A_{8}$          &    \multicolumn{2}{c|}{0.0000}    &   $-0.0102\pm 0.0001$ &  $ -0.0205\pm 0.0001$ &\multicolumn{2}{c|}{0.0000} &\multicolumn{2}{c|}{ 0.0000 }    &  \multicolumn{2}{c|}{ 0.0000 } \\
\hline
$A_{9}$          &    \multicolumn{2}{c|}{0.0000 }   &   $-0.0042\pm 0.0001$ &$ -0.0083 \pm 0.0001 $ &\multicolumn{2}{c|}{  0.0000} & \multicolumn{2}{c|}{ 0.0000} &  \multicolumn{2}{c|}{0.0000}  \\
\hline
$R_{D_s^*}$  & \multicolumn{2}{c|}{$ 0.2844\,\pm \, 0.0018 $ }   & \multicolumn{2}{c|}{$ 0.2842\,\pm \, 0.0018 $} & \multicolumn{2}{c|}{$ 0.2405 \pm 0.0015 $ }
& \multicolumn{2}{c|}{$0.2486 \,\pm \,  0.0016$}             &   \multicolumn{2}{c|}{$0.2781\,\pm \,0.0022 $}   \\
\hline
$A_{FB}^{\tau}$    & \multicolumn{2}{c|}{$-0.0896\,\pm \,0.0020 $} & \multicolumn{2}{c|}{$-0.0310\,\pm \,0.0016$} & \multicolumn{2}{c|}{$-0.1170 \,\pm \,  0.0020 $}
&\multicolumn{2}{c|}{$-0.0708  \,\pm \, 0.0021$}   & \multicolumn{2}{c|}{$-0.0229\,\pm \,0.0007$}  \\
\hline
$A_{FB}^{T}$       & \multicolumn{2}{c|}{$-0.3842  \,\pm \, 0.0026$} & \multicolumn{2}{c|}{$-0.2790\,\pm \,0.0019$} & \multicolumn{2}{c|}{ $-0.3842  \,\pm \, 0.0026$ }
&\multicolumn{2}{c|}{$-0.3842  \,\pm \, 0.0026$}   & \multicolumn{2}{c|}{$-0.1124\,\pm \,0.0002$ } \\
\hline
$F_L^{D_s^*}$      & \multicolumn{2}{c|}{$0.4482\,\pm \,0.0015$} & \multicolumn{2}{c|}{$0.4493\,\pm \,0.0015$}  & \multicolumn{2}{c|}{$ 0.4425  \,\pm \ 0.0015 $}
& \multicolumn{2}{c|}{$ 0.4606 \,\pm \,  0.0015$}    & \multicolumn{2}{c|}{$ 0.3237\,\pm \,0.0021$}  \\
\hline
$C_F^l$           & \multicolumn{2}{c|}{$-0.0550\,\pm \,0.0014 $}  & \multicolumn{2}{c|}{$-0.0558\,\pm \,0.0014$} & \multicolumn{2}{c|}{$-0.0555 \,\pm \  0.0014 $}
& \multicolumn{2}{c|}{$-0.0537\,\pm \,0.0014$}    & \multicolumn{2}{c|}{$-0.1316   \,\pm \,0.0004$}   \\
\hline
\end{tabular}}
 \caption{ Prediction of $B_s\to D_s^*(\to D_s\gamma,D_s\pi)\,\tau\,\nu$ decay observables in Scenerio - III.}
\label{dsg_sc3}
\end{table}

\begin{figure}[htbp]
\centering
\includegraphics[width=4cm,height=3cm]{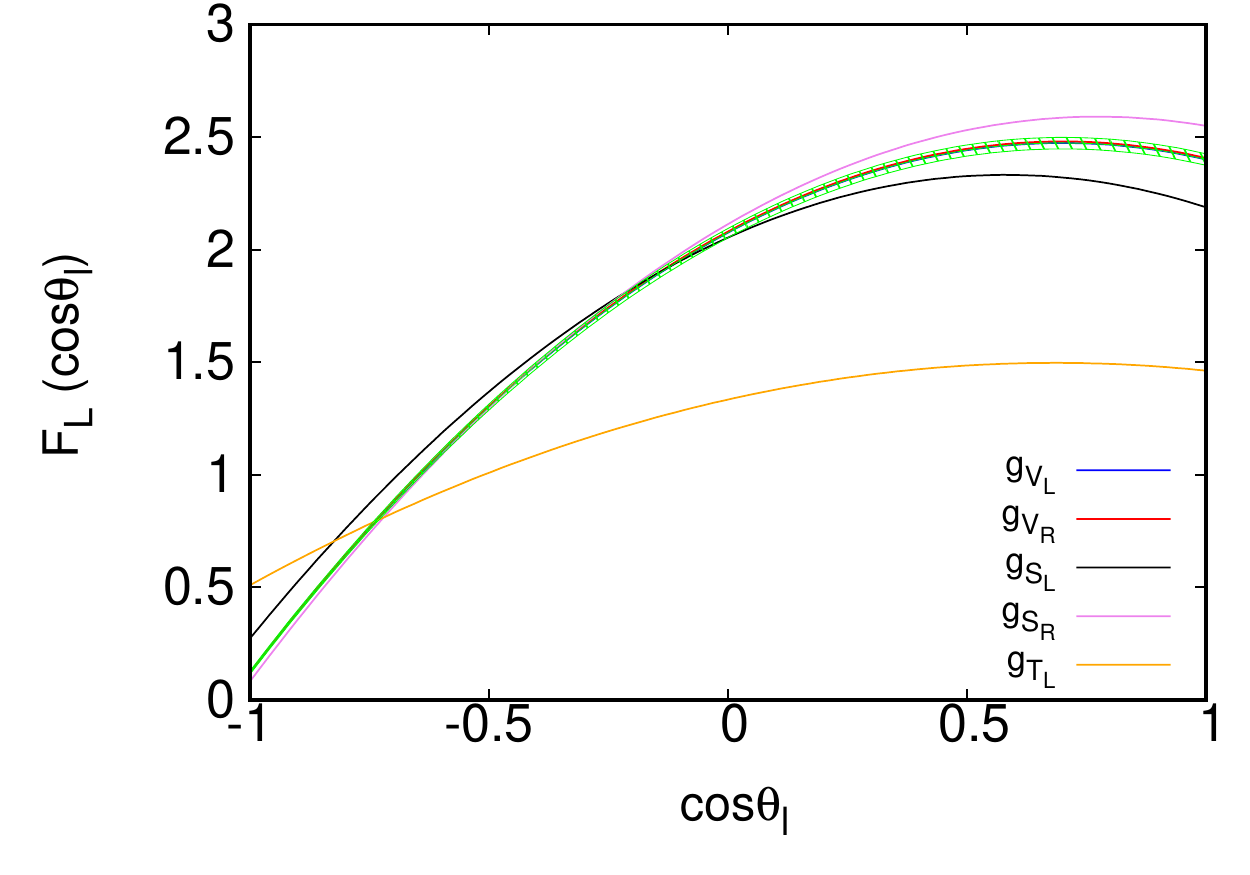}
\includegraphics[width=4cm,height=3cm]{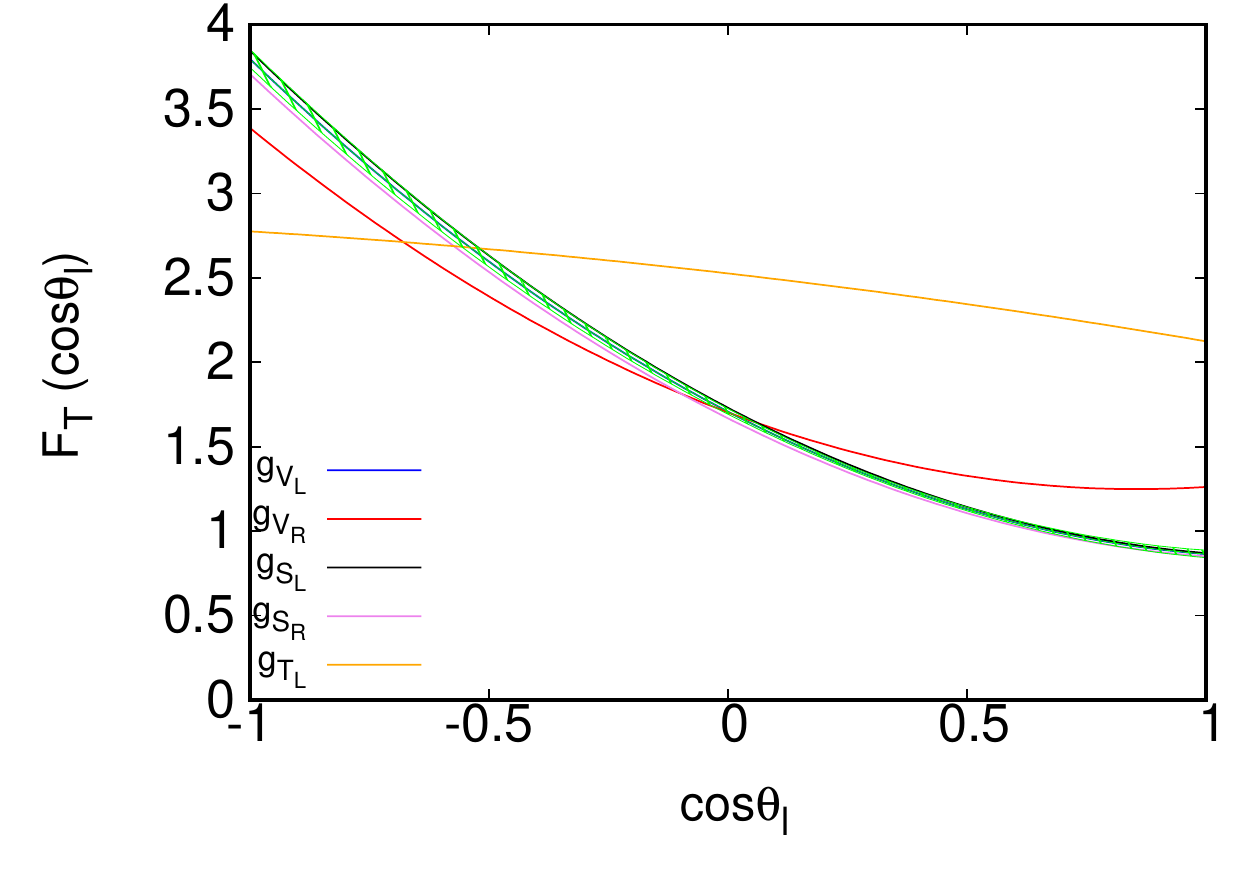}
\includegraphics[width=4cm,height=3cm]{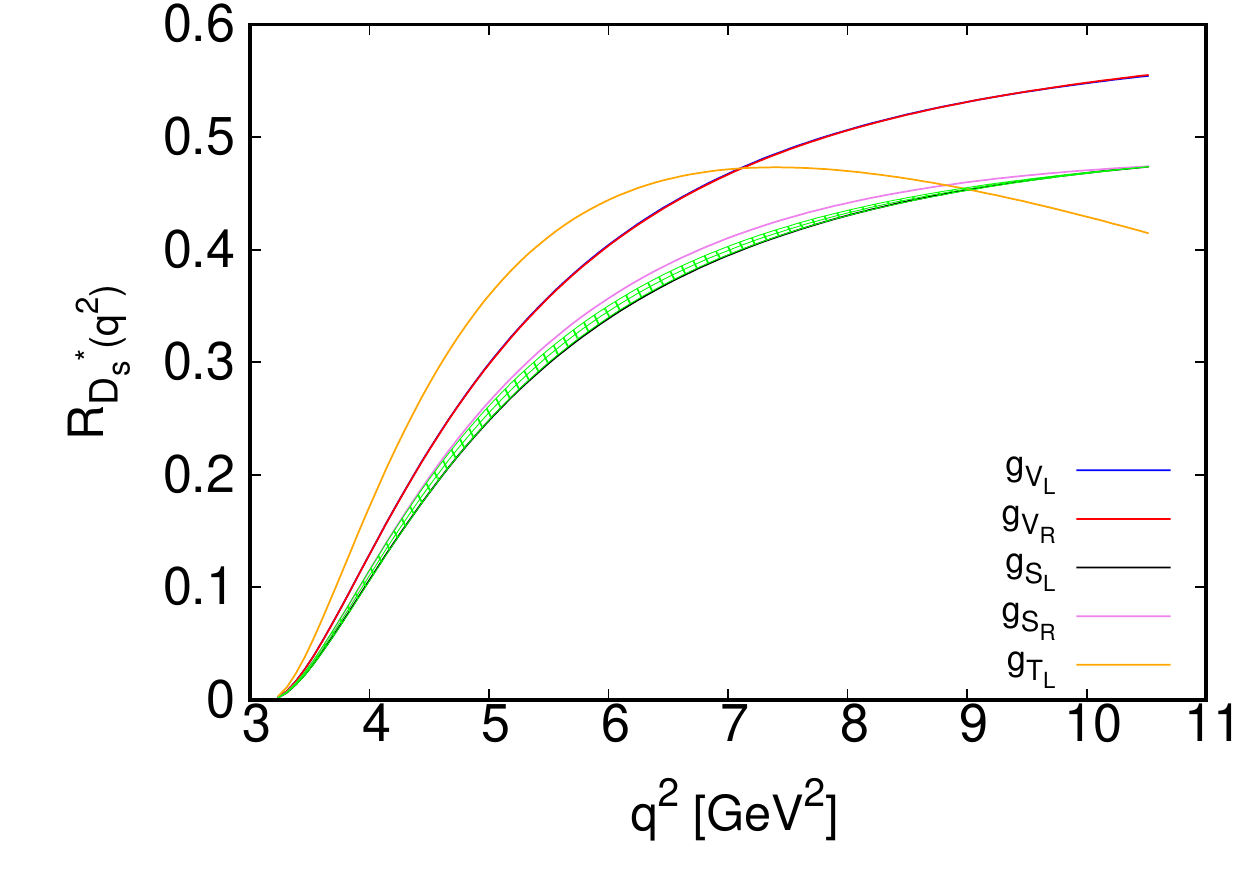}
\includegraphics[width=4cm,height=3cm]{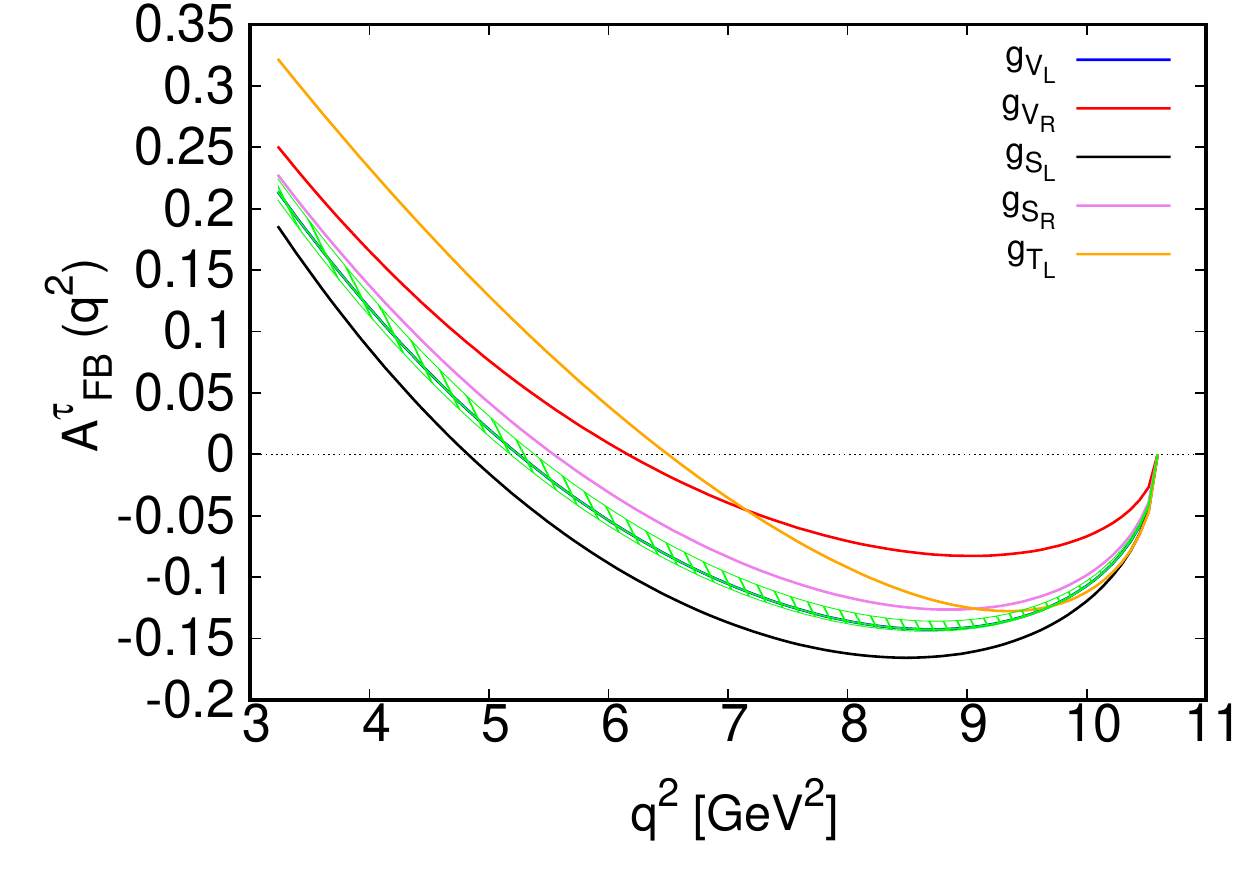}
\includegraphics[width=4cm,height=3cm]{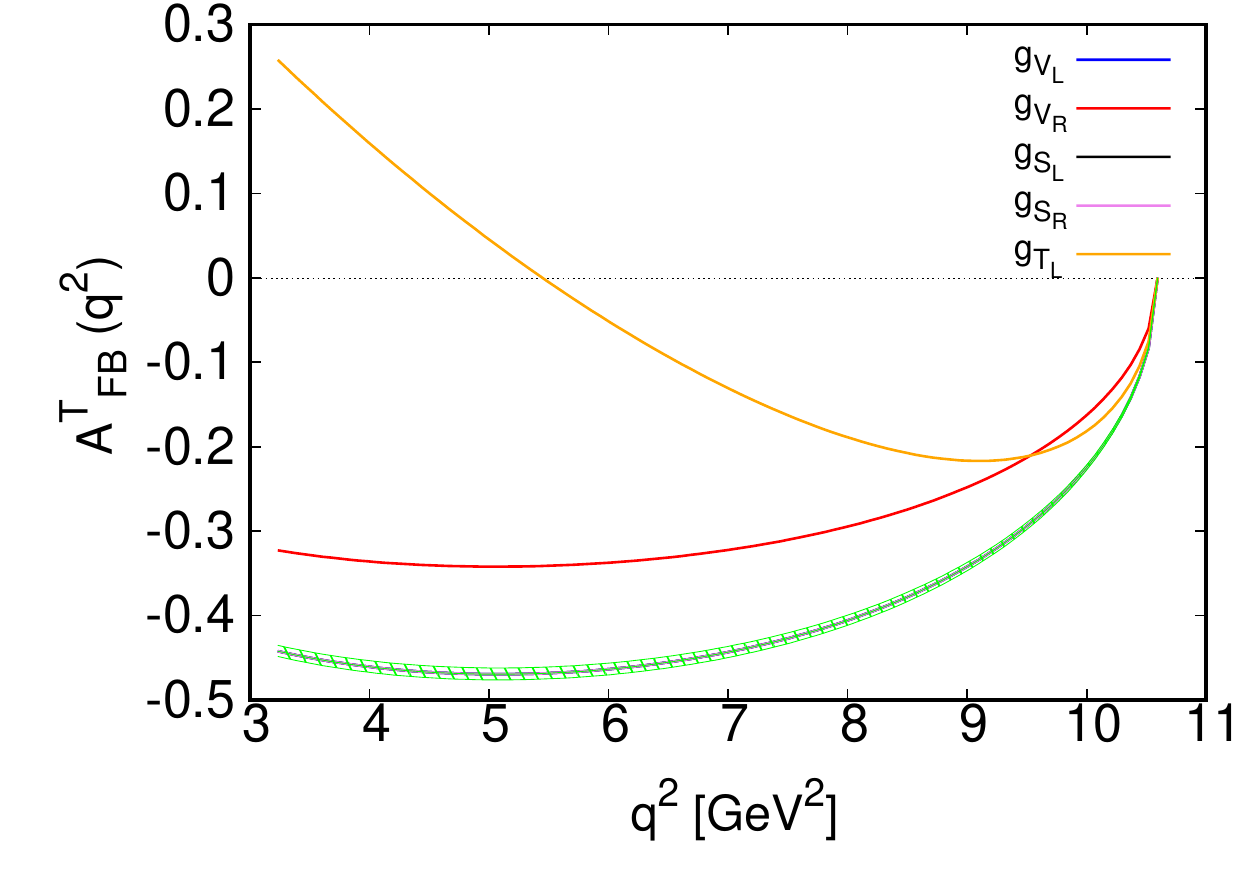}
\includegraphics[width=4cm,height=3cm]{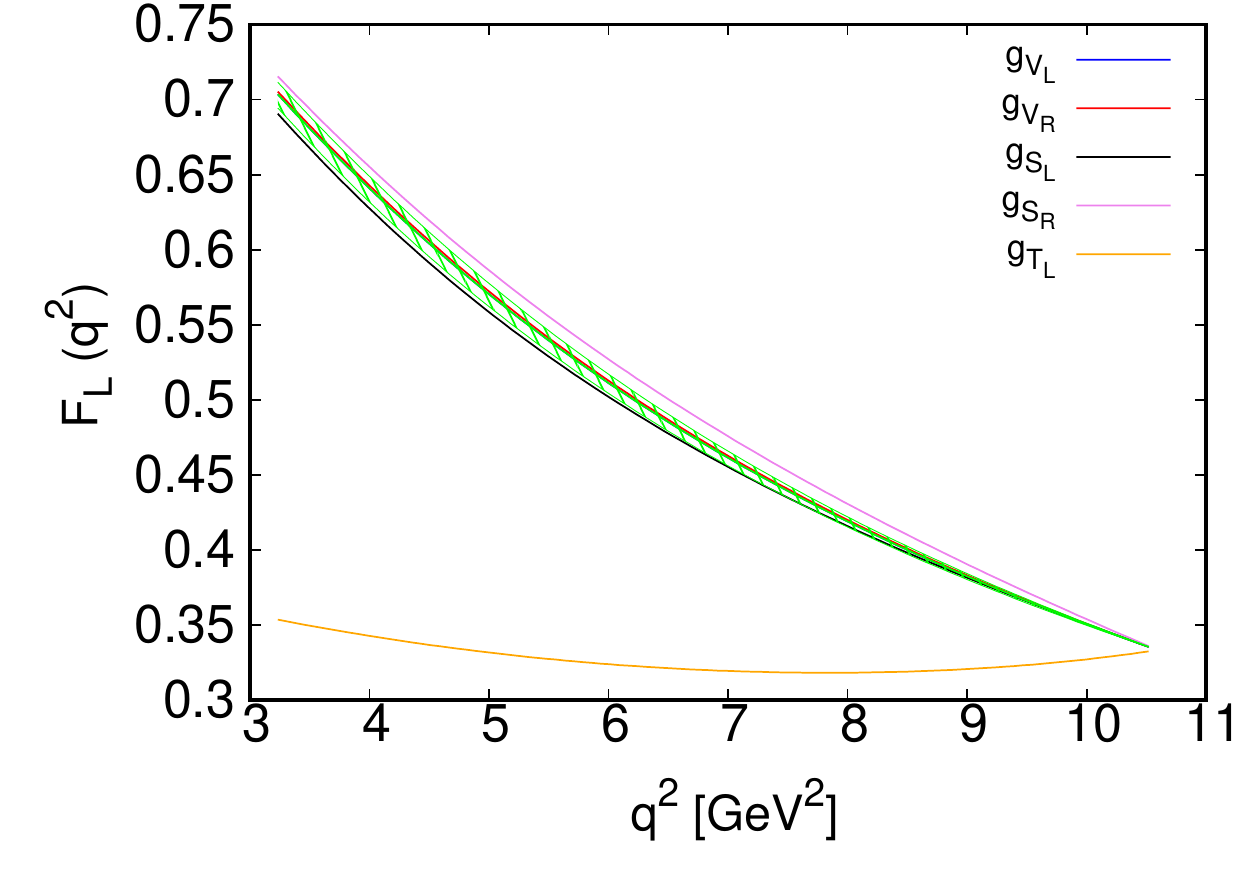}
\includegraphics[width=4cm,height=3cm]{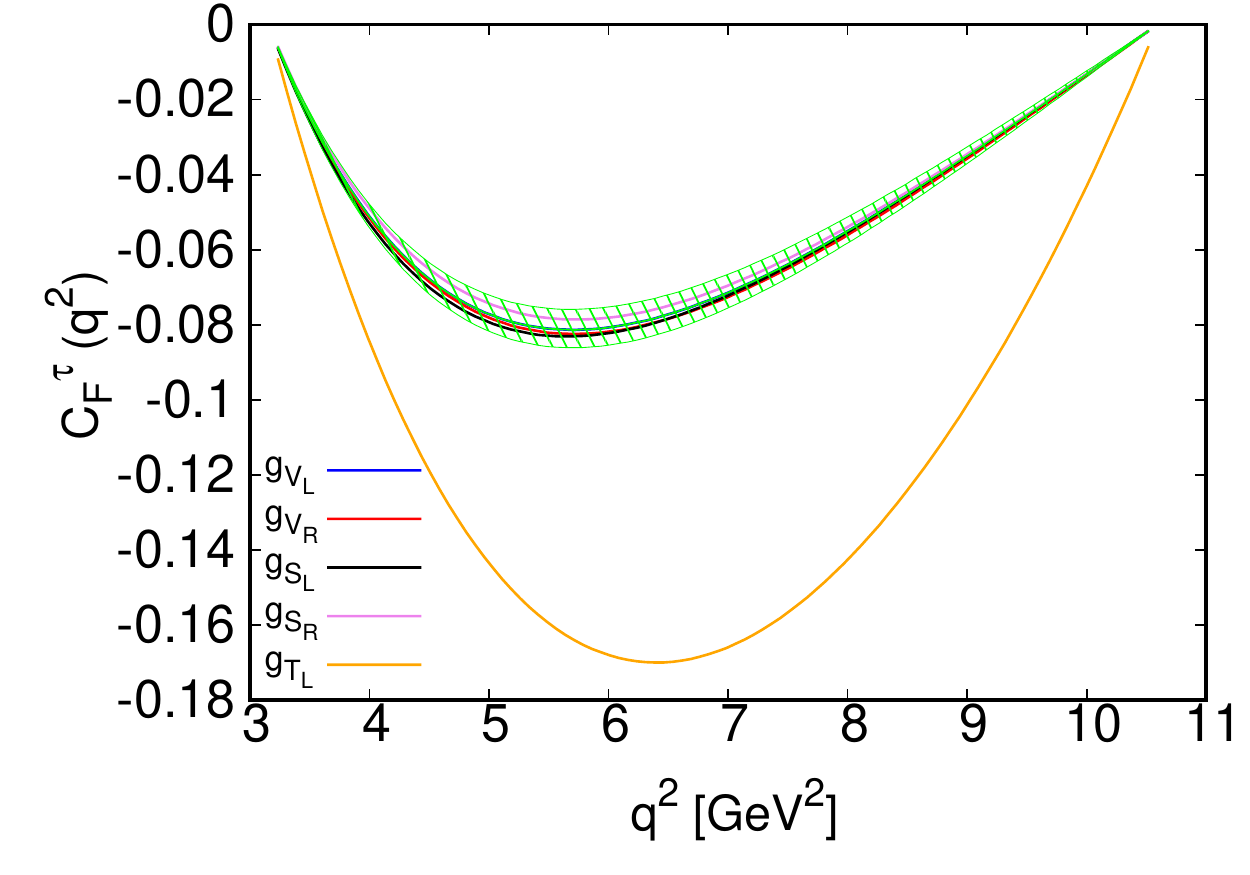}
\caption{The $q^2$ and $\cos\theta_{l}$ dependence of various physical observable of $B_s\to D_s^*(\to D_s \gamma, D_s\pi )\tau \nu$ in the SM and in the presence of the NP couplings of scenario - III. The SM central line and the corresponding error band are shown with green color. The  blue, red, black, violet and orange colors represents the effect of NP coupling $g_{V_L}$, $g_{V_R}$, $g_{S_{L}}$, $g_{S_R}$ and $g_{T_L}$  respectively.}
\label{dsg_sc3_same}
\end{figure}

\begin{figure}[htbp]
\centering
\includegraphics[width=4cm,height=3cm]{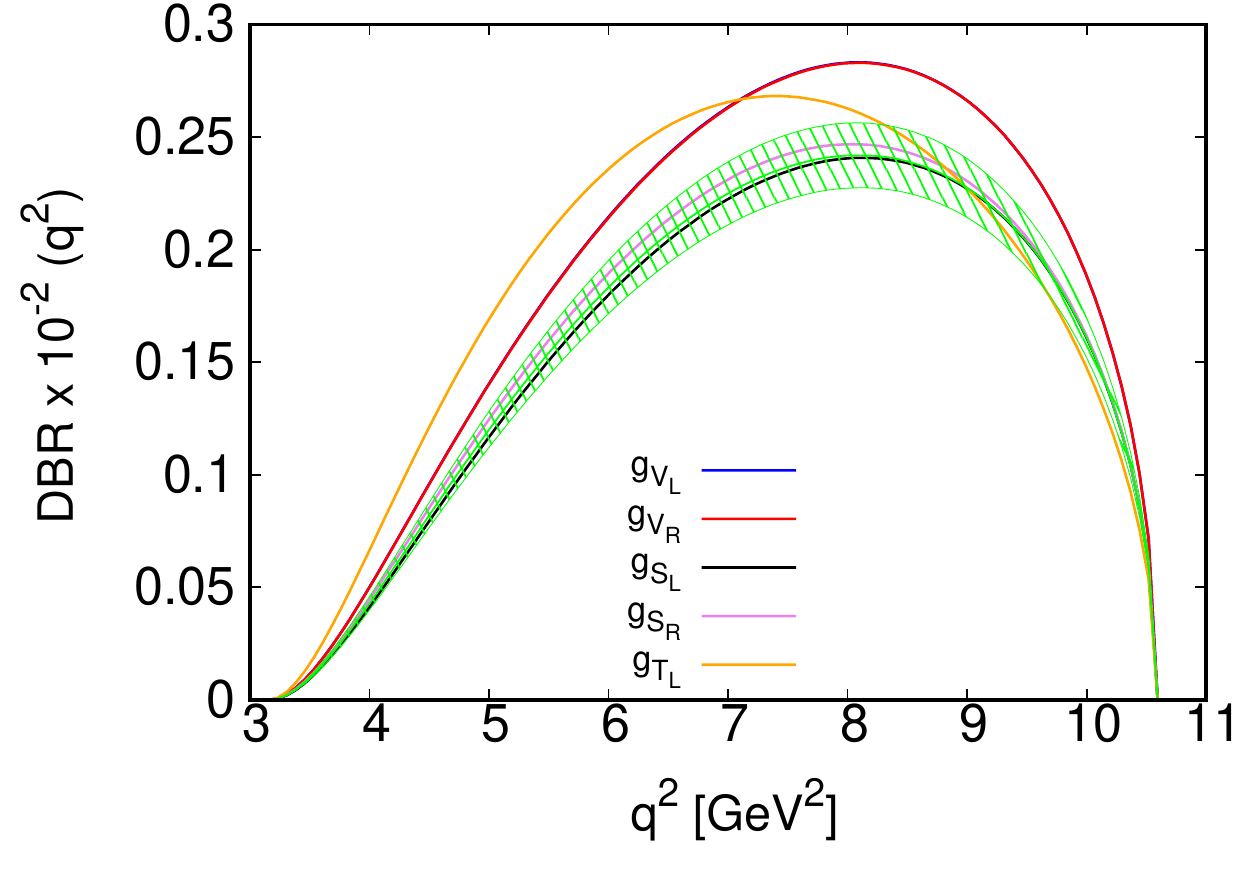}
\includegraphics[width=4cm,height=3cm]{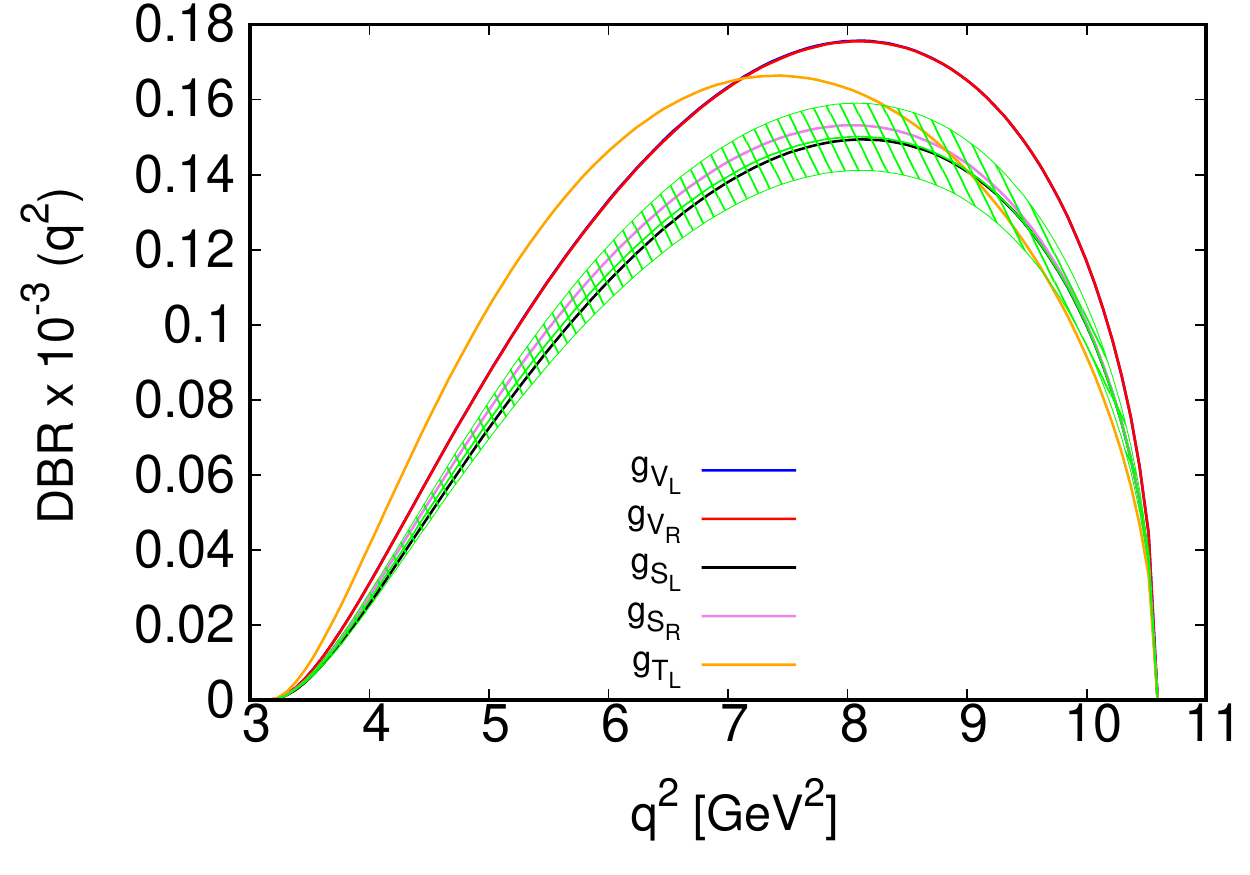}\hspace{1.5cm}
\includegraphics[width=4cm,height=3cm]{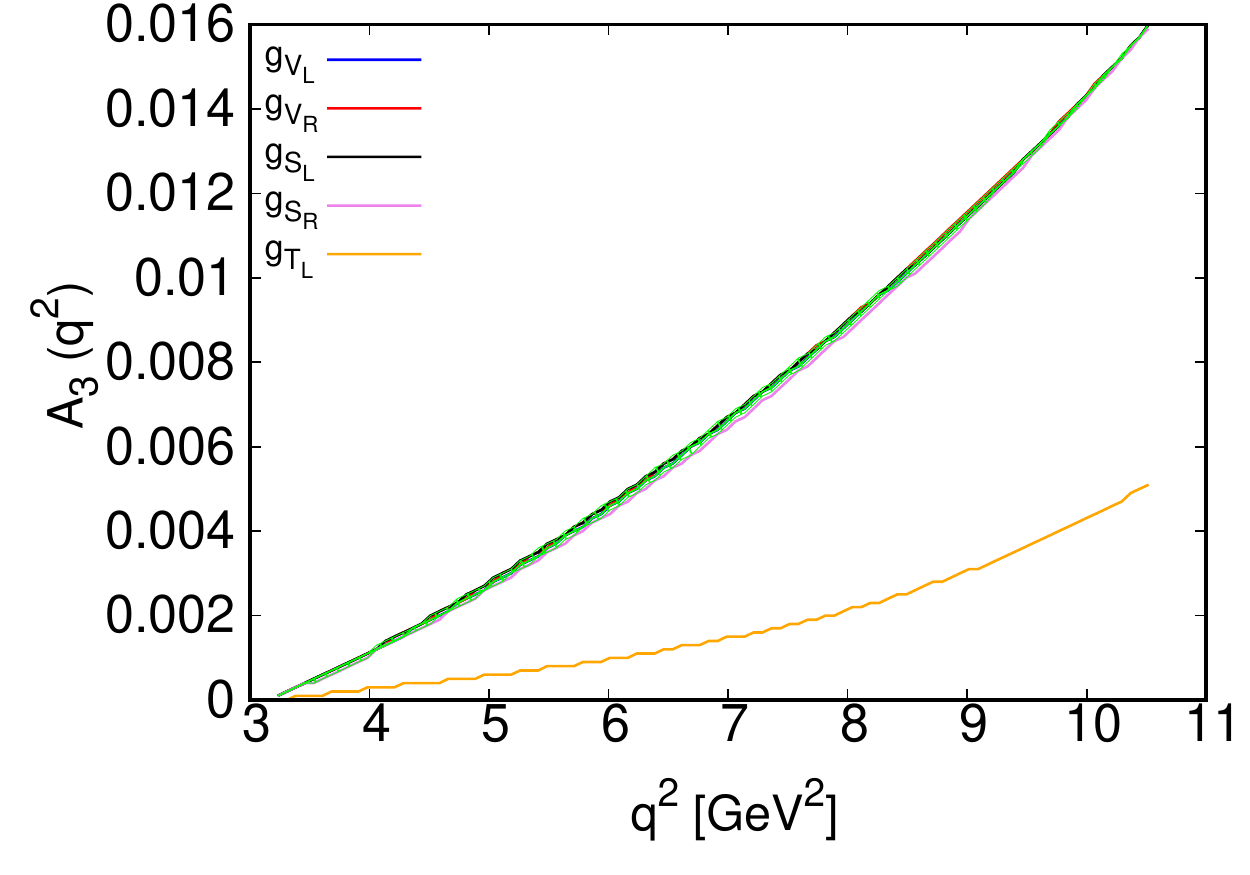}
\includegraphics[width=4cm,height=3cm]{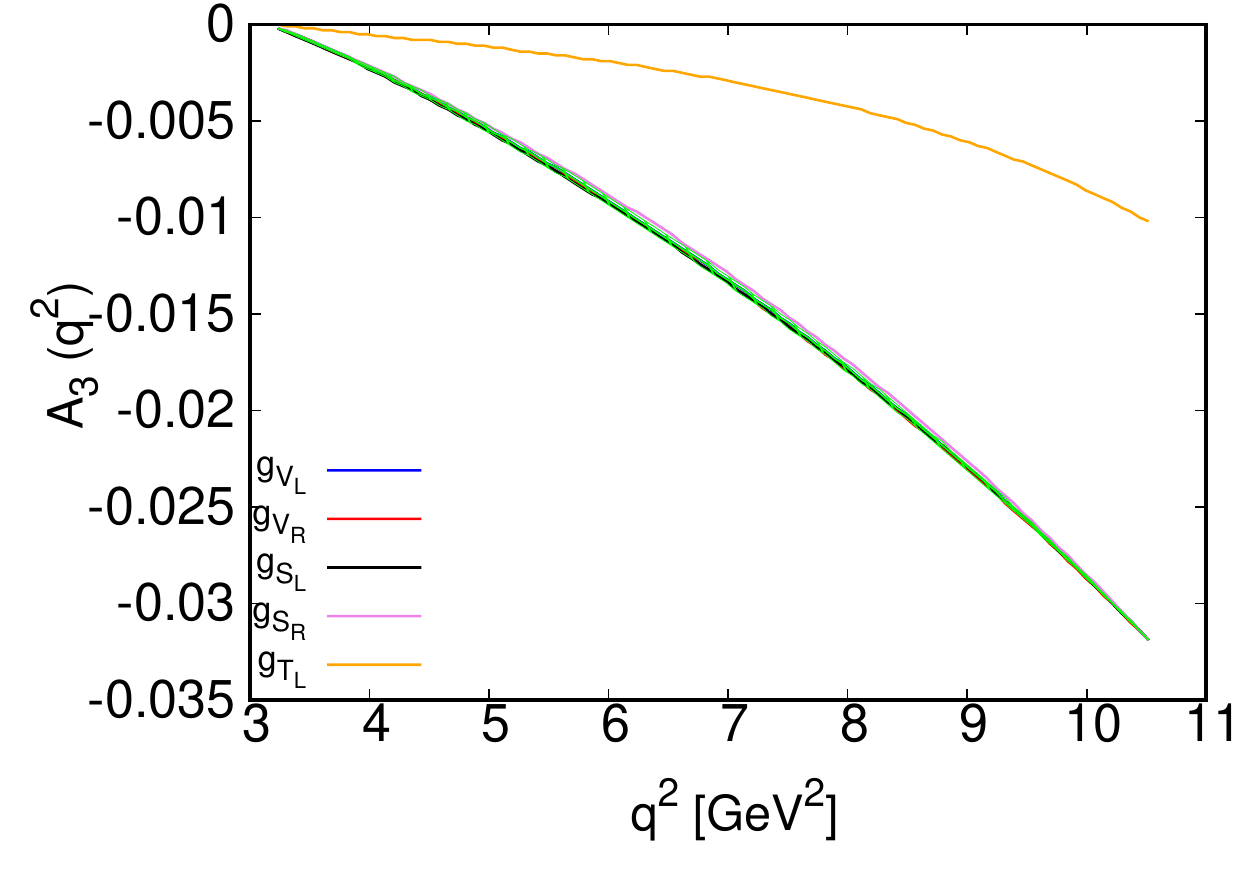}\hspace{1.5cm}
\includegraphics[width=4cm,height=3cm]{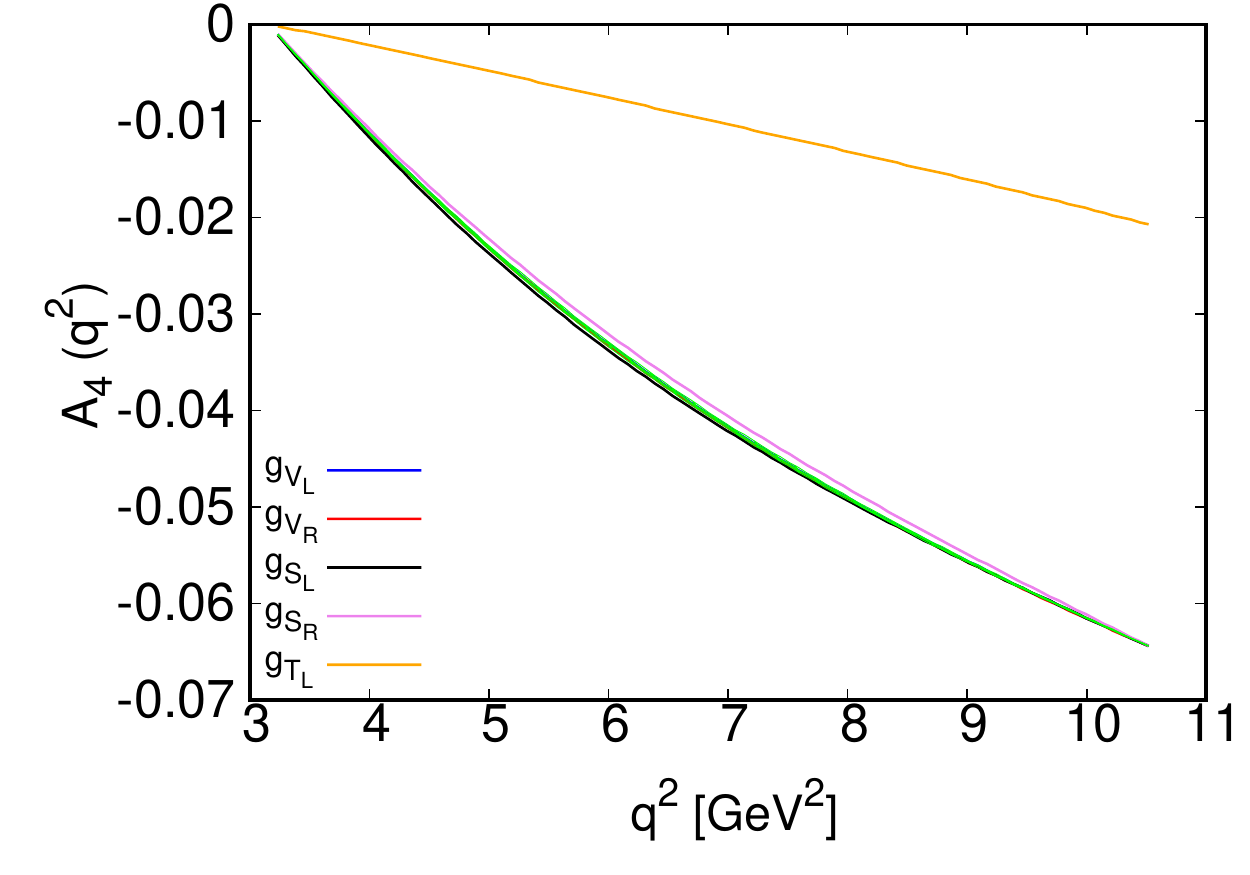}
\includegraphics[width=4cm,height=3cm]{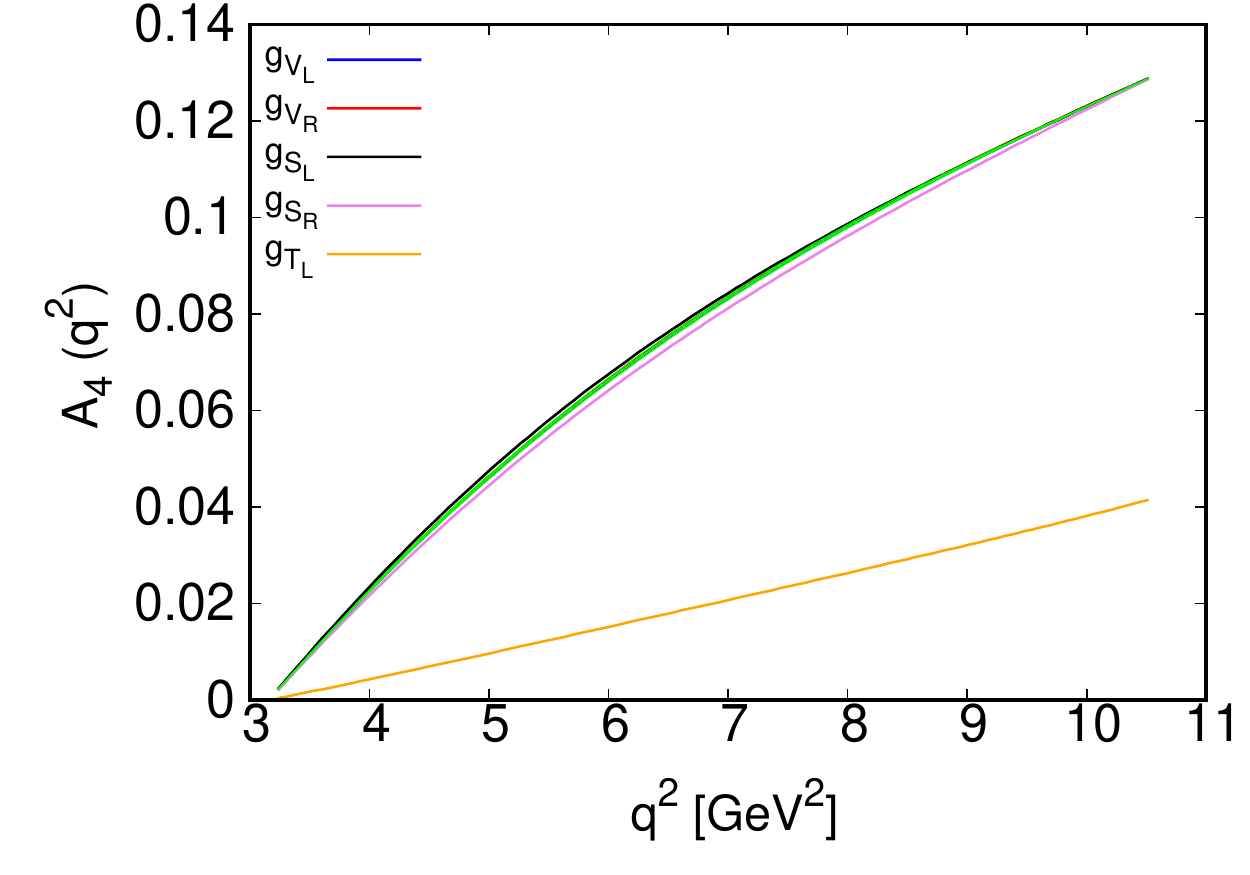}\hspace{1.5cm}
\includegraphics[width=4cm,height=3cm]{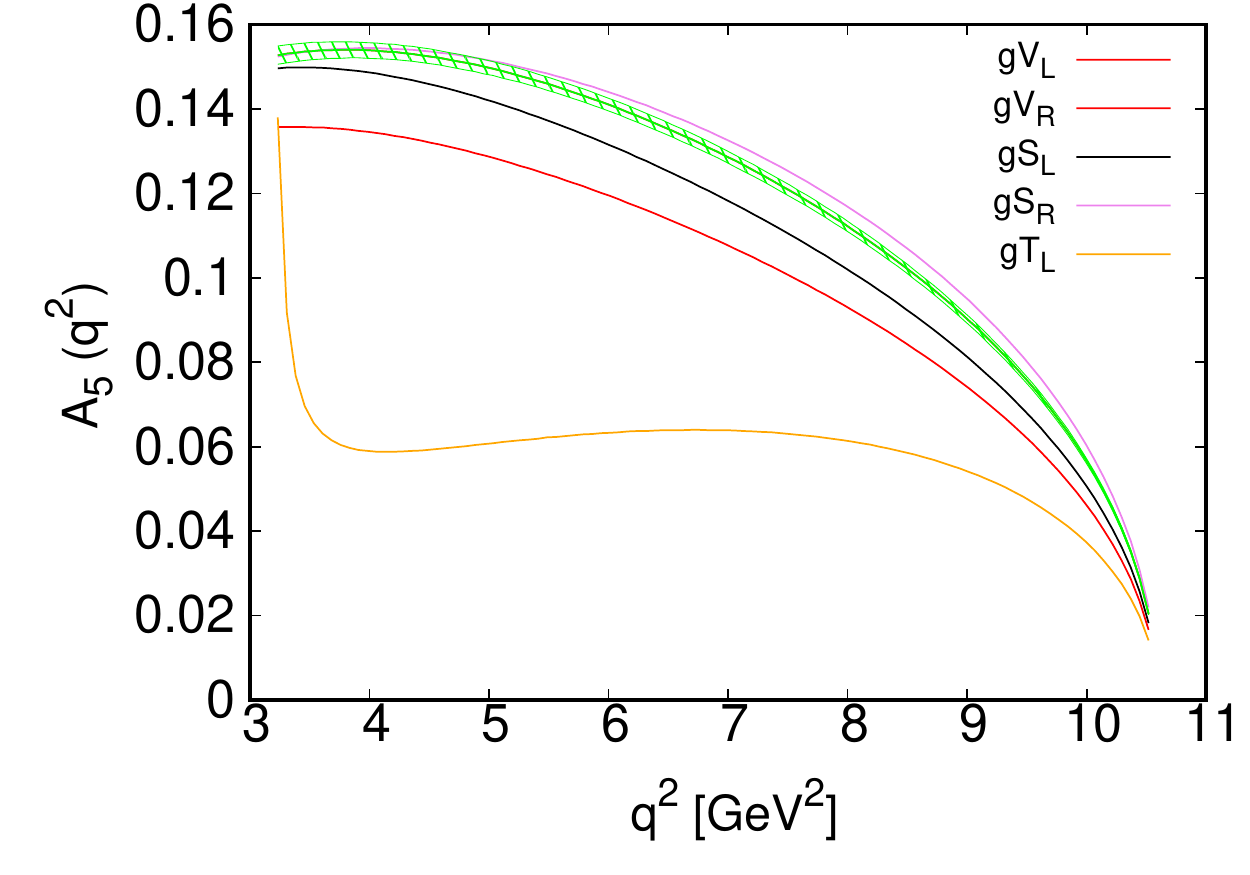}
\includegraphics[width=4cm,height=3cm]{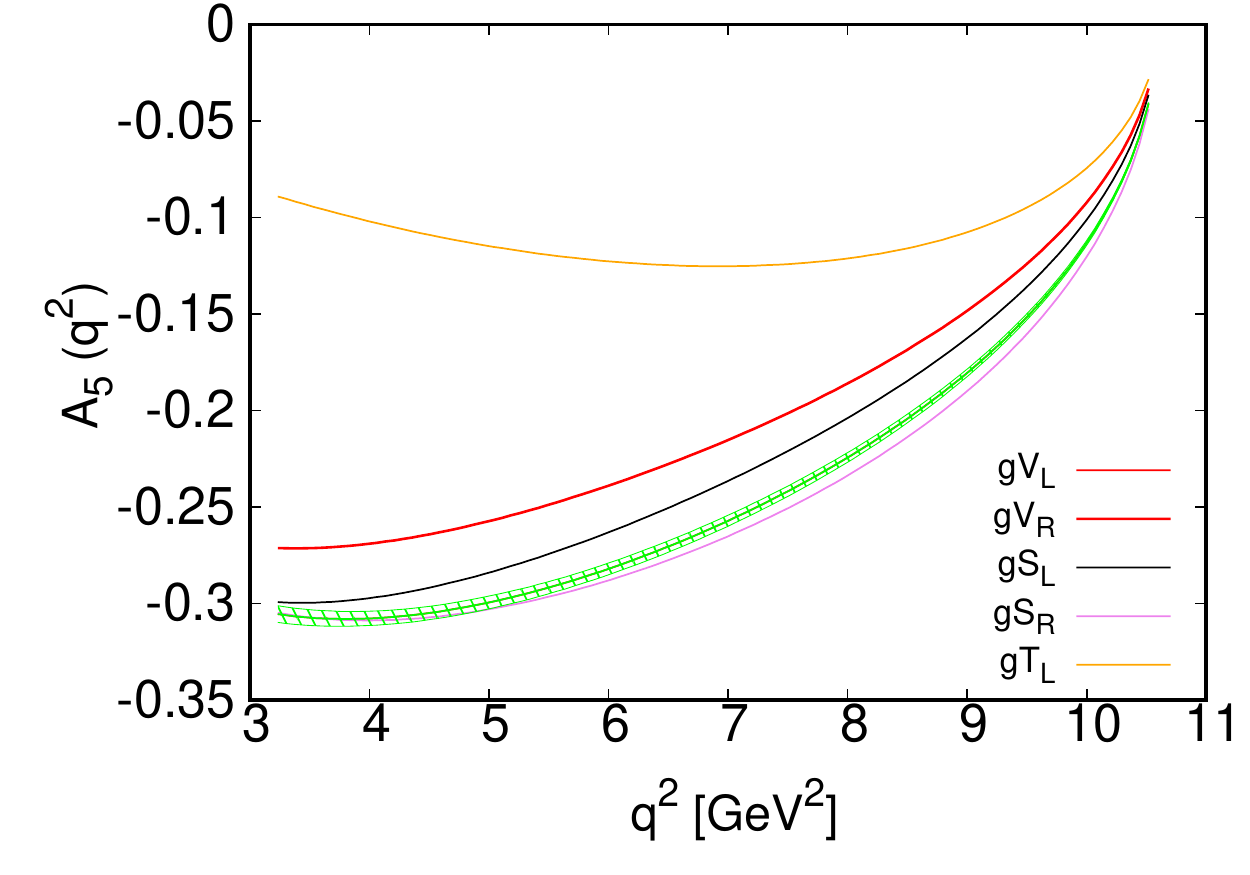}\hspace{1.5cm}
\includegraphics[width=4cm,height=3cm]{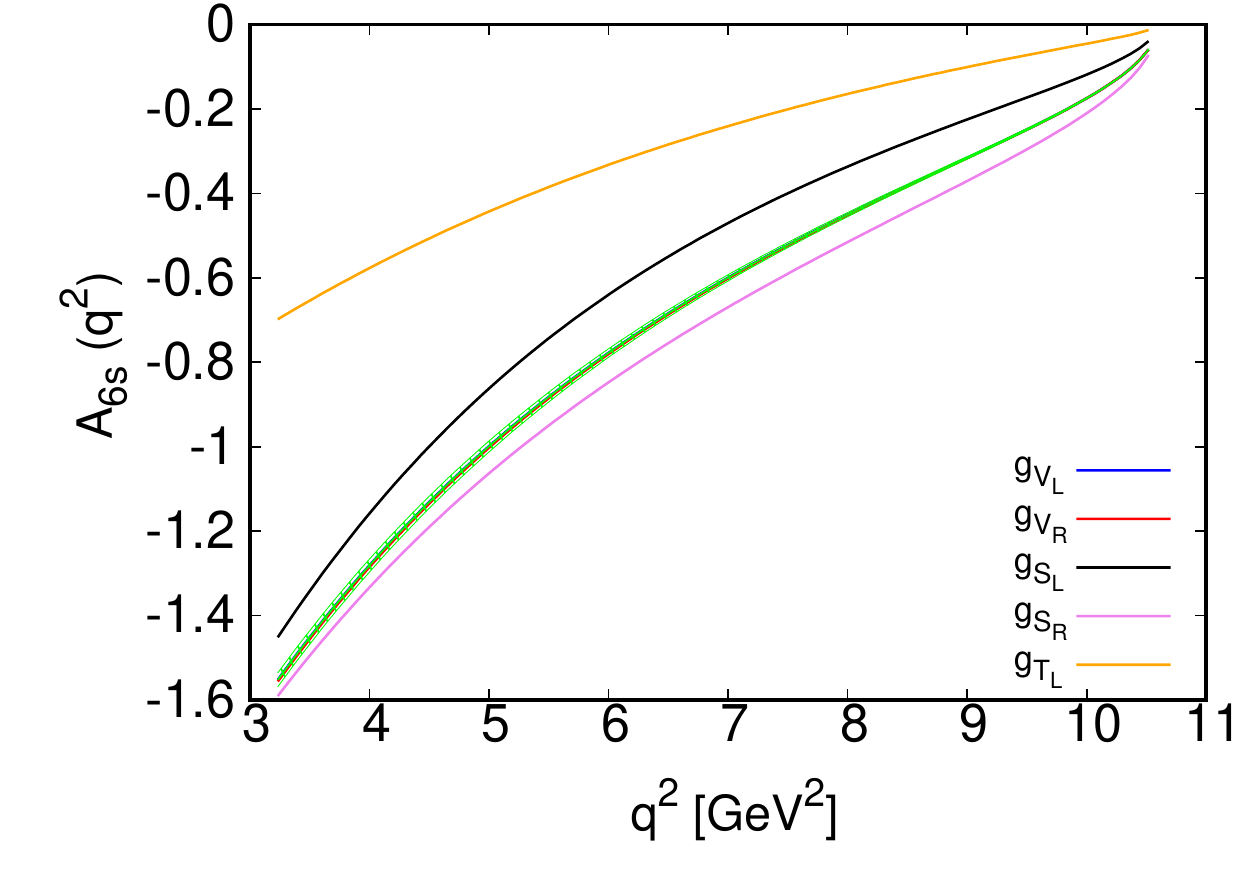}
\includegraphics[width=4cm,height=3cm]{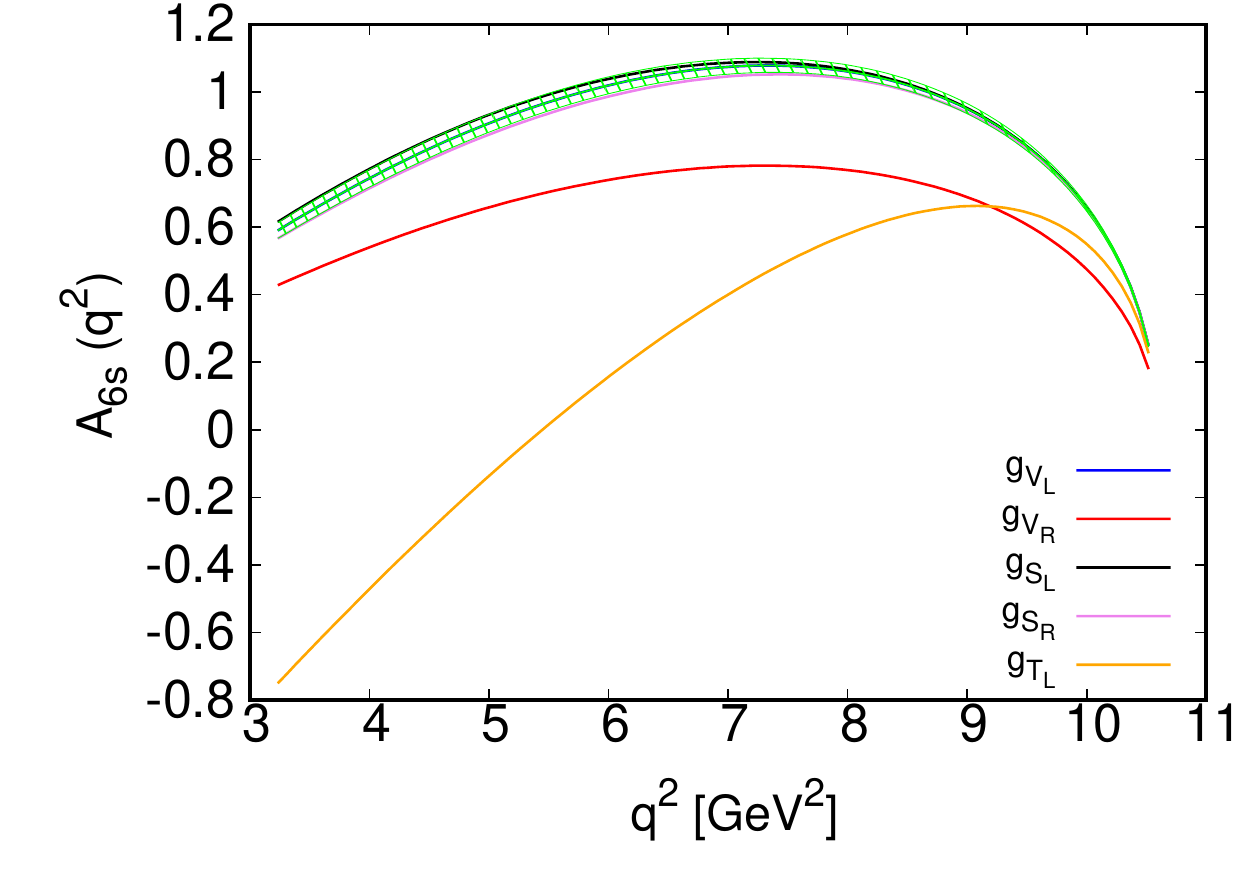}\hspace{1.5cm}
\includegraphics[width=4cm,height=3cm]{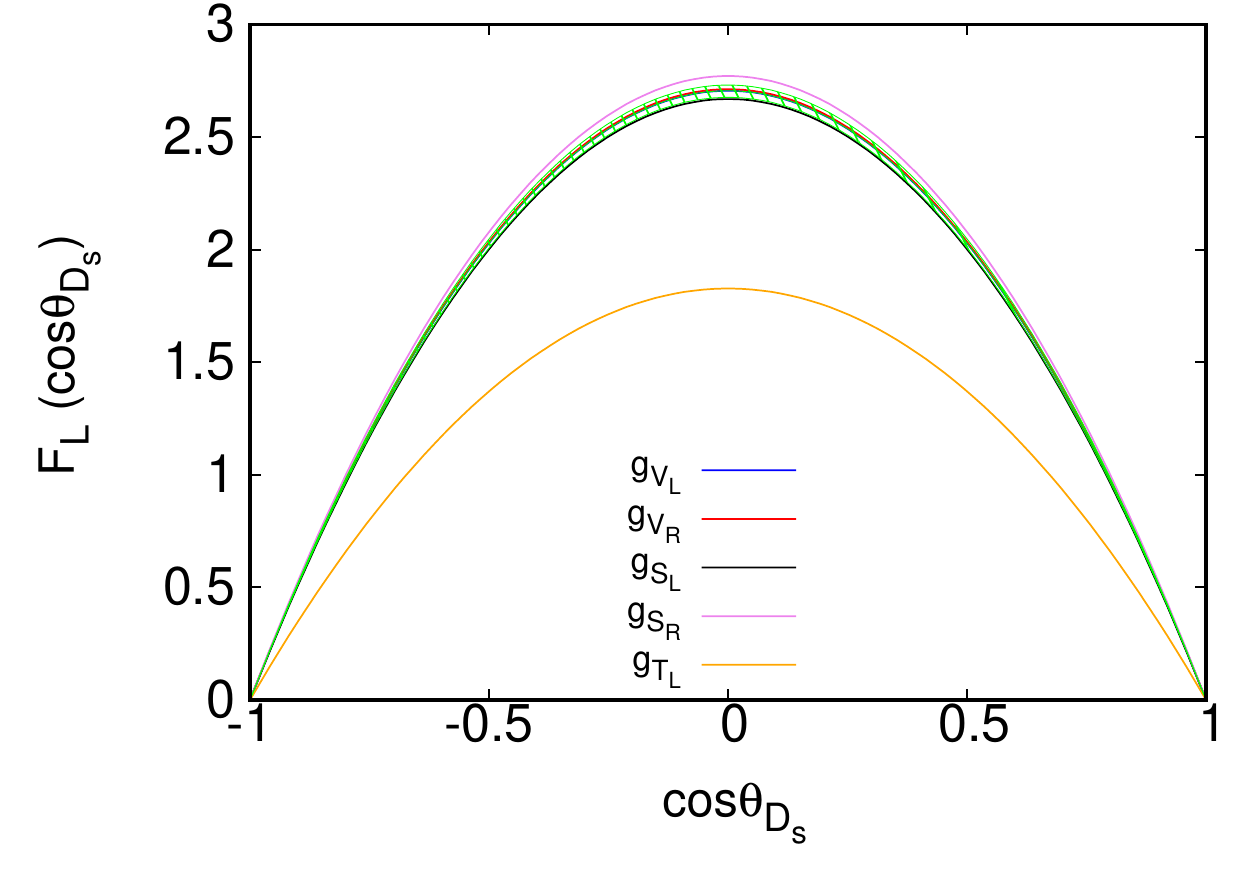}
\includegraphics[width=4cm,height=3cm]{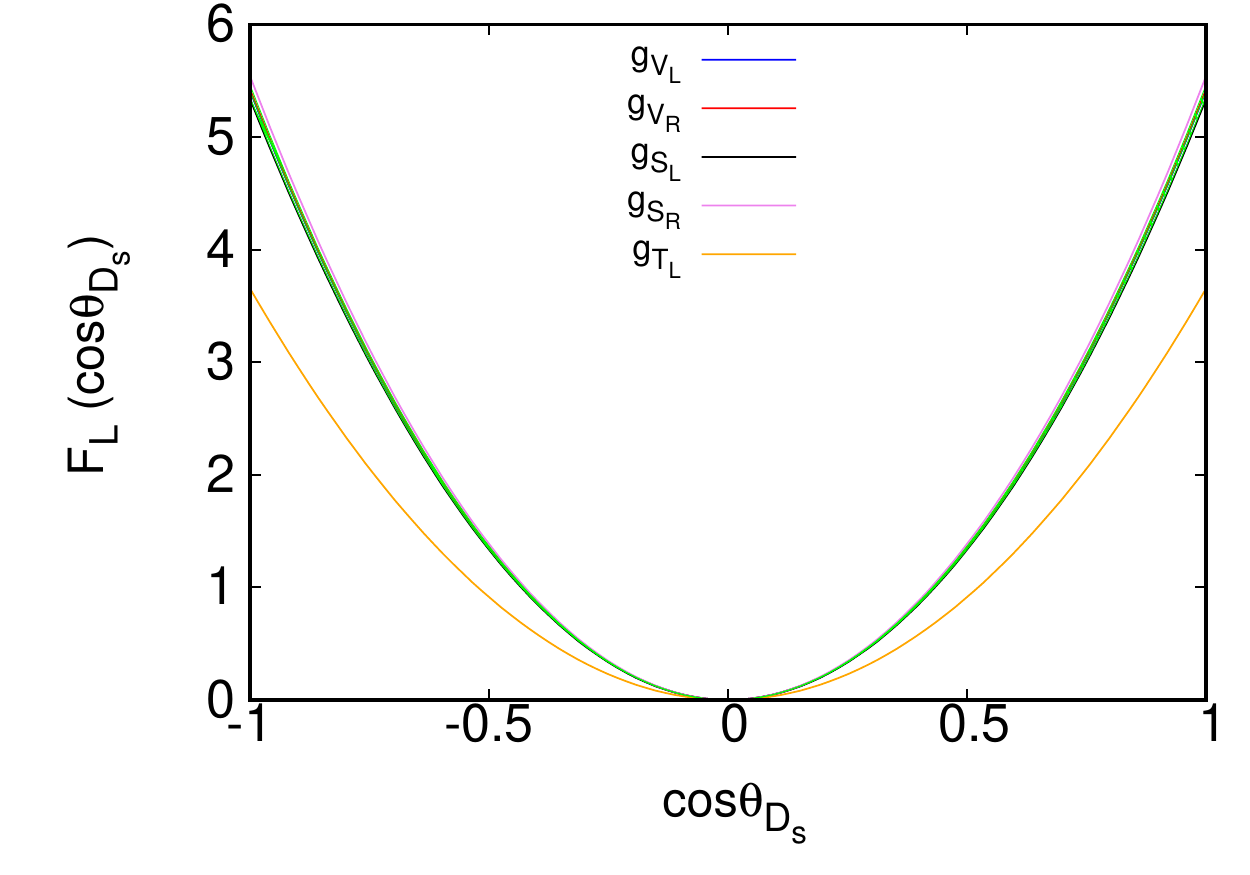}\hspace{1.5cm}
\includegraphics[width=4cm,height=3cm]{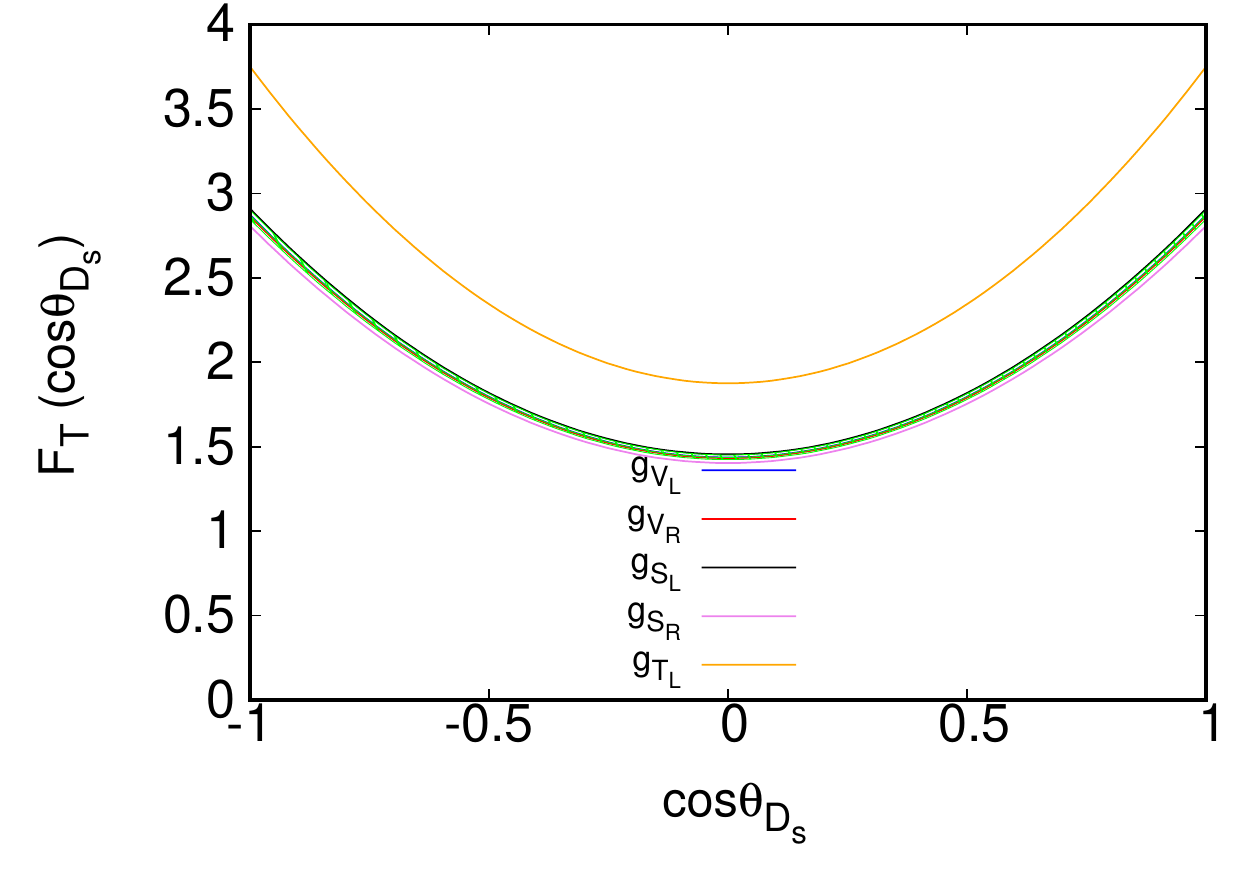}
\includegraphics[width=4cm,height=3cm]{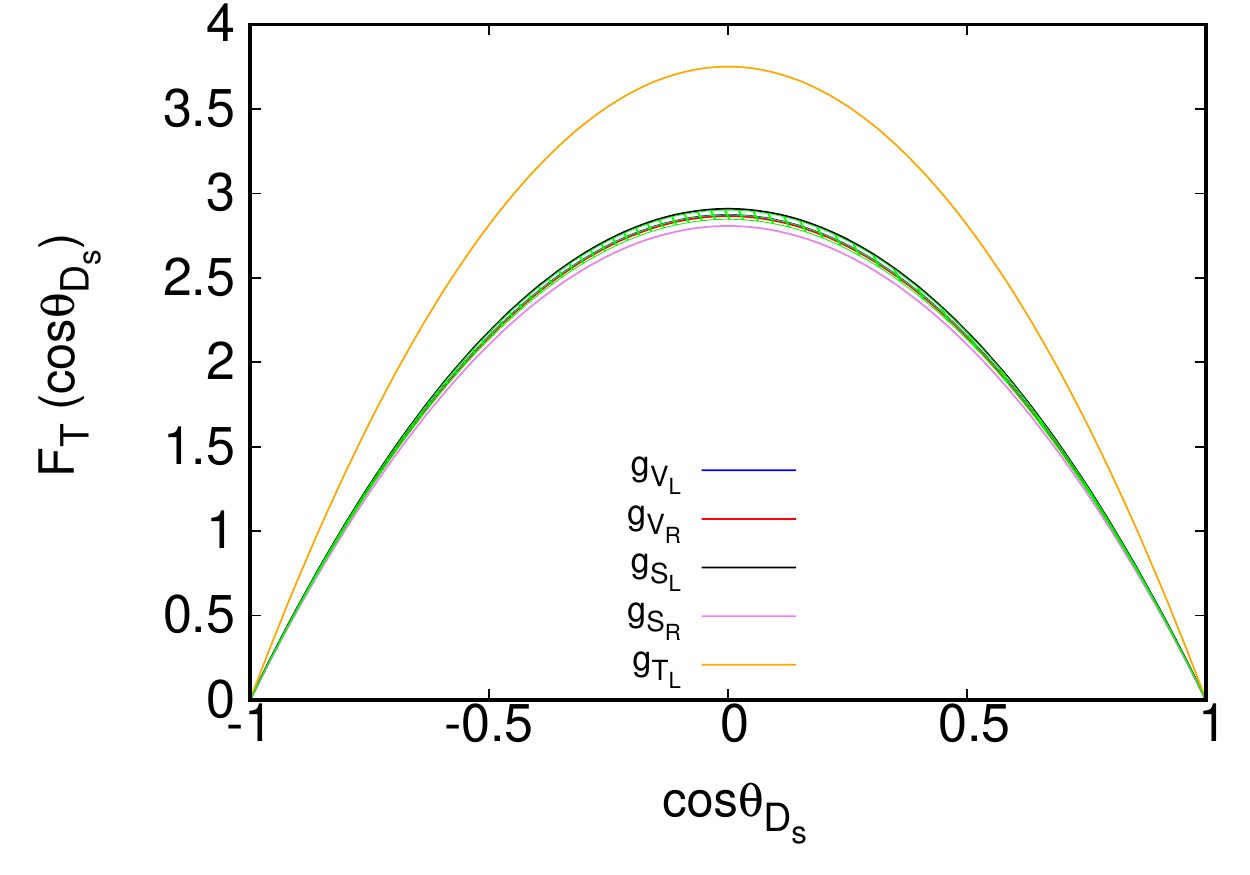}\hspace{1.5cm}
\includegraphics[width=4cm,height=3cm]{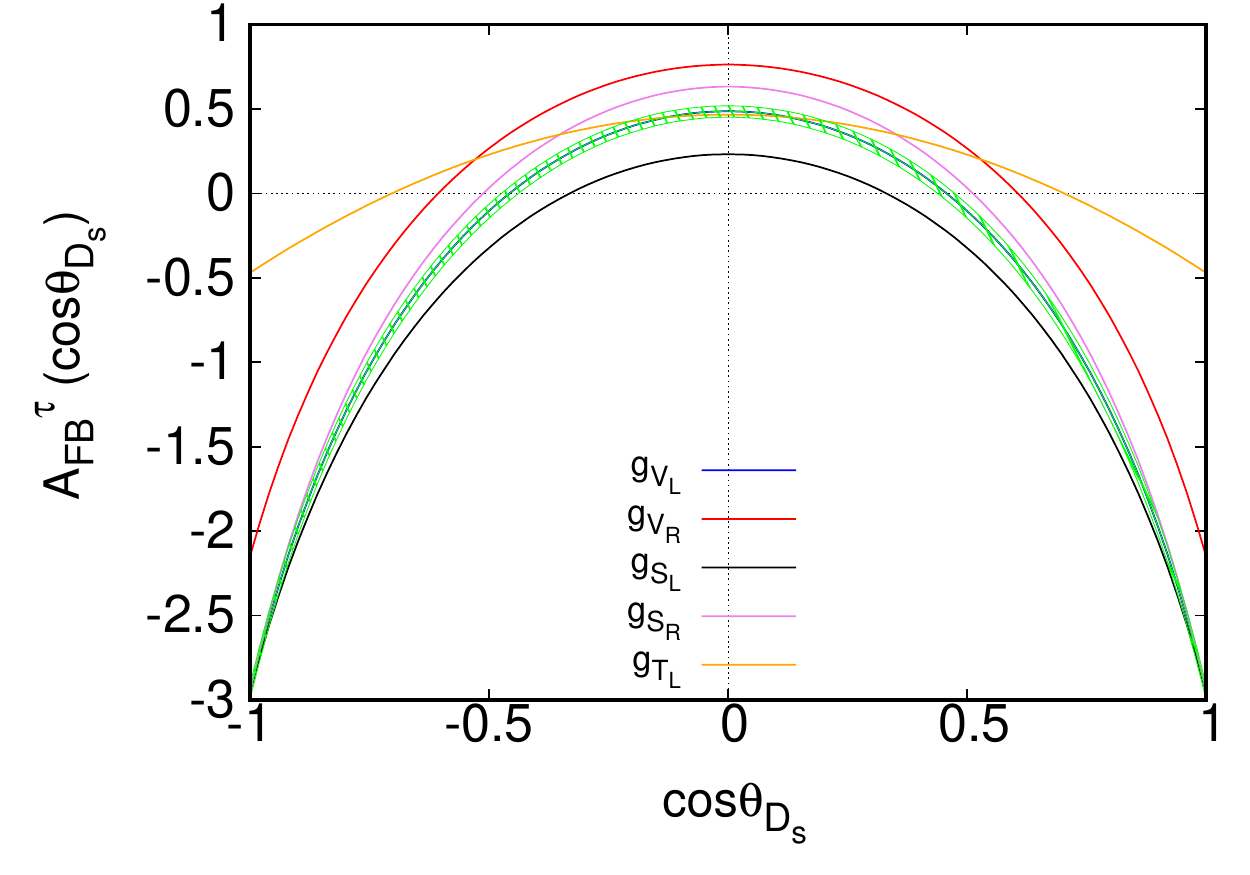}
\includegraphics[width=4cm,height=3cm]{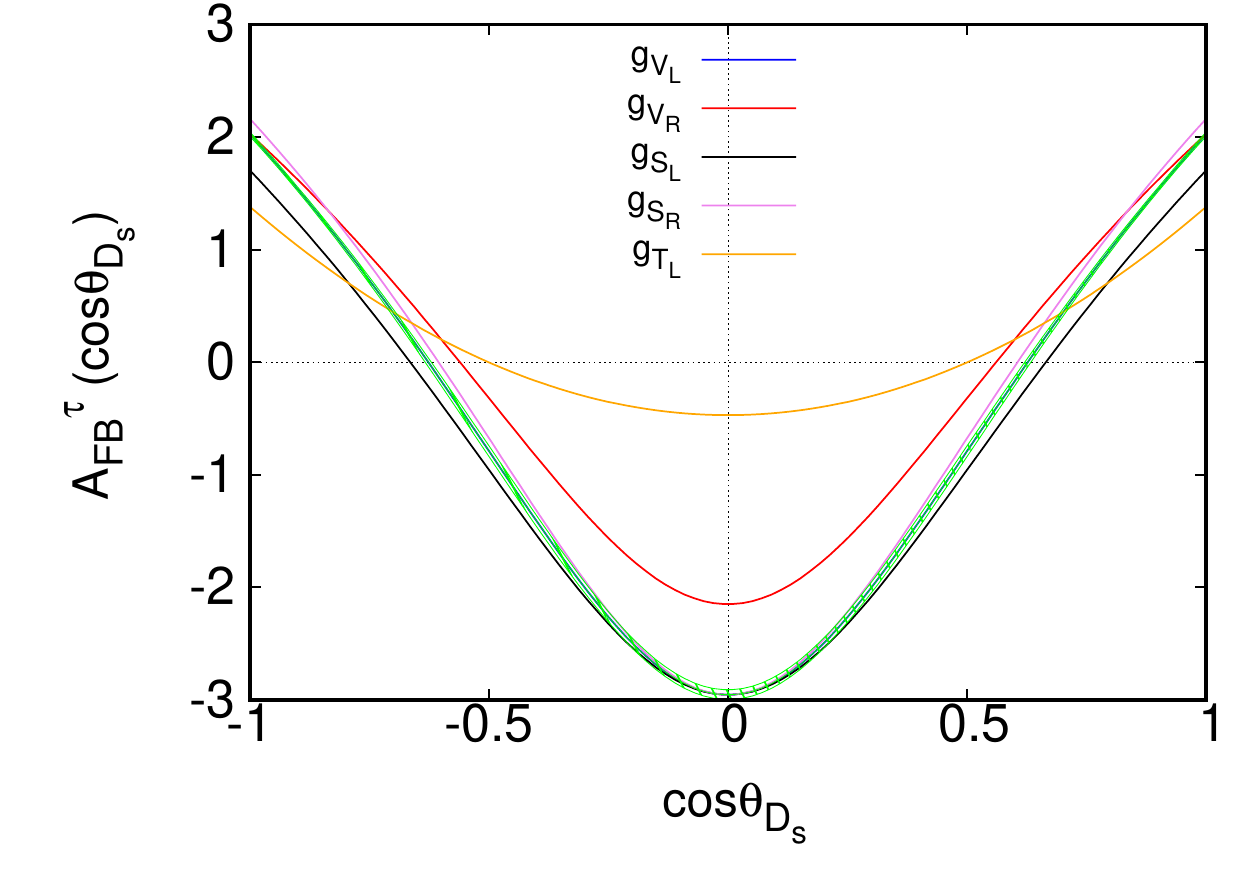}\hspace{1.5cm}
\includegraphics[width=4cm,height=3cm]{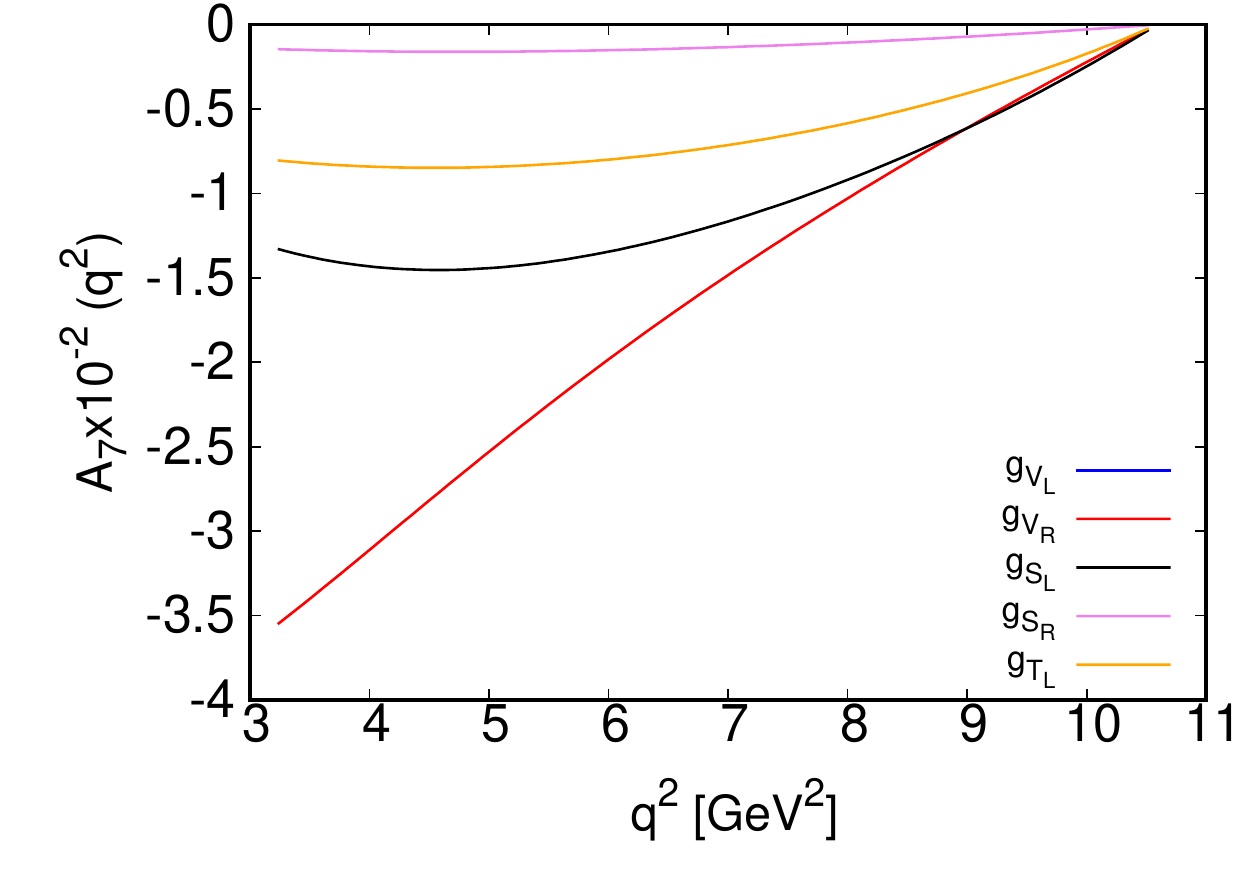}
 \includegraphics[width=4cm,height=3cm]{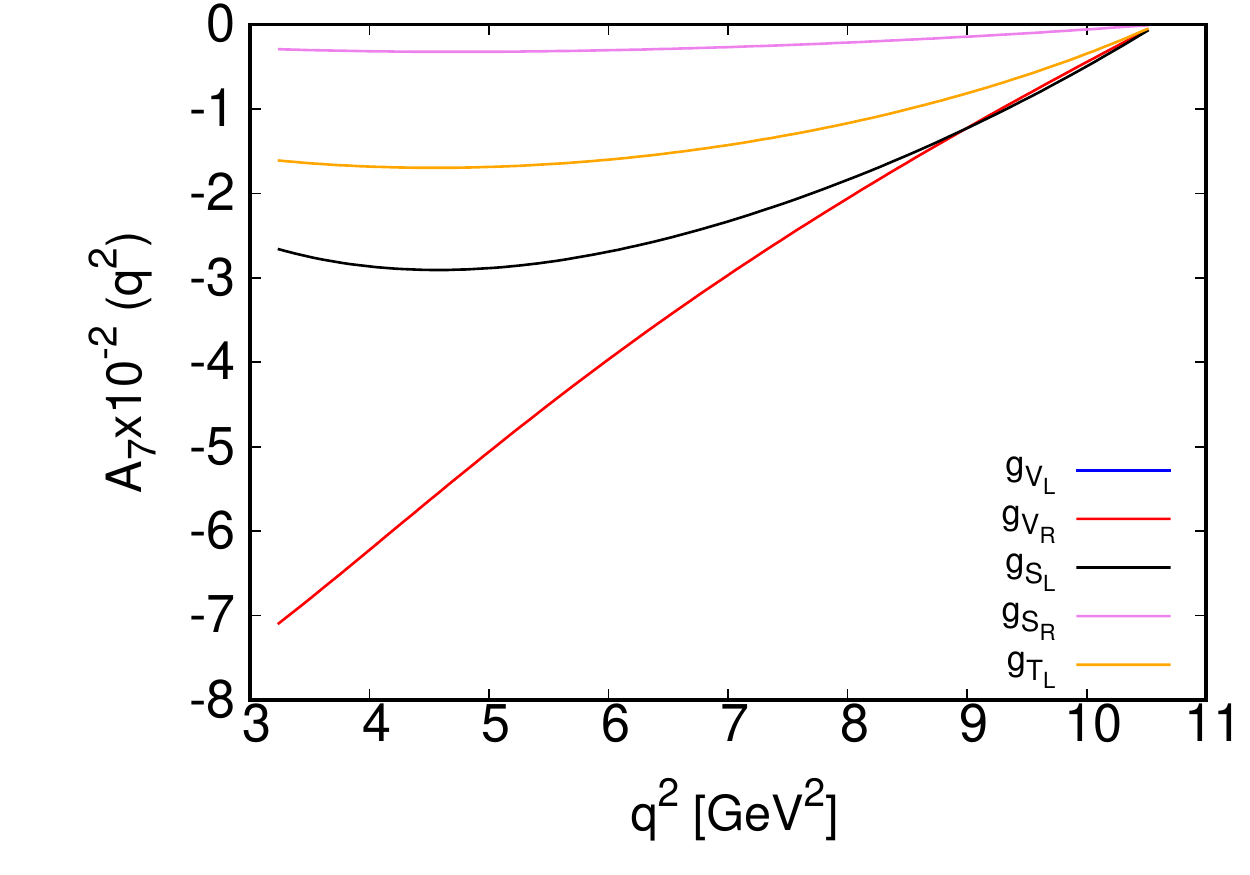}
 \hspace{1.5cm}
 \includegraphics[width=4cm,height=3cm]{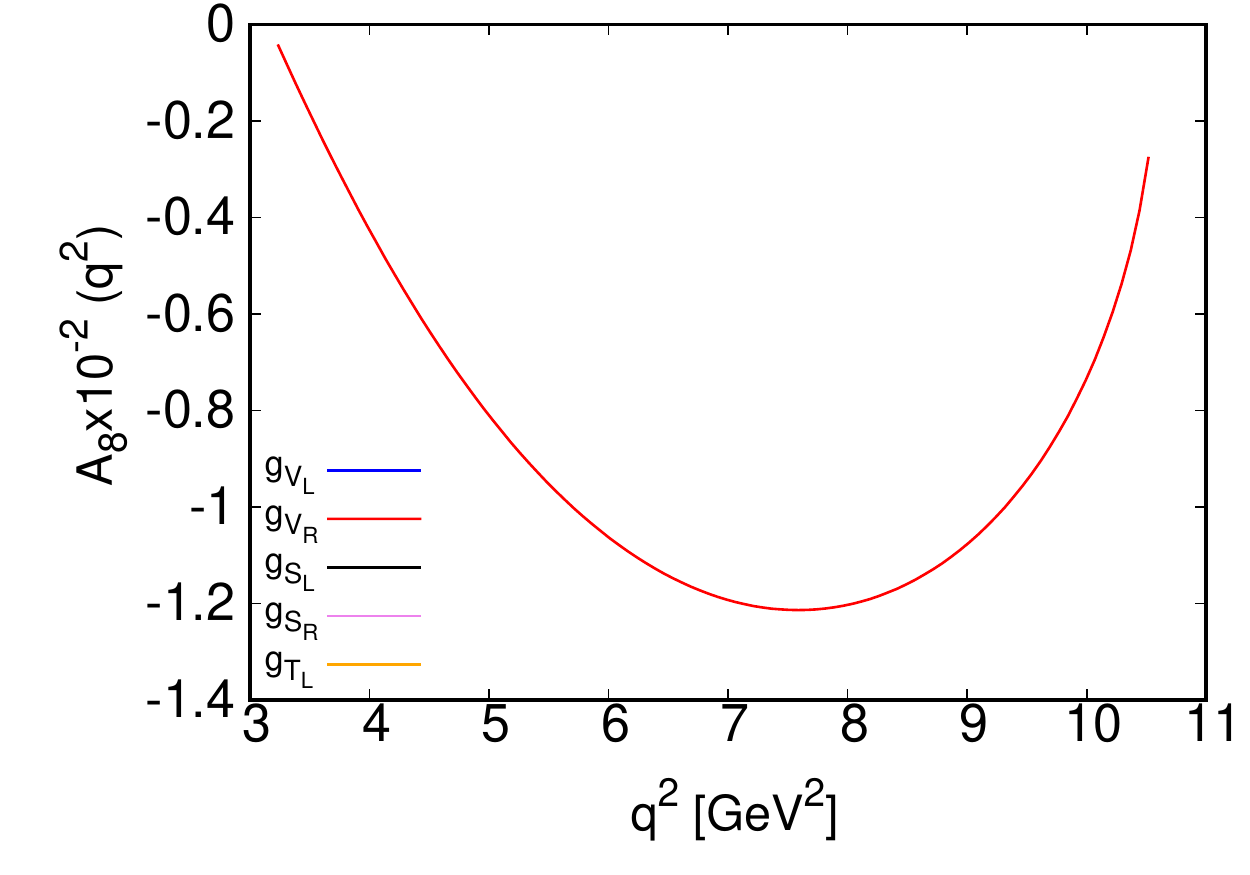}
 \includegraphics[width=4cm,height=3cm]{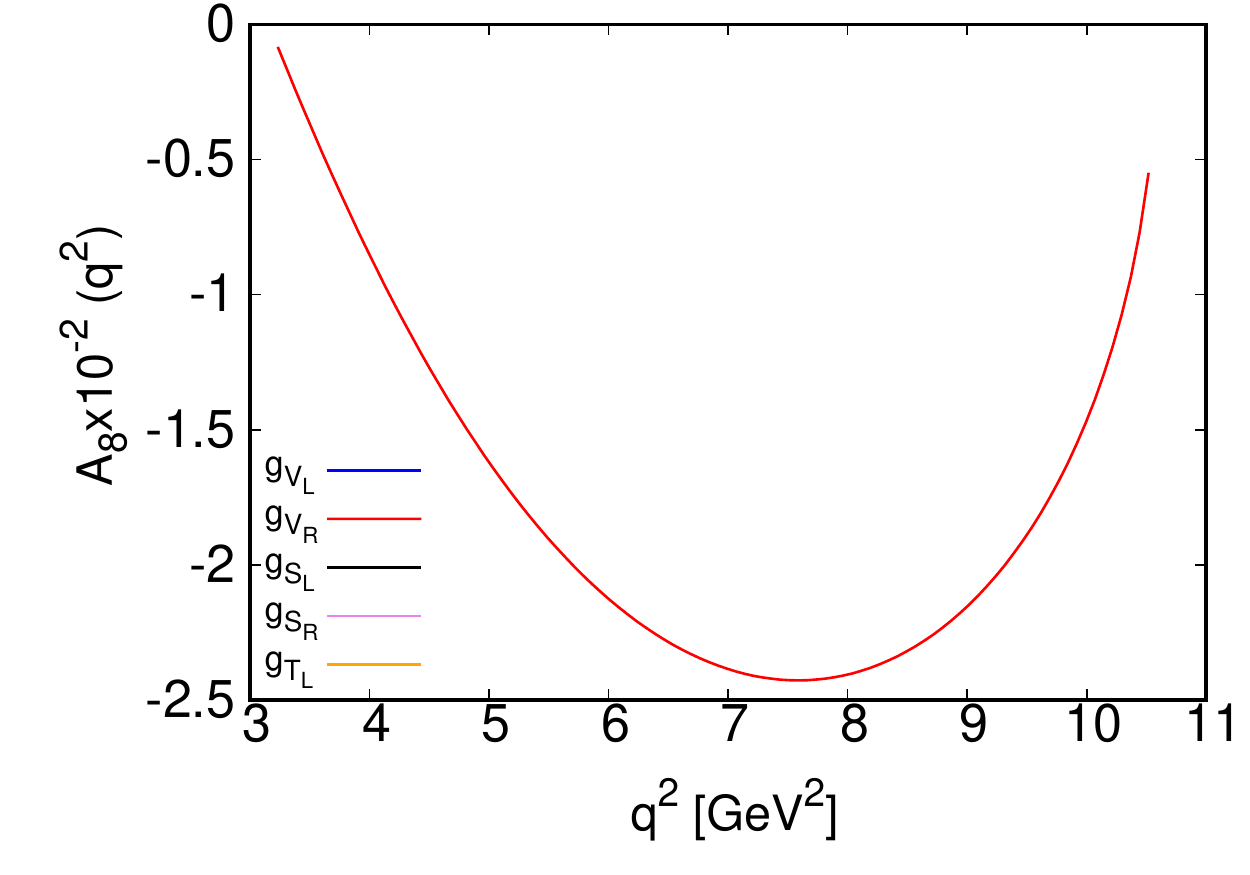}
 \hspace{1.5cm}
 \includegraphics[width=4cm,height=3cm]{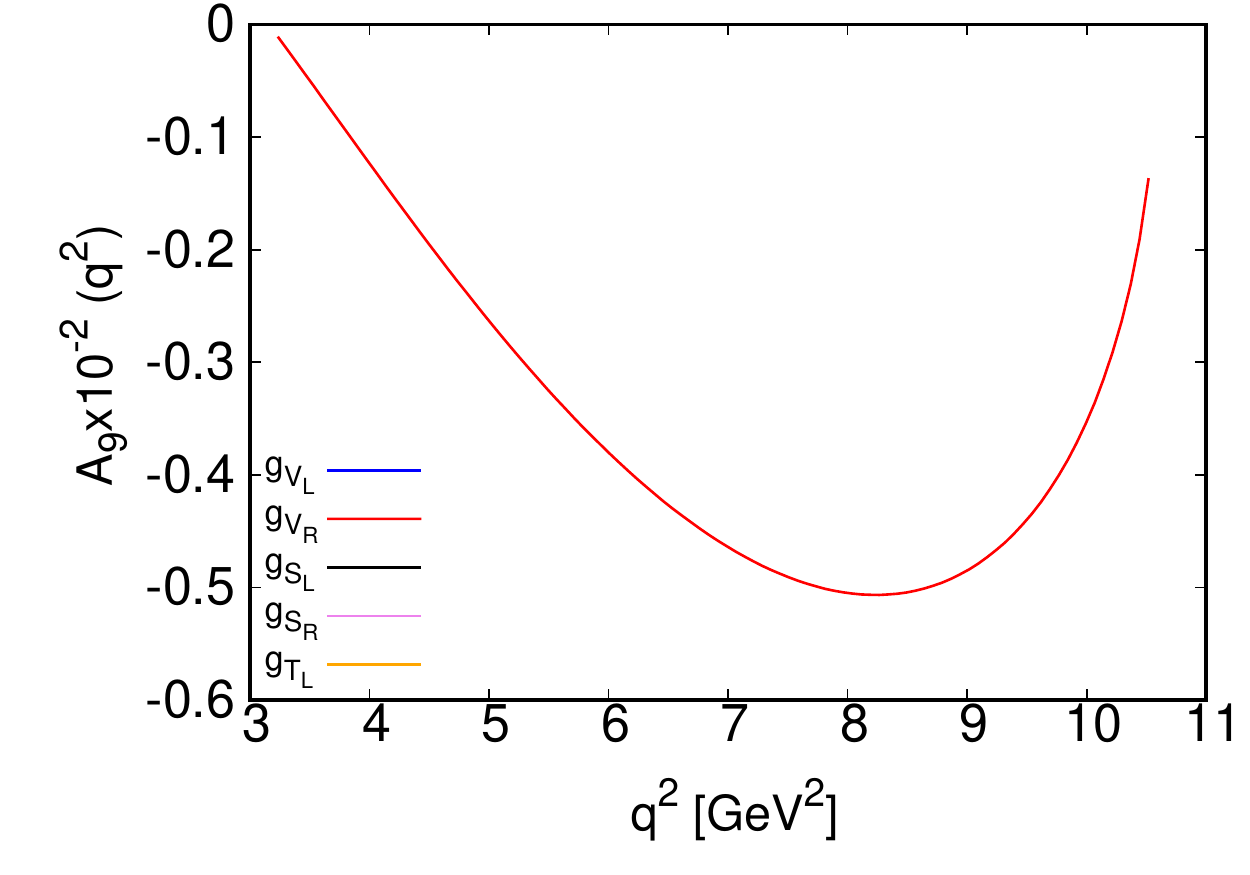}
 \includegraphics[width=4cm,height=3cm]{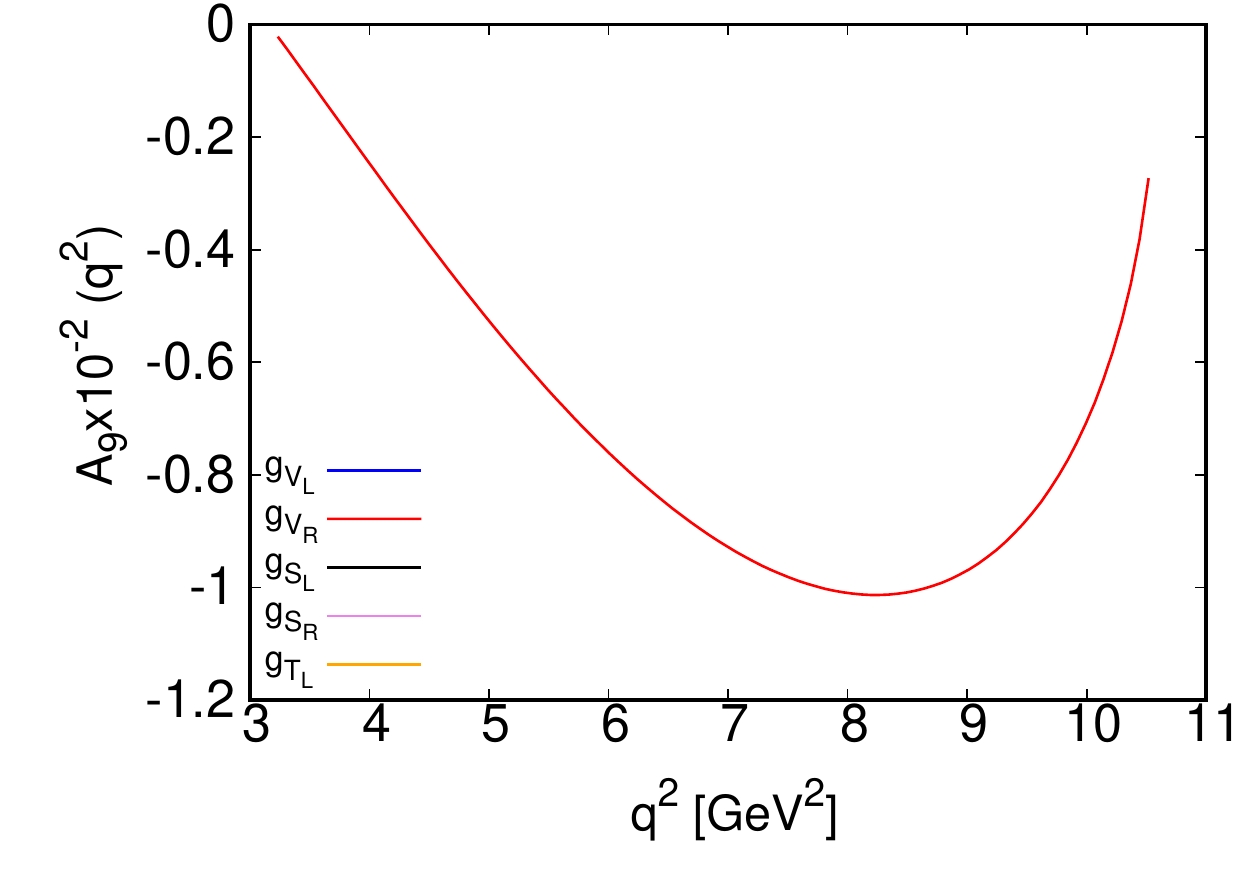}
\caption{The $q^2$ and $\cos\theta_{D_s}$ dependence of various physical observable of $B_s\to D_s^*(\to D_s \gamma, D_s\pi )\tau \nu$ in the SM and in the presence of the NP couplings of scenario - III. The SM central line and the corresponding error band are shown with green color. The  blue, red, black, violet and orange colors represents the effect of NP couplings of $g_{V_L}$, $g_{V_R}$, $g_{S_{L}}$, $g_{S_R}$ and $g_{T_L}$  respectively.}
\label{dsg_sc3_diff}
\end{figure}

In Fig~\ref{dsg_sc3_same} we show the $q^2$ and $\cos\theta_l$ dependence of various physical observables that exhibit same behaviour for 
the $D_s\gamma$ and $D_s\pi$ modes. NP contribution coming from $g_{V_L}$, $g_{V_R}$, $g_{S_{L}}$, $g_{S_R}$ and $g_{T_L}$ complex NP
couplings are shown with blue, red, black, violet and orange colored lines, respectively. Our observations are as follows.
\begin{itemize}
\item In case of $F_L(\cos\theta_l)$, a significant deviation from the SM prediction is observed due to $g_{T_L}$ NP coupling and it is quite
distinct from the rest of NP couplings. Similarly, we observe significant deviation in $F_T (\cos\theta_l)$ once $g_{V_{R}}$ and $g_{T_L}$ NP
couplings are switched on. Again, the behaviour of $F_T (\cos\theta_l)$ is quite distinct with $g_{T_L}$ NP coupling.

 \item In case of $R_{D_s^*}(q^2)$, maximum deviation from the SM prediction is observed with $g_{V_L}$, $g_{V_R}$ and $g_{T_L}$ NP 
couplings and they are clearly distinguishable from the SM prediction. Although the shape of the $q^2$ distribution is quite similar for 
$g_{V_L}$ and $g_{V_R}$ couplings, it is, however, quite distinct for $g_{T_L}$ NP coupling.  
 
\item In case of $A_{FB}^{\tau}(q^2)$, we observe a significant deviation from the SM due to $g_{V_R}$, $g_{S_{L}}$, $g_{S_R}$ and $g_{T_L}$
NP couplings. The zero crossing point is shifted to higher values of $q^2$ than in the SM for $g_{V_R}$, $g_{S_R}$ and $g_{T_L}$, 
whereas, it is shifted to a low value of $q^2$ for $g_{S_{L}}$ NP coupling. The observed zero crossings at $q^2=6.16\rm GeV^2$,
$q^2=4.82\rm GeV^2$, $q^2=5.54\rm GeV^2$ and $q^2=6.49\rm GeV^2$ in the presence of $g_{V_R}$, $g_{T_L}$, $g_{S_R}$ and $g_{S_{L}}$ are 
clearly distinguishable from the SM zero crossing of $q^2 = 5.25 \pm 0.10\,{\rm GeV^2}$ at the level of $9.1\sigma$, $4.3\sigma$, $2.9\sigma$ and
$12\sigma$ significance.

\item The observable $A_{FB}^T(q^2)$ shows a significant deviation from SM expectation once $g_{V_R}$ and $g_{T_L}$ NP couplings are 
switched on. We also observe a zero crossing in $A_{FB}^T(q^2)$ at $q^2 = 5.44{\rm GeV^2}$ with $g_{T_L}$ NP coupling. Similarly, a significant
deviation from the SM prediction is observed in $C_F^{\tau}(q^2)$ and $F_L(q^2)$ in the presence of $g_{T_L}$ NP coupling. The dip in $C_F^{\tau}(q^2)$
is shifted to a higher value of $q^2$ than in the SM.

\end{itemize}
In Fig~\ref{dsg_sc3_diff}, we display $q^2$ and $\cos\theta_{D_s}$ dependence of several 
observable for $D_s\gamma$~(left panel) and  $D_s\pi$~(right panel) modes. Our main observations are as follows. 

\begin{itemize}
\item In case of $DBR(q^2)$, we observe significant deviation from the SM prediction with $g_{T_L}$, $g_{V_L}$ and $g_{V_R}$ NP couplings
for both $D_s\gamma$ and $D_s\pi$ modes. The peak of the distribution, however, is shifted to a low value of $q^2$ than in the SM with 
$g_{T_L}$ NP coupling. 

\item The angular observable $A_3(q^2)$ and $A_4(q^2)$ show deviation from the SM in the presence of $g_{T_L}$ NP coupling for both $D_s\gamma$ 
and $D_s\pi$ modes. Similarly, in case of $A_5(q^2)$, deviation from the SM prediction is observed in the presence of $g_{V_{R}}$, $g_{S_{L}}$ 
and $g_{T_L}$ NP coupling in both the decay modes. The deviation in $A_5(q^2)$, however, is more pronounced with $g_{T_L}$ NP coupling.
 
\item Deviation from the SM prediction in $A_{6s}\,(q^2)$ is observed with  $g_{S_{L}}$, $g_{S_R}$ and $g_{T_L}$ NP couplings for the 
$D_s\gamma$ mode. The deviation is, however, more pronounced in case of $g_{T_L}$ NP coupling. Similarly, for $D_s\pi$ mode, we see 
significant deviation in $A_{6s}\,(q^2)$ in the presence of $g_{V_{R}}$ and $g_{T_L}$ NP couplings. We also observe a zero crossing in the
$A_{6s}\,(q^2)$ at $q^2=5.46\rm GeV^2$ with $g_{T_L}$ NP coupling. 
 
\item  The $A_7(q^2)$ is non-zero with $g_{V_{R}}$, $g_{S_{L}}$, $g_{S_R}$ and $g_{T_L}$ NP couplings for both $D_s\gamma$ and $D_s\pi$ 
decay mode. Similar conclusions can be made for $D_s\pi$ mode as well because of the strict
$A_7^{\pi} = 2\,A_7^{\gamma}$ relation.
 
\item The angular observables $A_8\,(q^2)$ and $A_9\,(q^2)$ are non-zero only in the presence of $g_{V_{R}}$ NP coupling for both $D_s\gamma$ 
and $D_s\pi$ modes. We observe a minimum of $A_8(q^2)$ and $A_9(q^2)$ at $q^2=7.5\,{\rm GeV^2}$ and $q^2=8.28\,{\rm GeV^2}$, respectively.

\item Although a slight deviation in $F_L(\cos\theta_{D_s})$ and $F_T(\cos\theta_{D_s})$ is observed with $g_{S_R}$ NP coupling, the 
deviation, however, is more pronounced with $g_{T_L}$ NP coupling for both $D_s\gamma$ and $D_s\pi$ modes and it is clearly distinguishable
from the SM prediction. 

\item Deviation from the SM prediction in $A_{FB}(\cos\theta_{D_s})$ is observed with $g_{V_{R}}$ $g_{S_{L}}$, $g_{S_R}$ and $g_{T_L}$ NP couplings for both $D_s\gamma$ and
$D_s\pi$ modes. In the $D_s\gamma$ mode, we observe that the zero crossing in $A_{FB}(\cos\theta_{D_s})$ shifts to lower value of 
$\cos\theta_{D_s}$ than in the SM with $g_{V_{R}}$, $g_{S_R}$ and $g_{T_L}$ NP couplings, whereas, it shifts to a higher value of 
$\cos\theta_{D_s}$ with $g_{S_{L}}$ NP coupling. The zero crossing points in $A_{FB}(\cos\theta_{D_s})$ at 
$\cos\theta_{D_s} = \pm 0.603\,, \pm 0.330\,, \pm 0.512\, {\rm and}\, \pm 0.703$ in the presence of $g_{V_{R}}$, $g_{S_{L}}$, $g_{S_R}$ and 
$g_{T_L}$ NP couplings are clearly distinguishable from the SM zero crossing of $\cos\theta_{D_s} = \pm 0.454 \pm 0.018$ at $8.16\sigma$,
$7\sigma$, $3.1\sigma$ and $13\sigma$ significance level, respectively. Similarly, for $D_s\pi$ mode, the zero crossing points in
$A_{FB}(\cos\theta_{D_s})$ at $\cos\theta_{D_s} = \pm 0.558\,, \pm 0.663\,, \pm 0.604\, {\rm and}\, \pm 0.500$ in the presence of these NP
couplings are clearly distinguishable from the SM zero crossing of $\cos\theta_{D_s} = \pm 0.626 \pm 0.0075$ at $9\sigma$, $4.9\sigma$,
$2.93\sigma$ and $16\sigma$ level of significance, respectively.

\end{itemize}

\section{Conclusions}
\label{conclusions}
Motivated by the anomalies present in several $b\to c\,l\,\nu$ quark level transition decays, we perform a detail angular analysis of 
$B_s\to D_s^*(\to D_s\gamma, D_s\pi)\,l\,\nu$ decays using the recent lattice QCD form factors. We use the latest global fit results of 
the possible NP couplings and estimate the effect of each NP couplings on several physical observables pertaining to $D_s\pi$ and $D_s\gamma$
modes in a model independent effective theory formalism. 

We first report the SM results. In the SM, we obtain the branching ratio to be of $\mathcal{O}(10^{-2})$ for $D_s\gamma$ channel 
and $\mathcal{O}(10^{-3})$ for $D_s\pi$ channel. The LHCb collaboration reported the first measurement of the 
branching ratio to be $\mathcal{B}(B_{s}^{0}\,\to\, D_s^{*-}\,\mu^+\, {\nu}_{\mu}) = (5.38\pm0.25\pm0.46)\times 10^{-2}$~
\cite{Aaij:2020hsi,Aaij:2020xjy} and it is in good agreement with our estimated results for the $D_s\gamma$ mode. The ratio of branching
ratio is found to be $R_{D_s^*} = 0.2430\pm 0.0015$ in the SM. 

For our NP analysis we work with three different NP scenarios with the best fit values obtained from various recent global fit results. 
We assume both real and complex NP couplings in our analysis. We study the underlying observables based on NP contribution coming from single
operators~($1D$) as well as from two different operators~($2D$). A brief summary of our results are as follows.
\begin{itemize}
\item In scenario - I, the observable $A_{FB}^{\tau}(q^2)$ is found to be interesting as the zero crossing point observed with $g_{S_{L}}$ and 
$g_{S_R}$ and $g_{S_{L}}=4g_{T_L}$ NP couplings stand at $1-2\sigma$ away from the SM zero crossing point. Similarly, the effect of $g_{V_L}$ NP coupling is found to 
be prominent for $DBR(q^2)$ and $R_{D_s^*}(q^2)$.

\item In scenerio-II, the deviation from the SM prediction observed for $DBR(q^2)$ and $R_{D_s^*}(q^2)$ is quite significant in the presence
of $(g_{V_L}$,\, $g_{S_{L}}=-4g_{T_L})$ and $(g_{V_L}$, $g_{S_R})$ NP couplings. The zero crossings in $A_{FB}^{\tau}(q^2)$ with 
$(g_{S_R}$, $g_{S_{L}})$ and $(g_{S_{L}}=4g_{T_L})$ NP couplings are clearly distinguishable from the SM zero crossing point at more than
 $8\sigma$ and $10\sigma$ significance. Similarly, the zero crossing in $A_{FB}^{\tau}(\cos\theta_{D_s})$ obtained with $(g_{S_R}$, $g_{S_{L}})$ 
and $(g_{S_{L}}=4g_{T_L})$ NP couplings are distinguishable from the SM zero crossing at more than $7\sigma$ for both the $D_s\gamma$ mode
and $D_s\pi$ mode. We find $A_7$ to be non-zero only in the presence of $g_{S_{L}}=4g_{T_L}$ NP 
coupling.

\item  In scenario-III, the zero crossings in $A_{FB}^{\tau}\,(q^2)$ in the presence of $g_{V_R}$, $g_{T_L}$, $g_{S_R}$ and $g_{S_{L}}$ NP
couplings are quite different from the SM zero crossing and they are clearly distinguishable from the SM prediction at the level of 
$9.1\sigma$, $4.3\sigma$, $2.9\sigma$ and $12\sigma$ significance. We also observe zero crossings in $A_{FB}^{T}\,(q^2)$ and $A_{6s}(q^2)$ 
with $g_{T_L}$ NP 
coupling that are absent in the SM. The angular observable $A_7$ is found to be non-zero in the presence of $g_{V_R}$, $g_{S_{L}}$, 
$g_{S_R}$ and $g_{T_L}$ NP couplings, whereas, $A_8$ and $A_9$ are found to be non-zero only for $g_{V_R}$ NP coupling. 
Moreover, the zero crossing points in $A_{FB}^{\tau}(\cos\theta_{D_s})$ obtained with $g_{V_R}$, $g_{S_{L}}$. $g_{S_R}$ and $g_{T_L}$ NP 
couplings are clearly distinguishable from the SM zero crossing at more than $8\sigma$, $7\sigma$, $3\sigma$ and $13\sigma$ significance 
level for the $D_s\gamma$ mode and they are distinguishable at more than $9\sigma$, $4\sigma$, $2\sigma$ and $15\sigma$ significance
for the $D_s\pi$ mode.
In general, the deviation from the SM prediction observed with complex tensor NP coupling $g_{T_L}$ is more pronounced for all the observables
in this scenario.
\end{itemize}
It should be noted that the angular observables $A_{FB}^{\tau}(q^2)$ and $A_{FB}^{\tau}(\cos\theta_{D_s})$ are quite interesting as they can
be used to distinguish between several NP scenarios. Similarly, presence of zero crossings in $A_{FB}^{T}\,(q^2)$ and $A_{6s}(q^2)$
would be a clear signal of complex tensor NP coupling. Moreover, the angular observables $A_7$, $A_8$ and $A_9$ will also play an important
role in identifying the exact NP Lorentz structures.
In conclusion, the results pertaining to $B_s\to D_s^*(\to D_s\gamma,D_s\pi)\,l\,\nu$ decay observables are very useful to explore ongoing 
flavor 
anomalies in $b\to cl\nu$ transitions and, in principle, it can provide us complementary information regarding NP in various $B$ meson decays. 
At the same time, it can also be useful in determining the value of the CKM matrix element $|V_{cb}|$. Moreover, study of these decay modes 
both theoretically and experimentally can act as a useful ingredient in maximizing future sensitivity to NP.

\FloatBarrier

\end{document}